%% file: paper.tex
\documentclass[12pt,titlepage,twoside]{article}
\usepackage{type1cm,amsmath,amssymb,bm,url}
\usepackage{aas_macros,braket,fancyhdr}
\usepackage{chapterbib}
\usepackage{cite} 
\input{colordvi.tex}
\usepackage[pdftex]{graphicx}
\usepackage{float}
\usepackage[top=2cm,bottom=2cm,left=2cm,right=2cm]{geometry}
\allowdisplaybreaks
\makeatletter
 
  \@addtoreset{equation}{section}
\makeatletter
 
  \@addtoreset{figure}{section}
\makeatletter
 
  \@addtoreset{table}{section}
 \@addtoreset{footnote}{section}
\makeatletter
\newcommand{\subsubsubsection}{\@startsection{paragraph}{4}{\z@}%
  {1.5\Cvs \@plus.5\Cdp \@minus.2\Cdp}%
  {.5\Cvs \@plus.3\Cdp}%
  {\reset@font\normalsize\sffamily}
}
\title{
{\Huge \it Probing the Early Universe \\ with the CMB Scalar, Vector and Tensor Bispectrum} \\
\vspace{3cm}
}
\author{
\Huge Maresuke Shiraishi}
\date{\today}

\begin{document}
\maketitle

\include{abst}
\include{acknow}

\tableofcontents
\newpage
\listoffigures
\newpage
\listoftables

\newcommand{\nnbe}{\stackrel{{\scriptscriptstyle (} -
{\scriptscriptstyle )}}{\nu}\!\!\!\! _{e}}
\newcommand{\nnbmu}{\stackrel{{\scriptscriptstyle (} -
{\scriptscriptstyle )}}{\nu}\!\!\!\! _{\mu}}

\def\up{\;\raise1.0pt\hbox{$'$}\hskip-6pt\partial\;}
\def\down{\;\overline{\raise1.0pt\hbox{$'$}\hskip-6pt
\partial}\;}
\def\bi#1{\hbox{\boldmath{$#1$}}}
\def\gsim{\raise2.90pt\hbox{$\scriptstyle
>$} \hspace{-6.4pt}
\lower.5pt\hbox{$\scriptscriptstyle
\sim$}\; }
\def\lsim{\raise2.90pt\hbox{$\scriptstyle
<$} \hspace{-6pt}\lower.5pt\hbox{$\scriptscriptstyle\sim$}\; }

\include{ch1_intro}
\include{ch2_inf}
\include{ch3_cmb}
\include{ch4_png}
\include{ch5_bis}
\include{ch6_mal}
\include{ch7_rot}
\include{ch8_pal}
\include{ch9_pmf}
\include{ch10_sum}


\appendix

\include{appendix}

\end{document}

%% file: abst.tex
\begin{abstract}
Although cosmological observations suggest that the fluctuations of seed fields are almost Gaussian, the possibility of a small deviation of their fields from Gaussianity is widely discussed. 
Theoretically, there exist numerous inflationary scenarios which predict large and characteristic non-Gaussianities in the primordial perturbations. These model-dependent non-Gaussianities act as sources of the Cosmic Microwave Background (CMB) bispectrum; therefore, the analysis of the CMB bispectrum is very important and attractive in order to clarify the nature of the early Universe. 
Currently, the impacts of the primordial non-Gaussianities in the scalar perturbations, where the rotational and parity invariances are kept, on the CMB bispectrum have been well-studied. However, for a complex treatment, the CMB bispectra generated from the non-Gaussianities, which originate from the vector- and tensor-mode perturbations and include the violation of the rotational or parity invariance, have never been considered in spite of the importance of this information. 

On the basis of our current studies \cite{Shiraishi:2010sm, Shiraishi:2010kd, Shiraishi:2010yk, Shiraishi:2011fi, Shiraishi:2011dh, Shiraishi:2011ph, Shiraishi:2011st}, this thesis provides the general formalism for the CMB bispectrum sourced by the non-Gaussianities not only in the scalar-mode perturbations but also in the vector- and tensor-mode perturbations. Applying this formalism, we calculate the CMB bispectrum from two scalars and a graviton correlation and that from primordial magnetic fields, and we then outline new constraints on these amplitudes. 
Furthermore, this formalism can be easily extended to the cases where the rotational or parity invariance is broken. We also compute the CMB bispectra from the non-Gaussianities of the curvature perturbations with a preferred direction and the graviton non-Gaussianities induced by the parity-violating Weyl cubic terms. We also present some unique impacts to the violation of these invariances on the CMB bispectrum.    

\end{abstract}

\bibliographystyle{JHEP}
\bibliography{paper}

%% file: acknow.tex
\section*{Acknowledgments}

First, I acknowledge the support of my supervisor, Prof. Naoshi Sugiyama. He backed my studies with much well-directed advice, enhancement of computing resources, and recruitment of high-caliber staff. In addition, owing to his powerful letters of recommendation, I could spend a fulfilling life in academia as a JSPS research fellow with ample opportunities to make presentations about our studies and communicate with top-level researchers in a variety of fields.

The work presented in this thesis is based on seven papers written in collaboration with Shuichiro Yokoyama, Kiyotomo Ichiki, Daisuke Nitta, and Keitaro Takahashi. Shuichiro Yokoyama supported and encouraged me in all seven papers. He mainly provided advice on nature in the inflationary era and on the cosmological perturbation theory. Kiyotomo Ichiki's comments focused mainly on the impacts of the primordial magnetic fields on CMB and on the technical side of numerical calculations. A preface to these works arose from a chat with Shuichiro Yokoyama and Kiyotomo Ichiki. They also corrected my drafts in a tactful way. Daisuke Nitta's main contribution was in telling me about several mathematical skills and fresh ideas necessary for calculating the primordial non-Gaussianities and the CMB bispectrum. Keitaro Takahashi gave me a polite lecture about the CMB scalar-mode bispectrum and discussed an application of our formalism to the primordial magnetic fields and cosmological defects. Owing to these four collaborators, I could gain a better understanding of the primordial non-Gaussianities and the CMB bispectrum. 

Financially, these individual works that comprise this thesis were supported in part by a Grant-in-Aid for JSPS Research under Grant No. 22-7477, a Grant-in-Aid for Scientific Research on Priority Areas No. 467 ``Probing the Dark Energy through an Extremely Wide and Deep Survey with the Subaru Telescope'', and a Grant-in-Aid for the Nagoya University Global COE Program ``Quest for Fundamental Principles in the Universe: from Particles to the Solar System and the Cosmos'' from the Ministry of Education, Culture, Sports, Science and Technology of Japan. 

I also acknowledge the Kobayashi-Maskawa Institute for the Origin of Particles and the Universe, Nagoya University, for providing computing resources useful in conducting the research reported in this thesis.

I was helped by many useful comments from Takahiko Matsubara, Chiaki Hikage, Kenji Kadota, Toyokazu Sekiguchi, and several researchers in other laboratories and institutions. My friends Masanori Sato and Shogo Masaki, my junior fellows Takahiro Inagaki and Yoshitaka Takeuchi, and other students provided advice on not only science and computers but also life. I owed my energy to them. Daichi Kashino provided a beautiful picture of the cosmic microwave background sky. Owing to Shohei Saga's careful reading, I could fix many typos in this thesis. 

Finally, I would like to thank my family for their concerns and hopes. 


%% file: ch1_intro.tex
\section{Introduction} \label{sec:intro}


\subsection{History of the Universe}

Several observational and theoretical studies on the cosmological phenomena such as the cosmic microwave background (CMB) radiation and matter clustering established the standard cosmological scenario that our Universe starts from microscopic scale and has been cooling down via the spatial expansion. Here, we summarize this scenario on the basis of Ref.~\cite{Baumann:2009ds}. 

In the primeval stage, the Universe may experience the accelerated spatial expansion, so-called inflation. In this stage, physics is determined in the quantum fluctuation. Via unknown reheating process, the energy of inflation is transformed into particles. Just after reheating, (strong), weak and electromagnetic interactions are unified and almost all of particles are relativistic. However, we believe that via cooling of the Universe and some symmetry breakings, particles become massive and decouple each other. Below $100 {\rm GeV}$, the electroweak symmetry breaking occurs and the weak interaction weakens as the temperature drops. At around $1 {\rm MeV}$, neutrinos decouple from electrons. Below $0.5 {\rm Mev}$, electrons become massive and $e^+$-$e^-$ annihilation frequently occurs. If the temperature reaches $0.1 \rm MeV$, the nucleus of light elements are produced from protons and neutrons. Observational abundance of these elements matches the theoretical estimation based on the Big Bang scenario. If the temperature becomes less than $1 \rm eV$ ($10^{11} \rm sec$), the energy density of matters dominates over that of radiations. At around $0.1 \rm eV$ ($380000 {\rm yrs}$), protons (and helium nucleus) and electrons combine into hydrogen (and helium) atoms. This process is called recombination. The CMB radiation is photon which decouples at that time and comes to us now. This is the black body radiation whose averaged temperature and its spatial anisotropies are, respectively, $2.725 {\rm K}$ and ${\cal O}(10^{-5}) {\rm K}$. The anisotropies of CMB intensity and polarizations reflect the density fluctuations in the primordial Universe. Resultant contrasts of matter distributions evolve observed large-scale structures in the balance between the gravitational force and pressure of radiations. Consequently, small-scale structures are produced earlier compared with large-scale structures. First stars arise at around $10^8 {\rm yrs}$. After these die, emitted photons ionize hydrogen atoms in the intergalactic medium until redshift $z \sim 6$. This phenomenon is called reionization. At latter half of the age of the Universe, the second accelerated expansion starts. This may be because an unknown energy with negative pressure, the so called dark energy. This expansion continues at the present epoch ($13.7 {\rm Gyrs}$).

\subsection{Access to the inflationary epoch}

At the inflationary era, the field values of physical quantities, such as metric and matters, quantum-mechanically fluctuate inside the horizon. However, the accelerated spatial expansion stretches these fluctuations beyond the horizon. Due to no causal physics, metric perturbations outside the horizon are preserved\footnote{This is valid only when there are no anisotropic stress fluctuations.}. These constant metric perturbations re-enter the horizon just before recombination and behave as seeds of the CMB fluctuations. In this sense, detailed analyses of the patterns of the CMB anisotropies will help explain the questions about the initial condition of our Universe, e.g., what kind of field there exists, what state gravity is in, and how strong the coupling is.

\subsection{Concept of this thesis}

Conventionally, the information of the primordial density fluctuations has been extracted from the two-point functions (power spectra) of the CMB fluctuations. 
There is a statistical property that 
although a non-Gaussian variable generates both even and odd-point correlations; a Gaussian variable generates only even-point correlations. 
Hence, it is hard to discriminate between the Gaussian and non-Gaussian signals in the CMB power spectrum. 
Theoretically, whether the primordial seed fluctuations are Gaussian depends completely on the inflationary models. Therefore, the check of the non-Gaussianity of the primordial fluctuations will lead to a more precise comprehension of the early Universe. 
To extract the non-Gaussian signals from the CMB anisotropy, we should focus on the higher-order correlations of the CMB fluctuations such as the CMB three-point correlations (bispectra). Owing to the recent precise observation of the Universe, the CMB bispectra are becoming detectable quantities. As a result, the CMB bispectra are good measures of the primordial non-Gaussianity. 

The primordial non-Gaussianities originating from the scalar components and their effects on the CMB bispectrum have been well-studied (Refs.~\cite{Bartolo:2004if, Komatsu:2010hc}). However, for some situations the vector components (vorticities) and tensor ones (gravitational waves) also act as non-Gaussian sources. This indicates that unknown signals, unlike the scalar case, may also appear in the CMB bispectra. To study these impacts in detail, we produced the general formulae for the CMB temperature and polarization bispectra from the scalar, vector and tensor non-Gaussianities \cite{Shiraishi:2010sm, Shiraishi:2010kd}. Next, utilizing these formulae and computing the practical CMB bispectra, we obtained new constraints on some primordial non-Gaussian sources and learned more about the nature of the early Universe \cite{Shiraishi:2010yk, Shiraishi:2011fi, Shiraishi:2011dh, Shiraishi:2011ph, Shiraishi:2011st}. 

This thesis aims to discuss the CMB bispectra induced by the primordial scalar, vector, and tensor non-Gaussianity on the basis of our recent studies \cite{Shiraishi:2010sm, Shiraishi:2010kd, Shiraishi:2010yk, Shiraishi:2011fi, Shiraishi:2011dh, Shiraishi:2011ph, Shiraishi:2011st}. More concrete organization of this thesis is as follows. 
In Secs.~\ref{sec:inflation} and \ref{sec:CMB_anisotropy}, we demonstrate how to generate the seed fluctuations in the inflationary era on the basis of some review papers and present formulae for the scalar, vector, and tensor modes of the CMB anisotropies as mentioned in Ref.~\cite{Shiraishi:2010sm}. We also review some observational findings obtained by the analysis of the CMB power spectra. 
In Secs.~\ref{sec:PNG} and \ref{sec:formula}, we describe the general formulae of the CMB bispectra generated from the primordial scalar, vector, and tensor non-Gaussianities \cite{Shiraishi:2010kd}. We then discuss the applications to the non-Gaussianities in two scalars and a graviton correlator \cite{Shiraishi:2010kd} (Sec.~\ref{sec:maldacena}), involving the violation of the rotational or parity invariance \cite{Shiraishi:2011ph, Shiraishi:2011st} (Secs.~\ref{sec:statistically_anisotropic} and \ref{sec:parity_violating}), and sourced by the primordial magnetic fields \cite{Shiraishi:2010yk, Shiraishi:2011fi, Shiraishi:2011dh} (Sec.~\ref{sec:PMF}). Finally, we summarize this thesis and discuss some future issues (Sec.~\ref{sec:sum}). In the appendices, we describe some mathematical tools and the detailed calculations required for the conduct of our formalism. 

\bibliographystyle{JHEP}
\bibliography{paper}

%% file: ch2_inf.tex
\section{Fluctuations in inflation}\label{sec:inflation}

Inflation expresses an exponential growth of the scale factor of the Universe in the early time, namely, $a \sim e^{H t}$. In Einstein gravity, this requires $p \sim -\rho$ with $p$ and $\rho$ being the pressure and energy density, and is often realized by the existence of a scalar field, inflaton. We believe that the small fluctuations of this field have created the curvature perturbations and the density contrasts of matters. Moreover, some vorticities and gravitational waves may also have evolved together. In this section, we briefly describe the physical treatment of these fluctuations in the inflationary era in accordance with Ref.~\cite{Baumann:2009ds}.

\subsection{Dynamics of inflation}%

As the action in the inflationary era, we consider the simple one including a scalar field $\phi$, which is called inflaton and minimally coupled with gravity as 
\begin{eqnarray}
S = \int d^4 x \sqrt{-g} 
\left[ \frac{1}{2}M_{\rm pl}^2 R - \frac{1}{2}g^{\mu \nu} \partial_\mu \phi
 \partial_\nu \phi - V(\phi) \right]~, 
\label{eq:phiaction}
\end{eqnarray} 
where $R$ denotes the Ricci scalar, $V(\phi)$ is the potential, and $M_{\rm pl} \equiv (8 \pi G)^{-1/2}$ is the reduced Planck mass. The energy momentum tensor and the field equation for $\phi$ are, respectively, given by 
\begin{eqnarray}
T_{\mu\nu} &\equiv& -\frac{2}{\sqrt{-g}} \frac{\delta S}{\delta g^{\mu \nu}} 
= \partial_\mu\phi\partial_\nu\phi-g_{\mu\nu}
\left({1\over 2}\partial^\sigma\phi\partial_\sigma\phi
+V(\phi)\right)~, \label{eq:Tscalarfield} \\ 
\frac{\delta S}{\delta \phi} &=& \frac{1}{\sqrt{-g}} \partial_\mu (\sqrt{-g} \partial^\mu \phi) + V_{\phi}= 0~,
\end{eqnarray}
where $V_{\phi} = dV / d\phi$. On the FLRW metric as 
\begin{eqnarray}
ds^2 = -dt^2 + a^2 dx^2 = a^2 (-d\tau^2 + dx^2)
\end{eqnarray}
with $\tau$ being the conformal time and under the assumption that $\phi(t, {\bf x}) \equiv \phi(t)$, the energy density and pressure of the scalar field are written as 
\begin{eqnarray}
\rho_\phi = \frac{1}{2}(\partial_t \phi)^2 + V(\phi) ~, \ \ 
p_\phi = \frac{1}{2}(\partial_t \phi)^2 - V(\phi) ~.
\end{eqnarray}
Thus, if $V$ exceeds $(\partial_t \phi)^2 / 2$ and the parameter $w_\phi \equiv p_\phi / \rho_\phi$ becomes less than $-1/3$ and the accelerated expansion can be realized.
 The Friedmann equation, the acceleration equation and the field equation are, respectively, given by
\begin{eqnarray}
\begin{split}
 H^2 &= \frac{1}{3 M_{\rm pl}^2} 
\left[ \frac{1}{2}(\partial_t \phi)^2 + V(\phi) \right] ~, \\
a^{-1} \frac{d^2 a}{dt^2} &= - \frac{1}{6 M_{\rm pl}^2} (\rho_\phi + 3 p_\phi) = H^2 (1-\epsilon_H)~, \\ 
\frac{d^2 \phi}{dt^2} + 3 H \frac{d \phi}{dt} + V_{\phi} &= 0 ~, \label{eq:inf_field_Friedmann}
\end{split}
\end{eqnarray}
where $H \equiv \partial_t a / a$ is the Hubble parameter and we have introduced the so-called Hubble slow-roll parameter as
\begin{eqnarray}
\epsilon_H \equiv - \frac{\partial_t H}{H^2} = - \frac{d \ln H}{d N} 
= \frac{3}{2} (w_\phi + 1) = \frac{1}{2}
\left( \frac{\partial_t \phi}{M_{\rm pl} H} \right)^2~, \label{eq:def_epsilon_H}
\end{eqnarray}
with $N$ being the $e$-folding number. For $w_\phi < -1/3$, $\epsilon_H < 1$ is realized and the Universe experiences an accelerated expansion. Moreover, this acceleration is kept stable if 
$\left| \frac{d^2 \phi}{dt^2} \right| \ll |3 H \partial_t \phi|, |V_\phi|$. This corresponds to 
\begin{eqnarray}
\eta_H \equiv - \frac{1}{H \partial_t \phi}\frac{d^2 \phi}{dt^2} = \epsilon_H
 - \frac{1}{2 \epsilon_H} \frac{d \epsilon_H}{d N} \ll 1 ~. \label{eq:def_eta_H}
\end{eqnarray}
Other slow-roll parameters are defined as the function of the potential: 
\begin{eqnarray} 
\epsilon(\phi) \equiv \frac{M_{\rm pl}^2}{2}
\left( \frac{V_\phi}{V} \right)^2 ~, \ \ 
\eta(\phi) \equiv M_{\rm pl}^2 \frac{V_{\phi \phi}}{V} ~.
\end{eqnarray}
Here, $\epsilon$ and $\eta$ are called the potential slow-roll
parameters, and in the slow-roll approximation, the Hubble and potential
slow-roll parameters are related as
\begin{eqnarray}
\epsilon_H \approx \epsilon~, \ \ \eta_H \approx \eta - \epsilon ~. 
\end{eqnarray} 
Hence, the slow-roll inflation occurs also for $\epsilon, |\eta| \ll 1$. When these slow-roll parameters reach unity as 
\begin{eqnarray}
\epsilon_H(\phi_{\rm end}) \equiv 1~, \ \ \epsilon(\phi_{\rm end}) \approx 1~.
\end{eqnarray}
inflation stops. 

The $e$-folding number as the function of given time during inflation is formulated as  
\begin{eqnarray}
N(\phi) \equiv \ln \frac{a_{\rm end}}{a} 
= \int_t^{t_{\rm end}} H dt 
&=& \int_\phi^{\phi_{\rm end}}
 \frac{H}{\partial_t \phi} d\phi 
\approx \int_{\phi_{\rm end}}^\phi
 \frac{V}{V_\phi} d\phi \nonumber \\
&=& \int^\phi_{\phi_{\rm end}}
 \frac{d \phi}{M_{\rm pl} \sqrt{2 \epsilon_H}} 
\approx \int^\phi_{\phi_{\rm end}}
 \frac{d \phi}{M_{\rm pl} \sqrt{2 \epsilon}} ~.
\end{eqnarray}
Note that $N(\phi) \gtrsim 60$ should be satisfied in order to solve the horizon and flatness problems.

\subsection{Curvature and tensor perturbations} 

Here, we summarize the analytical solutions of curvature and tensor perturbations in the de Sitter space-time, which is derived from the action (\ref{eq:phiaction}). For convenience, we adapt the comoving gauge as 
\begin{eqnarray} 
\delta \phi = 0 ~, \ \ g_{ij} = a^2[(1 + 2 {\cal R}) \delta_{ij} +
 h_{ij}] ~, \ \ 
\partial_i h^i{}_{j} = h^i{}_{i} = 0 ~. \label{eq:R_hij_def}
\end{eqnarray}
Comoving curvature perturbation ${\cal R}$ and the tensor perturbation $h_{ij}$ remain constant outside horizon if there exist no extra anisotropic stresses
\footnote{On superhorizon scales, this ${\cal R}$ is consistent with ${\cal R}$ in Refs.~\cite{Komatsu:2010fb, Motta:2012rn}, $\zeta$ in Refs.~\cite{Maldacena:2002vr, Lyth:2009zz}, $-{\cal R}$ in Refs.~\cite{Baumann:2009ds, Giovannini:2008yz}, and $- {\zeta}$ in Ref.~\cite{Shaw:2009nf}. In a numerical code CAMB \cite{Lewis:1999bs, Lewis:2004ef}, the primordial scalar-mode power spectrum is given by this ${\cal R}$.}
\footnote{In Sec.~\ref{sec:PMF}, we will show that due to the finite anisotropic stresses of the primordial magnetic field, the curvature perturbations (and gravitational waves) do not remain constant even on the superhorizon}. 


The quadratic actions of Eq.~(\ref{eq:phiaction}) for curvature and tensor perturbations are respectively given by 
\begin{eqnarray}
S_{\cal R}^{(2)} &=& M_{\rm pl}^2 \int d\tau d^3 x a^2 
\epsilon_H
\left[ \dot{\cal R}^2 - 
\left( \partial_i {\cal R} \right)^2 \right]~, \\ 
S_h^{(2)} &=& \frac{M_{\rm pl}^2}{8} \int d\tau d^3x a^2 
\left[ \dot{h}_{ij} \dot{h}_{ij} -
 \partial_l h_{ij} \partial_l h_{ij} \right] ~, 
\end{eqnarray}
where $~\dot{}~ \equiv d/d\tau$. Obeying the Fourier expansion as 
\begin{eqnarray}
{\cal R}({\bf x}, \tau) &=& \int \frac{d^3 {\bf k}}{(2\pi)^3} 
{\cal R} ({\bf k}, \tau) 
e^{i {\bf k} \cdot {\bf x}} ~, \\ 
h_{ij}({\bf x}, \tau) &=& \int \frac{d^3 {\bf k}}{(2\pi)^3} \sum_{\lambda = \pm 2}
 h^{(\lambda)} ({\bf k}, \tau)
e_{ij}^{(\lambda)}(\hat{\bf k})  e^{i {\bf
 k} \cdot {\bf x}} ~,
\end{eqnarray}
these are rewritten as 
\begin{eqnarray}
S_{\cal R}^{(2)} &=& \int d\tau 
\left( M_{\rm pl} a\right)^2 \epsilon_H
\int \frac{d^3 {\bf k}}{(2\pi)^3} 
\left[ |\dot{\cal R}({\bf k},
\tau) |^2 - k^2 |{\cal R}({\bf k}, \tau)|^2 \right] ~, \\ 
S_{h}^{(2)} &=& \sum_{\lambda = \pm 2}  
\int d\tau 
\left( \frac{M_{\rm pl} a}{2} \right)^2 
\int \frac{d^3 {\bf k}}{(2\pi)^3} 
\left[ |\dot{h}^{(\lambda)}({\bf k},
\tau) |^2 - k^2 |h^{(\lambda)}({\bf k}, \tau)|^2 \right] ~.
\end{eqnarray}
Here, $e_{ij}^{(\lambda)}$ is the transverse-traceless polarization
tensor which has two circular states $\lambda = \pm 2$ and is normalized as $e_{ij}^{(\lambda)}(\hat{\bf k}) e_{ij}^{(\lambda')}(-\hat{\bf k}) = 2
\delta_{\lambda, \lambda'}$. The convention
and useful properties of this tensor are described in Appendix
\ref{appen:polarization}. 
The variable transformation as $v^{(0)} \equiv z {\cal R}$, $z \equiv a \frac{\partial_t \phi}{H}$ (for scalar mode), $v^{(\pm 2)} \equiv \frac{a M_{\rm pl}}{\sqrt{2}} h^{(\pm 2)}$ (for tensor mode), and the variation principle as $\delta S / \delta v^{(\lambda)} = 0$ lead to the field equation as
\begin{eqnarray}
\ddot{v}_{\bf k}^{(\lambda)} 
+ \left( k^2 - \frac{2}{\tau^2} \right)v_{\bf k}^{(\lambda)} = 0 ~, \label{eq:mukhanov}
\end{eqnarray}
where we have used a relation in the de Sitter limit: $\ddot{z}/z = \ddot{a}/a = 2 / \tau^2$. 

To solve these equations, we perform the quantization of the field $v^{(\lambda)}$ as
\begin{eqnarray}
v_{\bf k}^{(\lambda)} 
= v_k(\tau) \hat{a}_{\bf k}^{(\lambda)} +
 v_k^*(\tau) \hat{a}_{- \bf k}^{(\lambda) \dagger} ~.
\end{eqnarray}
When we set the normalization of the mode functions as  
\begin{eqnarray}
\Braket{v_k, v_k } \equiv \frac{i}{\hbar} 
\left(v_k^* \dot{v}_k - \dot{v}_k^* v_k \right) =1~, \label{eq:norm_mode_func}
\end{eqnarray}
the canonical commutation relation between the creation ($\hat{a}_{\bf k}^{(\lambda) \dagger}$) and annihilation ($\hat{a}_{\bf k}^{(\lambda)}$) operators can be written as 
\begin{eqnarray}
\left[ \hat{a}_{\bf k}^{(\lambda)}, \hat{a}_{\bf k'}^{(\lambda') \dagger} \right] 
= (2\pi)^3 \delta({\bf k} - {\bf k'}) \delta_{\lambda, \lambda'}~.
\end{eqnarray}
A vacuum state is given by 
\begin{eqnarray}
\hat{a}_{\bf k}^{(\lambda)} \Ket{0} = 0 ~.
\end{eqnarray}
As a vacuum, one often choose the so-called Bunch-Davies Vacuum denoting the Minkowski vacuum in the far past. In this condition, i.e., $\tau \to - \infty$ or $|k \tau| \gg 1$, the field equation (\ref{eq:mukhanov}) is reduced to
\begin{eqnarray}
\ddot{v}_{\bf k}^{(\lambda)} + k^2 v_{\bf k}^{(\lambda)} = 0 ~.
\end{eqnarray}
This is equivalent to the equation for harmonic oscillators and hence easily solved as  
\begin{eqnarray}
v_k(\tau) = \frac{e^{-i k \tau}}{\sqrt{2k}} ~. \label{eq:bunch_davis_mode} 
\end{eqnarray}
Owing to two boundary conditions (\ref{eq:norm_mode_func}) and
(\ref{eq:bunch_davis_mode}), one can gain the solution of the mode function in the field equation (\ref{eq:mukhanov}) as
\begin{eqnarray}
v_{k}(\tau) = \frac{e^{-i k \tau}}{\sqrt{2k}} \left(1-\frac{i}{k \tau} \right)
 ~ . \label{eq:solution_mode_function}
\end{eqnarray} 
Using this, we can express the time evolution of the primordial curvature and tensor perturbations as 
\begin{eqnarray}
{\cal R}({\bf k}, \tau) &=& - \frac{H^2}{\partial_t \phi} \tau 
\left[ v_k(\tau) \hat{a}_{\bf k}^{(0)} 
+ v_k^*(\tau) \hat{a}_{- \bf k}^{(0) \dagger} \right] ~, \\ 
h^{(\pm 2)}({\bf k}, \tau) &=& - \sqrt{2} \frac{ H}{M_{\rm pl}} \tau
\left[ v_k(\tau) \hat{a}_{\bf k}^{(\pm 2)}
+ v_k^*(\tau) \hat{a}_{- \bf k}^{(\pm 2) \dagger} \right]~. 
\end{eqnarray}

Finally, we summarize these power spectra on superhorizon scales ($|k\tau| \ll 1$) as 
\begin{eqnarray}
\begin{split}
\Braket{\prod_{n=1}^2 {\cal R}({\bf k_n})} 
&\equiv (2 \pi)^3 P_{\cal R}(k_1) 
\delta\left( \sum_{n=1}^2  {\bf k_n} \right) ~, \\
 P_{\cal R}(k) &= \frac{H_*^2}{2k^3} 
\left( \frac{H_*}{(\partial_t \phi)_*}  \right)^2 
= \left( \frac{H_*}{M_{\rm pl}} \right)^2 \frac{1}{4 \epsilon_{H *} k^3}
\approx \left( \frac{H_*}{M_{\rm pl}} \right)^2 \frac{1}{4 \epsilon_{*} k^3}
 ~. \label{eq:ini_scal_power} 
\end{split}
\end{eqnarray}
and 
\begin{eqnarray}
\begin{split}
\Braket{ \prod_{n=1}^2 h^{(\lambda_n)}({\bf k_n}) }
&\equiv (2\pi)^3 \frac{P_h(k_1)}{2} 
\delta\left( \sum_{n=1}^2 {\bf k_n} \right) \delta_{\lambda_1,
\lambda_2} ~, \\
P_h (k) &= \left( \frac{H_*}{M_{\rm pl}} \right)^2 \frac{2}{k^3}~. \label{eq:ini_tens_power}
\end{split}
\end{eqnarray}
Here, we have evaluated all quantities at horizon crossing, namely $\tau_* =
-1/k$. Note that since ${\cal R}$ and $h^{(\pm 2)}$ are constant on superhorizon scales, these power spectra become the initial conditions for the CMB power spectra of the scalar and tensor modes. 

\subsection{Consistency relations in the slow-roll limit}

As a measure of the amplitude of the primordial gravitational wave, one often use the tensor-to-scalar ratio as 
\begin{eqnarray}
r \equiv
\frac{2 P_h(k)}{P_{\cal R}(k)}~. \label{eq:inf_def_tensor-to-scalar}
\end{eqnarray}
Comparing Eq.~(\ref{eq:ini_scal_power}) with
Eq.~(\ref{eq:ini_tens_power}), we find a consistency relation 
\begin{eqnarray}
r = 16 \epsilon_{H *} \approx 16 \epsilon_{*} ~. \label{eq:consistency_relarion_r_epsilon}
\end{eqnarray}

Using $Hdt = dN$, we find that $r$ is a measure of the evolution of the inflaton as
\begin{eqnarray}
r = \frac{8}{M_{\rm pl}^2} 
\left( \frac{d \phi}{d N} \right)^2 ~. 
\end{eqnarray}
By performing an integral over $N$ and an approximation as $r \sim {\rm const}$ during inflation, we obtain the so called Lyth bound \cite{Lyth:1996im}: 
\begin{eqnarray}
\frac{\Delta \phi}{M_{\rm pl}} \sim 
\left( \frac{r}{0.01} \right)^{1/2}~.
\end{eqnarray}
Therefore, if we observe $r > 0.01 (< 0.01)$, we may conclude that large-field (small-field) inflation, namely, $\Delta \phi > M_{\rm pl} (< M_{\rm pl})$ occurred. 

As measures for the shapes of the spectra, we often use the spectral indices, which are defined by 
\begin{eqnarray}
n_s - 4 \equiv \frac{d \ln P_{\cal R}}{d \ln k} ~, \ \ 
n_t - 3 \equiv \frac{d \ln P_h}{d \ln k} ~.
\end{eqnarray}
From Eqs.~(\ref{eq:ini_scal_power}) and (\ref{eq:ini_tens_power}), the
right-hand sides are expanded as
\begin{eqnarray}
\begin{split}
 \frac{d \ln P_{\cal R}}{d \ln k} 
&= \left( 2 \frac{d \ln H_*}{d N} 
- \frac{d \ln \epsilon_{H *}}{d N} \right)
 \frac{dN}{d \ln k} - 3 ~, \\  
\frac{d \ln P_h}{d \ln k} 
&= 2 \frac{d \ln H_*}{d N}
 \frac{dN}{d \ln k} - 3 ~. 
\end{split}
\end{eqnarray} 
From the definition of the Hubble slow-roll parameters
(\ref{eq:def_epsilon_H}) and (\ref{eq:def_eta_H}), we obtain 
\begin{eqnarray}
\frac{d \ln H_*}{d N} = - \epsilon_{H*}~, \ \ 
\frac{d \ln \epsilon_{H *}}{d N} = 2 (\epsilon_{H*} - \eta_{H*}) ~.
\end{eqnarray}
By using $k = a_* H_*$ and $d \ln k = d N + d \ln H_*$, we have
\begin{eqnarray}
\frac{dN}{d \ln k} 
= \left[ 1 + \frac{d \ln H_*}{dN} \right]^{-1}
\approx 1 + \epsilon_{H*} ~.
\end{eqnarray}
Consequently, we can summarize the consistency relations: 
\begin{eqnarray}
\begin{split}
n_s - 1 &= 2 \eta_{H *} - 4 \epsilon_{H *} 
\approx 2 \eta_{*} - 6 \epsilon_{*}  ~, \\
n_t &= - 2 \epsilon_{H *} \approx - 2 \epsilon_{*}  ~.
\end{split}
\end{eqnarray} 
From Eq.~(\ref{eq:consistency_relarion_r_epsilon}), we also find the
consistency relation between $r$ and $n_t$ as
\begin{eqnarray}
r = - 8 n_t ~.
\end{eqnarray}
As shown above, $r, n_s$ and $n_t$ depend on the slow-roll parameters and hence are observables which reflect the nature of inflation. 

\bibliographystyle{JHEP}
\bibliography{paper}

%% file: ch3_cmb.tex
\section{Fluctuations in cosmic microwave background radiation} \label{sec:CMB_anisotropy}

\begin{figure}[t]
  \centering \includegraphics[height=9cm,clip]{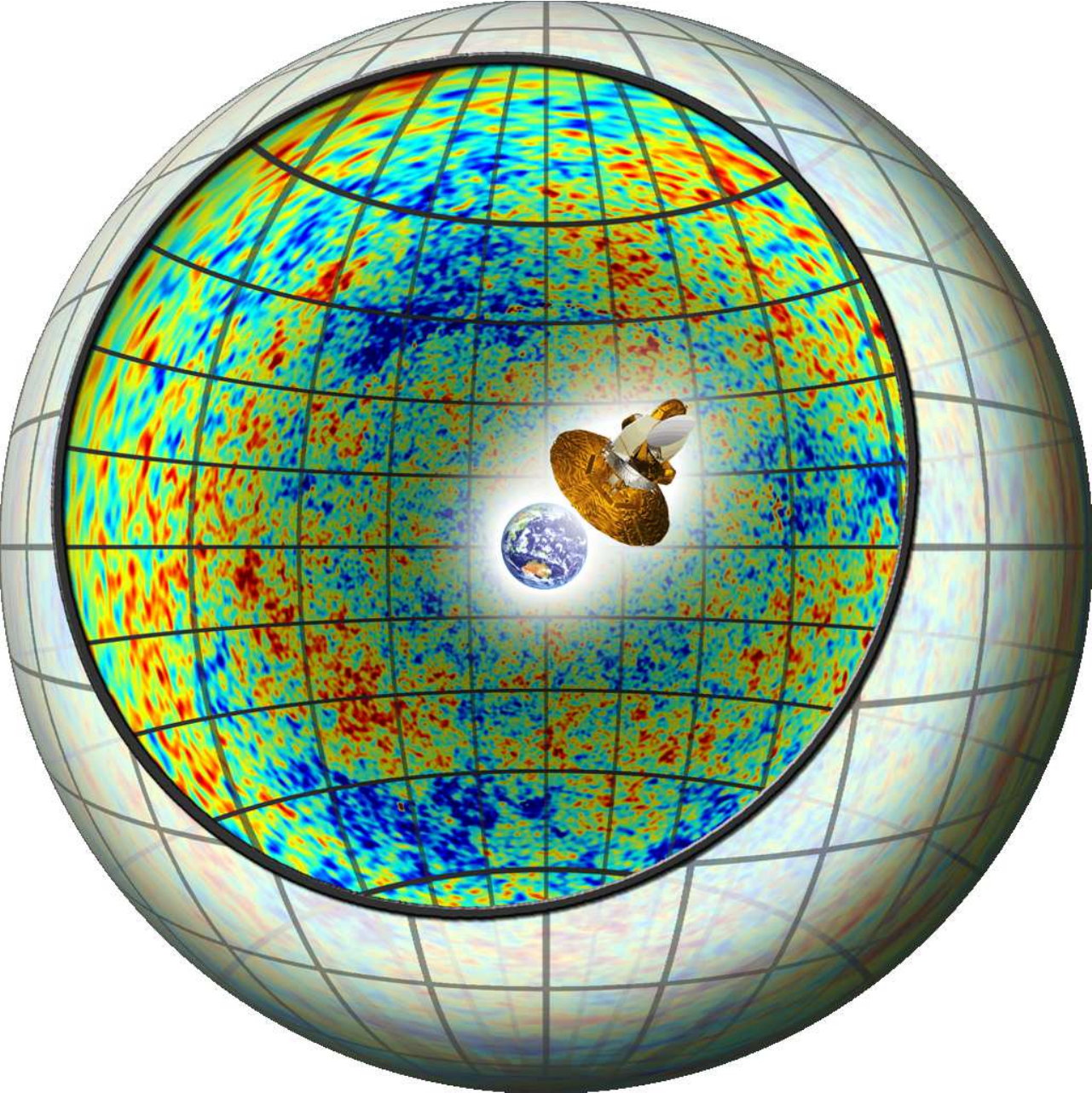}
  \caption{CMB anisotropy on the last scattering surface. The red (blue) parts correspond to the hot (cold) spots (Copyright 2011 by Daichi Kashino).}
  \label{fig:CMB_map}
\end{figure}

Cosmic microwave background (CMB) radiation is composed of photons which have decoupled from electrons in the epoch of the hydrogen and helium recombination at $z = 1089$ and it is observed as the perfectly black body radiation whose averaged temperature is $2.725 \rm K$. Historically, in 1949, Alpher and Herman predicted its existence as relics of the big bang Universe and its first detection came in 1964. More precisely, however, the CMB involves the spatial fluctuations of ${\cal O}(10^{-5}) \rm K$ (see Fig.~\ref{fig:CMB_map}). We had to wait the detection of the CMB anisotropy until the data of the COBE experiment were released in the 1990s.

Theoretically, the density contrast of the CMB is computed in the system where photons, neutrinos, baryons, dark matters and dark energy exist in the gravitational potential. Compared with the observational data, the values of several key parameters have been well-determined. The WMAP experiment established the facts that the Universe is close to spatially flat and the present structure grew from the nearly scale-invariant primordial fluctuations. These consequences are almost consistent with the prediction of the standard slow-roll inflation. Furthermore, we have a compelling evidence that the Universe is dominated by dark energy and
dark matter, which implies that $96\%$ of the total energy of the Universe remains unknown. Nowadays, some bare anomalies such as the preferred direction and the parity violation are furthermore being discussed \cite{Bennett:2010jb}, and we expect to extract more detailed information from the new precise measurements \cite{:2006uk}. 

In addition to the intensity of the CMB, the polarizations also lead to better understandings. The curl-free component of the polarizations, $E$ mode, reflects the recombination history, in particular, the reionization of the Universe. The curl component of the polarizations, $B$ mode, is generated from the primordial vector and tensor perturbations. Hence, the detection of the $B$-mode polarization will provide clues as to inflation and the physics beyond the standard model of the particle physics.

In this section, we describe the original formalism of CMB fluctuations including intensity and polarization anisotropies from the scalar, vector, and tensor modes partially on the basis of our paper \cite{Shiraishi:2010sm}, some publications \cite{Zaldarriaga:1996xe, Landriau:2002fx, Pritchard:2004qp, Weinberg:2008zzc} and some academic dissertations \cite{Zaldarriaga:1998rg, Brown:2006wv, Saito:2007mt}, and summarize current outputs from the analysis of the CMB power spectra. 

\subsection{Einstein equations}

Here, we derive the zeroth and first-order Einstein equations. 
Let us consider the flat (K = 0) FLRW metric and small perturbations in the synchronous
gauge (for open and closed cases, see \cite{Kamionkowski:1994sv, Spergel:1994tm, Lewis:1999bs}):
\begin{eqnarray}
ds^2 = a^2[-d\tau^2 + ( \delta_{ij} + h_{ij} ) dx^i dx^j ] ~.
\end{eqnarray}
We have the inverse metric to first order in perturbations as 
\begin{eqnarray}
g^{00} = -\frac{1}{a^2}~, \ \ 
g^{0i} = 0 ~, \ \  
g^{ij} = \frac{1}{a^2} (\delta^{ij} - h^{ij})~.
\end{eqnarray}

The Einstein equation with the cosmological constant $\Lambda$ can be written as 
\begin{eqnarray}
G^\mu{}_\nu = R^\mu{}_\nu - \frac{1}{2} \delta^\mu{}_\nu R = 8\pi G
 T^\mu{}_\nu - \Lambda \delta^\mu{}_\nu ~, \label{eq:einstein_general}
\end{eqnarray}
where the left-hand side denotes the curvature of space-time and the right-hand one is the energy
momentum tensor. The Ricci tensor $R_{\mu \nu}$ and Ricci scalar, namely
a contracted form of the Ricci tensor, $R$, are expressed with the
Christoffel symbols as  
\begin{eqnarray}
R_{\mu \nu} = R^\alpha{}_{\mu \alpha \nu}
= \Gamma^\alpha{}_{\mu \nu, \alpha} - \Gamma^\alpha{}_{\mu \alpha, \nu}
+ \Gamma^\alpha{}_{\beta \alpha} \Gamma^\beta{}_{\mu \nu} 
- \Gamma^\alpha{}_{\beta \nu} \Gamma^\beta{}_{\mu \alpha}~, 
\end{eqnarray}
where ${}_{, \alpha} \equiv \partial_\alpha$. 
The Christoffel symbols in a metric space without torsion are given by 
\begin{eqnarray}
\Gamma^\lambda{}_{\mu \nu} = \frac{1}{2} g^{\lambda \kappa}
(g_{\mu \kappa, \nu} + g_{\nu \kappa, \mu} - g_{\mu \nu, \kappa})~.
\end{eqnarray}
 Up to first order, we can express as 
\begin{eqnarray}
\begin{split}
\Gamma^{0}{}_{00} &= {\cal H} ~, \\
\Gamma^{i}{}_{00} &= \Gamma^{0}{}_{i0}= 0 ~,\\
\Gamma^{0}{}_{ij} &= {\cal H}(\delta_{ij} + h_{ij}) + \frac{1}{2}
 \dot{h}_{ij} ~,\\
\Gamma^i{}_{j0} &= {\cal H}\delta^i {}_{j} 
+ \frac{1}{2} \dot{h}^i {}_{j}  ~,\\
\Gamma^i{}_{jk} &= \frac{1}{2}(\partial_k h^i{}_{j} + \partial_j
 h^i{}_{k} - \partial^i h_{jk})~,
\end{split}
\end{eqnarray}
therefore each component of the Ricci tensor is calculated as
\begin{eqnarray}
\begin{split}
a^2 R^0{}_{0} &= 3 
\left( \frac{\ddot{a}}{a} - {\cal H}^2 \right) + \frac{1}{2} 
\left( \ddot{h}^i{}_i + {\cal H} \dot{h}^i{}_i \right) ~, \\
a^2 R^i{}_{0} &= - \frac{1}{2} 
\left( \partial^i \dot{h}^j{}_j - \partial^j \dot{h}^i{}_{j} \right)
 ~, \\
a^2 R^i{}_{j} &= 
\left( \frac{\ddot{a}}{a} + {\cal H}^2 \right) \delta^i{}_j +
 \frac{1}{2} \ddot{h}^i{}_j + {\cal H}\dot{h}^i{}_j + \frac{1}{2}
 {\cal H} \dot{h}^k{}_k \delta^i{}_j \\ 
&\quad - \frac{1}{2} 
\left( \partial^i\partial_j h^k{}_k + \nabla^2 h^i{}_j -
 \partial^i \partial_k h^k{}_j - \partial^k\partial_j h^i{}_k \right)~. \label{eq:ricci_tensor}
\end{split}
\end{eqnarray}
Here, ${\cal H} \equiv \dot{a}/a = aH$ is the Hubble parameter in terms of conformal time with $H$ being the observable Hubble parameter. Then the Ricci scalar is also given by
\begin{eqnarray}
R = R^\mu{}_\mu = \frac{1}{a^2} 
\left( 6 \frac{\ddot{a}}{a} + \ddot{h}^i{}_i + 3 {\cal H} \dot{h}^i{}_i -
 \nabla^2 h^i{}_i + \partial_i \partial^j h^i{}_j  \right)~.
\end{eqnarray}
Contracting the Einstein equation (\ref{eq:einstein_general}) allows one
to eliminate the Ricci scalar and reduce the Einstein equation to 
\begin{eqnarray}
R^\mu{}_\nu = 8 \pi G 
\left( T^\mu{}_\nu - \frac{1}{2} \delta^\mu{}_\nu T^\sigma{}_\sigma
\right) + \Lambda \delta^\mu{}_\nu ~. \label{eq:einstein_general_2}
\end{eqnarray}
Hence, in vacuum, we have $R^\mu {}_\nu = 0$. 

\subsubsection{Homogeneous contribution}

At zeroth order, the $00$ and $ii$ components of
Eq.~(\ref{eq:einstein_general}) lead to the Friedmann constraint
equation and the Raychaudhuri evolution equation, respectively. Substituting
Eq.~(\ref{eq:ricci_tensor}) into Eq.~(\ref{eq:einstein_general_2}), these are
obtained as 
\begin{eqnarray}
\begin{split}
{\cal H}^2 &= -\frac{8 \pi G}{3} a^2 \bar{T}^0{}_0 + \frac{a^2}{3}
 \Lambda ~, \\
2 \frac{\ddot{a}}{a} - {\cal H}^2 &= -\frac{8 \pi G}{3} a^2
 \bar{T}^i{}_i + a^2 \Lambda~.
\end{split}
\end{eqnarray}
The physical meaning of these equations can be illustrated with the
perfect fluid form as follows. The energy momentum tensor of the perfect
fluid is given by 
\begin{eqnarray}
T^\mu{}_\nu = (\rho + p) u^\mu u_\nu + p \delta^\mu {}_\nu ~,
\end{eqnarray}
hence the above equations change to  
\begin{eqnarray}
\begin{split}
{\cal H}^2 &= \frac{8 \pi G}{3} a^2 
\left( \bar{\rho} + \frac{\Lambda}{8 \pi G} \right) ~,\\
2 \frac{\ddot{a}}{a} - {\cal H}^2 
&= -8\pi G a^2 
\left( \bar{p} - \frac{\Lambda}{8 \pi G} \right)~.
\end{split}
\end{eqnarray}
Note that we may identify the cosmological constant as a component of
the perfect fluid as
\begin{eqnarray}
\bar{T}_{\Lambda}^\mu {}_\nu = \frac{\Lambda}{8\pi G} {\rm diag}(1,-1,-1,-1) ~.
\end{eqnarray}
To use a different phrase, an unperturbed perfect fluid of density
and pressure are given by $\rho_\Lambda = \Lambda / (8\pi G), p_\Lambda
= - \rho_\Lambda$. If we use $w = p / \rho$, then $w_\Lambda = -1$.  

For convenience, we change the Friedmann equation to 
\begin{eqnarray}
1 = \frac{8 \pi G}{3 {\cal H}^2} a^2 \bar{\rho} 
= \frac{8 \pi G}{3 H^2} \bar{\rho} 
= \sum_i \frac{8 \pi G}{3 H^2} \bar{\rho}_i~,
\end{eqnarray}
where in third equality, we decompose the total energy density in the
Universe into individual species $\bar{\rho}_i$. Introducing a quantity 
which means the ratio between the energy density of each species and
the critical density in the Universe at the present time, $\Omega_i$,
and which is expressed as $\Omega_i \equiv 8 \pi G \bar{\rho}_{i0} / (3
H_0^2)$, and using the scaling relation as $\bar{\rho}_i =
\bar{\rho}_{i0} / a^{n_i}$, this equation is rewritten as  
\begin{eqnarray}
\sum_i \frac{\Omega_i}{a^{n_i}} = \left( \frac{H}{H_0} \right)^2 ~. \label{eq:friedmann}
\end{eqnarray}
For radiations, matters and cosmological constant, we have $n_i = 4, 3,
0$, respectively. In this notation, we can also include a curvature term
as a component of $n_i = 2$. In Table~\ref{tab:FLRW_solutions}, we summarize the solutions of Eq.~(\ref{eq:friedmann}) if the cosmological fluid consists of a single component. 

\begin{table}[t]
\begin{center}
\begin{tabular}{| l |  c | c | c | c | c|}
\hline
 & $w$ & $\rho(a) $ & $a(t)$ & $a(\tau)$ & $\tau_{\rm i}$ \\
\hline
\hline
rad dom & $ 1/3$ & $a^{-4}$ & $t^{1/2}$ & $\tau$ &0  \\
mat dom & 0 & $a^{-3}$ & $t^{2/3}$ & $\tau^2$ & 0 \\
curv dom & - & $a^{-2}$ & $t$ & $e^{H_0 \Omega_k^{1/2} \tau}$ & $- \infty$  \\
$\Lambda$ dom & $-1$ & $a^0$ & $e^{Ht}$ & $- \tau^{-1}$ & $- \infty$ \\
\hline
\end{tabular}
\end{center}
\caption{FLRW solutions dominated by radiation, matter, curvature, or a cosmological constant.}
\label{tab:FLRW_solutions}
\end{table}

\subsubsection{Perturbed contribution}

At first order, $00$ and $ij$ components of
Eq.~(\ref{eq:einstein_general_2}) generate the evolution equations as   
\begin{eqnarray}
\begin{split}
&\ddot{h}^i{}_i + {\cal H} \dot{h}^i{}_i = 8 \pi G a^2 
\left( \delta T^0 {}_0 - \delta T^i{}_i  \right) ~, \\
& \ddot{h}^i{}_j + 2 {\cal H} \dot{h}^i{}_j + {\cal H} \dot{h}^k{}_k
\delta^i{}_j - (\partial^i \partial_j h^k{}_k + \nabla^2 h^i{}_j -
\partial^k \partial_j h^i{}_k - \partial_k \partial^i h^k{}_j) \\
 &\quad\qquad\quad = 16 \pi G a^2 
\left( \delta T^i{}_j - \frac{1}{2} \delta^i{}_j \delta T^\mu{}_\mu
 \right) ~, \label{eq:einstein_perturb_evolution}
\end{split}
\end{eqnarray}
and $00$ and $i0$ components of
Eq.~(\ref{eq:einstein_general}) generate the constraint equations as  
\begin{eqnarray}
\begin{split}
2 {\cal H} \dot{h}^i{}_{i} + \partial^j\partial_i h^i{}_j - \nabla^2
 h^i{}_i &= - 16 \pi G a^2 \delta T^0{}_0 ~, \\
\partial^j \dot{h}^i{}_j - \partial^i \dot{h}^j{}_j 
&= 16 \pi G a^2 \delta T^i{}_0~. \label{eq:einstein_perturb_constraint}
\end{split}
\end{eqnarray}
From here, let us express these equations with the variables in the helicity states. To do it, 
we decompose all kind of vectors and tensors, such as metric, velocities and energy momentum tensors, into each helicity part in accordance with the formulae: 
\begin{eqnarray}
\begin{split}
\omega_i ({\bf x}, \tau) &= \int \frac{d^3 {\bf k}}{(2\pi)^3} 
\left( \omega^{(0)} O^{(0)}_{i} + \sum_{\lambda = \pm 1} \omega^{(\lambda)}
O^{(\lambda)}_{i} \right) e^{i {\bf k} \cdot {\bf x}} ~, \\
\chi_{ij}({\bf x}, \tau) &= \int \frac{d^3 {\bf k}}{(2\pi)^3} 
\left( - \frac{1}{3} \chi_{\rm iso} \delta_{ij}
+ \chi^{(0)} O^{(0)}_{ij} + \sum_{\lambda = \pm 1} \chi^{(\lambda)}
O^{(\lambda)}_{ij} + \sum_{\lambda = \pm 2} \chi^{(\lambda)}
O^{(\lambda)}_{ij} \right) e^{i {\bf k} \cdot {\bf x}} ~,
\label{eq:decompose_fourier}
\end{split}
\end{eqnarray}
where we define the projection vectors and tensors as 
\begin{eqnarray}
\begin{split}
O_{a}^{(0)}(\hat{\bf k}) &\equiv i \hat{k}_a   ~, \\
O_{a}^{(\pm 1)}(\hat{\bf k}) &\equiv - i \epsilon^{(\pm 1)}_a(\hat{\bf k})  ~, \\
O_{ab}^{(0)}(\hat{\bf k}) &\equiv - \hat{k}_a \hat{k}_b + \frac{1}{3}\delta_{a,b} ~, \\
O_{ab}^{(\pm 1)}(\hat{\bf k}) &\equiv \hat{k}_a \epsilon_b^{(\pm 1)}(\hat{\bf k}) + \hat{k}_b
 \epsilon_a^{(\pm 1)}(\hat{\bf k})  ~, \\
O_{ab}^{(\pm 2)}(\hat{\bf k}) &\equiv e_{ab}^{(\pm 2)}(\hat{\bf k}) ~.
\end{split}
\end{eqnarray}
The polarization vector and tensor, $\epsilon_i^{(\pm 1)}, e_{ij}^{(\pm
2)}$, satisfy the divergenceless and transverse-traceless conditions as  
\begin{eqnarray}
\hat{k}_i \epsilon_i^{(\pm 1)}(\hat{\bf k}) = \hat{k}_i e_{ij}^{(\pm
 2)}(\hat{\bf k}) = e_{ii}^{(\pm 2)}(\hat{\bf k}) = 0 ~.
\end{eqnarray}
The prescription for the
scalar-vector-tensor decomposition and explicit forms of the
polarization vector and tensor are presented in Appendix
\ref{appen:polarization}. 
Then, from Eqs.~(\ref{eq:einstein_perturb_evolution}) and (\ref{eq:einstein_perturb_constraint}), we can rewrite the evolution equations as
\begin{eqnarray}
\begin{split}
\ddot{h}_{\rm iso} + {\cal H} \dot{h}_{\rm iso} 
&= - 8 \pi G a^2 
\left(\delta T^0{}_0 + \delta T_t^{\rm iso} \right) ~, \\
\ddot{h}^{(0)} +  2 {\cal H} \dot{h}^{(0)} + \frac{1}{3} k^2 (h_{\rm iso} -
 h^{(0)}) &= 16 \pi G a^2 \delta T_t^{(0)} ~, \\
\ddot{h}^{(\pm 1)} + 2 {\cal H} \dot{h}^{(\pm 1)} &= 16 \pi G a^2 \delta
 T_t^{(\pm 1)} ~, \\
\ddot{h}^{(\pm 2)} + 2 {\cal H} \dot{h}^{(\pm 2)} + k^2 h^{(\pm 2)} &=
 16 \pi G a^2 \delta T_t^{(\pm 2)} ~, \label{eq:decomposed_einstein_perturb_evolution}
\end{split}
\end{eqnarray}
and the constraint equations as 
\begin{eqnarray}
\begin{split}
{\cal H} \dot{h}_{\rm iso} + \frac{1}{3} k^2 (h_{\rm iso} - h^{(0)}) 
&= 8 \pi G a^2 \delta T^0{}_0 ~, \\
k \left( \dot{h}_{\rm iso} - \dot{h}^{(0)} \right) 
&= 24 \pi G a^2 \delta T_v^{(0)}  ~, \\
k \dot{h}^{(\pm 1)} &= - 16 \pi G a^2 \delta T_v^{(\pm 1)} ~. \label{eq:decomposed_einstein_perturb_constraint}
\end{split}
\end{eqnarray}
Here, we have obeyed the convention as
\begin{eqnarray}
 \begin{split}
\delta T^i{}_0 ({\bf x}, \tau) &= \int \frac{d^3 {\bf k}}{(2\pi)^3} 
\left( \delta T_v^{(0)} O^{(0)}_{i} 
+ \sum_{\lambda = \pm 1} \delta T_v^{(\lambda)}
O^{(\lambda)}_{i} \right) e^{i {\bf k} \cdot {\bf x}} ~, \\
\delta T^i{}_j ({\bf x}, \tau) &= \int \frac{d^3 {\bf k}}{(2\pi)^3} 
\left( - \frac{1}{3} \delta T_t^{\rm iso} \delta_{ij}
+ \delta T_t^{(0)} O^{(0)}_{ij} + \sum_{\lambda = \pm 1} \delta T_t^{(\lambda)}
O^{(\lambda)}_{ij} + \sum_{\lambda = \pm 2} \delta T_t^{(\lambda)}
O^{(\lambda)}_{ij} \right) e^{i {\bf k} \cdot {\bf x}} ~,
\label{eq:decompose_fourier}
\end{split}
\end{eqnarray}

Our perturbation quantities are related to the variables for the scalar
mode in the synchronous gauge of Ref.~\cite{Ma:1995ey}, namely $h$ and $\eta$, as 
\begin{eqnarray}
h_{\rm iso} = - h, \ \ h^{(0)} = - (h + 6 \eta)  ~.
\end{eqnarray}
Hence, we understand the correspondence to the gauge-invariant variables by Bardeen ($\Phi_A, \Phi_H$) \cite{Bardeen:1980kt} and Kodama-Sasaki ($\Psi, \Phi$) \cite{Kodama:1985bj}:
\begin{eqnarray}
\begin{split}
\Phi_A &= \Psi = - \frac{1}{2k^2} 
\left( \ddot{h}^{(0)} + {\cal H}\dot{h}^{(0)} \right) ~, \\
\Phi_H &= \Phi = \frac{1}{6} 
\left( h^{(0)} - h_{\rm iso} \right) - \frac{1}{2k^2} {\cal H} \dot{h}^{(0)}~.
\end{split}
\end{eqnarray}

\subsection{Boltzmann equations}

The distribution function of several species evolves in accordance with the Boltzmann equation as 
\begin{eqnarray}
\frac{d f}{d \tau} = \frac{\partial f}{\partial \tau} + \frac{\partial f}{\partial x^i} \frac{\partial x^i}{\partial \tau} + 
\frac{\partial f}{\partial p^\mu} \frac{\partial p^\mu}{\partial \tau} 
= \left(\frac{\partial f}{\partial \tau}\right)_C ~, \label{eq:boltzmann_general}
\end{eqnarray}
where $\tau$ is the conformal time, $p^\mu$ is the proper momentum of species, and a subscript $C$ denotes the collision term. In this Boltzmann equation, there exist two contributions: the gravitational redshift and the effect of scattering, which correspond to the third term of the first equality and the term of the second equality, respectively. 
For convenience, we introduce the comoving momentum and energy as $q^i \equiv a p^i, \epsilon \equiv a \sqrt{p^2 + m^2}$. Setting a unit vector parallel to the fluid momentum as ${\bf q} = q \hat{\bf n}$ and expanding the distribution function up to first order:
\begin{eqnarray}
f({\bf x}, q, \hat{\bf n}, \tau) = f^{(0)}(q)
\left[ 1 + f^{(1)}({\bf x}, q,
 \hat{\bf n}, \tau) \right] ~, \label{eq:distribution_expand}
\end{eqnarray}
the above Boltzmann equation is rewritten as 
\begin{eqnarray}
\frac{d f}{d \tau} &=& 
f^{(0)} 
\left( \frac{\partial f^{(1)}}{\partial \tau} + 
\frac{\partial f^{(1)}}{\partial x^i} 
 \frac{d x^i}{d \tau} \right) \nonumber \\
&&
+ \frac{\partial f^{(0)}}{\partial q}\frac{dq}{d\tau} +
f^{(0)} 
\frac{\partial f^{(1)}}{\partial q}\frac{dq}{d\tau}
+ f^{(1)} 
\frac{\partial f^{(0)}}{\partial q}\frac{dq}{d\tau}
 + f^{(0)} \frac{\partial f^{(1)}}{\partial \hat{n}^i}\frac{d \hat{n}^i}{d\tau}  
= \left(\frac{\partial f}{\partial \tau}\right)_C ~.
\end{eqnarray}
To estimate $dq / d\tau, d \hat{n}^i / d\tau$, we consider the geodesic equation as
\begin{eqnarray}
P^0 \frac{dP^\mu}{d\tau} + \Gamma^\mu {}_{\alpha \beta} P^\alpha P^\beta
 = 0 ~,
\end{eqnarray}
where $P^\mu$ is the canonical momentum as 
\begin{eqnarray}
P^\mu = \frac{1}{a^2}
\left( \epsilon , q_j 
\left( \delta^{ij} - \frac{1}{2} h^{ij} \right) \right) ~, \ \ 
P_\mu = 
\left( - \epsilon , q^j 
\left( \delta_{ij} + \frac{1}{2} h_{ij} \right) \right)~. 
\label{eq:p0pi}
\end{eqnarray}
The contraction is given by $P^\mu P_\mu = p^2 - (\epsilon/a)^2 = - m^2$. From $\mu = 0$ component, we obtain 
\begin{eqnarray}
\frac{dq}{d\tau} = - \frac{1}{2} q \hat{n}^i \hat{n}^j \frac{\partial h_{ij}}{\partial \tau}~.
\end{eqnarray}
Similarly, from the spatial components, we have
\begin{eqnarray}
 2 \frac{d \hat{n}^i}{d \tau} = \hat{n}^i \hat{n}_m \hat{n}_n \frac{\partial h^{mn}}{\partial \tau} - \hat{n}_j \frac{\partial h^{ij}}{\partial \tau} - \frac{2 q}{\epsilon} \hat{n}_m \hat{n}_n \partial^m h^{in} + \frac{q}{\epsilon} \hat{n}_m \hat{n}_n \partial^i h^{mn} \approx {\cal O}(h) ~.
\end{eqnarray}
Furthermore, since we have the zeroth order expression as $dx^i / d \tau = (q / \epsilon) \hat{n}^i$, the Boltzmann equations up to first order are expressed as
\begin{eqnarray}
\frac{\partial f^{(1)}}{\partial \tau} 
+ \frac{q}{\epsilon}\hat{n}^i \frac{\partial f^{(1)}}{\partial x^i} 
- \frac{1}{2} \hat{n}^i \hat{n}^j \frac{\partial h_{ij}}{\partial \tau} \frac{\partial \ln f^{(0)}}{\partial \ln q} = \frac{1}{f^{(0)}} 
\left(\frac{\partial f}{\partial \tau}\right)_C ~. \label{eq:boltzmann_general_1st}
\end{eqnarray} 

\begin{figure}[t]
  \centering \includegraphics[height=12cm,clip]{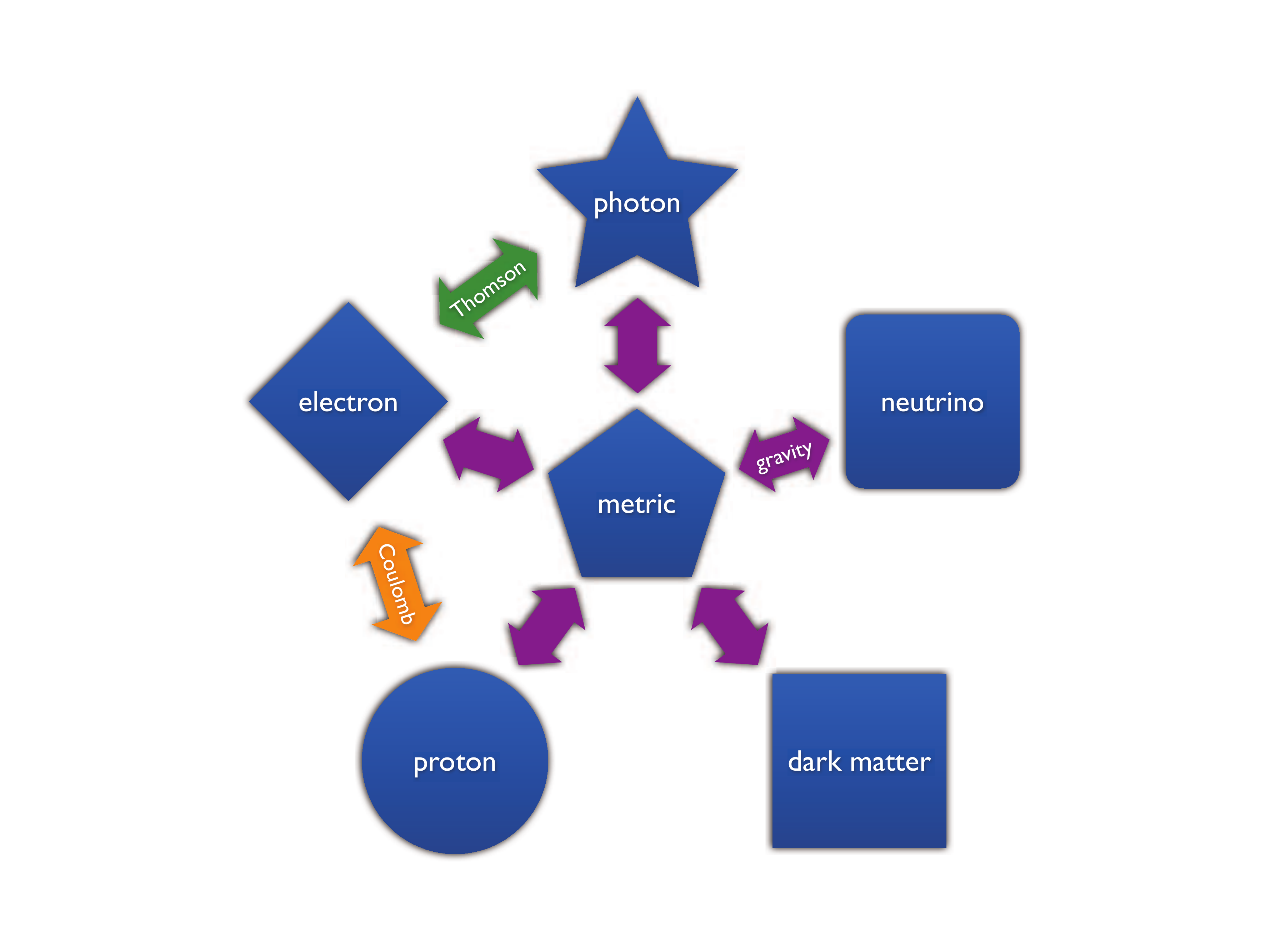}
  \caption{Interaction between several components in the Universe.}
  \label{fig:CMB}
\end{figure}

The general expression for the energy momentum tensor is given by 
\begin{eqnarray}
T^\mu{}_\nu = g_{\rm deg} \int (-g)^{-1/2} \frac{dP_1 dP_2 dP_3}{(2\pi)^3} \frac{P^\mu P_\nu}{P^0} f ~, 
\end{eqnarray}
where $g_{\rm deg}$ denotes the degree of freedom. 
Substituting the relations: 
\begin{eqnarray}
dP_1 dP_2 dP_3 = \left( 1 + \frac{1}{2} h^i{}_i \right) q^2 dq d\Omega_n ~, \ \
(-g)^{-1/2} = a^{-4} \left( 1 - \frac{1}{2} h^i{}_i \right) ~, 
\end{eqnarray}
and noting that
\begin{eqnarray}
\int \hat{n}^i \hat{n}_j d\Omega_n = \frac{4\pi}{3} \delta^i{}_j ~, \ \
\int \hat{n}_i d\Omega_n = \int \hat{n}_i \hat{n}_j \hat{n}_k d\Omega_n = 0 ~,
\end{eqnarray}
the homogeneous and linearized components of the energy momentum tensor are obtained as 
\begin{eqnarray}
\begin{split}
T^0{}_0 &= - \rho = -\frac{1}{(2\pi)^3 a^4} \int \int \epsilon 
f^{(0)}(1 + f^{(1)}) q^2 dq d\Omega_n  \\
&= - \frac{1}{2\pi^2 a^4} \int \epsilon f^{(0)} q^2 dq  - \frac{1}{(2\pi)^3 a^4} \int \int \epsilon f^{(0)} f^{(1)} q^2 dq d\Omega_n \equiv \bar{T}^0{}_0 + \delta T^0{}_0
~, \\
T^i{}_0 &= -\frac{1}{(2\pi)^3 a^4} \int \int q 
\left( \hat{n}^i - \frac{1}{2}\hat{n}_j h^{ij} \right) 
f^{(0)}(1 + f^{(1)}) q^2 dq d\Omega_n  \\
&= -\frac{1}{(2\pi)^3 a^4} \int \int q \hat{n}^i f^{(0)} f^{(1)} q^2 dq d\Omega_n 
\equiv  \delta T^i{}_0 ~, \\
T^i{}_j &=  \frac{1}{(2\pi)^3 a^4} \int \int \frac{q^2}{\epsilon} 
\left( \hat{n}^i - \frac{1}{2}\hat{n}_a h^{ia} \right)
 \left( \hat{n}_j + \frac{1}{2}\hat{n}^b h_{jb} \right)
f^{(0)}(1 + f^{(1)}) q^2 dq d\Omega_n  \\
&= \frac{1}{6\pi^2 a^4} \delta^i{}_j  
\int \frac{q^2}{\epsilon} f^{(0)} q^2 dq 
+ \frac{1}{(2\pi)^3 a^4} \int \int \frac{q^2}{\epsilon} 
\hat{n}^i \hat{n}_j  f^{(0)} f^{(1)} q^2 dq d\Omega_n  
\equiv \bar{T}^i{}_j + \delta T^i{}_j ~. \label{eq:EMT_any_component}
\end{split}
\end{eqnarray}
These components correspond to the density contrast $\delta$, velocity $v^i$ and anisotropic stress $\Pi^i{}_j$ of fluid as
\footnote{The anisotropic stress of the magnetic field is often normalized by photon's energy density as Eq.~(\ref{eq:EMT_PMF}).}
\begin{eqnarray}
\bar{T}^0{}_0 = - \bar{\rho} ~, \ \
\frac{\delta T^0{}_0}{\bar{\rho}} = - \delta ~, \ \
\frac{\delta T^i{}_0}{\bar{\rho} + \bar{p}} = - \frac{\delta T^0{}_i}{\bar{\rho} + \bar{p}} = - v^i ~, \ \ 
\bar{T}^i{}_j = \bar{p} \delta^i{}_j ~, \ \
\frac{\delta T^i{}_j}{\bar{p}} = \Pi^i{}_j ~. \label{eq:perturbation_variables}
\end{eqnarray}
Therefore, equating the integral of Eq.~(\ref{eq:boltzmann_general_1st}) over ${\bf q}_1, {\bf q}_2, {\bf q}_3$ with Eqs.~(\ref{eq:EMT_any_component}) and (\ref{eq:perturbation_variables}), we can see that the Boltzmann equation (\ref{eq:boltzmann_general_1st}) becomes the differential equations with respect to $\delta, v^i, \Pi^i{}_j$ for each species. These equations correspond to the Euler and continuity equations. Generally, as the species of the cosmological fluid which mainly generate the inhomogeneity of the cosmological structure, there exist baryon, photon, neutrino and cold dark matter (CDM), hence we can trace the evolution of their fluctuations due to solving these Boltzmann equations coupled with the Einstein equations (\ref{eq:decomposed_einstein_perturb_evolution}) and (\ref{eq:decomposed_einstein_perturb_constraint}). Between baryons and photons, Thomson scattering is effective, so that their Boltzmann equations have the collision term. On the other hand, for neutrinos and CDMs, since there are no short-length interactions, the right-hand side of the Boltzmann equations vanishes. All these species couple with the metric via gravity. This relation is illustrated in Fig.~\ref{fig:CMB}.

\subsection{Stokes parameters} 

Here, we introduce the Stokes parameters to characterize the
polarization states of the radiation field. For simplicity, at first, we consider a plane electromagnetic wave propagating along the $z$ axis. The Fourier decomposition of the radiation field is expressed as 
\begin{eqnarray}
{\bf E}(z, t) = \int_{-\infty}^\infty dk 
\left( \hat{\bf x} E_x e^{i \phi_x} +
 \hat{\bf y} E_y e^{i \phi_y} \right) e^{i(kz - \omega t)}~,
\end{eqnarray}
where $E_x, E_y$ and $\phi_x, \phi_y$ are the real quantities describing the amplitudes and phases in the $\hat{\bf x} - \hat{\bf y}$ plane, respectively, and $\omega = kc$ denotes the frequency of the wave. 

The Stokes parameters are given by
\begin{eqnarray}
\begin{split}
I &\equiv |E_x|^2 + |E_y|^2 ~, \\
Q &\equiv |E_x|^2 - |E_y|^2 ~, \\
U &\equiv - 2 {\rm Re} [E_x^* E_y] ~, \\
V &\equiv - 2 {\rm Im} [E_x^* E_y] ~, \label{eq:stokes_def}
\end{split}
\end{eqnarray}
where these are all real quantities. For the monochromatic wave, $I^2 = Q^2 + U^2 + V^2$ is satisfied. $I$ measures the intensity of radiation and is always positive. The other parameters represent the polarization states and can take ether positive or negative values. $Q$ and $U$ quantify the magnitude of the linear polarization, and $V$ parametrizes the circular polarization. While $I$ and $Q$ are parity-even quantities, $U$ and $V$ are parity-odd ones.

In order to see the transformation rule of $Q$ and $U$ under rotation of axes, let us introduce the new coordinate $(x', y')$, which is related to
the original coordinate $(x, y)$ by the rotation around the $z$ axis as
\begin{eqnarray}
\left(
  \begin{array}{cc}
 x' \\
 y'
  \end{array}
 \right)
= 
 \left(
  \begin{array}{cc}
  \cos\psi & \sin\psi \\
   -\sin\psi & \cos\psi 
  \end{array}
 \right)
\left(
  \begin{array}{cc}
 x \\
 y
  \end{array}
 \right) ~. \label{eq:pol_rotation}
\end{eqnarray}
Then, the radiation field is converted into
\begin{eqnarray}
\begin{split}
{\bf E}(z, t) &= \int_{-\infty}^\infty dk 
\left( \hat{\bf x'} E_x' e^{i \phi_x} +
 \hat{\bf y'} E_y' e^{i \phi_y} \right) e^{i(kz - \omega t)} ~, \\
\left( E_x' \pm i E_y' \right) &= e^{\pm i \psi} \left( E_x \pm i E_y \right)~; 
\end{split}
\end{eqnarray}
hence we have 
\begin{eqnarray}
Q' \pm i U' = e^{\mp 2 i \psi} (Q \pm i U) ~.
\end{eqnarray}
This implies that the linear combination, $Q \pm iU$, are the
spin-$\pm 2$ quantities. Therefore, the anisotropy of the linear
polarization should be expanded with the spin-$2$ spherical harmonics. 

\subsection{Boltzmann equations for photons}

Here, for quantifying the CMB anisotropy, let us focus on the linearized Boltzmann equation for photons. 
The distribution function of photons is given by 
\begin{eqnarray}
f = \left[ {\rm exp} 
\left\{ \frac{p}{T[1 + \Theta({\bf x}, \hat{\bf n}, t)]} \right\}
 - 1 \right]^{-1} ~,
\end{eqnarray}
hence we have \cite{Ma:1995ey} 
\begin{eqnarray}
f^{(0)} = \frac{1}{e^{p/T} - 1}~, \ \
f^{(1)}({\bf x}, q, \hat{\bf n}, \tau) 
= - \frac{q}{f^{(0)}} \frac{\partial f^{(0)}}{\partial q} \Theta
= - \frac{\partial \ln f^{(0)}}{\partial \ln q} \Theta ~, 
\label{eq:distribution_photon_expand}
\end{eqnarray}
where $\Theta \equiv \Delta T / T$. 
Substituting Eq.~(\ref{eq:distribution_photon_expand}) into Eq.~(\ref{eq:boltzmann_general_1st}), the Boltzmann equation in terms of $\Theta$ is expressed as 
\begin{eqnarray}
-q \frac{\partial f^{(0)}}{\partial q}
\left( \frac{\partial \Theta}{\partial \tau} 
+ \hat{n}^i \frac{\partial \Theta}{\partial x^i}
+ \frac{1}{2} \hat{n}^i \hat{n}^j \frac{\partial h_{ij}}{\partial \tau} 
\right) = \left( \frac{\partial f}{\partial \tau} \right)_C ~. 
\label{eq:boltzmann_photon_general_1st}
\end{eqnarray}
The first two terms in the bracket denote the free-streaming of the photon, 
whereas the remaining third term in the bracket account for the
gravitational redshift. As for the CMB polarization, there exist no
gravitational effects because $Q,U$ and $V$ themselves are first-order
quantities and the third term in the left-hand side of Eq.~(\ref{eq:boltzmann_general_1st}) does not appear. Hence, we summarize the Boltzmann equations of $\Theta, Q, U$ and $V$:
\begin{eqnarray}
\begin{split}
\frac{\partial \Theta}{\partial \tau} 
+ \hat{n}^i \frac{\partial \Theta}{\partial x^i}
+ \frac{1}{2} \hat{n}^i \hat{n}^j \frac{\partial h_{ij}}{\partial \tau}
&= \dot{\Theta}_{T} ~, \\
 \frac{\partial Q}{\partial \tau} 
+ \hat{n}^i \frac{\partial Q}{\partial x^i} 
&= \dot{Q}_T ~, \\
 \frac{\partial U}{\partial \tau} 
+ \hat{n}^i \frac{\partial U}{\partial x^i} 
&= \dot{U}_T ~, \\
 \frac{\partial V}{\partial \tau} 
+ \hat{n}^i \frac{\partial V}{\partial x^i} 
&= \dot{V}_T ~, \label{eq:boltzmann_photon_general_IQUV}
\end{split}
\end{eqnarray}
where $\Theta_T, Q_T, U_T$ and $V_T$ denote the collision terms of Thomson
scattering and $~\dot{}~ \equiv \partial / \partial \tau$ is the derivative with respect to the conformal time. Next we consider the contribution of these terms. 

\subsection{Thomson scattering}

\begin{figure}[t]
  \centering \includegraphics[height=9cm,clip]{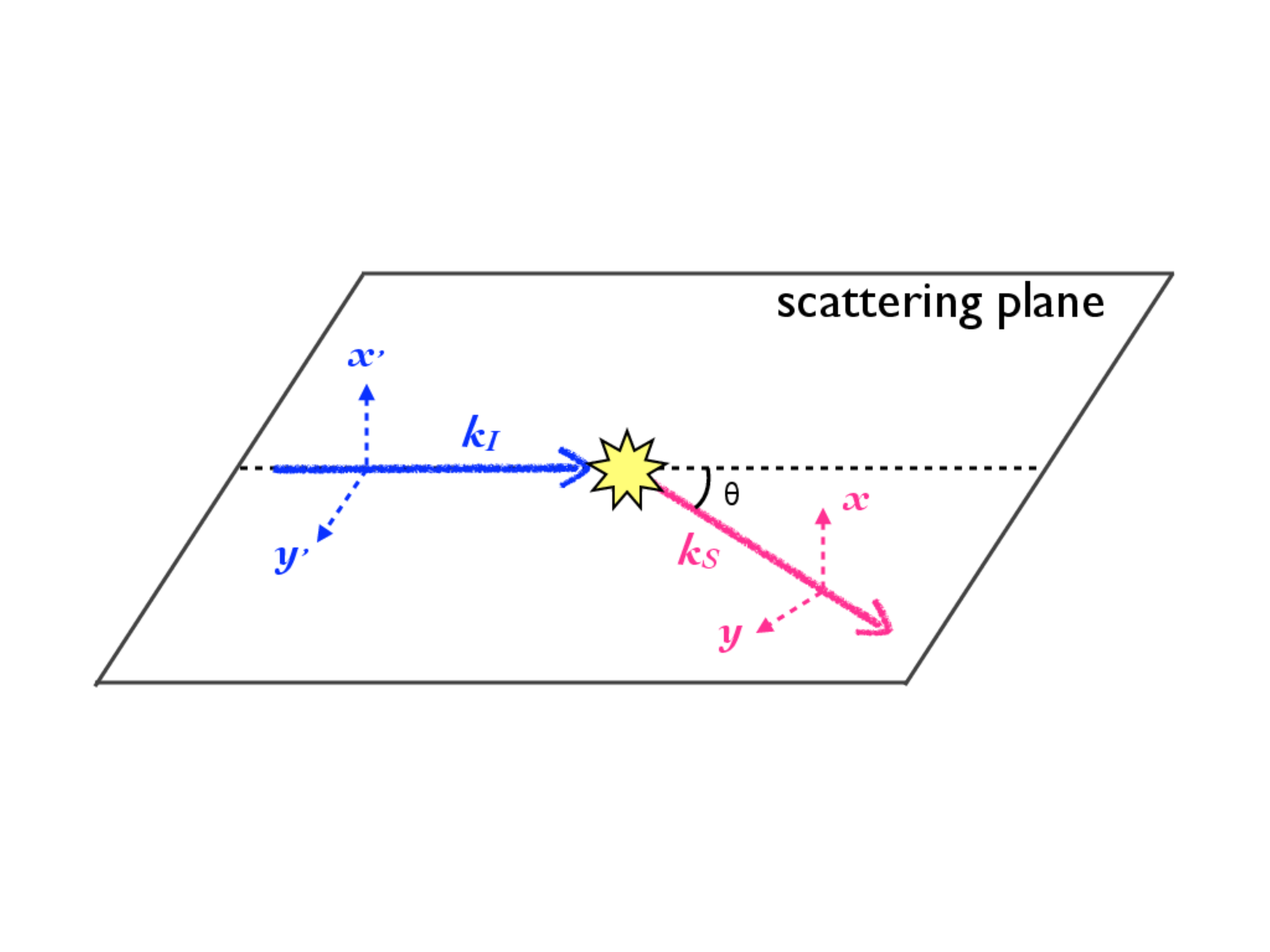}
  \caption{Geometry of Thomson scattering. Blue (Red) solid and two
 dashed arrows denote the incident (scattered) wave number vector and its orthogonal unit vectors, respectively. We set that $\hat{\bf x'} = \hat{\bf x}$}
  \label{fig:Thomson}
\end{figure}

The process of scattering off a photon by a charged particle without the energy exchange of photons is called the Rayleigh scattering. In particular, when the charged particle is an electron, the process is known as Thomson scattering. During the epoch of recombination, electrons scattered off photons by Thomson scattering. Here, we consider an incoming plane wave of the radiation with a wave number
vector ${\bf k_I}$ parallel to the $z$ axis
and an outgoing radiation scattered off by an electron with a wave number vector
${\bf k_S}$. We take the plane spanned by ${\bf k_I}$ and ${\bf k_S}$ as the scattering plane as shown in Fig.~\ref{fig:Thomson}.

The differential cross section of Thomson scattering is given by 
\begin{eqnarray}
\frac{d \sigma}{d \Omega} = \frac{3 \sigma_T}{8\pi} 
|\hat{\bf k_I} \cdot \hat{\bf k_S}|^2 ~, \label{eq:thomson_cross_sec}
\end{eqnarray}
where $d \Omega = d(\cos\theta) d\phi$ and $\sigma_T$ is the cross section of Thomson scattering. This equation quantifies the change of the intensity
by the scattering. For simplicity, we consider the case of the $x'$ axis parallel to the $x$ axis. Here, we suppose the incident radiation with the polarization states 
$I' = (I'_{y'}, I'_{x'}, U',V')$, where $I' = I'_{x'} + I'_{y'}$ and 
$Q' = I'_{y'} - I'_{x'}$. When there is no dependence on the azimuthal
angle, namely $\phi' = 0$, from the notation of Stokes parameters
(\ref{eq:stokes_def}) and (\ref{eq:thomson_cross_sec}), we obtain 
\begin{eqnarray}
I_{y} = \frac{3 \sigma_T}{16\pi} \cos^2 \theta I'_{y'} ~, \ \
I_{x} = \frac{3 \sigma_T}{16\pi} I'_{x'} ~, \ \
U = \frac{3 \sigma_T}{16\pi} \cos\theta U' ~,
\end{eqnarray} 
where we have normalized these equations so that the number of photons is
conserved during a single scattering. For a general case with a
non-vanishing azimuthal angle $\phi$, $Q'$ and $U'$ are replaced as
\begin{eqnarray}
Q' \pm i U' \Rightarrow e^{\mp 2i \phi} (Q' \pm i U')~.
\end{eqnarray}
Then, the changes of the Stokes parameters between the incident radiation
from $\hat{\bf n'} = \hat{\bf z} ~ (\theta' = 0, \phi' = 0)$, and the scattered radiation with
$\hat{\bf n} = (\theta, \phi)$ are given by \cite{Hu:1997hp}
\begin{eqnarray}
\begin{split}
\Delta \Theta(\hat{\bf n'} = \hat{\bf z}, \hat{\bf n}) 
&= \frac{1}{4\pi} 
\left[ \frac{3}{4}(1+\cos^2\theta) \Theta' 
- \sum_{s = \pm 2} \frac{3}{8} \sin^2 \theta 
 e^{- s i \phi} 
\left( Q' + \frac{s}{2} iU' \right) \right] ~, \\
\Delta(Q \pm iU)(\hat{\bf n'} = \hat{\bf z}, \hat{\bf n}) 
&= \frac{1}{4\pi} 
\left[ - \frac{3}{4}\sin^2\theta \Theta' 
+ \sum_{s = \pm 2} \frac{3}{8} 
\left( 1 \pm \frac{s}{2} \cos\theta \right)^2 e^{- s i \phi}
\left( Q' + \frac{s}{2} iU' \right) \right] ~, \label{eq:scattered_fields_special}
\end{split}
\end{eqnarray}
where we use $\Theta = \Delta I / I/ 4$. Using the explicit formulae of
the spin-$0$ and spin-$2$ spherical harmonics described in
Table~\ref{tab:spin2}, we can extend this expression to the form
corresponding to an arbitrary direction of $\hat{\bf n'}$: 
\begin{eqnarray}
\begin{split}
\Delta \Theta(\hat{\bf n'}, \hat{\bf n}) 
&= \sum_m  
\left[ 
\left\{ \frac{1}{10} Y_{2m}(\hat{\bf n}) Y_{2m}^* (\hat{\bf n'}) 
+ Y_{0m}(\hat{\bf n}) Y_{0m}^* (\hat{\bf n'}) 
\right\} \Theta' \right.  \\
&\qquad\quad \left. - 
\sum_{s = \pm 2}
\frac{3}{20} \sqrt{\frac{2}{3}} Y_{2m}(\hat{\bf n}) {}_{s}Y_{2m}^*
(\hat{\bf n'}) 
\left( Q' + \frac{s}{2} iU' \right)  \right] ~, \\
(\Delta Q \pm i \Delta U)(\hat{\bf n'}, \hat{\bf n})
&= \sum_m \frac{3}{10} {}_{\pm 2} Y_{2m}(\hat{\bf n})
\left[ 
- \sqrt{\frac{2}{3}} 
 Y_{2m}^* (\hat{\bf n'})  \Theta' +
\sum_{s = \pm 2}
{}_{s}Y_{2m}^*
(\hat{\bf n'}) 
\left( Q' + \frac{s}{2} iU' \right)  \right] ~. \label{eq:scattered_fields_general}
\end{split}
\end{eqnarray}
We will use Eq.~(\ref{eq:scattered_fields_general}) in the frame
satisfying ${\bf k} \parallel \hat{\bf z}$, where ${\bf k}$ is the wave
number vector. 
Integrating these equations
over all directions $\hat{\bf n'}$, we express the scattered fields as
\footnote{The Stokes parameter, $V$, which means the circular
polarization of photon, can be ignored because it cannot be generated
through Thomson scattering if this is initially absent.}
\begin{eqnarray}
\begin{split}
\dot{\Theta}_T  (\hat{\bf n}) 
&= -\dot{\kappa}   
\left[ \Theta(\hat{\bf n}) - \int d\Omega' 
\Delta \Theta(\hat{\bf n'}, \hat{\bf n}) 
 - {\bf v_b} \cdot \hat{\bf n} \right]~, \\ 
\left( \dot{Q} \pm i \dot{U} \right)_T  (\hat{\bf n}) 
&= -\dot{\kappa}   
\left[ (Q \pm iU)(\hat{\bf n}) - \int d\Omega' (\Delta Q \pm i \Delta U)(\hat{\bf n'}, \hat{\bf n}) 
\right] ~, \label{eq:line-of-sight_integral_general}
\end{split}
\end{eqnarray}
where we define the differential optical
depth as $\dot{\kappa} \equiv a \sigma_T n_e x_e$ with $n_e x_e$ being the density of ionized electrons, and its total value at time $\tau$ is given by
\begin{eqnarray}
\kappa(\tau) \equiv \int_\tau^{\tau_0} \dot{\kappa} (\tau') d\tau' ~,
\end{eqnarray} 
with $\tau_0$ being the present conformal time.

From here, we discuss the polarization property in more detail.
One of the key points in Eq.~(\ref{eq:scattered_fields_general}) is that
the temperature anisotropy generates the polarization of the CMB
photons. Then, what mode of the temperature anisotropy is related to the generation of the polarization? 
For simplicity, we suppose that the incident radiation
field is unpolarized, $Q' = U' = V' = 0$ and consider the case for $\hat{\bf n} = \hat{\bf z}$.
Integrating Eq.~(\ref{eq:scattered_fields_general}) over all incident
radiation, we gain
\begin{eqnarray}
Q \pm iU (\hat{\bf z}) = \frac{3 \sigma_T}{4\pi} \sqrt{\frac{2 \pi}{15}} 
\int d\Omega' Y_{22}(\theta', \phi') \Theta'(\theta', \phi')~. \label{eq:QU_n=z}
\end{eqnarray}
When the incident temperature (intensity) anisotropy is expanded with the spherical harmonics as
$\Theta'(\theta',\phi') = \sum_{\ell m} a'_{\ell m} Y_{\ell m}(\theta', \phi')$, Eq.~(\ref{eq:QU_n=z}) is replaced with 
\begin{eqnarray}
Q \pm iU (\hat{\bf z}) = \frac{3 \sigma_T}{4\pi} 
\sqrt{\frac{2 \pi}{15}} a'_{22} ~.
\end{eqnarray}
Thus, if there exists no quadrupole moment ($\ell = 2$) in the unpolarized radiation field, the total scattered radiation along the $z$ direction
would be never polarized. Long before recombination, in the thermal
equilibrium, the polarization states of photons are equally populated
and the incident radiation should not have any polarization. Therefore, there are only the unpolarized radiations before recombination. 
Allowing the polarization at the last scatters just before the photons
begin to stream freely, the polarized emission can lead to the multipole anisotropy and one has polarized quadrupole and octupole and so on. These effects are automatically involved in the Boltzmann equation.

\subsection{Transfer functions} 

Here, we derive the CMB anisotropy sourced from scalar-, vector- and tensor-mode perturbations. We obey a Fourier transformation as
\begin{eqnarray}
X(\hat{\bf n}, \tau) = \int \frac{d^3{\bf k}}{(2\pi)^3} \Delta_{X}(\tau,
 {\bf k}, \hat{\bf n}) ~,
\end{eqnarray}
where $X = \Theta, Q \pm iU$ and $\Delta_X$
is called the transfer function. Note that we include the factor $e^{i
{\bf k} \cdot \hat{\bf n}}$ in $\Delta_X$. 

At first, for convenience, we derive the transfer functions of photons when ${\bf k} \parallel \hat{\bf z}$. 
For ${\bf k} \parallel \hat{\bf z}$, the scalar-vector-tensor decomposition of the gravitational redshift and Doppler term in the Boltzmann equation of photons (\ref{eq:boltzmann_photon_general_IQUV}) and (\ref{eq:line-of-sight_integral_general}) are given by
 \begin{eqnarray}
\begin{split}
\frac{1}{2} \hat{n}^i \hat{n}^j \dot{h}_{ij}({\bf k} \parallel \hat{\bf
 z}, \tau) 
&= \frac{1}{2} 
\left[ - \frac{1}{3} \dot{h}_{\rm iso}({\bf k} \parallel \hat{\bf z}, \tau) 
+ \dot{h}^{(0)}({\bf k} \parallel \hat{\bf z}, \tau) \left(- \cos^2 \theta_{k,n} + \frac{1}{3}\right) \right] \\
&\quad + \sum_{\lambda = \pm 1} \frac{1}{\sqrt{2}} \sin \theta_{k,n} \cos \theta_{k,n} 
e^{\lambda i \phi_{k,n}} \dot{h}^{(\lambda)}({\bf k} \parallel \hat{\bf z}, \tau) \\
&\quad + \sum_{\lambda = \pm 2} \frac{1}{2\sqrt{2}} \sin^2 \theta_{k,n} e^{\lambda i
\phi_{k,n}} \dot{h}^{(\lambda)}({\bf k} \parallel \hat{\bf z}, \tau) \\
&\equiv \xi^{(0)}({\bf k} \parallel \hat{\bf z})
\left[ \frac{1}{3} \dot{h}_{\rm iso}^{(S)}(k, \tau) 
 + \dot{h}^{(S)}(k, \tau) 
\left(\cos^2 \theta_{k,n} - \frac{1}{3}\right) \right] \\
&\quad + \sum_{\lambda = \pm 1}  \sin \theta_{k,n} \cos \theta_{k,n}  
e^{\lambda i \phi_{k,n}} \xi^{(\lambda)} ({\bf k} \parallel \hat{\bf z})  \dot{h}^{(V)}(k, \tau)  \\
&\quad + \sum_{\lambda = \pm 2} \sin^2 \theta_{k,n} e^{\lambda i
\phi_{k,n}} \xi^{(\lambda)} ({\bf k} \parallel \hat{\bf z})  \dot{h}^{(T)}(k, \tau)
 ~, \\
{\bf v_b}({\bf k} \parallel \hat{\bf z}, \tau) \cdot \hat{\bf n} 
&= i \cos\theta_{k,n} v_b^{(0)}({\bf k} \parallel \hat{\bf z}, \tau) 
+ \sum_{\lambda = \pm 1} 
\frac{-i}{\sqrt{2}}  \sin\theta_{k,n} e^{\lambda i \phi_{k,n}} v_b^{(\lambda)} ({\bf k} \parallel \hat{\bf z}, \tau)  \\
& \equiv i \cos\theta_{k,n}
\xi^{(0)} ({\bf k} \parallel \hat{\bf z}) v_b^{(S)}(k,\tau)  \\
&\quad + \sum_{\lambda = \pm 1} 
-i \sin\theta_{k,n} e^{\lambda i \phi_{k,n}}
\xi^{(\lambda)} ({\bf k} \parallel \hat{\bf z}) v_b^{(V)}(k,\tau)
~, \label{eq:ISW_Doppler}
\end{split}
\end{eqnarray}
where $\xi^{(0)}, \xi^{(\pm 1)}$ and $\xi^{(\pm 2)}$ are the initial stochastic variables of scalar, vector and tensor modes, and we use the calculation results from Appendix \ref{appen:polarization} as
\begin{eqnarray}
\begin{split}
\epsilon_j^{(\pm 1)}(\hat{\bf z}) &= 
\frac{1}{\sqrt{2}}  
\left(
  \begin{array}{c}
   1 \\
   \pm i \\
   0 
  \end{array}
 \right)~, \\
O^{(\pm 1)}_{ij}(\hat{\bf z}) &=  
\frac{1}{\sqrt{2}}  
\left(
  \begin{array}{ccc}
   0 & 0 & 1 \\
   0 & 0 & \pm i \\
   1 & \pm i & 0
  \end{array}
 \right)~, \\
O^{(\pm 2)}_{ij}(\hat{\bf z}) 
&= e_{ij}^{(\pm 2)}(\hat{\bf z}) 
= \frac{1}{\sqrt{2}}  
\left(
  \begin{array}{ccc}
   1 & \pm i & 0 \\
   \pm i & -1 & 0 \\
   0 & 0 & 0
  \end{array}
 \right) ~.
\end{split}
\end{eqnarray}
According to Refs.~\cite{Ma:1995ey, Giovannini:2008yz, Lewis:2004ef}, if we equate $\xi^{(0)}$ with the comoving curvature perturbation on superhorizon scales ${\cal R}$, the initial conditions of the metric perturbations and the baryon velocity are
\footnote{Then, the parameters in Ref.~\cite{Ma:1995ey} are given by 
\begin{eqnarray}
\begin{split}
\eta(k, \tau_{\rm ini}) &= - {\cal R} 
\left[ 1 - \frac{5 + 4 R_\nu}{12(15 + 4 R_\nu)} (k \tau_{\rm ini})^2 \right] ~, \\ 
h(k, \tau_{\rm ini}) &= - \frac{1}{2} {\cal R} (k \tau_{\rm ini})^2 ~.
\end{split}
\end{eqnarray}
} 
\begin{eqnarray}
\begin{split}
h_{\rm iso}^{(S)}(k, \tau_{\rm ini}) &= - \frac{1}{4} (k \tau_{\rm ini})^2 ~, \\
h^{(S)}(k, \tau_{\rm ini}) &= - 3 - \frac{5}{2(15 + 4 R_\nu)} (k \tau_{\rm ini})^2 
~, \\
v_b^{(S)}(k, \tau_{\rm ini}) &= \frac{1}{36} (k \tau_{\rm ini})^3 ~.
\end{split}
\end{eqnarray}
In the tensor mode, equating $\xi^{(\pm 2)}$ with the primordial gravitational wave on superhorizon scales $h^{(\pm 2)}$, it is satisfied that
\begin{eqnarray}
h^{(T)}(k, \tau_{\rm ini}) = \frac{1}{2\sqrt{2}} 
\left[1 - \frac{5}{2(15 + 4 R_\nu)} (k \tau_{\rm ini})^2 \right]~.
\end{eqnarray}
Note that the Doppler effect does not affect only the tensor-mode perturbation. 
We introduce the transfer function in the Fourier space as \cite{Zaldarriaga:1996xe, Kosowsky:1994cy}
\begin{eqnarray}
\begin{split}
\Delta_{I}^{(S)}(\tau, {\bf k} \parallel \hat{\bf z}, \hat{\bf n})
&= \xi^{(0)}({\bf k} \parallel \hat{\bf z}) \tilde{\Delta}_I^{(S)}(\tau, k, \mu_{k,n}) ~, \\
(\Delta_{Q}^{(S)} \pm i \Delta_U^{(S)})
(\tau, {\bf k} \parallel \hat{\bf z}, \hat{\bf n})
&= 
\xi^{(0)}({\bf k} \parallel \hat{\bf z}) \tilde{\Delta}_P^{(S)}(\tau, k, \mu_{k,n})~, \\
\Delta_{I}^{(V)}(\tau, {\bf k} \parallel \hat{\bf z}, \hat{\bf n})
&= \sum_{\lambda = \pm 1} -i \sqrt{1 - \mu_{k,n}^2} e^{\lambda i \phi_{k,n}}
\xi^{(\lambda)}({\bf k} \parallel \hat{\bf z}) \tilde{\Delta}_I^{(V)}(\tau, k, \mu_{k,n}) ~, \\
(\Delta_{Q}^{(V)} \pm i \Delta_U^{(V)})
(\tau, {\bf k} \parallel \hat{\bf z}, \hat{\bf n})
&= \sum_{\lambda = \pm 1} 
\mp \lambda (1 \mp \lambda \mu_{k,n}) \sqrt{1 - \mu_{k,n}^2} 
e^{\lambda i \phi_{k,n}}
\xi^{(\lambda)}({\bf k} \parallel \hat{\bf z}) \tilde{\Delta}_P^{(V)}(\tau, k, \mu_{k,n})~, \\
\Delta_{I}^{(T)}(\tau, {\bf k} \parallel \hat{\bf z}, \hat{\bf n}) 
&=  (1 - \mu_{k,n}^2) 
\sum_{\lambda = \pm 2} e^{\lambda i \phi_{k,n}}
 \xi^{(\lambda)}({\bf k} \parallel \hat{\bf z}) 
\tilde{\Delta}_I^{(T)}(\tau, k, \mu_{k,n}) ~, \\
(\Delta_{Q}^{(T)} \pm i \Delta_{U}^{(T)})(\tau, {\bf k} \parallel \hat{\bf z}, \hat{\bf n}) 
&= \sum_{\lambda = \pm 2}  
\left(1 \mp \frac{\lambda}{2} \mu_{k,n} \right)^2  e^{\lambda i
 \phi_{k,n}} \xi^{(\lambda)}({\bf k} \parallel \hat{\bf z}) 
\tilde{\Delta}_{P}^{(T)}(\tau, k, \mu_{k,n}) ~,
\end{split}
\end{eqnarray}
where $\mu_{k,n} \equiv \hat{\bf k} \cdot \hat{\bf n}$. Then, from Eqs.~(\ref{eq:scattered_fields_general}) and (\ref{eq:line-of-sight_integral_general}), we can write the collision term of Thomson scattering for the scalar mode: 
\begin{eqnarray}
\begin{split}
\int d \Omega' \Delta\Theta^{(S)}(\hat{\bf n'}, \hat{\bf n})
&= 
\int \frac{d^3 {\bf k}}{(2\pi)^3}\int d \Omega'
\sum_m 
\frac{1}{10} \xi^{(0)}({\bf k} \parallel \hat{\bf z}) \\
&\quad \times
 \left[ 
\left\{  Y_{2m}(\hat{\bf n}) Y_{2m}^* (\hat{\bf n'}) 
+ 10 Y_{0m}(\hat{\bf n}) Y_{0m}^* (\hat{\bf n'}) 
\right\}  
\tilde{\Delta}_I^{(S)}(\tau, k, \mu') \right.  \\
&\left.\qquad 
- \sqrt{\frac{3}{2}} Y_{2m}(\hat{\bf n}) 
\left\{ \sum_{s = \pm 2} 
{}_{s}Y_{2m}^*(\hat{\bf n'}) \right\} 
\tilde{\Delta}_{P}^{(S)}(\tau, k, \mu')
\right] ~,\\
\int d \Omega' (\Delta Q^{(S)} \pm i \Delta U^{(S)})(\hat{\bf n'}, \hat{\bf n})
&= - \int \frac{d^3 {\bf k}}{(2\pi)^3}\int d \Omega'
\sum_m 
\frac{\sqrt{6}}{10} 
{}_{\pm 2} Y_{2 m}(\hat{\bf n})
\xi^{(0)}({\bf k} \parallel \hat{\bf z}) \\
&\quad \times
 \left[ Y_{2m}^* (\hat{\bf n'})   
\tilde{\Delta}_I^{(S)}(\tau, k, \mu') \right.  \\
&\left.\qquad 
- \sqrt{\frac{3}{2}} 
\left\{ \sum_{s = \pm 2} 
{}_{s}Y_{2m}^*(\hat{\bf n'}) \right\} 
\tilde{\Delta}_{P}^{(S)}(\tau, k, \mu')
\right]
 ~, 
\end{split}
\end{eqnarray}
for the vector mode: 
\begin{eqnarray}
\begin{split}
\int d \Omega' \Delta\Theta^{(V)}(\hat{\bf n'}, \hat{\bf n})
&= 
\int \frac{d^3 {\bf k}}{(2\pi)^3}\int d \Omega'
\sum_m 
\sum_{\lambda = \pm 1} 
\frac{1}{5}\sqrt{\frac{2\pi}{3}}
\lambda \xi^{(\lambda)}({\bf k} \parallel \hat{\bf z}) \\
&\quad \times
 \left[ i
\left\{  Y_{2m}(\hat{\bf n}) Y_{2m}^* (\hat{\bf n'}) 
+ 10 Y_{0m}(\hat{\bf n}) Y_{0m}^* (\hat{\bf n'}) 
\right\} \right. \\
&\left. \qquad\quad \times
Y_{1 \lambda}(\hat{\bf n'}) 
\tilde{\Delta}_I^{(V)}(\tau, k, \mu') \right.  \\
&\left.\qquad 
- \frac{3}{\sqrt{5}} Y_{2m}(\hat{\bf n}) 
\left\{ \sum_{s = \pm 2} 
{}_{s}Y_{2m}^*(\hat{\bf n'}) {}_{s}Y_{2 \lambda}(\hat{\bf n'})  
\right\}
\tilde{\Delta}_{P}^{(V)}(\tau, k, \mu')
\right] ~,\\
\int d \Omega' (\Delta Q^{(V)} \pm i \Delta U^{(V)})(\hat{\bf n'}, \hat{\bf n})
&= - \int \frac{d^3 {\bf k}}{(2\pi)^3}\int d \Omega' \sum_m  
\frac{2 \sqrt{\pi}}{5} {}_{\pm 2} Y_{2m}(\hat{\bf n})
\sum_{\lambda = \pm 1} \lambda \xi^{(\lambda)}({\bf k} \parallel \hat{\bf z})
\\
&\quad \times
\left[ i Y_{2m}^* (\hat{\bf n'}) Y_{1 \lambda}(\hat{\bf n'}) 
\tilde{\Delta}_I^{(V)} (\tau, k, \mu')
\right. \\
&\left.\qquad  
- \frac{3}{\sqrt{5}} \left\{
\sum_{s = \pm 2}
{}_{s}Y_{2m}^* (\hat{\bf n'}) 
{}_{s}Y_{2 \lambda}(\hat{\bf n'}) \right\}  
\tilde{\Delta}_P^{(V)} (\tau, k, \mu')
\right] ~, 
\end{split}
\end{eqnarray}
for the tensor mode: 
\begin{eqnarray}
\begin{split}
\int d \Omega' \Delta\Theta^{(T)}(\hat{\bf n'}, \hat{\bf n})
&= 
\int \frac{d^3 {\bf k}}{(2\pi)^3}\int d \Omega'
\sum_m \frac{2}{5} \sqrt{\frac{2\pi}{15}}  
\sum_{\lambda = \pm 2} \xi^{(\lambda)}({\bf k} \parallel \hat{\bf z}) \\
&\quad \times
 \left[ 
\left\{  Y_{2m}(\hat{\bf n}) Y_{2m}^* (\hat{\bf n'}) 
+ 10 Y_{0m}(\hat{\bf n}) Y_{0m}^* (\hat{\bf n'}) 
\right\} \right. \\
& \left. \qquad\quad \times
Y_{2 \lambda}(\hat{\bf n'}) 
\tilde{\Delta}_I^{(T)}(\tau, k, \mu') \right.  \\
&\left.\qquad - 
3 Y_{2m}(\hat{\bf n}) 
\left\{ \sum_{s = \pm 2} 
{}_{s}Y_{2m}^*(\hat{\bf n'}) {}_{s}Y_{2 \lambda}(\hat{\bf n'})  
\right\}
\tilde{\Delta}_{P}^{(T)}(\tau, k, \mu')
\right] ~,\\
\int d \Omega' (\Delta Q^{(T)} \pm i \Delta U^{(T)})(\hat{\bf n'}, \hat{\bf n})
&= - \int \frac{d^3 {\bf k}}{(2\pi)^3}\int d \Omega' \sum_m  
\frac{4}{5} \sqrt{\frac{\pi}{5}} {}_{\pm 2} Y_{2m}(\hat{\bf n})
\sum_{\lambda = \pm 2} \xi^{(\lambda)}({\bf k} \parallel \hat{\bf z})
\\
&\quad \times
\left[  Y_{2m}^* (\hat{\bf n'}) Y_{2 \lambda}(\hat{\bf n'}) 
\tilde{\Delta}_I^{(T)} (\tau, k, \mu')
\right. \\
&\left.\qquad - 
3 \left\{
\sum_{s = \pm 2}
{}_{s}Y_{2m}^* (\hat{\bf n'}) 
{}_{s}Y_{2 \lambda}(\hat{\bf n'}) \right\}  
\tilde{\Delta}_P^{(T)} (\tau, k, \mu')
\right] ~.
\end{split}
\end{eqnarray}
Here, we use 
\begin{eqnarray}
\begin{split}
\sqrt{1 - \mu^2} e^{\lambda i \phi} 
&= - \lambda \sqrt{8\pi \over 3} Y_{1 \lambda} ~, \ \
\sqrt{1 - \mu^2}(1 \mp \lambda \mu)e^{\lambda i
\phi} = \mp \sqrt{16 \pi \over 5} {}_{\pm 2} Y_{2 \lambda} \ \ ({\rm for} \ \lambda = \pm 1) ~, \\
(1 - \mu^2) e ^{\lambda i \phi} &= 4 \sqrt{2\pi \over 15} Y_{2 \lambda} ~, \ \ 
\left( 1 \mp \frac{\lambda}{2} \mu \right)^2 e^{\lambda i \phi} = 8 \sqrt{\pi \over 5} {}_{\pm 2} Y_{2 \lambda} \ \ ({\rm for} \ \lambda = \pm 2) ~.
\end{split}
\end{eqnarray}
Using the multipole expansion as 
\begin{eqnarray}
\tilde{\Delta}_{I/P}^{(S/V/T)} (\tau, k, \mu') = \sum_{l}(-i)^l
\sqrt{4\pi(2l+1)} Y_{l 0}(\hat{\bf n'})
 \tilde{\Delta}_{I/P,l}^{(S/V/T)} (\tau, k)
\end{eqnarray}
and the $\Omega'$-integrals for $\lambda = 0$:
\begin{eqnarray}
\begin{split}
\int d\Omega' Y_{2 m}^* Y_{l 0} 
&= \delta_{l,2} \delta_{m,0} ~, \\ 
\int d\Omega' Y_{0 m}^* Y_{l 0} 
&= \delta_{l,0} \delta_{m,0} ~, \\
\int d\Omega' {}_{\pm 2} Y_{2 m}^* Y_{l 0}  
&= \sqrt{\frac{5}{6}} \delta_{m, 0}
\left( \delta_{l,0} -  \frac{1}{\sqrt{5}} \delta_{l,2}  \right) ~,
\end{split}
\end{eqnarray}
for $\lambda = \pm 1$:
\begin{eqnarray}
\begin{split}
\int d\Omega' Y_{2 m}^* Y_{1 \lambda} Y_{l 0} 
&= \sqrt{\frac{3}{20\pi}} \delta_{m, \lambda}
\left( \delta_{l,1} - \sqrt{\frac{3}{7}} \delta_{l,3} \right) ~, \\ 
\int d\Omega' Y_{0 m}^* Y_{2 \lambda} Y_{l 0} 
&= 0 ~, \\
\int d\Omega' {}_{\pm 2} Y_{2 m}^* {}_{\pm 2} Y_{2 \lambda} Y_{l 0}  
&= \frac{1}{\sqrt{4\pi}} \delta_{m, \lambda}
\left( \delta_{l,0} \mp  \frac{\sqrt{3}}{3} \lambda \delta_{l,1} 
- \frac{\sqrt{5}}{7} \delta_{l,2}
\pm \frac{\sqrt{7}}{7} \lambda \delta_{l,3} 
- \frac{2}{21} \delta_{l,4} \right) ~,
\end{split}
\end{eqnarray}
and for $\lambda = \pm 2$:
\begin{eqnarray}
\begin{split}
\int d\Omega' Y_{2 m}^* Y_{2 \lambda} Y_{l 0} 
&= \frac{1}{\sqrt{4\pi}} \delta_{m, \lambda}
\left( \delta_{l,0} -  \frac{2 \sqrt{5}}{7} \delta_{l,2} + \frac{1}{7} \delta_{l,4} \right) ~, \\ 
\int d\Omega' Y_{0 m}^* Y_{2 \lambda} Y_{l 0} 
&= 0 ~, \\
\int d\Omega' {}_{\pm 2} Y_{2 m}^* {}_{\pm 2} Y_{2 \lambda} Y_{l 0}  
&= \frac{1}{\sqrt{4\pi}} \delta_{m, \lambda}
\left( \delta_{l,0} \mp  \frac{\sqrt{3}}{3} \lambda \delta_{l,1} + 
\frac{2 \sqrt{5}}{7} \delta_{l,2}
\mp \frac{\sqrt{7}}{28} \lambda \delta_{l,3} 
+ \frac{1}{42} \delta_{l,4} \right) ~,
\end{split}
\end{eqnarray}
we can obtain the anisotropies generated via Thomson scattering for the scalar mode:
\begin{eqnarray}
\begin{split}
\int d \Omega' \Delta\Theta^{(S)}(\hat{\bf n'}, \hat{\bf n})
& = 
\int \frac{d^3 {\bf k}}{(2\pi)^3}  \xi^{(0)}({\bf k} \parallel \hat{\bf z})
\left[ \tilde{\Delta}_{I,0}^{(S)}
- \sqrt{\frac{\pi}{5}}
Y_{2 0}(\hat{\bf n}) \psi^{(S)}(k, \tau)  \right] 
~, \\
\int d \Omega' (\Delta Q^{(S)} \pm i \Delta U^{(S)})(\hat{\bf n'}, \hat{\bf n})
& = \int \frac{d^3 {\bf k}}{(2\pi)^3}  \xi^{(0)}({\bf k} \parallel \hat{\bf z})
\left[ \sqrt{\frac{6 \pi}{5}}
{}_{\pm 2}Y_{2 0}(\hat{\bf n}) \psi^{(S)}(k, \tau)  \right] 
~, \\
\psi^{(S)}(k, \tau) 
&\equiv \tilde{\Delta}_{I,2}^{(S)} + \tilde{\Delta}_{P,0}^{(S)} + \tilde{\Delta}_{P,2}^{(S)} ~, \label{eq:scal_Thomson}
\end{split}
\end{eqnarray}
for the vector mode: 
\begin{eqnarray}
\begin{split}
\int d \Omega' \Delta\Theta^{(V)}(\hat{\bf n'}, \hat{\bf n})
& = 
\int \frac{d^3 {\bf k}}{(2\pi)^3}  
\left[\sum_{\lambda = \pm 1} \lambda \sqrt{\frac{8\pi}{15}}
Y_{2 \lambda}(\hat{\bf n}) \xi^{(\lambda)}({\bf k} \parallel \hat{\bf z}) \right] 
 \psi^{(V)}(k, \tau)
~, \\
\int d \Omega' ( \Delta Q^{(V)} \pm i \Delta U^{(V)})(\hat{\bf n'}, \hat{\bf n})
& = 
- \int \frac{d^3 {\bf k}}{(2\pi)^3}  
\left[\sum_{\lambda = \pm 1} \lambda \sqrt{\frac{16 \pi}{5}}
{}_{\pm 2}Y_{2 \lambda}(\hat{\bf n}) \xi^{(\lambda)}({\bf k} \parallel \hat{\bf z}) \right] 
 \psi^{(V)}(k, \tau), \\
\psi^{(V)}(k, \tau) 
&\equiv \frac{3}{10} \tilde{\Delta}_{I,1}^{(V)} + \frac{3}{10} \tilde{\Delta}_{I,3}^{(V)} - \frac{3}{5} \tilde{\Delta}_{P,0}^{(V)} - \frac{3}{7} \tilde{\Delta}_{P,2}^{(V)} 
+ \frac{6}{35} \tilde{\Delta}_{P,4}^{(V)}  ~, \label{eq:vec_Thomson}
\end{split}
\end{eqnarray}
and for the tensor mode: 
\begin{eqnarray}
\begin{split}
\int d \Omega' \Delta\Theta^{(T)}(\hat{\bf n'}, \hat{\bf n})
& = 
\int \frac{d^3 {\bf k}}{(2\pi)^3}  
\left[\sum_{\lambda = \pm 2}  \sqrt{\frac{32\pi}{15}}
Y_{2 \lambda}(\hat{\bf n}) \xi^{(\lambda)}({\bf k} \parallel \hat{\bf z}) \right] 
 \psi^{(T)}(k, \tau)
~, \\
\int d \Omega' (\Delta Q^{(T)} \pm i \Delta U^{(T)})(\hat{\bf n'}, \hat{\bf n})
& = 
- \int \frac{d^3 {\bf k}}{(2\pi)^3}  
\left[\sum_{\lambda = \pm 2}  \sqrt{\frac{64\pi}{5}}
{}_{\pm 2}Y_{2 \lambda}(\hat{\bf n}) \xi^{(\lambda)}({\bf k} \parallel \hat{\bf z}) \right] 
 \psi^{(T)}(k, \tau)
~, \\
\psi^{(T)}(k, \tau) 
&\equiv \frac{1}{10} \tilde{\Delta}_{I,0}^{(T)} +  \frac{1}{7} \tilde{\Delta}_{I,2}^{(T)} + \frac{3}{70} \tilde{\Delta}_{I,4}^{(T)} 
- \frac{3}{5} \tilde{\Delta}_{P,0}^{(T)} + \frac{6}{7} \tilde{\Delta}_{P,2}^{(T)} 
- \frac{3}{70} \tilde{\Delta}_{P,4}^{(T)}  ~. \label{eq:tens_Thomson}
\end{split}
\end{eqnarray}
Thus, from the Boltzmann equation (\ref{eq:boltzmann_photon_general_IQUV}),
the gravitational redshift and the Doppler terms (\ref{eq:ISW_Doppler}), and the collision term of Thomson scattering (\ref{eq:line-of-sight_integral_general}) and (\ref{eq:scal_Thomson}) - (\ref{eq:tens_Thomson}), we derive the Boltzmann equation for the scalar mode:  
\begin{eqnarray}
\begin{split}
\dot{\tilde{\Delta}}_{I}^{(S)} + i k\mu_{k,n}
 \tilde{\Delta}_{I}^{(S)} &= 
- \left[ \frac{1}{3} \dot{h}_{\rm iso}^{(S)} 
 + \left(\mu_{k,n}^2 - \frac{1}{3}\right)
\dot{h}^{(S)} 
\right] \\
&\quad  
- \dot{\kappa} 
\left[ \tilde{\Delta}_{I}^{(S)} - \tilde{\Delta}_{I,0}^{(S)}
+ \frac{3}{4}
\left( \mu_{k,n}^2 - \frac{1}{3} \right)  \psi^{(S)} 
- i\mu_{k,n}v_b^{(S)} \right] ~, \\
\dot{\tilde{\Delta}}_{P}^{(S)} + i k\mu_{k,n}
 \tilde{\Delta}_{P}^{(S)} &= - \dot{\kappa} 
\left[\tilde{\Delta}_{P}^{(S)} - \frac{3}{4} (1 - \mu_{k,n}^2) 
\psi^{(S)}  \right] ~, \label{eq:scal_boltzmann}
\end{split}
\end{eqnarray}
for the vector mode:
\begin{eqnarray}
\begin{split}
\dot{\tilde{\Delta}}_{I}^{(V)} + i k\mu_{k,n}
 \tilde{\Delta}_{I}^{(V)} &= - i \mu_{k,n} \dot{h}^{(V)} - \dot{\kappa} 
\left[ \tilde{\Delta}_{I}^{(V)} + i \mu_{k,n} \psi^{(V)} - v_b^{(V)} \right] ~, \\
\dot{\tilde{\Delta}}_{P}^{(V)} + i k\mu_{k,n}
 \tilde{\Delta}_{P}^{(V)} &= - \dot{\kappa} 
\left[ \tilde{\Delta}_{P}^{(V)} + \psi^{(V)} \right] ~, \label{eq:vec_boltzmann}
\end{split}
\end{eqnarray}
and for the tensor mode:
\begin{eqnarray}
\begin{split}
\dot{\tilde{\Delta}}_{I}^{(T)} + i k\mu_{k,n}
 \tilde{\Delta}_{I}^{(T)} &= -\dot{h}^{(T)} - \dot{\kappa} 
\left[ \tilde{\Delta}_{I}^{(T)} - \psi^{(T)} \right] ~, \\
\dot{\tilde{\Delta}}_{P}^{(T)} + i k\mu_{k,n}
 \tilde{\Delta}_{P}^{(T)} &= - \dot{\kappa} 
\left[ \tilde{\Delta}_{P}^{(T)} + \psi^{(T)} \right] ~. \label{eq:tens_boltzmann}
\end{split}
\end{eqnarray}
The multipole expansion of these equations gives
\begin{eqnarray}
\dot{\tilde{\Delta}}_{I/P, l}^{(S/V/T)} + \frac{k}{2 l + 1} 
\left[(l + 1) \tilde{\Delta}_{I/P, l}^{(S/V/T)} - l \tilde{\Delta}_{I/P, l-1}^{(S/V/T)} \right] 
&= V_{I/P,l}^{(S/V/T)}~,
\end{eqnarray}
where 
\begin{eqnarray}
\begin{split}   
V_{I,l}^{(S)}
&= 
- \left[ \frac{1}{3} \dot{h}_{\rm iso}^{(S)} \delta_{l, 0} 
 - \frac{2}{15} \dot{h}^{(S)} \delta_{l,2} 
\right] 
- \dot{\kappa} 
\left[ \tilde{\Delta}_{I,l}^{(S)} (1 - \delta_{l,0}) 
- \frac{1}{10} \psi^{(S)} \delta_{l,2} 
+ \frac{1}{3} v_b^{(S)} \delta_{l,1} \right] ~, \\
V_{P,l}^{(S)} &= - \dot{\kappa} 
\left[\tilde{\Delta}_{P,l}^{(S)} 
- \frac{1}{10} \psi^{(S)} 
\left( \delta_{l,2} + 5 \delta_{l,0} \right) \right] ~, \\
V_{I,l}^{(V)} &= \frac{1}{3} \dot{h}^{(V)} \delta_{l,1} - \dot{\kappa} 
\left[ \tilde{\Delta}_{I,l}^{(V)} - \frac{1}{3} \psi^{(V)} \delta_{l,1} - v_b^{(V)}\delta_{l,0} \right] ~, \\
V_{P,l}^{(V)} &= - \dot{\kappa} 
\left[ \tilde{\Delta}_{P,l}^{(V)} + \psi^{(V)} \delta_{l,0} \right] ~, \\ 
V_{I,l}^{(T)} &= - \dot{h}^{(T)} \delta_{l,0} - \dot{\kappa} 
\left[ \tilde{\Delta}_{I,l}^{(T)} - \psi^{(T)} \delta_{l,0} \right] ~, \\
V_{P,l}^{(T)} &= - \dot{\kappa} 
\left[ \tilde{\Delta}_{P,l}^{(T)} + \psi^{(T)} \delta_{l,0} \right]
~. 
\end{split}
\end{eqnarray}  

Following the line of sight integration \cite{Seljak:1996is}, we can
give the explicit solution of this Boltzmann equation as
follows. At first, let us present the derivation for tensor mode. 
Using $de^{-\kappa}/d\tau = \dot{\kappa}e^{-\kappa}$ and
multiplying $e^{i k \mu_{k,n} \tau - \kappa}$ in both sides of
Eq.~(\ref{eq:tens_boltzmann}), we have
\begin{eqnarray}
\begin{split}
\frac{d}{d\tau}
\left( \tilde{\Delta}_{I}^{(T)} e^{ik\mu_{k,n} \tau - \kappa} \right)
&= e^{ik\mu_{k,n}}[-\dot{h}^{(T)} e^{-\kappa} + g \psi^{(T)}] ~, \\
\frac{d}{d\tau}
\left( \tilde{\Delta}_{P}^{(T)} e^{ik\mu_{k,n} \tau - \kappa} \right)
&= - e^{ik\mu_{k,n}} g \psi^{(T)} ~.
\end{split}
\end{eqnarray}
where $g(\tau) \equiv \dot{\kappa} e^{-\kappa}$ is the visibility
function which describes the probability that a given CMB photon last
scattered at a given time. Through the integral over conformal time and
some treatments, we obtain each from of the transfer function at $\tau = \tau_0$ as
\begin{eqnarray}
\begin{split}
\tilde{\Delta}_{I}^{(T)} &= \int_0^{\tau_0} d\tau e^{- i \mu_{k,n} x}
\left[ -\dot{h}^{(T)} e^{-\kappa} + g \psi^{(T)} \right]
~, \\
\tilde{\Delta}_{P}^{(T)} &= - \int_0^{\tau_0} d\tau e^{- i \mu_{k,n} x} g \psi^{(T)} ~,
\end{split}
\end{eqnarray} 
Here we use $\kappa(\tau_0) = 0, \kappa(\tau =
0) \rightarrow \infty$ and $x \equiv k(\tau_0 - \tau)$. In the same manner, we also obtain the scalar-mode function:
\begin{eqnarray}
\begin{split}
\tilde{\Delta}_{I}^{(S)} &= \int_0^{\tau_0} d\tau e^{- i \mu_{k,n} x}
 \left[ - \left\{ \frac{1}{3} \dot{h}_{\rm iso}^{(S)} 
 + \left(\mu_{k,n}^2 - \frac{1}{3}\right)
\dot{h}^{(S)} 
\right\} e^{- \kappa}  
\right. \\
&\left. \qquad\qquad\qquad\qquad
+ g 
\left\{ \tilde{\Delta}_{I,0}^{(S)}
- \frac{3}{4}
\left( \mu_{k,n}^2 - \frac{1}{3} \right) \psi^{(S)}
+ i\mu_{k,n}v_b^{(S)} \right\} \right] 
 ~, \\
&= \int_0^{\tau_0} d\tau e^{- i \mu_{k,n} x}
 \left[ 
\left( - \frac{1}{3} \dot{h}_{\rm iso}^{(S)} 
+ \frac{\dddot{h}^{(S)}}{k^2}
+ \frac{1}{3} \dot{h}^{(S)} 
\right) e^{- \kappa} 
\right. \\
&\left. \qquad\qquad\qquad\qquad
+ g 
\left( \frac{2 \ddot{h}^{(S)}}{k^2} 
+ \tilde{\Delta}_{I,0}^{(S)}
+ \frac{3}{4} \frac{\ddot{\psi}^{(S)}}{k^2} 
+ \frac{1}{4} \psi^{(S)}
- \frac{\dot{v}_b^{(S)}}{k} \right) 
\right. \\
&\left. \qquad\qquad\qquad\qquad
+ \dot{g}
\left( \frac{\dot{h}^{(S)} }{k^2} 
+ \frac{3}{2} \frac{\dot{\psi}^{(S)}}{k^2}
- \frac{v_b^{(S)}}{k}
\right)
+ \frac{3}{4} \frac{\ddot{g}\psi^{(S)}}{k^2}
\right] 
 ~, \\
\tilde{\Delta}_{P}^{(S)} 
&= \int_0^{\tau_0} d\tau e^{- i \mu_{k,n} x} 
(1 - \mu_{k,n}^2) 
\frac{3}{4} g \psi^{(S)}  ~,
\end{split}
\end{eqnarray}
and the vector-mode one:
\begin{eqnarray}
\begin{split}
\tilde{\Delta}_{I}^{(V)} &= \int_0^{\tau_0} d\tau e^{- i \mu_{k,n} x}
\left[ - i \mu_{k,n} 
\left( \dot{h}^{(V)} e^{-\kappa} + g \psi^{(V)} \right) + g v_b^{(V)} \right]
 \\
&= \int_0^{\tau_0} d\tau e^{- i \mu_{k,n} x}
\left[  
\frac{\ddot{h}^{(V)}}{k} e^{-\kappa} 
+ g 
\left(\frac{\dot{h} + \dot{\psi}^{(V)}}{k} + v_b^{(V)} \right) + \dot{g} 
\frac{\psi^{(V)}}{k} \right]
~, \\
\tilde{\Delta}_{P}^{(V)} &= - \int_0^{\tau_0} d\tau e^{- i \mu_{k,n} x}
 g \psi^{(V)}  ~.
\end{split}
\end{eqnarray}
where in the second equality of each equation, we neglect the topological terms.

Consequently, we can summarize the transfer functions when ${\bf k} \parallel \hat{\bf z}$ for the scalar mode:
\begin{eqnarray}
\begin{split}
\Delta_I^{(S)}(\tau_0, {\bf k} \parallel \hat{\bf z}, \hat{\bf n})
& = \xi^{(0)}({\bf k} \parallel \hat{\bf z})
\int_0^{\tau_0} d\tau e^{- i \mu_{k,n} x}
 S_I^{(S)} (k,\tau)
~, \\
(\Delta_Q^{(S)} \pm i \Delta_U^{(S)}) (\tau_0, {\bf k} \parallel \hat{\bf z}, \hat{\bf n})
& = \frac{4}{3} \sqrt{\frac{6 \pi}{5}} {}_{\pm 2} Y_{2 0}(\hat{\bf n})
\xi^{(0)}({\bf k} \parallel \hat{\bf z}) 
\int_0^{\tau_0} d\tau e^{- i \mu_{k,n} x}
S_P^{(S)} (k,\tau) ~, \\
S_I^{(S)}(k,\tau) 
&\equiv \left( - \frac{1}{3} \dot{h}_{\rm iso}^{(S)} 
+ \frac{\dddot{h}^{(S)}}{k^2}
+ \frac{1}{3} \dot{h}^{(S)} 
\right) e^{- \kappa} 
\\
&\quad + g 
\left( \frac{2 \ddot{h}^{(S)}}{k^2} 
+ \tilde{\Delta}_{I,0}^{(S)}
+ \frac{3}{4} \frac{\ddot{\psi}^{(S)}}{k^2} 
+ \frac{1}{4} \psi^{(S)}
- \frac{\dot{v}_b^{(S)}}{k} \right) 
\\
&\quad 
+ \dot{g}
\left( \frac{\dot{h}^{(S)} }{k^2} 
+ \frac{3}{2} \frac{\dot{\psi}^{(S)}}{k^2}
- \frac{v_b^{(S)}}{k}
\right)
+ \frac{3}{4} \frac{\ddot{g}\psi^{(S)}}{k^2} 
~, \\
S_P^{(S)}(k,\tau) 
&\equiv \frac{3}{4} g \psi^{(S)}(k, \tau) ~,
\label{eq:transfer_scal_general}
\end{split}
\end{eqnarray}
for the vector mode: 
\begin{eqnarray}
\begin{split}
\Delta_I^{(V)}(\tau_0, {\bf k} \parallel \hat{\bf z}, \hat{\bf n})
& = 
\left[\sum_{\lambda = \pm 1} i \lambda \sqrt{\frac{8\pi}{3}}
Y_{1 \lambda}(\hat{\bf n}) \xi^{(\lambda)}({\bf k} \parallel \hat{\bf
 z}) \right] 
\int_0^{\tau_0} d\tau e^{- i \mu_{k,n} x} S_I^{(V)} (k,\tau)
~, \\
(\Delta_Q^{(V)} \pm i \Delta_U^{(V)}) (\tau_0, {\bf k} \parallel \hat{\bf z}, \hat{\bf n})
& = 
\left[\sum_{\lambda = \pm 1} \lambda \sqrt{\frac{16 \pi}{5}}
{}_{\pm 2}Y_{2 \lambda}(\hat{\bf n}) \xi^{(\lambda)}({\bf k} \parallel \hat{\bf z}) \right] 
\int_0^{\tau_0} d\tau e^{- i \mu_{k,n} x}
S_P^{(V)} (k,\tau) ~, \\
S_I^{(V)}(k,\tau)
&\equiv
\frac{\ddot{h}^{(V)}}{k} e^{-\kappa} 
+ g 
\left(\frac{\dot{h} + \dot{\psi}^{(V)}}{k} + v_b^{(V)} \right) + \dot{g} 
\frac{\psi^{(V)}}{k} ~, \\
S_P^{(V)}(k,\tau) 
&\equiv - g \psi^{(V)} ~, \label{eq:transfer_vec_general}
\end{split}
\end{eqnarray}
and for the tensor mode:
\begin{eqnarray}
\begin{split}
\Delta_I^{(T)}(\tau_0, {\bf k} \parallel \hat{\bf z}, \hat{\bf n})
&= \left[\sum_{\lambda = \pm 2}  \sqrt{\frac{32\pi}{15}}
Y_{2 \lambda}(\hat{\bf n}) \xi^{(\lambda)}({\bf k} \parallel \hat{\bf z}) \right]  
 \int_0^{\tau_0} d\tau e^{- i \mu_{k,n} x}S_I^{(T)} (k, \tau) ~,  \\
(\Delta_Q^{(T)} \pm i \Delta_U^{(T)})(\tau_0, {\bf k} \parallel \hat{\bf z}, \hat{\bf n})
&= \left[\sum_{\lambda = \pm 2}  \sqrt{\frac{64\pi}{5}}
{}_{\pm 2}Y_{2 \lambda}(\hat{\bf n}) \xi^{(\lambda)}({\bf k} \parallel \hat{\bf z}) \right] 
\int_0^{\tau_0} d\tau e^{- i \mu_{k,n} x}S_P^{(T)} (k, \tau) ~, \\
S_I^{(T)} (k,\tau) &\equiv -\dot{h}^{(T)} e^{-\kappa} + g \psi^{(T)} ~, \\
S_P^{(T)} (k,\tau) &\equiv - g \psi^{(T)} ~. \label{eq:transfer_tens_general}
\end{split}
\end{eqnarray}
Here, we have introduced the source function, $S_{I/P}^{(S/V/T)}(k,\tau)$. 


\begin{figure}[t]
  \centering \includegraphics[height=12cm,clip]{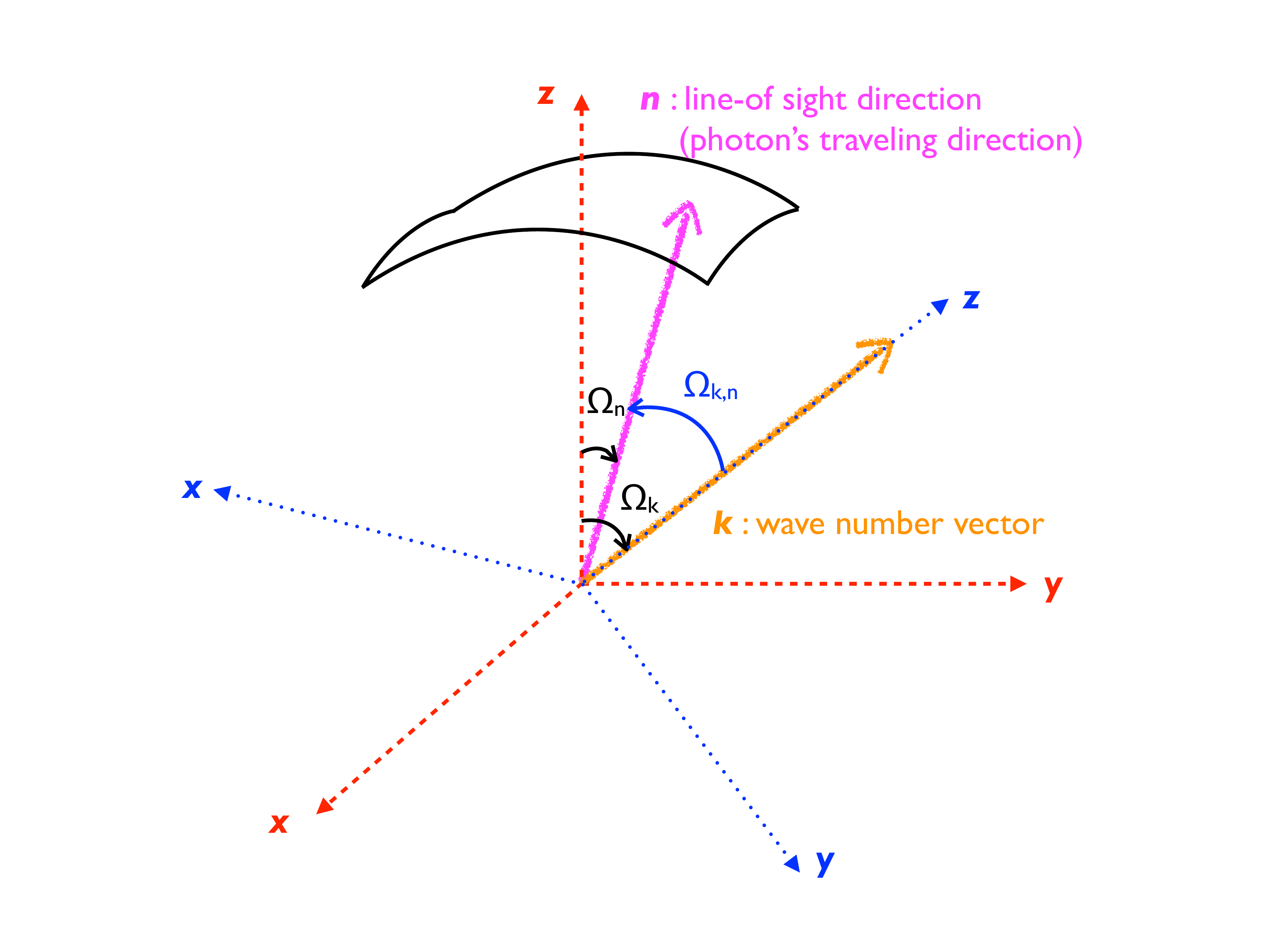}
  \caption{Geometry for the line-of-sight direction. }
  \label{fig:line-of-sight}
\end{figure}

\subsection{All-sky formulae for the CMB scalar-, vector- and tensor-mode anisotropies}

In this subsection, let us formulate the all-mode CMB coefficients $a_{\ell m}$ in the all-sky analysis on the basis of the derivation in
 Refs.~\cite{Shiraishi:2010sm, Weinberg:2008zzc}. 

Since the CMB anisotropy is described in the spherical coordinate system, its intensity ($I$) and two polarization ($Q$ and $U$) fields should be expanded by the spin-0 and spin-2 spherical harmonics, respectively, as
\begin{eqnarray}
\begin{split}
\Theta^{(Z)}(\hat{\bf n}) &= \sum_{\ell, m} a_{I, \ell m}^{(Z)} Y_{\ell m}(\hat{\bf n}) ~, \\
(Q^{(Z)} \pm i U^{(Z)})(\hat{\bf n}) &= \sum_{\ell, m} a_{\pm 2, \ell m}^{(Z)} ~ {}_{\pm 2}Y_{\ell m}(\hat{\bf n}) ~. \label{eq:field_expand_define}
\end{split}
\end{eqnarray}
Here, the index $Z$ denotes the mode of perturbations: $Z = S$
(scalar), $= V$ (vector) or $=T$ (tensor). The main difficulty when computing the spectrum of polarization arises from the variance under rotations in the plane perpendicular to $\hat{\bf n}$. While $Q$ and $U$ are easily calculated in a coordinate system where ${\bf k} \parallel \hat{\bf z}$, the superposition of the different modes is complicated by the behavior of $Q$ and $U$ under rotations. However, using the spin raising and lowering operators $\up, \down$ defined in Appendix \ref{appen:spherical_harmonics}, we can obtain spin-0 quantities. This leads to the rotational invariant fields like the intensity one and there are no ambiguities connected with the rotation of coordinate system arise. Acting these operators on $Q \pm i U$ in Eq.~(\ref{eq:field_expand_define}), we have 
\begin{eqnarray}
\begin{split}
\down^2 (Q^{(Z)} + iU^{(Z)})(\hat{\bf n}) &= \sum_{\ell m} 
\left[ \frac{(\ell + 2)!}{(\ell - 2)!} \right]^{1/2} a_{2, \ell m}^{(Z)} Y_{\ell m}(\hat{\bf n}) ~, \\
\up^2 (Q^{(Z)} - iU^{(Z)})(\hat{\bf n}) &= \sum_{\ell m} 
\left[ \frac{(\ell + 2)!}{(\ell - 2)!} \right]^{1/2} a_{-2, \ell m}^{(Z)} Y_{\ell m}(\hat{\bf n}) ~, \label{eq:Q_pm_iU_updown}
\end{split}
\end{eqnarray}
 Instead of $a_{\pm 2, \ell m}^{(Z)}$, it is convenient to introduce their linear combinations as \cite{Newman:1966ub}
\begin{eqnarray}
\begin{split}
a_{E, \ell m}^{(Z)} &\equiv -\frac{1}{2} 
\left( a_{2, \ell m}^{(Z)} + a_{-2, \ell m}^{(Z)} \right) ~, \\
a_{B, \ell m}^{(Z)} &\equiv \frac{i}{2} 
\left( a_{2, \ell m}^{(Z)} - a_{-2, \ell m}^{(Z)} \right) ~. \label{eq:EB_generate}
\end{split} 
\end{eqnarray}
These fields $E$ and $B$ have parity-even and odd properties, respectively, in analogy with the electric and magnetic fields. Then, from Eqs.~(\ref{eq:Q_pm_iU_updown}) and (\ref{eq:EB_generate}), we can express  
\begin{eqnarray}
\begin{split}
a_{X, \ell m}^{(Z)} &= \int d\Omega_n Y_{\ell m}^*(\Omega_n) 
\int \frac{d^3{\bf k}}{(2 \pi)^3}
 \Delta_{X}^{(Z)}(\tau_0, {\bf k}, \hat{\bf n}) ~, \\ 
\Delta_E^{(Z)} (\tau_0, {\bf k}, \hat{\bf n}) &\equiv - \frac{1}{2} 
\left[ \frac{(\ell-2)!}{(\ell+2)!}\right]^{1/2} 
\left[ \down^2 (\Delta_{Q}^{(Z)} + i \Delta_U^{(Z)})
+ \up^2 (\Delta_{Q}^{(Z)} - i \Delta_U^{(Z)}) \right] (\tau_0, {\bf k}, \hat{\bf n}) ~, \\
\Delta_B^{(Z)} (\tau_0, {\bf k}, \hat{\bf n}) &\equiv \frac{i}{2} \left[ \frac{(\ell-2)!}{(\ell+2)!}\right]^{1/2} 
 \left[ \down^2 (\Delta_{Q}^{(Z)} + i \Delta_U^{(Z)})
- \up^2 (\Delta_{Q}^{(Z)} - i \Delta_U^{(Z)}) \right] (\tau_0, {\bf k}, \hat{\bf n}) ~. 
\label{eq:alm_fourier}
\end{split}
\end{eqnarray}
Here, $X$ discriminates between intensity and two polarization (electric and magnetic) modes, respectively, as $X = I,E,B$.
When ${\bf k} \parallel \hat{\bf z}$, using Eqs.~(\ref{eq:transfer_scal_general}) - (\ref{eq:transfer_tens_general}) and the operations derived from Eq.~(\ref{eq:operators1}) as 
\begin{eqnarray} 
\begin{split}
\down^2 \left[ {}_{2}Y_{2 \lambda}(\hat{\bf n}) e^{-i \mu_{k,n} x} \right] &= 
\left(-\partial_{\mu_{k,n}} + {\lambda \over 1-\mu_{k,n}^2}\right)^2 
\left[(1-\mu_{k,n}^2) {}_{2}Y_{2 \lambda}(\Omega_{k,n}) e^{-i \mu_{k,n} x} \right] ~, \\ 
\up^2 \left[ {}_{- 2}Y_{2 \lambda}(\hat{\bf n}) e^{-i \mu_{k,n} x} \right] &=
\left(-\partial_{\mu_{k,n}} - {\lambda \over 1-\mu_{k,n}^2}\right)^2 
\left[(1-\mu_{k,n}^2) {}_{- 2}Y_{2 \lambda}(\Omega_{k,n}) e^{-i \mu_{k,n} x} 
\right]~, 
\end{split}
\end{eqnarray}
we obtain more explicit expressions as
\begin{eqnarray}
\begin{split}
\Delta_E^{(S)} (\tau_0, {\bf k} \parallel \hat{\bf z}, \hat{\bf n}) 
&= 
\left[ \frac{(\ell-2)!}{(\ell+2)!}\right]^{1/2} 
\xi^{(0)} ({\bf k} \parallel \hat{\bf z})
\int_0^{\tau_0} d\tau
 S_P^{(S)}(k,\tau) \hat{\mathcal{E}}^{(S)}(x) e^{-i \mu_{k,n} x}
 ~, \\
\Delta_E^{(V)} (\tau_0, {\bf k} \parallel \hat{\bf z}, \hat{\bf n}) 
&= \left[ \frac{(\ell-2)!}{(\ell+2)!}\right]^{1/2} 
\left[ \sum_{\lambda = \pm 1} 
- \lambda i \sqrt{\frac{8\pi}{3}} Y_{1 \lambda} (\hat{\bf n})
\xi^{(\lambda)}({\bf k} \parallel \hat{\bf z} ) \right] \\
&\quad \times
 \int_0^{\tau_0} d\tau S_P^{(V)}(k,\tau) \hat{\mathcal{E}}^{(V)}(x) e^{ - i \mu_{k,n} x} ~, \\
\Delta_B^{(V)} (\tau_0, {\bf k} \parallel \hat{\bf z}, \hat{\bf n}) 
&= \left[ \frac{(\ell-2)!}{(\ell+2)!}\right]^{1/2}  
\left[ \sum_{\lambda = \pm 1} 
- i \sqrt{\frac{8\pi}{3}} Y_{1 \lambda} (\hat{\bf n})
\xi^{(\lambda)}({\bf k} \parallel \hat{\bf z} ) \right]
\\
&\quad\times
 \int_0^{\tau_0} d \tau S_P^{(V)}(k,\tau) \hat{\mathcal{B}}^{(V)}(x) e^{- i \mu_{k,n} x}  ~, \\
\Delta_E^{(T)} (\tau_0, {\bf k} \parallel \hat{\bf z}, \hat{\bf n}) 
&= \left[ \frac{(\ell-2)!}{(\ell+2)!}\right]^{1/2} 
 \left[ \sum_{\lambda = \pm 2} \sqrt{\frac{32\pi}{15}} Y_{2 \lambda}(\hat{\bf n}) \xi^{(\lambda)}({\bf k} \parallel \hat{\bf z}) \right] \\ 
&\quad\times\int_0^{\tau_0} d\tau
 S_P^{(T)}(k, \tau) \hat{\mathcal{E}}^{(T)}(x) e^{- i \mu_{k,n} x} ~, \\
\Delta_B^{(T)} (\tau_0, {\bf k} \parallel \hat{\bf z}, \hat{\bf n}) 
&= \left[ \frac{(\ell-2)!}{(\ell+2)!}\right]^{1/2} 
 \left[ \sum_{\lambda = \pm 2} - \frac{\lambda}{2} \sqrt{\frac{32\pi}{15}} Y_{2 \lambda}(\hat{\bf n}) 
\xi^{(\lambda)}({\bf k} \parallel \hat{\bf z}) \right]  \\
&\quad\times \int_0^{\tau_0} d\tau
 S_P^{(T)}(k, \tau) \hat{\mathcal{B}}^{(T)}(x) e^{- i \mu_{k,n} x}
\end{split}
\end{eqnarray} 
with the operators $\hat{\mathcal{E}}, \hat{\mathcal{B}}$ defined as 
\begin{eqnarray}
\begin{split}
\hat{\mathcal{E}}^{(S)}(x) &\equiv (1 + \partial_x^2)^2 x^2 ~, \\
\hat{\mathcal{E}}^{(V)}(x) &\equiv 4x + (12+x^2)
 \partial_x + 8x \partial_x^2 + x^2 \partial_x^3  ~, \\
\hat{\mathcal{B}}^{(V)}(x) &\equiv x^2 + 4x\partial_x + x^2 
 \partial_x^2~, \\
\hat{\mathcal{E}}^{(T)}(x) &\equiv -12 + x^2 (1-\partial_x^2) - 8x
 \partial_x  ~, \\
\hat{\mathcal{B}}^{(T)}(x) &\equiv 8x + 2x^2 \partial_x  ~. \\
 \label{eq:x_operator}
\end{split}
\end{eqnarray} 

From here, we want to show analytical expressions of $a_{\ell m}$'s.
 In the above discussion, we have analytical formulae of the transfer functions when ${\bf k} \parallel \hat{\bf z}$. This implies that we consider the physics in the blue basis of Fig.~\ref{fig:line-of-sight} and their transfer functions are completely determined by not the angle between $\hat{\bf z}$ and ${\hat{\bf n}}$, namely $\Omega_n$, but the angle between ${\bf k}$ and $\hat{\bf n}$, namely $\Omega_{k,n}$. However, as shown in Eq.~(\ref{eq:alm_fourier}), to obtain $a_{\ell m}$'s, we have to consider the physics in the red basis Fig.~\ref{fig:line-of-sight} and perform the $\Omega_n$-integral. Here, instead of the transformation of the transfer functions, it is a better way to transform the integration variable in the $a_{\ell m}$ as $\Omega_n \rightarrow \Omega_{k,n}$. This can be done by using the Wigner $D$-matrix $D_{m m'}^{(\ell)}$, which is the unitary
irreducible matrix of rank $2\ell+1$ that forms a representation of the
rotational group. The property of this matrix and the relation with
spin-weighted spherical harmonics are explained in Appendix \ref{appen:wigner_d_matrix}. If we consider the rotational matrix
\begin{eqnarray}
S(\Omega_k) 
\equiv \left( 
  \begin{array}{ccc}
  \cos\theta_k \cos\phi_k & -\sin\phi_k  & \sin\theta_k \cos\phi_k \\
 \cos\theta_k \sin\phi_k  &  \cos\phi_k & \sin\theta_k \sin\phi_k \\
 -\sin\theta_k & 0 & \cos\theta_k
  \end{array}
 \right)
\end{eqnarray}
 corresponding to the configuration ($\alpha = \phi_k, \beta = \theta_k, \gamma = 0$) of Eq.~(\ref{eq:rotational_matrix_zyz}) and satisfying 
\begin{eqnarray}
\Omega_n = S(\Omega_k) \Omega_{k,n}~, \label{eq:transform_rotate}
\end{eqnarray}
 the transformation equation (\ref{eq:D_transform}) can be equated with 
\begin{eqnarray}
Y_{\ell m}^*(\Omega_n)  = \sum_{m'} D_{m m'}^{(\ell)}
\left( S(\Omega_k) \right) Y_{\ell m'}^*(\Omega_{k,n})~. \label{eq:wig_D_cmb}
\end{eqnarray}
Using this equation and the relation of the coordinate transformation as $d \Omega_{n} = d \Omega_{k,n}$, the $a_{\ell m}$ of arbitrary mode is written as
\begin{eqnarray}
a_{X, \ell m}^{(Z)} 
&=& \int \frac{d^3{\bf k}}{(2 \pi)^3} \left[ \sum_{m'} D^{(\ell)}_{m m'}
\left( S(\Omega_k) \right) \int d \Omega_{k,n} Y^*_{\ell m'}(\Omega_{k,n})
\Delta^{(Z)}_{X}(\tau_0, {\bf k} \parallel \hat{\bf z}, \hat{\bf n}) \right] \nonumber \\
&=&  \left[ \frac{4 \pi}{2\ell + 1} \right]^{1/2} 
\int \frac{d^3{\bf k}}{(2 \pi)^3} \left[ \sum_{m'} (-1)^{m'}
{}_{-m'}Y_{\ell m}^*(\Omega_{k}) \int d \Omega_{k,n} Y^*_{\ell m'}(\Omega_{k,n})
\Delta^{(Z)}_{X}(\tau_0, {\bf k} \parallel \hat{\bf z}, \hat{\bf n}) \right] ~. \nonumber \\
\end{eqnarray}
In the second equality, we have obeyed the relation of Eq.~(\ref{eq:D_Y}) in this case as
\begin{eqnarray}
D_{m m'}^{(\ell)} \left( S(\Omega_{k}) \right) = 
\left[ \frac{4 \pi}{2\ell + 1} \right]^{1/2} (-1)^{m'}
{}_{-m'}Y_{\ell m}^*(\Omega_{k})~. 
\end{eqnarray}
Using the mathematical results of the $\Omega_{k,n}$-integrals
\begin{eqnarray}
\begin{split}
e^{- i \mu_{k,n} x} &= \sum_{L} 4\pi  (-i)^L j_L(x) \sqrt{\frac{2L+1}{4\pi}} Y_{L 0}(\Omega_{k,n}) ~, \\
\int d\Omega_{k,n} Y_{\ell m'}^* e^{- i \mu_{k,n} x}
&= (-i)^{\ell} \delta_{m', 0} \sqrt{4\pi (2\ell + 1)} j_\ell(x) ~, \\
\int d\Omega_{k,n} Y_{\ell m'}^* Y_{1 \pm 1} e^{- i \mu_{k,n} x}
&= (-i)^{\ell - 1} \delta_{m', \pm 1} \sqrt{\frac{3}{2} (2\ell + 1) \frac{(\ell+1)!}{(\ell-1)!}} \frac{j_\ell(x)}{x} ~, \\
\int d\Omega_{k,n} Y_{\ell m'}^* Y_{2 \pm 2} e^{- i \mu_{k,n} x}
&= (-i)^{\ell - 2} \delta_{m', \pm 2} \sqrt{\frac{15}{8} (2\ell + 1) \frac{(\ell+2)!}{(\ell-2)!} } \frac{j_\ell(x)}{x^2} ~, \\
\end{split}
\end{eqnarray}
we can find the general formulae of the $a_{\ell m}$ for all-mode perturbations:
\begin{eqnarray}
a^{(Z)}_{X, \ell m} = 
4\pi (-i)^{\ell} \int \frac{d^3{\bf k}}{(2 \pi)^3} 
\sum_{\lambda} [{\rm sgn}(\lambda)]^{\lambda + x} {}_{-\lambda} Y^*_{\ell m}(\Omega_k) \xi^{(\lambda)}({\bf k})
\mathcal{T}^{(Z)}_{X, \ell}(k)
\label{eq:all_alm_general} ~ ,
\end{eqnarray}
where the helicity of the perturbations is expressed
by $\lambda$: $\lambda = 0$ for $(Z = S)$, $= \pm 1$ for $(Z = V)$ or $=
\pm 2$ for $(Z = T)$, the index $x$ discriminates between the two parity states: $x = 0$ for $X=I,E$, $x = 1$ for $X=B$, and the time-integrated transfer functions $\mathcal{T}^{(Z)}_{X, \ell}(k)$ are expressed as
\footnote{In Ref.~\cite{Shiraishi:2010sm}, there are three typos: right-hand sides of Eqs.~(B21), (B22) and (B23) must be multiplied by a factor $-1$, respectively.}
\begin{eqnarray}
\begin{split}
 \mathcal{T}_{I,\ell}^{(S)}(k) &=  \int_0^{\tau_0} d\tau S_I^{(S)}(k,\tau)j_\ell(x)
  ~, \\
 \mathcal{T}_{E,\ell}^{(S)}(k) &= \left[ \frac{(\ell-2)!}{(\ell+2)!} \right] ^{1/2} \int_0^{\tau_0} d\tau S_P^{(S)} \hat{\mathcal{E}}^{(S)}(x) j_\ell(x) ~, \\ 
\mathcal{T}_{I,\ell}^{(V)}(k) &= \left[ \frac{(\ell+1)!}{(\ell-1)!} \right]^{1/2}
\int_0^{\tau_0} d\tau S_I^{(V)}(k, \tau) \frac{j_\ell(x)}{x}
~, \\
 \mathcal{T}_{E,\ell}^{(V)}(k) &=   
- \left[ \frac{(\ell+1)!}{(\ell-1)!} \frac{(\ell-2)!}{(\ell+2)!} \right]^{1/2}
\int_0^{\tau_0} d\tau S_P^{(V)}(k, \tau) \hat{\mathcal{E}}^{(V)}(x)
\frac{j_\ell(x)}{x} ~, \\
 \mathcal{T}_{B,\ell}^{(V)}(k) &= 
- \left[ \frac{(\ell+1)!}{(\ell-1)!} \frac{(\ell-2)!}{(\ell+2)!} \right]^{1/2}
\int_0^{\tau_0} d\tau S_P^{(V)}(k, \tau) \hat{\mathcal{B}}^{(V)}(x)
\frac{j_\ell(x)}{x} ~, \\
\mathcal{T}_{I,\ell}^{(T)}(k) &= - \left[ \frac{(\ell+2)!}{(\ell-2)!} \right]^{1/2} \int_0^{\tau_0} d\tau S_I^{(T)}(k, \tau) \frac{j_\ell(x)}{x^2} ~, \\
 \mathcal{T}_{E,\ell}^{(T)}(k) &= - \int_0^{\tau_0} d\tau S_P^{(T)}(k, \tau) \hat{\mathcal{E}}^{(T)}(x) \frac{j_\ell(x)}{x^2} ~, \\
 \mathcal{T}_{B,\ell}^{(T)}(k) &= \int_0^{\tau_0} d\tau S_P^{(T)}(k, \tau) \hat{\mathcal{B}}^{(T)}(x) \frac{j_\ell(x)}{x^2} \label{eq:transfer_explicit}~.
\end{split}
\end{eqnarray}  
Note that in the all-sky analysis, due to the dependence of
transfer functions on $\phi_{k,n}$, $a_{\ell m}$'s depend on the helicity state through the spin spherical harmonics. In the above discussion, we take the synchronous gauge and derive the $a_{\ell m}$. However, in the same manner, we can obtain the identical form of Eq.~(\ref{eq:all_alm_general}) even in another gauge. In this case, the different points can be confined only in the transfer function (\ref{eq:transfer_explicit}). In a numerical code CAMB \cite{Lewis:1999bs,Lewis:2004ef}, these transfer functions of polarization modes are expanded as, e.g., 
\begin{eqnarray}
\begin{split}
{\cal T}_{E, \ell}^{(T)}(k) 
&= - \int_0^{\tau_0} d\tau 
\left[ \frac{\ddot{g}\psi^{(T)} + 2 \dot{g} \dot{\psi}^{(T)} + g \ddot{\psi}^{(T)}}{k^2} 
+ \frac{4}{k}\frac{\dot{g} \psi^{(T)} + g \dot{\psi}^{(T)}}{x}
- g \psi^{(T)} \left( 1 - \frac{6}{x^2} \right)
\right] j_{\ell}(x) ~,\\
{\cal T}_{B, \ell}^{(T)}(k) 
&= - 2 \int_0^{\tau_0} d\tau 
\left[ g \left( \frac{2 \psi^{(T)}}{x} + \frac{\dot{\psi}^{(T)}}{k} \right) 
+ \frac{\dot{g}\psi^{(T)}}{k} 
\right] j_{\ell}(x)~. 
\end{split}
\end{eqnarray}

\subsection{Flat-sky formulae for the CMB scalar-, vector- and tensor-mode anisotropies}

The flat-sky approximation uses the (2D) plane wave expansion of the CMB fluctuation instead of the spherical harmonics one, and it is valid if we restrict observed direction (parallel to $\hat{\bf n}$) only close to the $z$ axis \cite{Zaldarriaga:1996xe,Seljak:1996ti}. As confirmed in Ref.~\cite{Zaldarriaga:1996xe}, the flat-sky
power spectra of $E$- and $B$-mode polarizations sourced from the
primordial scalar and tensor perturbations are in good agreement with the all-sky ones for $\ell \gtrsim 40$. 
On the basis of these studies, we have also compared the all-sky power spectra
with the flat-sky ones for the $I,E$, and $B$ modes from scalar, vector, and tensor perturbations and found their consistencies at $\ell \gtrsim 40$.  

As mentioned in the previous subsection, in order to estimate the $a_{\ell m}$, one must construct the transfer functions for arbitrary ${\bf k}$.  
 In other words, we want to obtain the transfer functions expressed
 by arbitrary ${\bf k}$ (whose direction is denoted by $\Omega_k$) and
 $\hat{\bf n}$ (denoted by $\Omega_n$) instead of $\Omega_{k,n}$ in Fig.~\ref{fig:line-of-sight}. 
In the $I$ modes, only by changing $\Omega_{k,n}$ to $\Omega_k$ and $\Omega_n$ with the relation (\ref{eq:transform_rotate}), the transfer
 functions for arbitrary $\bf k$ can be obtained.
In the $E$ and $B$ modes, in addition to this treatment, one must consider the
 mixing between $\Delta_Q$ and $\Delta_U$ under the transformation
 $S(\Omega_k)$ as described in Ref.~\cite{Zaldarriaga:1996xe}. 
This effect is expressed as
\begin{eqnarray} 
({\Delta_Q^{(Z)}}' \pm i{\Delta_U^{(Z)}}')(\tau_0, {\bf k}, \hat{\bf n}) 
= e^{\mp 2i \psi} (\Delta_Q^{(Z)} \pm i\Delta_U^{(Z)})(\tau_0, {\bf k} \parallel \hat{\bf z},
\hat{\bf n})~.
\end{eqnarray}
with the mixing angle $\psi$. The angle $\psi$ represents the rotation
angle between $\hat{\bf \theta}_{k,n}$ and $\hat{\bf \theta}_{n}$, where
$\hat{\bf \theta}_{k,n}$ and $\hat{\bf \theta}_{n}$ are the unit vectors orthogonal to $\hat{\bf n}$ in 
a particular basis in which ${\bf k}\parallel \hat{\bf z}$ and a general basis, respectively (see Fig.~\ref{fig:line-of-sight}). 

In the flat-sky analysis, i.e., $\theta_n \rightarrow 0$, by using
 Eqs. (\ref{eq:transfer_scal_general}) - (\ref{eq:transfer_tens_general}) and by using the limit of $\psi$ as $\psi \rightarrow \phi_n -\phi_k + \pi$, the
 transfer functions for arbitrary ${\bf k}$ are derived as
\begin{eqnarray}
\begin{split}
\Delta_{I}^{(S)} (\tau_0, {\bf k}, \hat{\bf n}) 
&\rightarrow \xi ({\bf k})  \int_0^{\tau_0} d\tau  S_I^{(S)}(k,\tau)
e^{- i {\bf k} \cdot \hat{\bf n} D} ~, \\ 
(\Delta_Q^{(S)} \pm i\Delta_U^{(S)})(\tau_0, {\bf k}, \hat{\bf n})
&\rightarrow e^{\mp 2i (\phi_n -\phi_k)} \sin^2 \theta_k  \xi({\bf k}) 
\int_0^{\tau_0} d\tau S_P^{(S)}(k,\tau) e^{- i {\bf k} \cdot \hat{\bf n} D} ~,
 \\
\Delta_{I}^{(V)} (\tau_0, {\bf k}, \hat{\bf n}) 
&\rightarrow i \sin \theta_k
\left( \sum_{\lambda = \pm 1} \xi^{(\lambda)}({\bf k}) \right ) \int_0^{\tau_0} d\tau
 S_I^{(V)}(k,\tau) e^{-i {\bf k} \cdot \hat{\bf n} D}~, \\
(\Delta_Q^{(V)} \pm i\Delta_U^{(V)})(\tau_0, {\bf k}, \hat{\bf n})
&\rightarrow e^{\mp 2i (\phi_n - \phi_k)} 
\sin\theta_k  \\
&\quad\times
\left[- \cos\theta_k 
\left( \sum_{\lambda = \pm 1} \xi^{(\lambda)}({\bf k}) \right )
\pm \left( \sum_{\lambda = \pm 1} \lambda \xi^{(\lambda)}({\bf k}) \right ) \right]  \\
&\quad \times \int_0^{\tau_0} d\tau S_P^{(V)}(k,\tau) e^{-i {\bf k} \cdot
 \hat{\bf n} D} ~, \\
\Delta_{I}^{(T)} (\tau_0, {\bf k}, \hat{\bf n}) 
&\rightarrow \sin^2 \theta_k
\left( \sum_{\lambda = \pm 2} \xi^{(\lambda)}({\bf k}) \right )  \int_0^{\tau_0} d\tau
S_I^{(T)}(k,\tau) e^{-i {\bf k} \cdot \hat{\bf n} D}~, \\
(\Delta_Q^{(T)} \pm i\Delta_U^{(T)})(\tau_0, {\bf k}, \hat{\bf n})
&\rightarrow e^{\mp 2i (\phi_n - \phi_k)} 
\left[ (1 + \cos^2 \theta_k) 
\left( \sum_{\lambda = \pm 2} \xi^{(\lambda)}({\bf k}) \right ) \right. \\
&\qquad\qquad\qquad\quad \left. \mp 2 \cos \theta_k
 \left( \sum_{\lambda = \pm 2} \frac{\lambda}{2} \xi^{(\lambda)}({\bf k}) \right )
 \right] 
\int_0^{\tau_0} d\tau S_P^{(T)}(k,\tau) e^{- i {\bf k} \cdot
 \hat{\bf n} D} ~.  \label{eq:transfer_flat}
\end{split}
\end{eqnarray} 
It is important to note that the $\phi_k$ dependence which are inherent
in the vector and tensor perturbations vanishes in the flat-sky
approximation, besides a trivial $\phi_k$ dependence due to a spin-2
nature of the Stokes $Q$ and $U$ parameters. One may explicitly see that
$\phi_{k,n}$ dependence vanishes in the transfer functions when taking
$\theta_n \rightarrow 0$ because the $S(\Omega_k)$ matrix rotates the basis with the new $z$ axis always being on the $x-z$ plane in a particular basis in which ${\bf k} \parallel \hat{z}$ (see Fig.~\ref{fig:line-of-sight}). 
This approximation means that 
for $\theta_n \ll 1$, it is valid to calculate the CMB fluctuation on
the basis of vector and tensor perturbations fixed as $\theta_n = 0$,
namely, $\phi_{k,n} = \pi$. 

In the flat-sky limit, $a_{\ell m}$ in the all-sky analysis
described as Eq.~(\ref{eq:alm_fourier}) is modified by using the plane wave as 
\begin{eqnarray}
\begin{split}
a_{X, \ell m}^{(Z)} &\rightarrow 
\int d^2{\bf \Theta} e^{-i {\boldsymbol \ell} \cdot {\bf \Theta}} \int \frac{d^3{\bf k}}{(2 \pi)^3} \Delta_{X}^{(Z)}(\tau_0, {\bf k}, \hat{\bf n})  \equiv a_X^{(Z)}({\boldsymbol \ell})~,  \\
\Delta_{E}^{(Z)}(\tau_0, {\bf k}, \hat{\bf n})
 &\rightarrow  
 \frac{1}{2} 
\left[ \sum_{s = \pm 2} 
\left(\Delta_{Q}^{(Z)} + \frac{s}{2} i \Delta_U^{(Z)} \right)(\tau_0, {\bf k}, \hat{\bf n})
 e^{- s i(\phi_\ell - \phi_n)} \right]  ~, \\
 \Delta_{B}^{(Z)}(\tau_0, {\bf k}, \hat{\bf n}) 
&\rightarrow - \frac{i}{2} 
\left[ \sum_{s = \pm 2} 
\frac{s}{2} \left( \Delta_{Q}^{(Z)} + \frac{s}{2} i \Delta_U^{(Z)} \right)(\tau_0, {\bf k}, \hat{\bf n})
 e^{- s i(\phi_\ell - \phi_n)} \right] ~, \label{eq:form_flat_al}
\end{split}
\end{eqnarray} 
where $\bf \Theta$ is the 2D vector projecting $\hat{\bf n}$ to the flat-sky
plane expressed as ${\bf \Theta} = (\Theta {\rm cos}\phi_n, \Theta
{\rm sin}\phi_n)$
\footnote{Not confuse $\bf \Theta$ with the CMB temperature fluctuation.}.

For example, in order to obtain $a_{I, \ell m}^{(T)}$, we substitute Eq. (\ref{eq:transfer_flat}) into
Eq. (\ref{eq:form_flat_al}) and calculate as follows:   
\begin{eqnarray}
a_I^{(T)}({\boldsymbol \ell}) &=& \int \frac{d^3{\bf k}}{(2 \pi)^3} 
\sin^2 \theta_k
\left( \sum_{\lambda = \pm 2}\xi^{(\lambda)}({\bf k}) \right) 
\int_0^{\tau_0} d\tau \int d^2{\bf \Theta} e^{- i({\bf k}^{\parallel} D + {\boldsymbol \ell}) \cdot {\bf \Theta}} S_I^{(T)}(k,\tau) e^{-i k_z D} \nonumber \\
&=& \int \frac{d^3{\bf k}}{(2\pi)^3} \sin^2 \theta_k 
\left( \sum_{\lambda = \pm 2}\xi^{(\lambda)}({\bf k}) \right) 
\int_0^{\tau_0} d\tau (2\pi)^2 \delta({\bf k}^{\parallel}D + {\boldsymbol \ell}) S_I^{(T)}(k,\tau) e^{-i k_z D} \nonumber \\
&=& \int_0^{\tau_0} d\tau \int_{-\infty}^{\infty} \frac{dk_z}{2\pi}
 \left( \sum_{\lambda = \pm 2}\xi^{(\lambda)}({\bf k}^{\parallel} = -{\boldsymbol \ell}/D, k_z) \right) \nonumber \\
&&\times \frac{\ell^2}{(k_z D)^2 + \ell^2} S_I^{(T)}(k = \sqrt{k_z^2 + (\ell/D)^2},\tau)
 \frac{1}{D^2}e^{-i k_z D} \label{eq:flat_alt}~,
\end{eqnarray} 
where $D=\tau_0-\tau$ is the conformal distance and we have decomposed
$\bf{k}$ into two-dimensional vector parallel to the flat sky and that
orthogonal to it, ${\bf k}=({\bf k}^{\parallel},k_z)$.
In order to obtain the last equation, we use following relations which are
satisfied under ${\bf k}^{\parallel} = -{\boldsymbol \ell}/D$ as
\begin{eqnarray}
\begin{split}
k &= \sqrt{k_z^2 + 
\left(\frac{\ell}{D} \right)^2}  ~, \\
\sin \theta_k &= \frac{\ell}{kD} = \frac{\ell}{\sqrt{(k_zD)^2 + \ell^2}}  ~, \\
\cos \theta_k &= {\rm sgn}(k_z)
\sqrt{1 - 
\left( \frac{\ell}{kD} \right)^2 }  ~, \\
\phi_k &= \phi_\ell + \pi ~. \label{eq:delta2_condi}
\end{split}
\end{eqnarray}
The other-mode $a_{\ell m}$'s can be calculated in the same manner. Thus, we summarize the all-mode $a({\boldsymbol \ell})$'s: 
\begin{eqnarray}
\begin{split}
a_X^{(Z)}({\boldsymbol \ell}) &= 
 \int_0^{\tau_0} d\tau \int_{-\infty}^{\infty}
 \frac{dk_z}{2 \pi}  
\sum_{\lambda} [{\rm sgn}(\lambda)]^x \xi^{(\lambda)}({\bf k}^{\parallel} = -{\boldsymbol \ell}/D, k_z)
 \frac{1}{D^2}e^{- i k_z D} {\cal S}_{X, \ell}^{(Z)}(k_z, \tau) 
 \\
{\cal S}_{I, \ell}^{(S)}(k_z, \tau) &= S_I^{(S)}(k = \sqrt{k_z^2 + (\ell/D)^2},\tau)  ~, \\
{\cal S}_{E, \ell}^{(S)}(k_z, \tau) &= 
\frac{\ell^2}{(k_z D)^2 + \ell^2}
S_P^{(S)}(k = \sqrt{k_z^2 + (\ell/D)^2},\tau)~, \\
{\cal S}_{I, \ell}^{(V)}(k_z, \tau) &= i \frac{\ell}{\sqrt{(k_z D)^2 + \ell^2}} S_I^{(V)}(k = \sqrt{k_z^2 +
(\ell/D)^2},\tau) ~, \\ 
{\cal S}_{E, \ell}^{(V)}(k_z, \tau) &= - {\rm sgn}(k_z) \frac{\ell}{\sqrt{(k_z D)^2 + \ell^2}} \sqrt{1 -
 \frac{\ell^2}{(k_z D)^2 + \ell^2}} S_P^{(V)}(k = \sqrt{k_z^2 + (\ell/D)^2},\tau)
  ~, \\ 
{\cal S}_{B, \ell}^{(V)}(k_z, \tau) &= -i \frac{\ell}{\sqrt{(k_z D)^2 + \ell^2}} S_P^{(V)}(k = \sqrt{k_z^2 + (\ell/D)^2},\tau) ~, \\
{\cal S}_{I, \ell}^{(T)}(k_z, \tau) &= 
\frac{\ell^2}{(k_z D)^2 + \ell^2} S_I^{(T)}(k = \sqrt{k_z^2 + (\ell/D)^2},\tau)
 ~,  \\
{\cal S}_{E, \ell}^{(T)}(k_z, \tau) &= 
\left( 2 - \frac{\ell^2}{(k_z D)^2 + \ell^2} \right) S_P^{(T)}(k =
 \sqrt{k_z^2 + (\ell/D)^2},\tau)  ~,  \\
{\cal S}_{B, \ell}^{(T)}(k_z, \tau) &= 2 i ~{\rm sgn}(k_z) \sqrt{ 1 - \frac{\ell^2}{(k_z D)^2 + \ell^2} }
 S_P^{(T)}(k = \sqrt{k_z^2 + (\ell/D)^2},\tau)  ~,
\end{split}
\end{eqnarray} 
where we label ${\cal S}_{X, \ell}^{(Z)}$ as the modified source function. 
One can formulate the flat-sky CMB power spectrum and bispectrum by using these formulae \cite{Babich:2004gb, Pitrou:2008ak, Boubekeur:2009uk, Shiraishi:2010sm}.   
\subsection{CMB power spectrum}

To extract several information about the Universe from the observational
data, the two-point correlation function of the CMB fluctuations (called CMB power spectrum) is often-used. 
Here, we formulate the CMB power spectrum and summarize the
constraints on several model parameters from the current observational
data. 

If we assume the Gaussianity and the symmetry under the parity and
rotational transformations in the initial stochastic variables, their power spectrum can be expressed as
\begin{eqnarray}
\Braket{\xi^{(\lambda_1)}({\bf k_1}) \xi^{(\lambda_2) *}({\bf k_2})} 
=  (2\pi)^3 P_Z(k_1) \delta({\bf k_1} - {\bf k_2}) \delta_{\lambda_1,
    \lambda_2} 
\times 
\begin{cases}
 1 & (Z = S) \\
1 / 2 & (Z = V, T)
 \end{cases}
\end{eqnarray} 
This implies that the couplings between the different modes of the
perturbation vanish in the power spectrum. 
Then, from the formula of the all-sky $a_{\ell m}$ (\ref{eq:all_alm_general}),
the CMB power spectra of all modes are
derived as
\begin{eqnarray}
\begin{split}
\Braket{\prod_{n=1}^2 a_{X_n, \ell_n m_n}^{(Z_n)}} 
&\equiv C_{X_1 X_2, \ell_1}^{(Z_1)} 
(-1)^{m_1} \delta_{\ell_1, \ell_2}
\delta_{m_1, -m_2} \delta_{Z_1, Z_2} \delta_{x_1, x_2} ~, \\
C_{X_1 X_2, \ell_1}^{(Z_1)} &= \frac{2}{\pi} 
\int k_1^2 dk_1 P_{Z_1}(k_1) {\cal T}_{X_1, \ell_1}^{(Z_1)}(k_1) 
{\cal T}_{X_2, \ell_1}^{(Z_1)}(k_1) \label{eq:cmb_power_general}
\end{split} 
\end{eqnarray} 
where we use the relation derived by the reality condition of
the metric perturbation:
\begin{eqnarray}
\xi^{(\lambda)}({\bf k}) = (-1)^\lambda \xi^{(\lambda)
*}(- {\bf k}) ~.
\end{eqnarray}

\begin{figure}[t]
  \begin{tabular}{cc}
    \begin{minipage}{0.5\hsize}
  \begin{center}
    \includegraphics[width=8cm,clip]{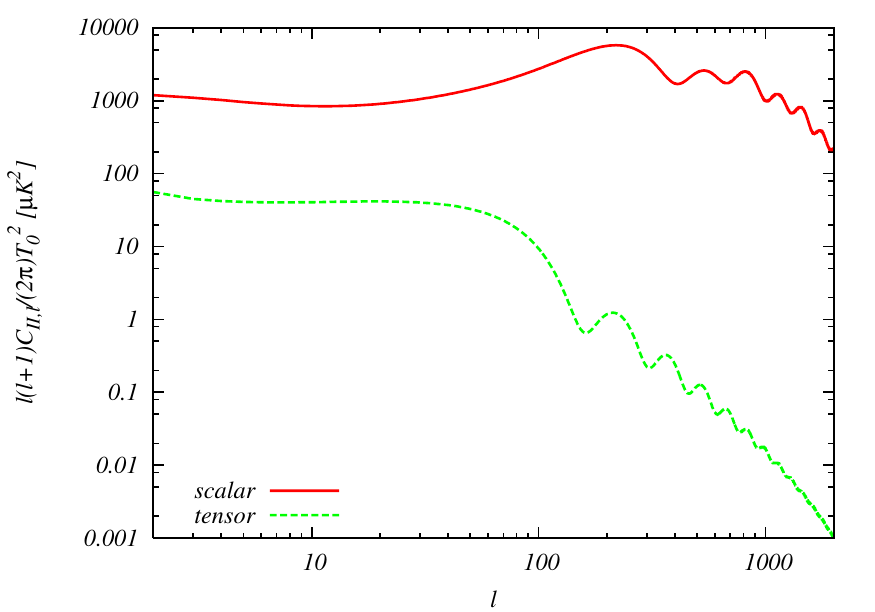}
  \end{center}
\end{minipage}
\begin{minipage}{0.5\hsize}
  \begin{center}
    \includegraphics[width=8cm,clip]{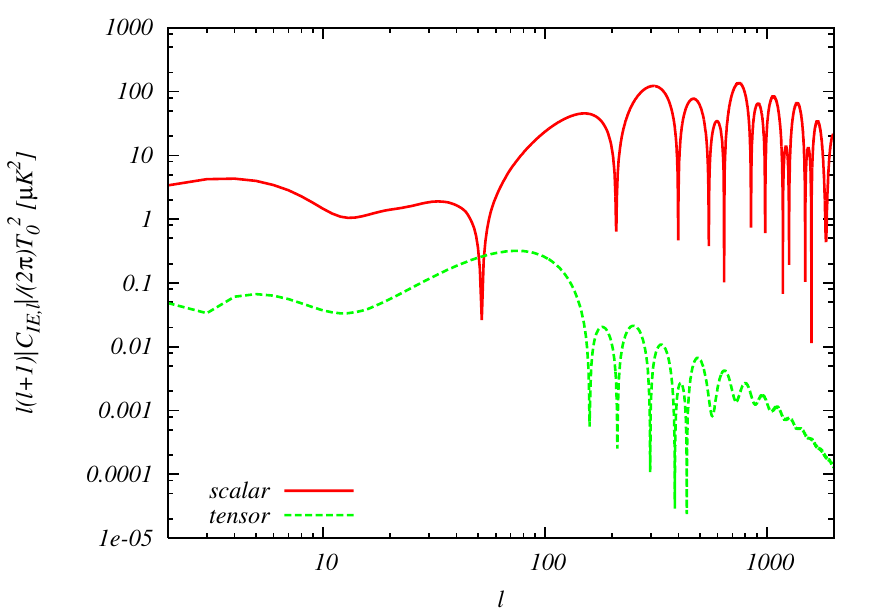}
  \end{center}
\end{minipage}
\\
    \begin{minipage}{0.5\hsize}
  \begin{center}
    \includegraphics[width=8cm,clip]{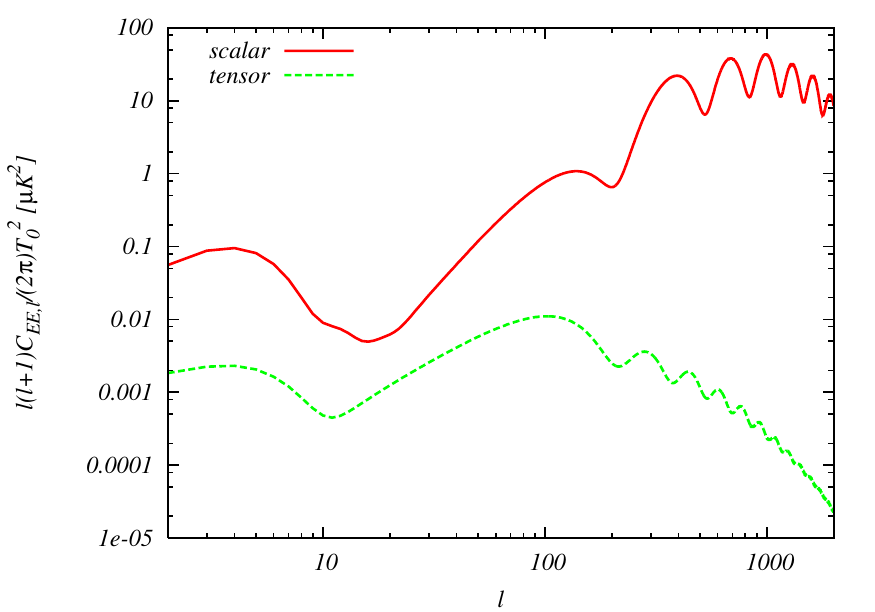}
  \end{center}
\end{minipage}
\begin{minipage}{0.5\hsize}
  \begin{center}
    \includegraphics[width=8cm,clip]{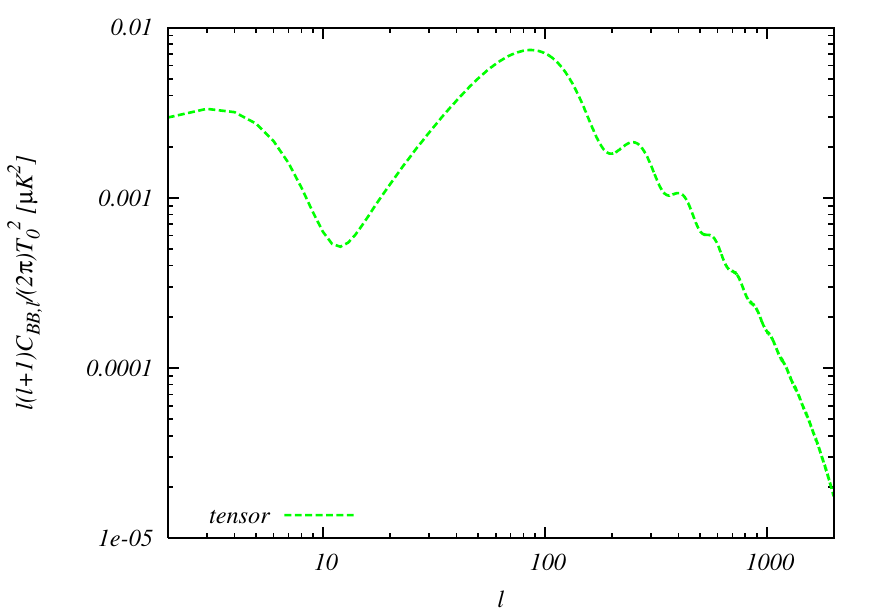}
  \end{center}
\end{minipage}
\end{tabular}
  \caption{The CMB spectra of the $II$ (left top panel), $IE$ (right top
 one), $EE$ (left bottom one) and $BB$ (right bottom one) modes. Here we
  consider a power-law flat $\Lambda$CDM model and fix
  the tensor-to-scalar ratio as $r = 0.1$. The other
  cosmological parameters are fixed to the mean values reported in Ref.~\cite{Komatsu:2010fb}.} \label{fig:WMAP_theory_XX.pdf}
\end{figure}

In Fig.~\ref{fig:WMAP_theory_XX.pdf}, we plot the CMB intensity and
polarization power spectra of
the scalar and tensor modes. Here we think that the scalar and tensor
perturbations are sourced from the primordial curvature perturbations
($\xi^{(0)} = {\cal R}$) 
and primordial gravitational waves ($\xi^{(\pm 2)} = h^{(\pm 2)}$),
respectively. The ratio between these power spectra, called the tensor-to-scalar ratio
$r$, is defined as
\footnote{This definition is consistent with
Eq.~(\ref{eq:inf_def_tensor-to-scalar}), and the notation of
Ref.~\cite{Komatsu:2010fb} and CAMB \cite{Lewis:1999bs}}
\begin{eqnarray}
r \equiv \frac{2 P_T(k)}{P_S(k)}~.  \label{eq:tensor-to-scalar_def} 
\end{eqnarray}
At first, let us focus on the $II$ spectra in the left top panel. In
the scalar spectrum, the dominant signal is generated from the acoustic
oscillation of the photon and baryon fluid. The first acoustic peak is located at $\ell_1 \sim 220$. This scale is corresponding to the angle of the
sound horizon at the recombination epoch as $\ell_1 \sim 2 \pi \tau_0 /
r_s(z_*)$. At small scales corresponding to $\ell \gg \ell_1$, due to the
difference between the photon and baryon speeds, the coupling between photons
and baryons is ineffective and the acoustic oscillation is highly
damped. This effect is well-known as the Silk damping \cite{Silk:1967kq}. On
the other hand, the gravitational blue shift due
to the potential decay in the late time affects the fluctuations for $\ell \ll \ell_1$. This is called the
integrate Sachs Wolfe (ISW) effect \cite{Hu:1993xh} and arises from the terms
proportional to $e^{-\kappa}$ of the source function $S_I^{(Z)}$ in
Eqs.~(\ref{eq:transfer_scal_general}) and (\ref{eq:transfer_tens_general}). In the tensor spectrum, the ISW effect
leads to the dominant signals and the scattering effect (the second term
of the source function $S_I^{(T)}$ in Eq.~(\ref{eq:transfer_tens_general})) is
subdominant \cite{Pritchard:2004qp}. As shown in
Eqs.~(\ref{eq:transfer_scal_general}) and
(\ref{eq:transfer_tens_general}), the $EE$ and $BB$ spectra (the bottom
two panels) have no
gravitational redshift term and there is no ISW effect. Instead of it,
the scattering term generates the CMB fluctuation. The most interesting
signature in the polarization power spectra is the enhancement at $\ell
\lesssim 10$. This can be caused by Thomson scattering at reionization of hydrogen which may have occurred at $z \sim 10$ \cite{Zaldarriaga:1996xe}. 
Therefore, these signals can be important to determine the optical depth of the Universe, $\kappa$. The $IE$ spectrum described in the right top
panel seems to include the features of both the $II$ and $EE$ spectra and has both positive and negative values. Unlike the above three cases, the $BB$ spectrum never arises from the scalar
mode because the scalar mode has only one helicity, namely $\lambda =
0$
\footnote{The vector mode generates the $BB$ spectrum due to its two helicities. However, due to the decay of the vector potential via the Einstein equation, this becomes the subdominant signal. To avoid this, the sources such as cosmic strings and magnetic fields need to exist and support.}. 
Hence, we believe that the $BB$ spectrum directly tells us the amplitude of
the primordial gravitational waves depending on the energy scale of inflation.    

\begin{figure}[t]
  \centering \includegraphics[height=8cm,clip]{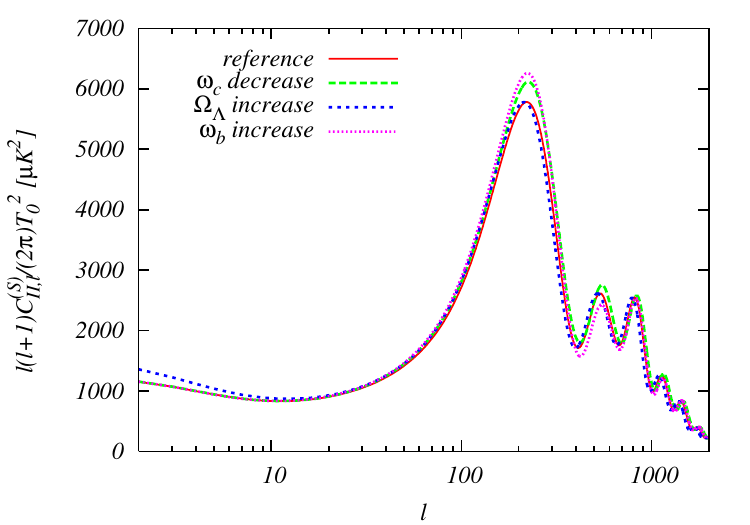}
  \caption{CMB $II$ power spectra sourced from the scalar-mode
 perturbations in a power-law flat $\Lambda$CDM model. The red solid
 line is plotted with $\omega_c = 0.112, \Omega_\Lambda = 0.728, \omega_b
 = 0.02249, n_s = 0.967, \kappa = 0.088$ \cite{Komatsu:2010fb}. The green dashed, blue dotted and magenta dot-dashed lines are calculated if $\omega_c$ decreases to $0.1$, $\Omega_\Lambda$ increases to $0.8$, and $\omega_b$ increases
 to $0.028$, respectively.}
  \label{fig:SS_II_paramdep_LCDM.pdf}
\end{figure}

Theoretically, the CMB power spectrum depends on the parameters which determine the dynamics of the Universe as the energy density of the cosmic fluids, curvature, and the Hubble constant $H_0$. Figure \ref{fig:SS_II_paramdep_LCDM.pdf} shows the
dependence of $C_{II, \ell}^{(S)}$ on the density parameters of cold dark matters, the cosmological constant and baryons as $\omega_c \equiv \Omega_c (H_0 / 100 {\rm sec \cdot Mpc/ km})^2, \Omega_\Lambda$ and $\omega_b \equiv \Omega_b (H_0 / 100 {\rm sec \cdot Mpc/ km})^2$, respectively. From this figure, one can observe that as $\omega_c$ decreases, the overall amplitude of $C_{II, \ell}^{(S)}$
enlarges. This behavior is understood as follows. If $\omega_c$
decreases, since the radiation dominated era is lengthened, the gravitational potential for smaller $k$ enters the
horizon and decays. Thus, $C_{II, \ell}^{(S)}$ at corresponding
multipoles as $\ell \sim k \tau$ is boosted due to the gravitational blue
shift. This is the so called early ISW effect
\cite{Hu:1993xh}. Next, focusing on the blue dotted line, one can find that if
$\Omega_\Lambda$ becomes large, $C_{II, \ell}^{(S)}$ is boosted for $\ell
\lesssim 10$. This is due to the late ISW effect, that is,
$\Lambda$ dominates the Universe earlier and the potential at larger
scales is destroyed, hence $C_{II, \ell}^{(S)}$ at corresponding $\ell$'s is
amplified. We also notice that when $\omega_b$ enlarges, the ratio of the amplitude between the first and second peaks of the magenta
dot-dashed line increases. 
Solving the coupled Boltzmann equations, the acoustic oscillation of the
baryon-photon fluid in the matter dominated era is roughly given by
\begin{eqnarray}
\Theta \sim \left(\frac{1}{3} + R\right)\Phi \cos(k r_s) - R \Phi~,
\end{eqnarray}
where $\Phi$ is the potential of the conformal Newtonian gauge, $r_s$
denotes the sound horizon and $R \equiv 3 \rho_b / (4 \rho_\gamma)$. 
Then if $\omega_b$ increases and $R$ becomes large, this equation
experiences increase in amplitude and suppression of the intercept. Hence, 
the difference of $C_{II, \ell}^{(S)} \propto \Theta^2$ between the odd-
and even-number peaks increases. These parameters
are limited with the others from the current observational data set as
Table \ref{tab:WMAP7_LCDM_bounds}. Other than these parameters, massive neutrinos and some relativistic components also make impacts on the CMB
fluctuation (e.g. Refs.~\cite{Lesgourgues:2006nd, Shiraishi:2009fu}). 

\begin{table}
\begin{center}
\begin{tabular}{ccc}
\hline 
Parameter & WMAP 7-yr & WMAP + BAO + $H_0$ 
 \\
\hline\hline
$100\omega_b$
&$2.249^{+0.056}_{-0.057}$ 
&$2.255\pm 0.054$ 
\\
$\omega_c$
&$0.1120\pm 0.0056$ 
&$0.1126\pm 0.0036$ 
\\
$\Omega_\Lambda$
&$0.727^{+0.030}_{-0.029}$ 
&$0.725\pm 0.016$ 
\\
$n_s$
&$0.967\pm 0.014$ 
&$0.968\pm 0.012$ 
\\
$\kappa$
&$0.088\pm 0.015$ 
&$0.088\pm 0.014$ 
\\
$A_S$
&$(2.43\pm 0.11)\times 10^{-9}$ 
&$(2.430\pm 0.091)\times 10^{-9}$ 
\\
$r$ & $< 0.36$ & $< 0.24$ \\ 
\hline
$H_0$
&$70.4\pm 2.5$~{km/s/Mpc} 
&$70.2\pm 1.4$~{km/s/Mpc} 
\\
$\Omega_b$
&$0.0455\pm 0.0028$ 
&$0.0458\pm 0.0016$ 
\\
$\Omega_c$
&$0.228\pm 0.027$ 
&$0.229\pm 0.015$ 
\\
$\omega_m$
&$0.1345^{+0.0056}_{-0.0055}$ 
&$0.1352\pm 0.0036$ 
\\
$z_{\rm reion}$
&$10.6\pm 1.2$ 
&$10.6\pm 1.2$ 
\\
$t_0$
&$13.77\pm 0.13$~Gyr 
&$13.76\pm 0.11$~Gyr 
\\
\hline
\end{tabular}
\caption{Summary of the cosmological parameters of $\Lambda$CDM with
 finite $r$ model from the WMAP 7-year data \cite{Komatsu:2010fb}, and the data set combined with the results of
 the galaxy survey \cite{Percival:2009xn} and Hubble constant
 measurement \cite{Riess:2009pu}, respectively. Here $z_{\rm reion}$
 denotes the redshift at the reionization epoch, $t_0$ is the present
 time of the Universe, and $\omega_m \equiv \omega_b + \omega_c$.} 
\label{tab:WMAP7_LCDM_bounds}
\end{center}
\end{table}

The CMB power spectrum also depends on the primordial curvature
perturbations and primordial gravitational waves. 
Conventionally, these spectra are parametrized as
\begin{eqnarray}
 \frac{k^3 P_S(k)}{2 \pi^2} =  A_S 
\left( \frac{k}{0.002 {\rm Mpc}^{-1}} \right)^{n_s - 1} ~,
\label{eq:wmap_def_power_curvature_perturbation}
\end{eqnarray}
and Eq.~(\ref{eq:tensor-to-scalar_def}). As shown in
Eq.~(\ref{eq:all_alm_general}), the magnitudes of the primordial curvature
perturbations $A_S$ and gravitational waves $r A_S$ simply determine the
overall amplitude of the $C_{X_1 X_2, \ell_1}^{(S)}$ and $C_{X_1 X_2, \ell_1}^{(T)}$, respectively. The spectral index of the power spectrum of the curvature perturbations $n_s$ changes in slope of $C_{X_1 X_2, \ell_1}^{(Z_1)}$. From the observational data, the constraints on $A_S, n_s, r$ are given as Table \ref{tab:WMAP7_LCDM_bounds}. Here, we
want to note that $n_s = 1$ is excluded at about 3-sigma level. This
implies the deviation from the exact de Sitter expansion in inflation. As shown in the bound on $r$, unlike the primordial curvature
perturbation, the primordial gravitational wave has not been detected
yet. However, some experimental groups aim to discover the $BB$ spectrum
through remove of some noisy foreground emission and improvement of the
instruments \cite{:2006uk, Bischoff:2010ud, Keating:2011iq,
litebird}. If these projects achieve, it will be possible to judge the
existence of the primordial gravitational waves of $r < 0.01$. 

So far, we discussed under the assumption that the parity and rotational
invariances are kept. On the other hand, there are a lot of studies which
probe the somewhat exotic scenarios where these invariances violate. 
As a theoretical prediction, if parity-violating action such as the Chern-Simon term exists in the early Universe, 
$\Braket{\xi^{(+2)}({\bf k}) \xi^{(+2) *} ({\bf k'})} \ne \Braket{\xi^{(-2)}({\bf
k}) \xi^{(-2) *}({\bf k'}) }$ and the $IB$ and $EB$ spectra
appear \cite{Alexander:2004wk, Saito:2007kt, Gluscevic:2010vv,
Gruppuso:2011ci}. Using the parametrization as
\begin{eqnarray}
C_{IB, \ell}^{\rm obs} \equiv C_{IE, \ell} \sin(2 \Delta \alpha) ~,
\end{eqnarray}
the parity violation is limited as $-5.0^\circ < \Delta \alpha < 2.8^\circ$ (95\%
CL) \cite{Komatsu:2010fb}. 
The rotational invariance is broken if the Universe has the preferred direction and this situation is realized by the anisotropic inflation
\cite{Yokoyama:2008xw, Watanabe:2010fh, Karciauskas:2011fp}. This leads
to the direction-depending power spectrum as
\begin{eqnarray}
P_S({\bf k}) = P_S^{\rm iso}(k) [1 + g (\hat{\bf k} \cdot \hat{\bf n}
 )^2 ]
~,
\end{eqnarray} 
and produces the off-diagonal components in the CMB power spectrum as
$\ell_1 \ne \ell_2$. From the CMB observational data, the magnitude of the statistical
anisotropy has been limited as $g = 0.15 \pm 0.039$ \cite{Groeneboom:2008fz}

Furthermore, owing to the progress of the observational accuracy, the
deviation of the Gaussianity can be measured. Beyond the power spectrum,
this is achieved by using the three-point function (bispectrum). 
In the next section, we discuss about how to extract the information on
the early Universe from the CMB scalar, vector and tensor bispectrum.

In addition, we can add other components of fluids in the analysis of
the CMB spectrum. From Sec.~\ref{sec:PMF}, we focus on the effect
of the primordial magnetic fields on the CMB. 

\bibliographystyle{JHEP}
\bibliography{paper}

%% file: ch4_png.tex
\section{Primordial non-Gaussianities}\label{sec:PNG}

The study of non-Gaussian impacts in the cosmological fluctuations provides an important information of the early Universe~\cite{Komatsu:2009kd}. The primordial non-Gaussianities are measures of the interactions in inflation, hence constraining this will lead to a great deal about the inflationary dynamics. 
It may also puts strong constraints on alternatives to the
inflationary paradigm (e.g., Refs.~\cite{Creminelli:2007aq, Koyama:2007if,
Buchbinder:2007at, Lehners:2007wc, Lehners:2008my, Koyama:2007ag,
Koyama:2007mg}). 

In the previous part, 
we expanded the inflationary action to second order in the comoving
curvature perturbation ${\cal R}$ and the gravitational waves $h^{(\pm 2)}$. These actions allowed us to compute the power spectra $P_{\cal
R}(k)$ and $P_h(k)$. If the fluctuations ${\cal R}$ and $h^{(\pm 2)}$ obey the exact Gaussian statistics, the power spectrum (or two-point correlation function) contains all the information 
\footnote{Odd-point correlation functions of Gaussian fluctuations vanish while their even-point functions can be expanded by two-point functions due to the Wick's theorem.}.
However, for the non-Gaussian fluctuations, higher-order correlation functions beyond the two-point function contain additional information about inflation.
 Estimating the leading non-Gaussian effects requires the expansion of the action to third order since we must take into account the leading non-trivial interaction terms. In this section, we review recent studies about the primordial non-Gaussianity based on e.g.,~Refs.~\cite{Baumann:2009ds, Bartolo:2004if, Komatsu:2010hc}.

\subsection{Bispectrum of the primordial fluctuations}

At first, we give the definition of the bispectrum of the initial
perturbations $\xi^{(\lambda)}$ of the scalar ($\lambda = 0$), vector
($\lambda = \pm 1$), and tensor ($\lambda = \pm 2$) modes.
The Fourier transformation of the two-point function is the power
spectrum 
\begin{eqnarray}
\Braket{\prod_{n=1}^2 \xi^{(\lambda_n)}({\bf k_n})} 
=  (2\pi)^3 P_Z(k_1) 
\delta\left( \sum_{n=1}^3 {\bf k_n} \right) 
\delta_{\lambda_1, \lambda_2} (-1)^{\lambda_1} 
\times 
\begin{cases}
 1 & (Z = S) \\
1 / 2 & (Z = V, T)
 \end{cases} ~.
\end{eqnarray}
Similarly, the Fourier mode of the three-point function is so called the
bispectrum 
\begin{eqnarray}
\Braket{\prod_{n=1}^3 \xi^{(\lambda_n)}({\bf k_n})} 
= (2\pi)^3 F^{\lambda_1 \lambda_2 \lambda_3}({\bf k_1}, {\bf k_2}, {\bf k_3}) 
\delta\left( \sum_{n=1}^3 {\bf k_n} \right)~. \label{eq:def_ini_bis}
\end{eqnarray} 
Here, the translation invariance of the background results in the delta function denoting the momentum conservation. If the scale invariance is kept, we have 
\begin{eqnarray}
F^{\lambda_1 \lambda_2 \lambda_3}(b {\bf k_1}, b {\bf k_2}, b {\bf k_3}) 
= b^{-6} F^{\lambda_1 \lambda_2 \lambda_3}({\bf k_1}, {\bf k_2}, {\bf k_3})~. 
\end{eqnarray}
Moreover, due to the rotational invariance, the independent variables are reduced to $k_2 / k_1$ and $k_3 / k_1$. 

In order to compute the primordial bispectrum, it is necessary to deal with the vacuum evolution under the finite interactions carefully. This is not the leading order effect in calculating the power spectrum. The in-in formalism is a powerful method to compute the primordial higher-order cosmological correlation \cite{Maldacena:2002vr, Schwinger:1960qe, Calzetta:1986ey, Jordan:1986ug, Weinberg:2005vy}. In Sec.~\ref{sec:parity_violating}, using this formalism, we actually discuss the computation for the bispectrum of the gravitational waves. 

\subsection{Shape of the non-Gaussianities} \label{subsec:NG_shape}

The delta function in Eq.~(\ref{eq:def_ini_bis}) results in a closed triangle in Fourier space. The triangle configuration at which the primordial bispectrum is amplified is dependent on the inflationary models; therefore the shape of the non-Gaussianity is a powerful clue to inflation \cite{Babich:2004gb, Fergusson:2008ra}. 

\begin{figure}[t]
\centering \noindent
\includegraphics[width=16cm]{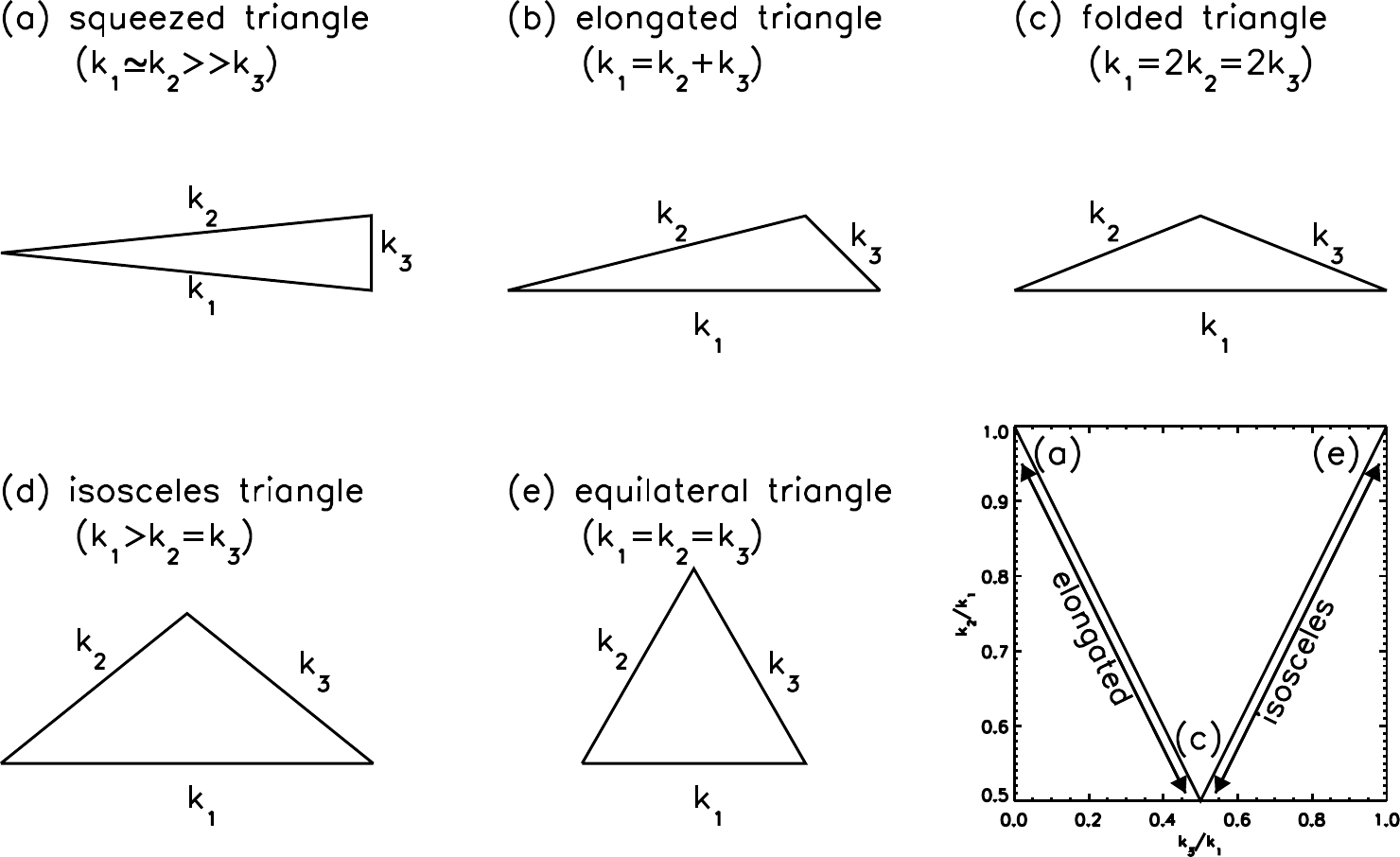}
\caption{%
 Representations of triangles forming the bispectrum. This figure is adopted from Ref.~\cite{Jeong:2008rj}. } 
\label{fig:triangle.pdf}
\end{figure}

We can study the bispectrum shape by plotting the magnitude of $ (k_1 k_2 k_3)^2 F^{\lambda_1 \lambda_2 \lambda_3}(k_1, k_2, k_3)$ as a function of $k_2 / k_1$ and $k_3 / k_1$ for $k_3 \leq k_2 \leq k_1$. For identification of each triangle, one often use the following names: squeezed ($k_1 \approx k_2 \gg k_3$), elongated ($k_1 = k_2 + k_3$), folded ($k_1 = 2 k_2 = 2 k_3$), isosceles ($k_2 = k_3$), and equilateral ($k_1 = k_2 = k_3$). In Fig.~\ref{fig:triangle.pdf}, the visual representations of these triangles are depicted.  

From here, we concentrate on three representative shapes of the primordial bispectrum: ``local'', ``equilateral'', and ``orthogonal''. 
Then, it may be convenient to decompose the non-Gaussianity of the curvature perturbations into the magnitude-depending and shape-depending parts:  
\begin{eqnarray}
F^{0 0 0}(k_1, k_2, k_3) =  \frac{3}{5} f_{\rm NL} (2 \pi^2 A_S)^2
 S(k_1, k_2, k_3)~, \label{eq:ini_bis_scalar}
\end{eqnarray}
where $A_S$ is the magnitude of curvature perturbations defined in
Eq.~(\ref{eq:wmap_def_power_curvature_perturbation}).

\subsubsection{Local type}

The simplest way to parametrize the non-Gaussianity of curvature perturbation is to expand by Gaussian fluctuation ${\cal R}_g$ \cite{Komatsu:2001rj} as
\begin{eqnarray}
{\cal R}({\bf x}) = {\cal R}_g({\bf x}) + \frac{3}{5} f_{\rm NL}^{\rm
 local}
\left[ {\cal R}_g({\bf x})^2 - \Braket{{\cal R}_g({\bf x})^2} \right] ~.
\end{eqnarray}
From this equation, we can see that the non-Gaussianity is localized at a given point in the real space. Therefore, we call this the local-type non-Gaussianity, and $f_{\rm NL}^{\rm local}$ is called the local-type nonlinearity parameter. Then, the bispectrum of the local-type non-Gaussianity of the curvature perturbations is derived as
\begin{eqnarray}
F^{0 0 0}_{\rm local}(k_1, k_2 ,k_3) = \frac{6}{5} f_{\rm NL}^{\rm local} 
\left[ P_{\cal R}(k_1) P_{\cal R}(k_2) + 2 \ {\rm perms.}\right] ~.
\end{eqnarray}
Fixing the spectral index as $n_s = 1$ and equating this equation with Eq.~(\ref{eq:ini_bis_scalar}), we can write 
\begin{eqnarray}
S^{\rm local}(k_1, k_2, k_3)  = 2 \left[ \frac{1}{(k_1 k_2)^3} + 2 \ {\rm perms.}\right] ~. \label{eq:S_loc}
\end{eqnarray}
This is boosted in the squeezed limit: $k_3 \ll k_1 \approx k_2$ as shown in the top-left panel of Fig.~\ref{fig:2D_loc_equil_orthog.pdf}. 
In this limit, the bispectrum reaches
\begin{eqnarray}
F^{000}_{\rm local}(k_1, k_2, k_3 \to 0) = \frac{12}{5} f_{\rm NL}^{\rm local} P_{\cal R}(k_1) P_{\cal R}(k_3) ~.
\end{eqnarray} 

In Refs.~\cite{Salopek:1990jq, Falk:1992sf, Gangui:1993tt, Maldacena:2002vr, Acquaviva:2002ud}, the authors found that the local-type non-linearity parameter is tiny in the single field slow-roll inflation as
\begin{eqnarray}
f_{\rm NL}^{\rm local} = \frac{5}{12} (1 - n_s) 
= \frac{5}{6} (2 \epsilon_H - \eta_H) 
= \frac{5}{6} (3 \epsilon - \eta)  ~,
\end{eqnarray}
which gives $f_{\rm NL}^{\rm local} = 0.015$ for $n_s = 0.963$. In contrast, large $f_{\rm NL}^{\rm local}$ can be realized in the models with multiple light fields during inflation \cite{Bartolo:2001cw, Bernardeau:2002jy,
Bernardeau:2002jf, Sasaki:2008uc, Naruko:2008sq, Byrnes:2008wi,
Byrnes:2006fr, Assadullahi:2007uw, Valiviita:2006mz, Vernizzi:2006ve,
Allen:2005ye}, the curvaton scenario \cite{Linde:1996gt, Moroi:2001ct,
Lyth:2002my}, and inhomogeneous reheating \cite{Dvali:2003em, Kofman:2003nx}.

\subsubsection{Equilateral  type}

The equilateral bispectrum is parametrized as \cite{Creminelli:2005hu}
\begin{eqnarray}
S^{\rm equil}(k_1, k_2, k_3) =
6 \left[ \left\{- \frac{1}{(k_1 k_2)^3} + 2 \ {\rm perms.} \right\} 
- \frac{2}{(k_1 k_2 k_3)^2} 
+ \left\{ \frac{1}{k_1 k_2^2 
k_3^3 } 
+ 5 \ {\rm perms.}
\right\}
\right] ~. \label{eq:S_eq}
\end{eqnarray}
This bispectrum is obtained in the inflationary models with non-canonical kinetic terms for the scalar field. For example, the so-called Dirac-Born-Infeld (DBI) inflation \cite{Alishahiha:2004eh, Silverstein:2003hf} predicts $f_{\rm NL}^{\rm equil} \propto - 1 / c_s^2$ for $c_s \ll 1$ with $c_s$ being the effective sound speed of the scalar field fluctuation. We can also find a lot of other large $f_{\rm NL}^{\rm equil}$ models \cite{ArkaniHamed:2003uz, Seery:2005wm, Chen:2006nt, Cheung:2007st, Li:2008qc}. This bispectrum has a peak at the equilateral limit, namely, $k_1 = k_2 = k_3$ as described in the bottom-left panel of Fig.~\ref{fig:2D_loc_equil_orthog.pdf}. Due to the orthogonality between the local- and equilateral-type bispectra, these can be measured almost independently. 

In Sec.~\ref{sec:parity_violating}, we confirm that the graviton non-Gaussianity originated from the Weyl cubic action is also categorized as the equilateral type.

\subsubsection{Orthogonal type}

\begin{figure}[t]
\centering \noindent
\includegraphics[width=17cm]{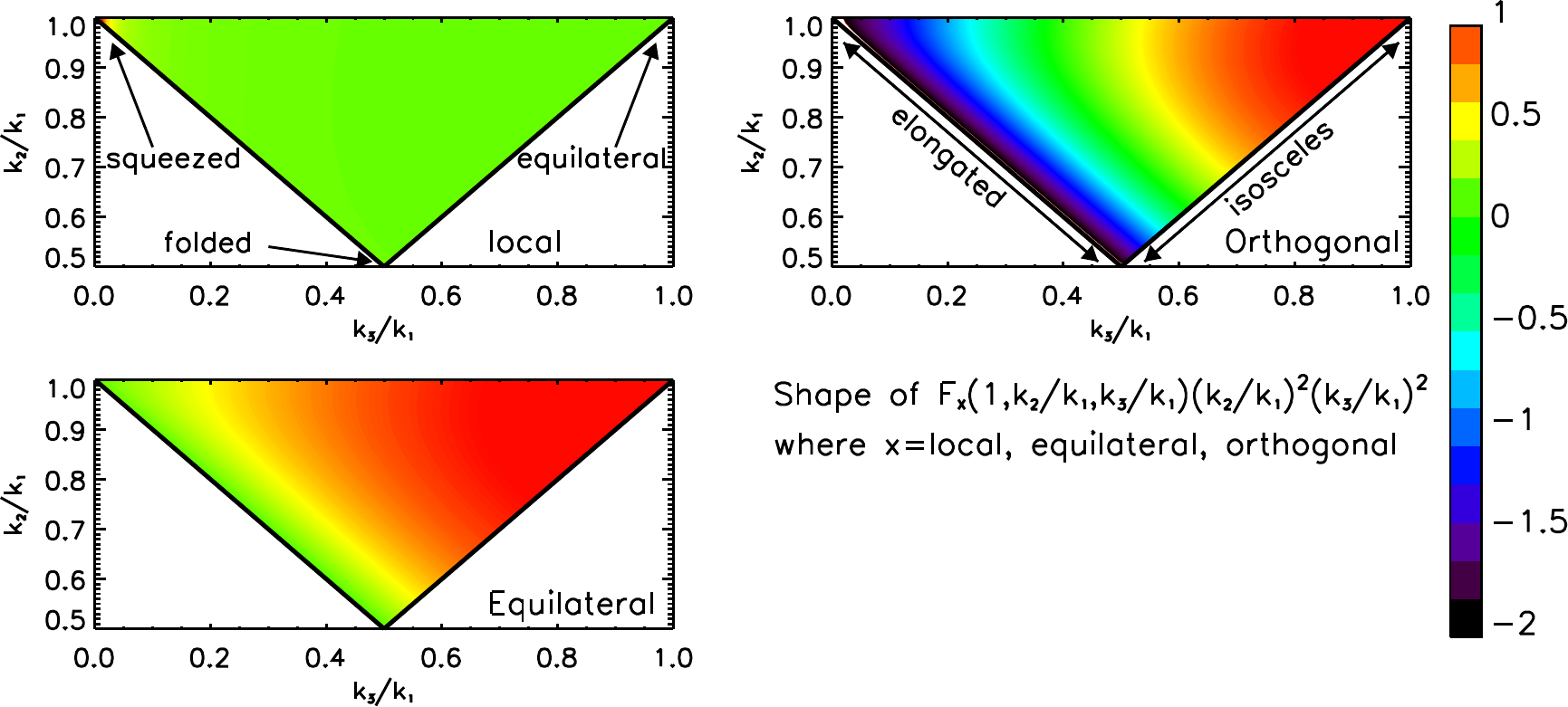}
\caption{%
 Two dimensional color contour for the shapes of the primordial bispectra.
 Each panel describes the normalized amplitude of $(k_1 k_2 k_3)^2 S(k_1,k_2,k_3)$ as a function of $k_2/k_1$ and $k_3/k_1$ for $k_3\le k_2\le k_1$. The amplitude is normalized by the maximum value of $(k_1 k_2 k_3)^2 S(k_1,k_2,k_3)$. 
The top-left panel is the local form given by Eq.~(\ref{eq:S_loc}), which diverges at the squeezed configuration. The bottom-left panel is the equilateral form given by Eq.~(\ref{eq:S_eq}), which is amplified at the equilateral configuration. The top-right panel is the orthogonal form given by Eq.~(\ref{eq:S_orthog}), which has a positive peak at the equilateral configuration, and a negative valley along the elongated configurations. This is quoted in Ref.~\cite{Komatsu:2010hc}.} 
\label{fig:2D_loc_equil_orthog.pdf}
\end{figure}

In the orthogonal type,  we use the following parametrization: 
\begin{eqnarray}
S^{\rm orthog}(k_1, k_2, k_3)  = 6 \left[ 
\left\{- \frac{3}{(k_1 k_2)^3} + 2 \ {\rm perms.} \right\} 
- \frac{8}{(k_1 k_2 k_3)^2 } 
+ \left\{ \frac{3}{k_1 k_2^2 
k_3^3 } 
+ 5 \ {\rm perms.}
\right\}
\right] \label{eq:S_orthog} ~.
\end{eqnarray}
This is nearly orthogonal to both the local-type and equilateral-type non-Gaussianities \cite{Senatore:2009gt}. This bispectrum can arise from a linear combination of higher-derivative scalar-field interaction terms which produce the equilateral bispectra. This function has a positive peak at the equilateral configuration and negative valley along the elongated configurations as seen in the top-right panel of Fig.~\ref{fig:2D_loc_equil_orthog.pdf}.  

\subsection{Observational limits}

Using the optimal estimators \cite{Yadav:2007rk, Yadav:2007ny, Smith:2009jr, Senatore:2009gt, Komatsu:2010hc}, the constraints on the nonlinearity parameters from the CMB data (WMAP 7-yr data) are obtained as \cite{Komatsu:2010fb}
\begin{eqnarray}
-10 < f_{\rm NL}^{\rm local} < 74  ~, \ \
-214 < f_{\rm NL}^{\rm equil} < 266 ~, \ \
-410 < f_{\rm NL}^{\rm orthog} < 6 \ \ (95\% \ {\rm CL}) ~. \label{eq:nonlinearity_param_limit}
\end{eqnarray} 
As another approach for extracting the non-Gausssinity from the CMB data, the methods with the Minkowski functionals have been developed \cite{Eriksen:2004df, Hikage:2008gy, Matsubara:2010te}. 

The PLANCK satelite \cite{:2006uk} and the proposed CMBPol mission \cite{Baumann:2008aq} will give tighter bounds as $\sigma(f_{\rm NL}^{\rm local}) \sim 5$ and $2$. At the level of $f_{\rm NL}^{\rm local} = {\cal O}(1)$, we need to be concerned about the contamination of the signals from late-time secondary effects. Studies on the gravitational non-linear evolution at late times can be seen in Refs.~e.g., \cite{Bartolo:2008sg, Pitrou:2008hy, Nitta:2009jp, Pitrou:2010sn}.   

In addition, the primordial non-Gaussianity also imprints its signatures on the large scale structure in the Universe. Estimating the primordial non-Gaussianity from the data of the matter distribution is hard due to large contamination of late-time gravitational nonlinear evolution. Regardless of it, the scale-dependence of the bias parameter between biased objects and linearized matter density fields is a good indicator for the primordial local-type non-Gaussianity \cite{Dalal:2007cu, Slosar:2008hx})~. From the luminous red galaxies (LRGs) sample of SDSS, Ref.~\cite{Slosar:2008hx} obtained a bound as
\begin{eqnarray}
-29 < f_{\rm NL}^{\rm local} < 70 \ \ (95\% \ {\rm CL}) ~.
\end{eqnarray}
This is comparable to bounds from the CMB data. The way to extract the information on the primordial non-Gaussianity from the matter distribution has continuously been studied (see, e.g., Refs.~\cite{Takeuchi:2010bc, Gong:2011gx, Yokoyama:2011qr}).

\subsection{Beyond the standard scalar-mode non-Gaussianities}

Historically, as described above, only the non-Gaussianity of curvature perturbations has been well-known studied. However, the non-Gaussianity of vector- or tensor-mode perturbation can be generated from the cosmological defects~\cite{Takahashi:2008ui,Hindmarsh:2009qk}, the magnetic fields~\cite{Brown:2005kr}, the nonlinear gravitational waves \cite{Maldacena:2002vr, Adshead:2009bz, Maldacena:2011nz, McFadden:2011kk, Gao:2011vs}, and so on. Furthermore, somewhat exotic non-Gaussianities including the violation of the rotatoional or parity invariance in the early Universe have recently been discussed (see, e.g., Refs.~\cite{Yokoyama:2008xw, Dimopoulos:2008yv, Maldacena:2011nz, Soda:2011am}). Hence, for precise comprehension of the early Universe, detailed analyses of these signals are necessary. 

This is the main motivation of our studies: construction of the general formulae for the CMB bispectrum with not only scalar- but also vector- and tensor-mode contributions, and computation and analysis of the CMB bispectrum sources from these novel non-Gaussianities. In the following sections, we focus on our studies. 

\bibliographystyle{JHEP}
\bibliography{paper}

%% file: ch5_bis.tex
\section{General formalism for the CMB bispectrum from primordial scalar, vector and tensor non-Gaussianities}\label{sec:formula}

In this section, on the basis of the formulation of the CMB anisotropy in Sec.~\ref{sec:CMB_anisotropy}, we develop the formulae for
the CMB bispectrum sourced from scalar-, vector-, and tensor-mode
non-Gaussianity. These results have been published in our paper \cite{Shiraishi:2010kd}. 

At first, we should remember an expression of CMB
fluctuation discussed in Sec.~\ref{sec:CMB_anisotropy}. In the all-sky analysis, the CMB fluctuations of intensity, and $E$ and $B$-mode polarizations are expanded with the spin-0
spherical harmonics, respectively. Then, the coefficients of CMB
fluctuations, called $a_{\ell m}$'s, are described as
\begin{eqnarray}
a_{X, \ell m}^{(Z)} = 4\pi (-i)^\ell \int \frac{d^3 {\bf k}}{(2 \pi)^3} 
\sum_\lambda [{\rm sgn}(\lambda)]^{\lambda+x} {}_{-\lambda}Y^*_{\ell
m}(\hat{\bf k}) \xi^{(\lambda)}({\bf k}) \mathcal{T}^{(Z)}_{X,\ell}(k)~, \label{eq:alm_ang}
\end{eqnarray}
where the index $Z$ denotes the mode of perturbations: $Z = S$
(scalar), $= V$ (vector) or $=T$ (tensor) and its helicity is expressed
by $\lambda$; $\lambda = 0$ for $Z = S$, $= \pm 1$ for $Z = V$ or $=
\pm 2$ for $Z = T$, $X$ discriminates between intensity and two
polarization (electric and magnetic) modes, respectively, as $X = I,E,B$
and $x$ is determined by it: $x = 0$ for $X = I,E$ or $=1$ for $X = B$,
$\xi^{(\lambda)}$ is the initial perturbation decomposed on each
helicity state and $\mathcal{T}_{X, \ell}^{(Z)}$ is the time-integrated
transfer function in each sector given by Eq.~(\ref{eq:transfer_explicit})\footnote{Here, we set $0^0 = 1$.}.

Next, we expand $\xi^{(\lambda)}$ with spin-$(-\lambda)$ spherical harmonics as
\begin{eqnarray}
\xi^{(\lambda)}({\bf k}) \equiv \sum_{\ell m} \xi^{(\lambda)}_{\ell m} (k)
{}_{-\lambda} Y_{\ell m}(\hat{\bf k})~, \label{eq:xi_def}
\end{eqnarray}
 and eliminate the angular dependence in Eq.~(\ref{eq:alm_ang}) by performing
$\hat{\bf k}$-integral: 
\begin{eqnarray}
a^{(Z)}_{X, \ell m} = 4 \pi (-i)^\ell \int_0^\infty \frac{k^2 dk}{(2 \pi)^3}
\sum_\lambda [{\rm sgn}(\lambda)]^{\lambda+x} \xi_{\ell m}^{(\lambda)}(k) 
\mathcal{T}^{(Z)}_{X, \ell}(k)~. \label{eq:alm}
\end{eqnarray}
Here, we use the orthogonality relation of the spin-$\lambda$ spherical
harmonics described in Appendix \ref{appen:spherical_harmonics} as
\begin{eqnarray}
\int d^2 \hat{\bf n} {}_{\lambda}Y^*_{\ell' m'}(\hat{\bf n})
{}_{\lambda}Y_{\ell m}(\hat{\bf n}) = \delta_{\ell, \ell'}\delta_{m, m'}~.
\end{eqnarray} 
Then the CMB bispectrum generated from the primordial non-Gaussianity of the scalar, vector and tensor perturbations is written down as 
\begin{eqnarray}
\begin{split}
\Braket{\prod_{n=1}^3 a_{X_n, \ell_n m_n}^{(Z_n)}} 
&= 
\left[ \prod_{n=1}^3 4\pi (-i)^{\ell_n} \int_0^\infty \frac{k_n^2 dk_n}{(2\pi)^3} 
\mathcal{T}^{(Z_n)}_{X_n, \ell_n}(k_n) 
\sum_{\lambda_n} [{\rm sgn}(\lambda_n)]^{\lambda_n + x_n}
\right] \\
& \times 
\Braket{\prod_{n=1}^3 \xi_{\ell_n m_n}^{(\lambda_n)}(k_n)} ~ , \\
\Braket{\prod_{n=1}^3 \xi_{\ell_n m_n}^{(\lambda_n)}(k_n)} 
&= 
\left[ \prod_{n=1}^3 \int d^2 \hat{\bf k_n} 
{}_{-\lambda_n} Y_{\ell_n m_n}^*(\hat{\bf k_n}) \right]
\Braket{\prod_{n=1}^3 \xi^{(\lambda_n)}({\bf k_n})}
 ~.
\label{eq:cmb_bis_general}
\end{split}
\end{eqnarray}
This formalism will be applicable to diverse sources of the scalar, vector and
tensor non-Gaussianities, such as, a scalar-graviton coupling \cite{Shiraishi:2010kd} (Sec.~\ref{sec:maldacena}), cosmic strings \cite{Takahashi:2008ui,
Hindmarsh:2009qk}, primordial magnetic fields \cite{Shiraishi:2010yk, Shiraishi:2011fi, Shiraishi:2011dh} (Sec.~\ref{sec:PMF}), and statistically-anisotropic and parity-violating interactions \cite{Shiraishi:2011ph, Shiraishi:2011st} (Secs.~\ref{sec:statistically_anisotropic} and \ref{sec:parity_violating}). 

To compute the CMB bispectrum composed of arbitrary perturbation modes, we have to reduce the expanded primordial bispectrum, $\Braket{\prod_{n=1}^3 \xi_{\ell_n m_n}^{(\lambda_n)}(k_n)}$, involving the contractions of the wave number vector and polarization vector and tensor, and the integrals over the angles of the wave number vectors. As shown later, this is elegantly completed by utilizing the Wigner symbols and spin-weighted spherical harmonics. 
  
If the initial bispectrum satisfies the rotational invariance, the CMB bispectrum is divided into the Wigner-$3j$ symbol depending on the azimuthal quantum numbers and the angle-averaged function as 
\begin{eqnarray}
\Braket{\prod_{n=1}^3 a_{\ell_n m_n}^{(Z_n)}} 
= \left(
  \begin{array}{ccc}
  \ell_1 & \ell_2 & \ell_3 \\
  m_1 & m_2 & m_3 
  \end{array}
 \right) B_{X_1 X_2 X_3, \ell_1 \ell_2 \ell_3}^{(Z_1 Z_2 Z_3)}~. \label{eq:cmb_bis_def_angle}
\end{eqnarray}


 \begin{figure}[t]
  \begin{tabular}{cc}
    \begin{minipage}{0.5\hsize}
  \begin{center}
    \includegraphics[width=8cm,clip]{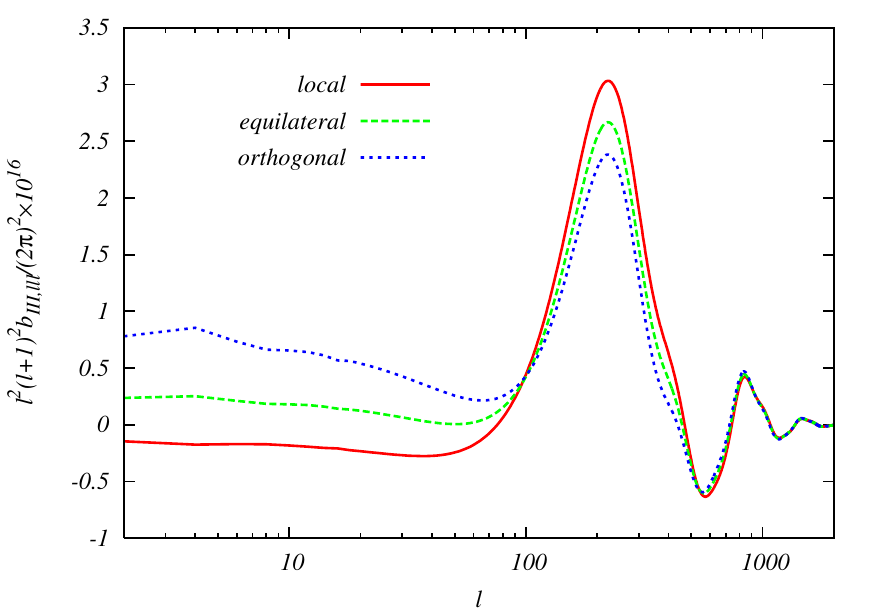}
  \end{center}
\end{minipage}
\begin{minipage}{0.5\hsize}
  \begin{center}
    \includegraphics[width=8cm,clip]{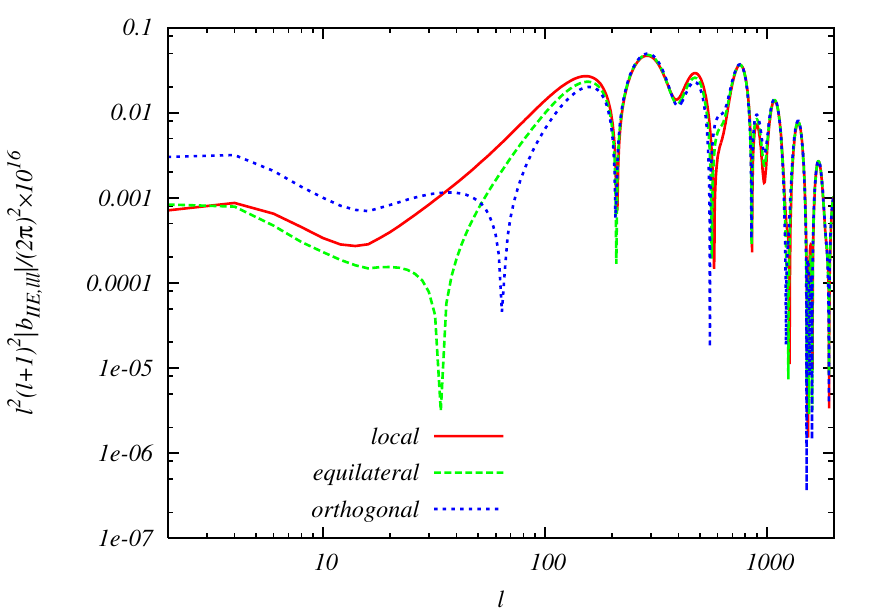}
  \end{center}
\end{minipage}
\\
    \begin{minipage}{0.5\hsize}
  \begin{center}
    \includegraphics[width=8cm,clip]{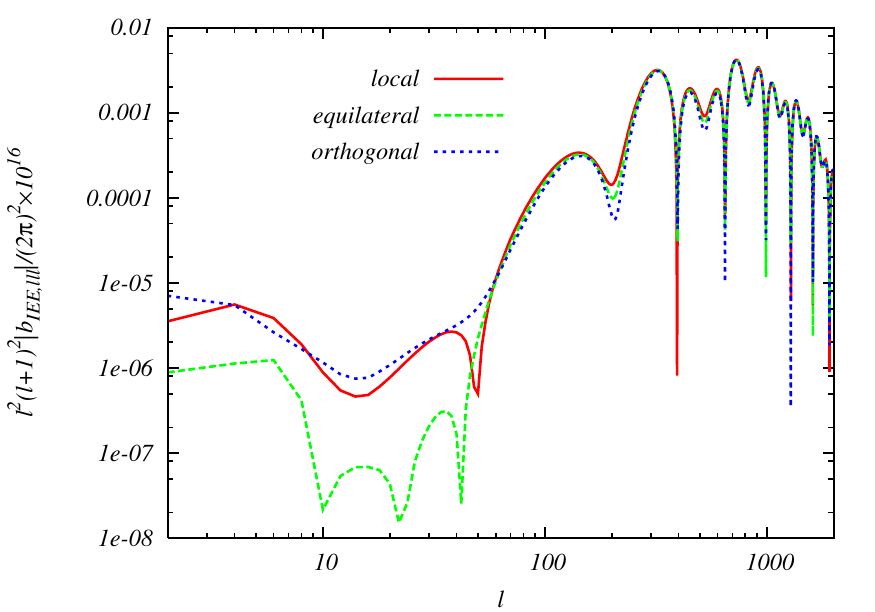}
  \end{center}
\end{minipage}
\begin{minipage}{0.5\hsize}
  \begin{center}
    \includegraphics[width=8cm,clip]{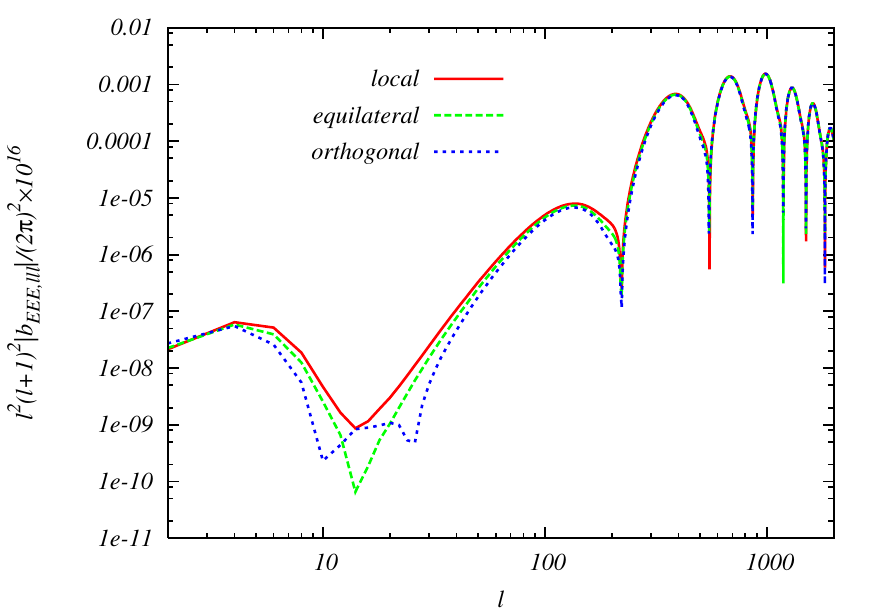}
  \end{center}
\end{minipage}
\end{tabular}
  \caption{The CMB $III$ (left top panel), $IIE$ (right top
 one), $IEE$ (left bottom one), and $EEE$ (right bottom one) bispectra induced from the local-type (red solid line), equilateral-type (green dashed one), and orthogonal-type (blue dotted one) non-Gaussianities of curvature perturbations. The three multipoles are fixed as $\ell_1 = \ell_2 = \ell_3 \equiv \ell$. Here we
  consider a power-law flat $\Lambda$CDM model and fix the nonlinear parameters as $f_{\rm NL}^{\rm local} = f_{\rm NL}^{\rm equil} = f_{\rm NL}^{\rm orthog}= 100$. The other cosmological parameters are fixed to the mean values reported in Ref.~\cite{Komatsu:2010fb}.} \label{fig:WMAP_theory_SSS_XXX_samel.pdf}
\end{figure}


 \begin{figure}[t]
  \begin{tabular}{cc}
    \begin{minipage}{0.5\hsize}
  \begin{center}
    \includegraphics[width=8cm,clip]{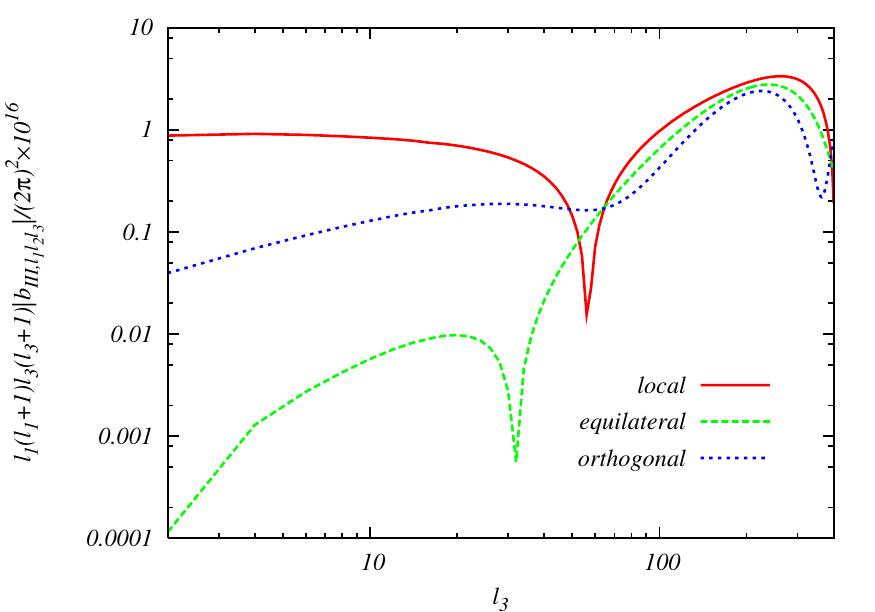}
  \end{center}
\end{minipage}
\begin{minipage}{0.5\hsize}
  \begin{center}
    \includegraphics[width=8cm,clip]{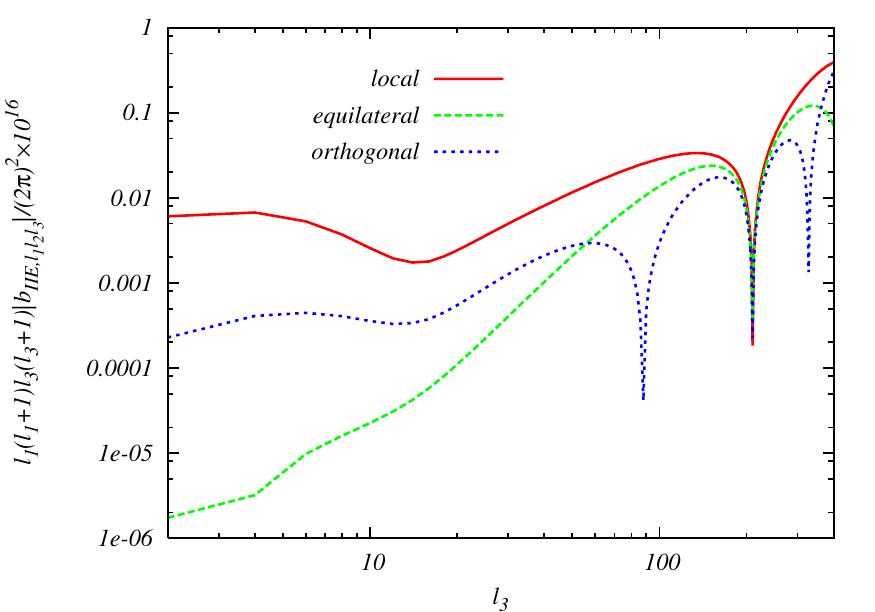}
  \end{center}
\end{minipage}
\\
    \begin{minipage}{0.5\hsize}
  \begin{center}
    \includegraphics[width=8cm,clip]{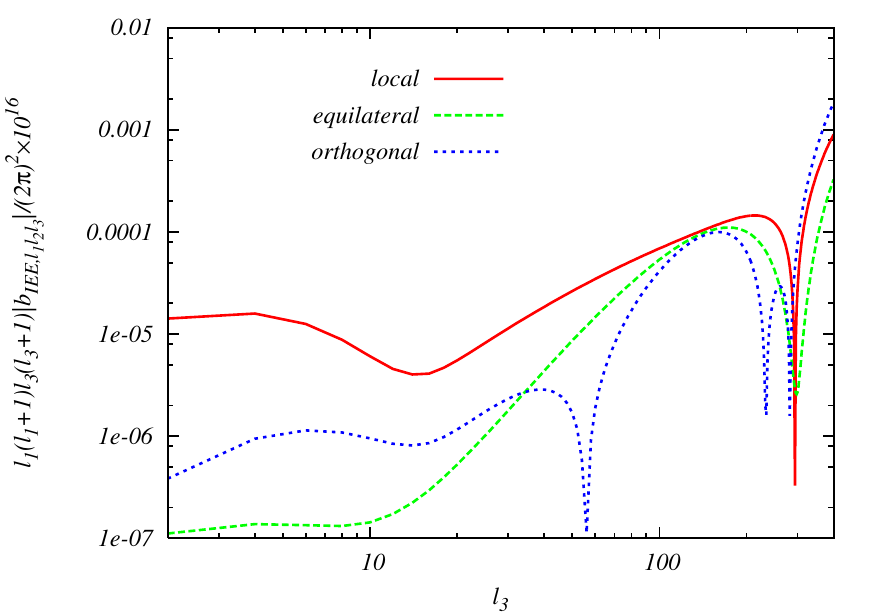}
  \end{center}
\end{minipage}
\begin{minipage}{0.5\hsize}
  \begin{center}
    \includegraphics[width=8cm,clip]{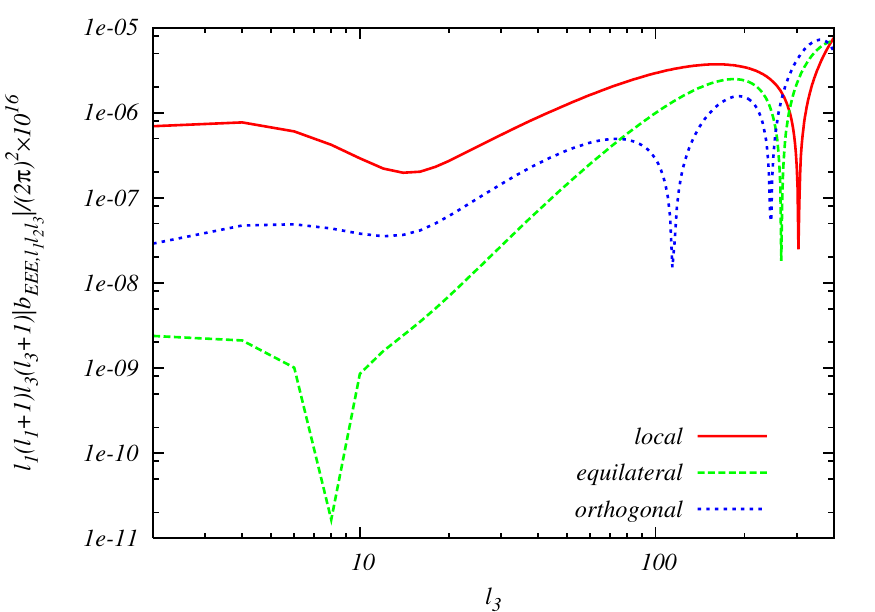}
  \end{center}
\end{minipage}
\end{tabular}
  \caption{The CMB $III$ (left top panel), $IIE$ (right top one), $IEE$ (left bottom one), and $EEE$ (right bottom one) bispectra induced from the local-type (red solid line), equilateral-type (green dashed one), and orthogonal-type (blue dotted one) non-Gaussianities of curvature perturbations with $f_{\rm NL}^{\rm local} = f_{\rm NL}^{\rm equil} = f_{\rm NL}^{\rm orthog}= 100$. Here, we fix the two multipoles as $\ell_1 = \ell_2 = 200$, and plot each curve as the function in terms of $\ell_3$. The cosmological parameters are identical to those in Fig.~\ref{fig:WMAP_theory_SSS_XXX_samel.pdf}.} \label{fig:WMAP_theory_SSS_XXX_difl.pdf}
\end{figure}

Let us focus on the CMB bispectrum from the curvature perturbations. Then,
from Eqs.~(\ref{eq:def_ini_bis}), (\ref{eq:ini_bis_scalar}), (\ref{eq:S_loc}), (\ref{eq:S_eq}), (\ref{eq:S_orthog}), (\ref{eq:cmb_bis_general}), (\ref{eq:cmb_bis_def_angle}) and the
knowledge of Appendix \ref{appen:wigner}, we derive
the reduced bispectra as 
\begin{eqnarray}
b_{X_1 X_2 X_3, \ell_1 \ell_2 \ell_3}^{(S S S)} 
&=& \left( I_{\ell_1 \ell_2 \ell_3}^{0~0~0} \right)^{-1} 
B_{X_1 X_2 X_3, \ell_1 \ell_2
\ell_3}^{(S S S)} \nonumber \\ 
&=&
\int_0^\infty y^2 dy 
\left[ \prod_{n=1}^3 \frac{2}{\pi} \int_0^\infty k_n^2 dk_n 
{\cal T}_{X_n,\ell_n}^{(S)}(k_n) j_{\ell_n}(k_n y) \right] 
F^{000}(k_1, k_2, k_3)~, \label{eq:cmb_bis_curvature}
\end{eqnarray}
where we have introduced the $I$ symbol as 
\begin{eqnarray}
I^{s_1 s_2 s_3}_{l_1 l_2 l_3}
\equiv \sqrt{\frac{(2 l_1 + 1)(2 l_2 + 1)(2 l_3 + 1)}{4 \pi}}
\left(
  \begin{array}{ccc}
  l_1 & l_2 & l_3 \\
  s_1 & s_2 & s_3
  \end{array}
 \right)~.
\end{eqnarray} 
In Figs.~\ref{fig:WMAP_theory_SSS_XXX_samel.pdf} and \ref{fig:WMAP_theory_SSS_XXX_difl.pdf}, we plot these CMB bispectra for each multipole configuration. From these, one can see that depending on the shape of the primordial non-Gaussianity, the overall magnitudes of the bispectra do not change, but the shapes in $\ell$ space change. Comparing these with the CMB data, the bounds on the nonlinearity parameters (\ref{eq:nonlinearity_param_limit}) are obtained. 

In this simple CMB bispectrum, there exists no dependence of the initial bispectrum on the polarization vector and tensor, hence we can derive the above formulae easily. Considering the vector- and tensor-mode contributions in the CMB bispectrum, however, the computation becomes so cumbersome due to the complicated angular dependence of the polarization vector and tensor. 
This difficulty is also true for the bispectrum where the rotational or parity invariance is broken. From the next section, we deal with these complicated bispectra depending on the several scenarios by applying the wonderful mathematical tools such as the Wigner symbols and the spin-weighted spherical harmonics. 

\bibliographystyle{JHEP}
\bibliography{paper}

%% file: ch6_mal.tex
\section{CMB bispectrum induced by the two scalars and a graviton correlator}\label{sec:maldacena}

In this section, adapting Eq.~(\ref{eq:cmb_bis_general}) to the primordial non-Gaussianity in two scalars and a graviton correlation \cite{Maldacena:2002vr}, we compute the CMB scalar-scalar-tensor bispectrum. This discussion is based on our paper \cite{Shiraishi:2010kd}. 

\subsection{Two scalars and a graviton interaction during inflation}

We consider a general single-field inflation model with Einstein-Hilbelt action
\cite{Garriga:1999vw}~:
\begin{eqnarray}
S=\int d^4x \sqrt{-g}\left[\frac{M_{\rm pl}^2}{2}R
+p(\phi,X)\right] ~,
\end{eqnarray}
where $g$ is the determinant of the metric, $R$ is the Ricci scalar, $M_{\rm
pl}^2\equiv 1/(8\pi G)$, $\phi$ is a scalar field, and $X \equiv - g^{\mu \nu}
\partial_{\mu}\phi\partial_\nu \phi/2$. Using the background equations, the
slow-roll parameter and the sound speed for perturbations are given by
\begin{eqnarray}
\epsilon \equiv -\frac{\dot{H}}{H^2}=\frac{X p_{,X}}{H^2M_{\rm pl}^2}~, \ \
c_s^2 \equiv \frac{p_{,X}}{2 X p_{,XX} + p_{,X}} ~,
\end{eqnarray}
where $H$ is the Hubble parameter, the dot means a derivative with respect to the
physical time $t$ and $p_{,X}$ denotes partial derivative of $p$ with respect to
$X$.
 We write a metric by ADM formalism
\begin{eqnarray}
ds^2 = -N^2dt^2+a^2e^{\gamma_{ab}}
(dx^a+N^adt)(dx^b+N^bdt) ~,
\end{eqnarray}
where $N$ and $N^a$ are respectively the lapse function and shift vector,
$\gamma_{ab}$ is a transverse and traceless tensor as
$\gamma_{aa}=\partial_a\gamma_{ab}=0$, and $e^{\gamma_{ab}} \equiv
\delta_{ab}+\gamma_{ab}+\gamma_{ac}\gamma_{cb}/2+\cdots$. On the 
flat hypersurface, the gauge-invariant curvature perturbation
$\zeta$ is related to the first-order fluctuation of the scalar field
$\varphi$ as $\zeta = - H\varphi/\dot{\phi}$
\footnote{Here, $\zeta$ and $\gamma_{ab}$ are equivalent to ${\cal R}$ and $h_{ab}$ in Eq.~(\ref{eq:R_hij_def}), respectively.}. 
Following the conversion equations (\ref{eq:scal_decompose}) and (\ref{eq:tens_decompose}), we decompose $\zeta$ and $\gamma_{ab}$ into the helicity states as 
\begin{eqnarray}
\xi^{(0)}({\bf k}) = {\zeta}({\bf k}) ~ , \ \
\xi^{(\pm 2)}({\bf k}) = \frac{1}{2} e_{ab}^{(\mp 2)}(\hat{\bf k}) \gamma_{ab}({\bf k}) ~.
\end{eqnarray}
Here, $e^{(\pm 2)}_{ab}$ is a transverse and traceless polarization
tensor explained in Appendix \ref{appen:polarization}.  
The interaction
parts of this action have been derived by Maldacena
\cite{Maldacena:2002vr} up to the third-order terms.  
In particular, we will focus on an interaction between two scalars and a graviton. This is because the correlation  between a small wave number of the tensor
mode and large wave numbers of the scalar modes will remain despite the
tensor mode decays after the mode reenters the cosmic horizon.
We find a leading term of the two scalars
and a graviton interaction in the action coming from the matter part of
the Lagrangian through $X$ as
\begin{eqnarray}
X|_{\rm 3rd-order}\supset a^{-2}\frac{p_{,X}}{2}\gamma_{ab}\partial_a \varphi 
\partial_b\varphi~,
\end{eqnarray}
therefore, the interaction part is given by
\begin{eqnarray}
S_{\rm int}\supset\int d^4x\, a g_{tss} \gamma_{ab}\partial_a \zeta \partial_b\zeta~.
\end{eqnarray}
Here, we introduce a coupling constant $g_{tss}$. 
From the definition of $\zeta, \gamma_{ab}$ and the slow-roll parameter, $g_{tss} = \epsilon$.  For a general
 consideration, let us deal with $g_{tss}$ as a free parameter.  In
 this sense, constraining this parameter may offer a probe of the
 nature of inflation and gravity in the early Universe.  The primordial
 bispectrum is then computed using in-in formalism in the next
 subsection.

\subsection{Calculation of the initial bispectrum}

\begin{figure}[t]
  \centering \includegraphics[height=8cm,clip]{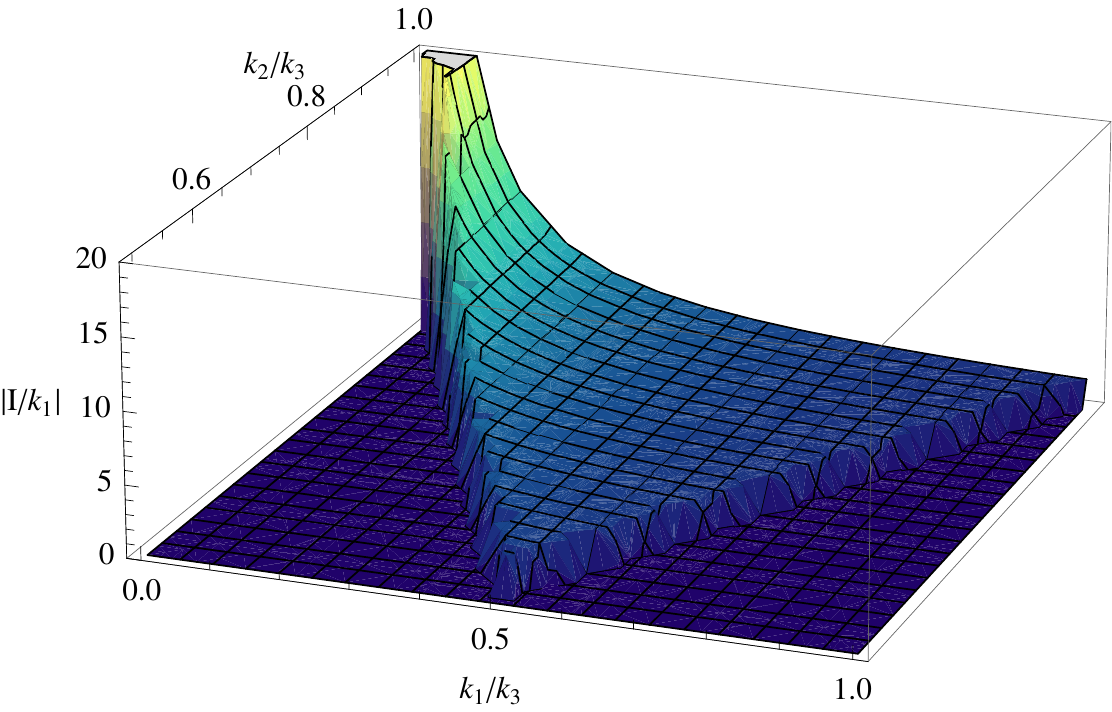}
  \caption{Shape of 
$I / k_1$. For the symmetric property and the triangle
 condition, we limit the plot range as $k_1 \leq k_2 \leq k_3$ and $|k_1 - k_2| \leq k_3 \leq k_1 + k_2$.}
  \label{fig:SST_Iok1}
\end{figure}

In the same manner as discussed in Ref.~\cite{Maldacena:2002vr}, we
calculate the primordial bispectrum generated from two scalars and a
graviton in the lowest order of the slow-roll parameter:
\begin{eqnarray}
\begin{split}
\Braket{\xi^{(\pm 2)}({\bf k_1}) \xi^{(0)}({\bf k_2}) \xi^{(0)}({\bf k_3})}
 &= (2 \pi)^3 F^{\pm 2 0 0}({\bf k_1}, {\bf k_2}, {\bf k_3})
\delta\left(\sum_{n=1}^3 {\bf k_n}\right)~, \\
F^{\pm 2 0 0}({\bf k_1}, {\bf k_2}, {\bf k_3}) 
&\equiv \frac{4 g_{tss} I(k_1,k_2,k_3) k_2 k_3}{\prod_i
(2k^3_i)}\frac{H_*^4}{2c^2_{s*}\epsilon^2_* M_{\rm pl}^4} 
e_{ab}^{(\mp 2)}(\hat{\bf k_1}) \hat{k_2}_a \hat{k_3}_b ~,  \\
I(k_1, k_2, k_3) &\equiv  -k_t + \frac{k_1 k_2 + k_2 k_3 + k_3
 k_1}{k_t} + \frac{k_1 k_2 k_3}{k_t^2}~, \label{eq:mal_bis}
\end{split}
\end{eqnarray}
where $k_t \equiv k_1 + k_2 + k_3$, and $*$ means that it is evaluated
at the time of horizon crossing, i.e., $a_*H_* = k$.  Here, we keep the
angular and polarization dependences, $e_{ab}^{(\mp 2)}(\hat{\bf k_1})
\hat{k_2}_a \hat{k_3}_b$, which have sometimes been omitted in the
literature for simplicity \cite{Mack:2001gc,Shiraishi:2010sm,Caprini:2009vk}. 
We show, however, that expanding this term with spin-weighted spherical harmonics enables us to formulate the
rotational-invariant bispectrum in an explicit way.
The statistically isotropic power spectra of $\xi^{(0)}$ and
$\xi^{(\pm 2)}$ are respectively given by
\begin{eqnarray}
\begin{split}
\Braket{\xi^{(0)}({\bf k}) \xi^{(0) *}({\bf k'})}
&\equiv (2\pi)^3 P_S(k) \delta({\bf k} - {\bf k'})~, \\ 
\frac{k^3 P_S(k)}{2 \pi^2} &= \frac{H_*^2}{8 \pi^2 c_{s*}\epsilon_* M_{\rm pl}^2} \equiv A_S~, \\
\Braket{\xi^{(\lambda)}({\bf k}) \xi^{(\lambda') *}({\bf k'})}
&\equiv (2\pi)^3 \frac{P_T(k)}{2} \delta({\bf k} - {\bf k'})
\delta_{\lambda, \lambda'} \ ({\rm for} \  \lambda, \lambda' = \pm 2)
~, \\
\frac{k^3 P_T(k)}{2 \pi^2} &= \frac{H_*^2}{ \pi^2 M_{\rm pl}^2} 
= 8 c_{s *}
 \epsilon_* A_S \equiv \frac{r}{2} A_S~, \label{eq:def_power}
\end{split}
\end{eqnarray}
where $r$ is the tensor-to-scalar ratio 
and $A_S$ is the amplitude of primordial curvature perturbations
\footnote{For $c_{s *} = 1$, these results are identical to Eqs.~(\ref{eq:ini_scal_power}) and (\ref{eq:ini_tens_power}).}. 
Note that the power spectra satisfy the scale invariance because we consider them in the lowest order of the slow-roll parameter.
Using these equations, we parametrize the initial bispectrum in this case from
Eq.~(\ref{eq:mal_bis}) as
\begin{eqnarray}
F^{\pm2 0 0}({\bf k_1}, {\bf k_2}, {\bf k_3}) 
&=& f^{(TSS)}(k_1, k_2, k_3) e_{ab}^{(\mp 2)}(\hat{\bf k_1}) \hat{{k_2}}_a
\hat{k_3}_b~, \label{eq:mal_F} \\
f^{(TSS)}(k_1, k_2, k_3) &\equiv& \frac{16 \pi^4 A_S^2 g_{tss}}{k_1^2 k_2^2 k_3^2} \frac{I(k_1, k_2, k_3)}{k_t} \frac{k_t}{k_1} ~. \label{mal_ftss}
\end{eqnarray}
Note that $f^{(TSS)}$ seems not to depend on the tensor-to-scalar
ratio. 
In Fig.~\ref{fig:SST_Iok1}~, we show the shape of $I/k_1$. From this, we confirm that the initial bispectrum $f^{(TSS)}$ (\ref{mal_ftss}) dominates in the squeezed limit as $k_1 \ll k_2 \simeq k_3$ like the local-type bispectrum of scalar modes given by Eq.~(\ref{eq:S_loc}).

In the squeezed limit, the ratio of $f^{(TSS)}$ to the
scalar-scalar-scalar counterpart $f^{(SSS)} = \frac{6}{5}f_{\rm NL}
P_S(k_1) P_S(k_2)$, which has been considered
frequently, reads
\begin{equation}
\frac{f^{(TSS)}}{f^{(SSS)}}
= \frac{10 g_{tss}}{3 f_{\rm NL}} \frac{I}{k_t} \frac{k_t k_2}{k_3^2}
\rightarrow \frac{20 g_{tss}}{3 f_{\rm NL}} \frac{I}{k_t}~.
\end{equation}
In the standard slow-roll inflation model, this ratio becomes ${\cal
O}(1)$ and does not depend on
the tensor-to-scalar ratio because $g_{tss}$ and $f_{\rm NL}$ are
proportional to the slow-roll parameter $\epsilon$, and 
$I/k_t$ has a nearly flat shape. The average of amplitude is evaluated as
$I/k_t \approx -0.6537$. Therefore, it manifests the comparable
importance of the higher order correlations of tensor modes to the scalar ones in the standard inflation scenario.

\subsection{Formulation of the CMB bispectrum}

Here, using Eqs.~(\ref{eq:cmb_bis_general}), (\ref{eq:mal_bis}) and (\ref{eq:mal_F}), we explicitly calculate the CMB tensor-scalar-scalar bispectrum as 
\begin{eqnarray}
\Braket{a_{X_1, \ell_1 m_1}^{(T)} a_{X_2, \ell_2 m_2}^{(S)} a_{X_3, \ell_3 m_3}^{(S)}} 
&=& 
\left[ \prod_{n=1}^3 4\pi (-i)^{\ell_n} \int_0^\infty \frac{k_n^2 dk_n}{(2\pi)^3} 
\mathcal{T}^{(Z_n)}_{X_n, \ell_n}(k_n)
\right] 
\nonumber \\
&& \times 
\sum_{\lambda_1 = \pm 2} \left( \frac{\lambda_1}{2} \right)^{x_1}
\left[ \prod_{n=1}^3 \int d^2 \hat{\bf k_n} \right]
{}_{-\lambda_1} Y_{\ell_1 m_1}^*(\hat{\bf k_1})  
 Y_{\ell_2 m_2}^*(\hat{\bf k_2}) 
 Y_{\ell_3 m_3}^*(\hat{\bf k_3}) \nonumber \\
&&\times (2 \pi)^3 
f^{(TSS)}(k_1, k_2, k_3) e_{ab}^{(\mp 2)}(\hat{\bf k_1}) \hat{{k_2}}_a
\hat{k_3}_b
\delta\left(\prod_{n=1}^3 {\bf k_n}\right)
~.
\end{eqnarray}

At first, we express all parts containing the angular dependence with
the spin spherical harmonics\footnote{Equations~(3.14) and (3.21) in Ref.~\cite{Shiraishi:2010kd} include typos.}:
\begin{eqnarray}
e^{(\mp 2)}_{ab}(\hat{\bf k_1})\hat{k_2}_a \hat{k_3}_b 
&=& \frac{(8\pi)^{3/2}}{6} \sum_{M m_a m_b}
{}_{\pm 2} Y^*_{2 M}(\hat{\bf k_1})
Y^*_{1 m_a}(\hat{\bf k_2}) Y^*_{1 m_b}(\hat{\bf k_3}) 
\left(
  \begin{array}{ccc}
  2 &  1 & 1 \\
   M & m_a & m_b
  \end{array}
 \right)~, \\
\delta\left( \sum_{i=1}^3 {\bf k_i} \right) 
&=& 8 \int_0^\infty y^2 dy 
\left[ \prod_{i=1}^3 \sum_{L_i M_i} 
 (-1)^{L_i/2} j_{L_i}(k_i y) 
Y_{L_i M_i}^*(\hat{\bf k_i}) \right] \nonumber \\
&&\times 
I_{L_1 L_2 L_3}^{0~0~0}
\left(
  \begin{array}{ccc}
  L_1 & L_2 & L_3 \\
  M_1 & M_2 & M_3 
  \end{array}
 \right)~, \label{eq:delta}
\end{eqnarray}
where we used the relations listed in  Appendices \ref{appen:wigner} and \ref{appen:polarization} and 
\begin{eqnarray}
I^{s_1 s_2 s_3}_{l_1 l_2 l_3}
\equiv \sqrt{\frac{(2 l_1 + 1)(2 l_2 + 1)(2 l_3 + 1)}{4 \pi}}
\left(
  \begin{array}{ccc}
  l_1 & l_2 & l_3 \\
  s_1 & s_2 & s_3
  \end{array}
 \right)~.
\end{eqnarray} 
Secondly, using Eq.~(\ref{eq:gaunt}), we replace all the integrals of spin spherical
harmonics with the Wigner symbols:
\begin{eqnarray}
\begin{split}
\int d^2 \hat{\bf k_1}~ {}_{\mp 2}Y_{\ell_1 m_1}^*(\hat{\bf k_1}) 
Y_{L_1 M_1}^*(\hat{\bf k_1})  {}_{\pm 2}Y_{2 M}^*(\hat{\bf k_1})
&= I^{\pm 2 0 \mp 2}_{\ell_1 L_1 2}
\left(
  \begin{array}{ccc}
  \ell_1 &  L_1 & 2 \\
   m_1 & M_1 & M
  \end{array}
 \right)~, \\
\int d^2 \hat{\bf k_2}~ Y_{\ell_2 m_2}^*(\hat{\bf k_2}) 
Y_{L_2 M_2}^*(\hat{\bf k_2}) Y_{1 m_a}^*(\hat{\bf k_2}) 
&= I^{0~ 0~ 0}_{\ell_2 L_2 1}
\left(
  \begin{array}{ccc}
  \ell_2 &  L_2 & 1 \\
   m_2 & M_2 & m_a
  \end{array}
 \right)~, \\
\int d^2 \hat{\bf k_3}~ Y_{\ell_3 m_3}^*(\hat{\bf k_3}) 
Y_{L_3 M_3}^*(\hat{\bf k_3}) Y_{1 m_b}^*(\hat{\bf k_3})
&= I^{0~ 0~ 0}_{\ell_3 L_3 1}
\left(
  \begin{array}{ccc}
  \ell_3 &  L_3 & 1 \\
   m_3 & M_3 & m_b
  \end{array}
 \right)~.
\end{split}
\end{eqnarray}  
Thirdly, using the summation formula of five Wigner-$3j$ symbols as Eq.~(\ref{eq:sum_9j_3j}), we sum up the Wigner-$3j$ symbols
with respect to azimuthal quantum numbers in the above equations and express
with the Wigner-$9j$ symbol as
\begin{eqnarray}
&& \sum_{\substack{M_1 M_2 M_3 \\ M m_a m_b}}
\left(
  \begin{array}{ccc}
  L_1 &  L_2 & L_3 \\
   M_1 & M_2 & M_3
  \end{array}
 \right)
\left(
  \begin{array}{ccc}
  2 &  1 & 1 \\
   M & m_a & m_b
  \end{array}
 \right) 
\nonumber \\
&&\qquad \times
\left(
  \begin{array}{ccc}
  \ell_1 &  L_1 & 2 \\
   m_1 & M_1 & M
  \end{array}
 \right)
\left(
  \begin{array}{ccc}
  \ell_2 &  L_2 & 1 \\
   m_2 & M_2 & m_a
  \end{array}
 \right)
\left(
  \begin{array}{ccc}
  \ell_3 &  L_3 & 1 \\
   m_3 & M_3 & m_b
  \end{array}
 \right) \nonumber \\
&& \qquad\qquad\qquad = 
\left(
  \begin{array}{ccc}
  \ell_1 & \ell_2 & \ell_3 \\
   m_1 & m_2 & m_3
  \end{array}
 \right)
\left\{
  \begin{array}{ccc}
  \ell_1 & \ell_2 & \ell_3 \\
   L_1 & L_2 & L_3 \\
   2 & 1 & 1 \\
  \end{array}
 \right\}~.
\end{eqnarray}
After these treatments and 
the summation over $\lambda_1 = \pm 2$ as
\begin{eqnarray}
\sum_{\lambda_1 = \pm 2} \left(\frac{\lambda_1}{2}\right)^{x_1} 
I^{\lambda_1 0 -\lambda_1}_{\ell_1 L_1 2} = 
\begin{cases}
2 I^{2 0 -2}_{\ell_1 L_1 2}&
({\rm for \ } x_1 + L_1 + \ell_1 = {\rm even} )~, \\
0 & ({\rm for \ } x_1 + L_1 + \ell_1 = {\rm odd} ) ~,
\end{cases} \label{eq:sum_lambda_mal}
\end{eqnarray}
we can obtain the CMB angle-averaged bispectrum induced from the nonlinear coupling between two scalars and a graviton as 
\begin{eqnarray}
B^{(TSS)}_{X_1 X_2 X_3, \ell_1 \ell_2 \ell_3} 
&=& \frac{(8 \pi)^{3/2}}{3} \sum_{L_1 L_2 L_3} 
(-1)^{\frac{L_1 + L_2 + L_3}{2}} I^{0~0~0}_{L_1 L_2 L_3} 
I^{2 0 -2}_{\ell_1 L_1 2} I^{0~0~0}_{\ell_2 L_2 1} I^{0~ 0~ 0}_{\ell_3 L_3 1} 
\left\{
  \begin{array}{ccc}
  \ell_1 & \ell_2 & \ell_3 \\
  L_1 & L_2 & L_3 \\
  2 & 1 & 1
  \end{array}
 \right\} \nonumber \\
&& \times \int_0^\infty y^2 dy  
\left[ \prod_{n=1}^3 \frac{2}{\pi} (-i)^{\ell_n} \int_0^\infty k_n^2 d
 k_n {\cal T}_{X_n, \ell_n}^{(Z_n)} j_{L_n}(k_n y)  \right]
f^{(TSS)}(k_1, k_2, k_3). 
\label{eq:mal_cmb_bis}   
\end{eqnarray} 
Note that the absence of the summation over $m_1, m_2$ and $m_3$ in this equation means that the tensor-scalar-scalar bispectrum maintains the
rotational invariance. As described above, this consequence is derived from
the angular dependence in the polarization tensor. Also in vector modes,
if their power spectra obey the statistical isotropy like
Eq.~(\ref{eq:def_power}), one can obtain the rotational invariant bispectrum by
considering the angular dependence in the polarization vector 
as Eq.~(\ref{eq:pol_vec}).
Considering Eq.~(\ref{eq:sum_lambda_mal}) and the selection rules of the Wigner symbols explained in Appendix~\ref{appen:wigner}, we can see that the values of $L_1, L_2$ and  $L_3$ are limited as 
\begin{eqnarray}
\begin{split}
& L_1 =
\begin{cases}
|\ell_1 \pm 2|, \ell_1 & ({\rm for \ } X_1 = I,E ) \\
|\ell_1 \pm 1| & ({\rm for \ } X_1 = B )
\end{cases}~, \ \ 
 L_2 = |\ell_2 \pm 1|~, \ \
 L_3 = |\ell_3 \pm 1|~, \\
& |L_1 - L_2| \leq L_3 \leq L_1 + L_2~, \ \ 
\sum_{i=1}^3 L_i = {\rm even}~, 
\end{split}
\end{eqnarray}
and the bispectrum (\ref{eq:mal_cmb_bis}) has nonzero value under the conditions: 
\begin{eqnarray}
|\ell_1 - \ell_2| \leq \ell_3 \leq \ell_1 + \ell_2~, \ \ 
\sum_{i=1}^3 \ell_i =
\begin{cases}
{\rm even} & ({\rm for \ } X_1 = I,E ) ~, \\
{\rm odd} & ({\rm for \ } X_1 = B ) ~.
\end{cases} 
\end{eqnarray} 

\begin{figure}[t]
  \centering \includegraphics[height=8cm,clip]{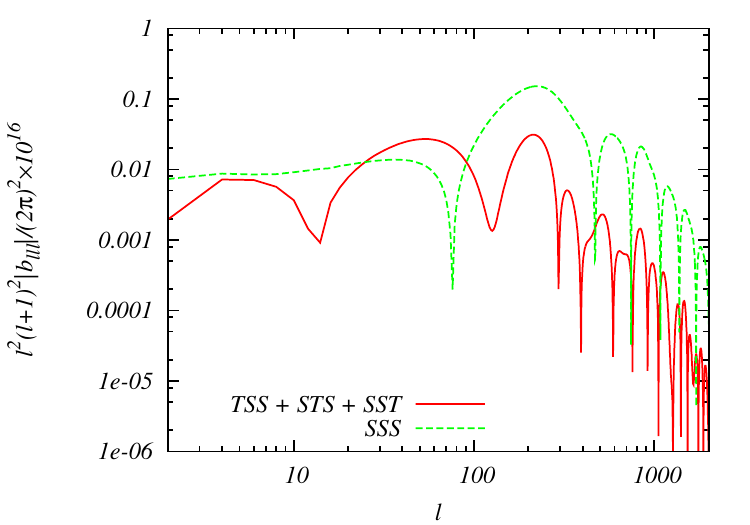}
  \caption{Absolute values of the CMB reduced bispectra of 
 temperature fluctuation for $\ell_1 = \ell_2 = \ell_3 \equiv \ell$. 
The lines correspond to the spectra 
  generated from tensor-scalar-scalar correlation given by
 Eq.~(\ref{eq:mal_cmb_red_bis}) with $g_{tss} = 5$ (red solid line) and
 the primordial non-Gaussianity in the scalar curvature perturbations
 with $f^{\rm local}_{\rm NL} = 5$ (green dashed line). 
 The other cosmological parameters are fixed to the mean values limited
 from WMAP-7yr data reported in Ref.~\cite{Komatsu:2010fb}.}
  \label{fig:SST_III_samel}
\end{figure}

\begin{figure}[t]
  \centering \includegraphics[height=8cm,clip]{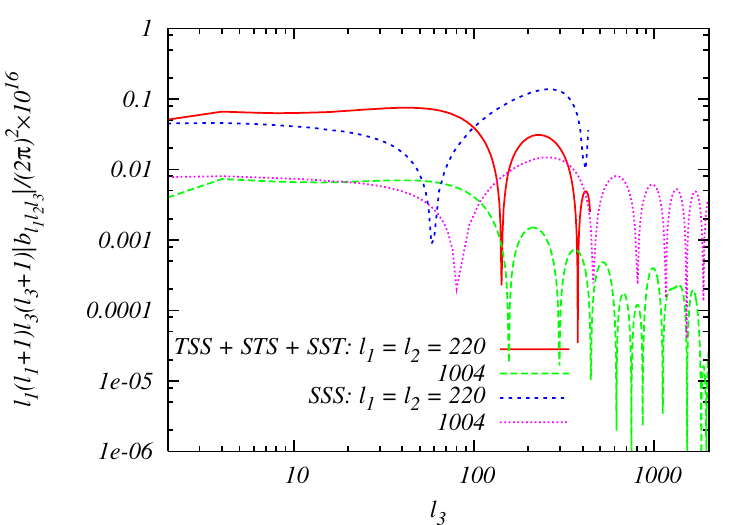}
  \caption{Absolute values of the CMB reduced bispectra of 
 temperature fluctuation generated from tensor-scalar-scalar correlation
 given by Eq.~(\ref{eq:mal_cmb_red_bis}) $(TSS + STS + SST)$ and the primordial non-Gaussianity in the scalar curvature perturbations $(SSS)$ as a function of $\ell_3$ with
 $\ell_1$ and $\ell_2$ fixed to some values as indicated. The
 parameters are fixed to the same values defined in
 Fig.~\ref{fig:SST_III_samel}.}
  \label{fig:SST_III_difl}
\end{figure}

In Figs.~\ref{fig:SST_III_samel} and \ref{fig:SST_III_difl}, we describe the reduced CMB bispectra of intensity mode sourced from two scalars and a graviton coupling: 
\begin{eqnarray}
b^{(TSS)}_{I I I, \ell_1 \ell_2 \ell_3} + b^{(STS)}_{I I I, \ell_1
 \ell_2 \ell_3} + b^{(SST)}_{I I I, \ell_1 \ell_2 \ell_3} 
= \left( I_{\ell_1 \ell_2 \ell_3}^{0~0~0}\right)^{-1}
\left(B^{(TSS)}_{I I I, \ell_1 \ell_2 \ell_3} 
+ B^{(STS)}_{I I I, \ell_1 \ell_2 \ell_3} 
+ B^{(SST)}_{I I I, \ell_1 \ell_2 \ell_3} 
\right)~,
\label{eq:mal_cmb_red_bis}
\end{eqnarray}
and primordial curvature perturbations (\ref{eq:cmb_bis_curvature}): 
\begin{eqnarray}
b^{(SSS)}_{III, \ell_1 \ell_2 \ell_3} 
= \left(I^{0~0~0}_{\ell_1 \ell_2 \ell_3}\right)^{-1} 
B^{(SSS)}_{III, \ell_1 \ell_2 \ell_3}~.
\end{eqnarray} 
 For the numerical computation, we modify the Boltzmann Code for
Anisotropies in the Microwave Background (CAMB) \cite{Lewis:2004ef,
Lewis:1999bs}
\footnote{The CMB bispectra generated from the two scalars and a graviton correlator in Figs.~\ref{fig:SST_III_samel} and \ref{fig:SST_III_difl} become slightly smaller than those in Ref.~\cite{Shiraishi:2010kd} due to the accuracy enhancement of the numerical calculation.}. 
In the calculation of the Wigner-$3j$ and $9j$ symbols,
we use the Common Mathematical Library SLATEC \cite{slatec} and the
summation formula of three Wigner-$6j$ symbols (\ref{eq:sum_9j_6j}). As
the radiation transfer functions of scalar and tensor modes, namely, ${\cal T}^{(S)}_{X_i, \ell_i}$ and ${\cal T}^{(T)}_{X_i, \ell_i}$, we use Eq.~(\ref{eq:transfer_explicit}). From the behavior of each line shown in Fig.~\ref{fig:SST_III_difl} at small $\ell_3$ that the reduced CMB bispectrum is roughly proportional to $\ell^{-2}$, we can confirm that the tensor-scalar-scalar bispectrum has a nearly squeezed-type
configuration corresponding to the shape of the initial bispectrum as discussed above. From Fig.~\ref{fig:SST_III_samel}, by comparing the green dashed line with the red solid line roughly estimated as  
\begin{eqnarray}
|b^{(TSS)}_{I I I, \ell \ell \ell} + b^{(STS)}_{I I I, \ell \ell \ell}
+ b^{(SST)}_{I I I, \ell \ell \ell}| 
\sim \ell^{-4} \times 8 \times 10^{-18}
 |g_{tss}|~, \label{eq:mal_cmb_bis_raugh}
\end{eqnarray}
we find that $|g_{tss}| \sim 5$ is comparable to
$f^{\rm local}_{\rm NL} = 5$ corresponding to the upper bound expected
from the PLANCK experiment.

\subsection{Estimation of the signal-to-noise ratio}

Here, we compute the signal-to-noise ratio by comparing the intensity bispectrum of
Eq.~(\ref{eq:mal_cmb_bis}) with the zero-noise (ideal) data and examine the bound on
the absolute value of $g_{tss}$. 
The formulation of (the square of) the signal-to-noise ratio $(S/N)$ is
reported in Refs.~\cite{Komatsu:2001rj} and \cite{Bartolo:2004if}. In our case, it can be
expressed as 
\begin{eqnarray}
\left( \frac{S}{N} \right)^2 = \sum_{2 \leq \ell_1 \leq \ell_2 \leq
 \ell_3 \leq \ell} 
\frac{ \left(B^{(TSS)}_{I I I \ell_1 \ell_2 \ell_3} 
+ B^{(STS)}_{I I I \ell_1 \ell_2 \ell_3} 
+ B^{(SST)}_{I I I \ell_1 \ell_2 \ell_3}\right)^2 }{\sigma^2_{\ell_1 \ell_2
\ell_3}}~, \label{eq:SN}
\end{eqnarray}
where $\sigma_{\ell_1 \ell_2 \ell_3}$ denotes the variance of the bispectrum. 
Assuming the weakly non-Gaussianity, the variance 
can be estimated as \cite{Spergel:1999xn, Gangui:2000gf} 
\begin{eqnarray}
\sigma^2_{\ell_1 \ell_2 \ell_3} \approx C_{\ell_1} C_{\ell_2} C_{\ell_3}
\Delta_{\ell_1 \ell_2 \ell_3}~, 
\end{eqnarray}
where $\Delta_{\ell_1 \ell_2 \ell_3}$ takes $1, 6$ or $2$ 
for $\ell_1 \neq \ell_2 \neq \ell_3, \ell_1 = \ell_2 = \ell_3$, or the
case that two $\ell$'s are the same, respectively. $C_\ell$ denotes that the CMB
angular power spectrum included the noise spectrum, which is
neglected in our case.   

\begin{figure}[t]
  \centering \includegraphics[height=8cm,clip]{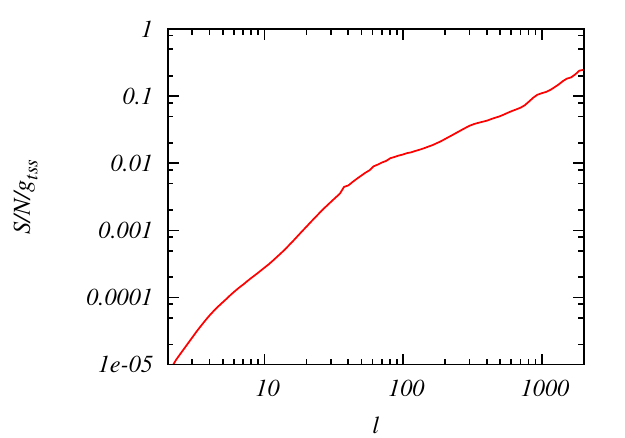}
  \caption{Signal-to-noise ratio normalized by
 $g_{tss}$ as a function
 of the maximum value between $\ell_1, \ell_2$ and $\ell_3$, namely, $\ell$. Each parameter is fixed to the same values defined in Fig. \ref{fig:SST_III_samel}.}
  \label{fig:SST_SN}
\end{figure}

In Fig.~\ref{fig:SST_SN}, the numerical result of Eq.~(\ref{eq:SN}) is
presented. We find that $(S/N)$ is a monotonically increasing function
roughly proportional to $\ell$ for $\ell < 2000$. It is
compared with the order estimation of Eq.~(\ref{eq:SN}) as
Ref.~\cite{Bartolo:2004if}
\begin{eqnarray}    
\left( \frac{S}{N} \right) &\sim& \sqrt{ \frac{\ell^3}{24} } 
\times \sqrt{\frac{(2 \ell)^3}{4 \pi}}
\left| \left(
  \begin{array}{ccc}
  \ell & \ell & \ell \\
   0 & 0 & 0
  \end{array}
 \right) \right| 
\frac{\ell^3 | b^{(TSS)}_{I I I \ell \ell \ell} + b^{(STS)}_{I I I \ell \ell
\ell} + b^{(SST)}_{I I I \ell \ell \ell} |}{(\ell^2 C_\ell)^{3/2}}
\nonumber \\
&\sim& \ell \times 5.4 \times 10^{-5} |g_{tss}| ~.
\end{eqnarray}
Here, we use Eq.~(\ref{eq:mal_cmb_bis_raugh}) and the approximations as $\sum \sim \ell^3/24$, 
$\ell^3 
\left(
  \begin{array}{ccc}
  \ell & \ell & \ell \\
   0 & 0 & 0
  \end{array}
 \right)^2 \sim 0.36 \times \ell$, and $\ell^2 C_\ell \sim 6 \times
 10^{-10}$. 
 We confirm that this is consistent with Fig. \ref{fig:SST_SN}, which justifies
 our numerical calculation in some sense.
This figure shows that from the WMAP and PLANCK
experimental data \cite{Komatsu:2010fb, :2006uk}, 
which are roughly noise-free at $\ell \lesssim 500$ and $1000$, respectively, 
expected $(S/N) / g_{tss}$ values are $0.05$ and $0.11$. Hence, to obtain $(S/N) > 1$, we need $|g_{tss}| > 20 $ and $9$. The latter value is close to a naive estimate $|g_{tss}| \lesssim 5$, which was discussed at the end of the previous subsection.

\subsection{Summary and discussion }\label{subsec:sum}

In this section, we present a full-sky formalism of the CMB bispectrum
 sourced from the primordial non-Gaussianity not only in the scalar 
 but also in the vector and tensor perturbations. 
As an extension of the previous formalism discussed in
 Ref.~\cite{Shiraishi:2010sm}, the new formalism contains the
 contribution of the polarization vector and tensor in the initial
 bispectrum.  In Ref.~\cite{Shiraishi:2010sm}, we have shown that
 in the all-sky analysis, the CMB bispectrum of vector or tensor mode
 cannot be formed as a simple angle-averaged bispectrum in the same
 way as that of scalar mode.  This is because the angular integrals over
 the wave number vectors have complexities for the non-orthogonality of
 spin spherical harmonics whose spin values differ from each other if
 one neglects the angular dependence of the polarization vector or
 tensor. In this study, however, we find that this difficulty vanishes
 if we maintain the angular dependence in the initial bispectrum.
 
 To present how to use our formalism, we
 compute the CMB bispectrum induced by the nonlinear mode-coupling
 between the two scalars and a graviton \cite{Maldacena:2002vr}. The
 typical value of the reduced bispectrum in temperature fluctuations is
 calculated as a function of the coupling constant between scalars and
 gravitons $g_{tss}$: $|b^{(TSS)}_{III, \ell \ell \ell} + b^{(STS)}_{III, \ell
 \ell \ell} + b^{(SST)}_{III, \ell \ell \ell}| \sim \ell^{-4} \times 8
 \times 10^{-18} |g_{tss}|$.  Through the computation of the
 signal-to-noise ratio, we find that the two scalars and a graviton coupling can be detected by the WMAP and PLANCK experiment if $|g_{tss}| \sim {\cal O}(10)$. Although we do not include the
 effect of the polarization modes in the estimation of $g_{tss}$ in
 this study, they will provide more beneficial information of the
 nonlinear nature of the early Universe.

\bibliographystyle{JHEP}
\bibliography{paper}

%% file: ch7_rot.tex
\section{Violation of the rotational invariance in the CMB bispectrum}\label{sec:statistically_anisotropic}

The current cosmological observations, particularly 
Cosmic Microwave Background (CMB), 
tell us that the Universe is almost isotropic, and 
primordial density fluctuations are almost Gaussian random fields.
However, in keeping with the progress of the experiments, 
 there have been many works that verify the possibility of the small deviation of the statistical isotropy, e.g., the so-called ``Axis of Evil''.
The analyses of the power spectrum by employing the current CMB data suggest that the deviation of the statistical isotropy is about $10 \%$ at most 
(e.g. \cite{Groeneboom:2008fz, Groeneboom:2009cb, Frommert:2009qw,
Hanson:2009gu, Bennett:2010jb, Hanson:2010gu}). 
Toward more precise measurements in future experiments,
there are a lot of theoretical discussions about
the effects of the statistical anisotropy on the CMB power spectrum,
\cite{Ackerman:2007nb, Boehmer:2007ut, Shtanov:2009wp, Watanabe:2010bu,
Gumrukcuoglu:2010yc},
e.g., the presence of the off-diagonal configuration
of the multipoles in the CMB power spectrum, which vanishes in the isotropic spectrum.

As is well known, it might be difficult to explain such statistical anisotropy
in the standard inflationary scenario.
However, recently, 
there have been several works about the possibility of generating the
statistically anisotropic primordial density fluctuations
in order to introduce nontrivial dynamics of the vector field. 
\cite{Dimopoulos:2006ms, Koivisto:2008xf, Yokoyama:2008xw,
Dimopoulos:2008yv, Karciauskas:2008bc, Bartolo:2009pa, Dimopoulos:2009am,
Dimopoulos:2009vu, ValenzuelaToledo:2009nq, 
ValenzuelaToledo:2010cs, Dimastrogiovanni:2010sm,
Dimopoulos:2010xq,Karciauskas:2011fp}. 
In Ref.~\cite{Yokoyama:2008xw}, the authors considered 
a modified hybrid inflation model where a waterfall field couples not only with an inflaton field 
but also with a massless vector field.
They have shown that, owing to the effect of fluctuations of the vector field, the primordial density fluctuations
may have a small deviation from the statistical isotropy and also
the deviation from the Gaussian statistics.
If the primordial density fluctuations deviate from the Gaussian statistics,
they produces the non-zero higher order spectra (corresponding to higher order correlation functions), e.g., the bispectrum (3-point function), the trispectrum (4-point function) and so on.
Hence, in the model presented in Ref.~\cite{Yokoyama:2008xw},
we can expect that there are characteristic signals not only in the CMB power spectrum but also in the CMB bispectrum. 

With these motivations, in this work, we calculate the CMB
statistically anisotropic bispectrum sourced from the curvature
perturbations generated in the modified hybrid inflation scenario
proposed in Ref.~\cite{Yokoyama:2008xw}, on the basis of the useful formula presented in Ref.~\cite{Shiraishi:2010kd}.
Then, we find the peculiar configurations of the multipoles which never appear in the isotropic bispectrum, like off-diagonal components in the CMB power spectrum. These discussions are based on Ref.~\cite{Shiraishi:2011ph}. 

This section is organized as follows. In the next subsection, we briefly review the inflation model where the scalar waterfall field couples with the vector field and calculate the bispectrum of curvature perturbations based on Ref.~\cite{Yokoyama:2008xw}. In Sec.~\ref{subsec:cmb_bis_RV}, we give an exact form of the CMB statistically anisotropic bispectrum and analyze its behavior by numerical computation. Finally, we devote the final subsection to the summary and discussion. 

Throughout this section, we obey the definition of the Fourier transformation as 
\begin{eqnarray}
f({\bf x}) \equiv \int \frac{d^3 {\bf k}}{(2 \pi)^3} \tilde{f}({\bf k}) e^{i {\bf k} \cdot {\bf x}}~,
\end{eqnarray}
and a normalization as $M_{\rm pl} \equiv (8\pi G)^{-1/2} = 1$.

\subsection{Statistically-anisotropic non-Gaussianity in curvature perturbations}

In this subsection, we briefly review the mechanism of generating the statistically anisotropic bispectrum induced by primordial curvature perturbations proposed in Ref.~\cite{Yokoyama:2008xw}, 
where the authors set the system like the hybrid inflation wherein there are two scalar fields:
inflaton $\phi$ and waterfall field $\chi$, and a vector field $A_\mu$ coupled with a waterfall field. 
The action is given by   
\begin{eqnarray}
S = \int dx^4 \sqrt{-g} 
\left[\frac{1}{2}R - \frac{1}{2}g^{\mu\nu}(\partial_{\mu} \phi
 \partial_\nu \phi + \partial_\mu \chi \partial_\nu \chi)
- V(\phi, \chi, A_\nu) 
- \frac{1}{4} g^{\mu\nu} g^{\rho \sigma}f^2(\phi) F_{\mu \rho} F_{\nu \sigma}
\right]. \label{eq:action_RV} 
\end{eqnarray}
Here, $F_{\mu \nu} \equiv \partial_\mu A_\nu - \partial_\nu A_\mu$ is the field strength of the vector field
$A_\mu$, $V(\phi, \chi, A_\mu)$
is the potential of fields and 
$f(\phi)$ denotes a gauge coupling. 
To guarantee the isotropy of the background Universe,
we need the condition that the energy density of the vector field
is negligible in the total energy of the Universe and we assume a small expectation value of the vector field. 
Therefore, we neglect the effect of the vector field on the background dynamics and also the evolution of
the fluctuations of the inflaton. 
 In the standard hybrid inflation (only with the inflaton and the waterfall field), inflation suddenly ends owing to the tachyonic instability of the waterfall field, which
is triggered when the inflaton reaches a critical value $\phi_{\rm e}$. 
In the system described using Eq.~(\ref{eq:action_RV}), 
however, $\phi_{\rm e}$ may fluctuate 
owing to the fluctuation of the vector field and it generates additional curvature perturbations. 

Using the $\delta N$ formalism \cite{Starobinsky:1982ee, Starobinsky:1986fxa, Sasaki:1995aw, Sasaki:1998ug, Lyth:2004gb, Lyth:2005fi}, 
the total curvature perturbation on the uniform-energy-density hypersurface at the end of inflation $t=t_{\rm e}$ can be estimated in terms of 
the perturbation of the $e$-folding number as  
\begin{eqnarray}
\zeta (t_{\rm e}) &=& \delta N (t_{\rm e}, t_*) \nonumber \\
&=& \frac{\partial N}{\partial \phi_*} \delta \phi_* 
+ \frac{1}{2} \frac{\partial^2 N}{\partial \phi_*^2} \delta \phi_*^2  
\nonumber \\
&& 
+ \frac{\partial N}{\partial \phi_{\rm e}} 
\frac{d \phi_{\rm e}(A)}{d A^\mu} \delta A_{\rm e}^\mu 
+ \frac{1}{2} 
\left[ \frac{\partial N}{\partial \phi_{\rm e}} 
\frac{d^2 \phi_{\rm e}(A)}{dA^\mu dA^\nu}
+ \frac{\partial^2 N}{\partial \phi_{\rm e}^2} 
\frac{d \phi_{\rm e}(A)}{d A^\mu} \frac{d \phi_{\rm e}(A)}{d A^\nu} \right] 
\delta A_{\rm e}^\mu \delta A_{\rm e}^\nu ~.
\end{eqnarray}
Here, $t_*$ is the time when the scale of interest crosses the
horizon during the slow-roll inflation. 
Assuming the sudden decay of all fields into radiations just after inflation, the curvature perturbations on the uniform-energy-density hypersurface become constant after inflation ends
\footnote{This $\zeta$ is consistent with ${\cal R}$ in Eq.~(\ref{eq:R_hij_def}).}. Hence, at the leading order, the power spectrum and the bispectrum of curvature perturbations are respectively derived as 
\begin{eqnarray}
\Braket{\prod_{n=1}^2 \zeta({\bf k_n})} 
&=& (2\pi)^3 N_*^2 P_{\phi} (k_1) \delta\left(\sum_{n=1}^2 {\bf k_n}\right)
 \nonumber \\
&& + N_{\rm e}^2 \frac{d \phi_{\rm e}(A)}{d A^\mu} \frac{d \phi_{\rm e}(A)}{d A^\nu} 
\Braket{\delta A^\mu_{\rm e}({\bf k_1}) \delta A^\nu_{\rm e}({\bf k_2})} ~, \label{eq:ini_power_deltaN} \\
\Braket{\prod_{n=1}^3 \zeta({\bf k_n})} 
&=& (2\pi)^3 N^2_* N_{**} [P_\phi(k_1) P_\phi(k_2) + 2 \ {\rm perms.}]
\delta\left(\sum_{n=1}^3 {\bf k_n}\right) \nonumber \\
&& + N_{\rm e}^3 \frac{d \phi_{\rm e}(A)}{d A^\mu} \frac{d \phi_{\rm e}(A)}{d A^\nu} \frac{d \phi_{\rm e}(A)}{d A^\rho} 
\Braket{\delta A^\mu_{\rm e}({\bf k_1}) \delta A^\nu_{\rm
e}({\bf k_2}) \delta A^\rho_{\rm e}({\bf k_3})} \nonumber \\
&& + N_{\rm e}^4 \frac{d \phi_{\rm e}(A)}{d A^\mu} \frac{d \phi_{\rm e}(A)}{d A^\nu}
\left( \frac{1}{N_{\rm e}} \frac{d^2 \phi_{\rm e}(A)}{d A^\rho d A^\sigma} 
+ \frac{N_{\rm e e}}{N_{\rm e}^2} \frac{d \phi_{\rm e}(A)}{dA^\rho} 
\frac{d \phi_{\rm e}(A)}{dA^\sigma} \right) \nonumber \\
&& \times 
\left[ \Braket{\delta A^\mu_{\rm e}({\bf k_1}) \delta A^\nu_{\rm e}({\bf k_2})
(\delta A^\rho \star \delta A^\sigma)_{\rm e}({\bf k_3})} + 2 \ {\rm perms.}\right]~, \label{eq:ini_bis_deltaN}
\end{eqnarray}
where $P_\phi(k) = H_*^2 / (2 k^3)$ is the power spectrum of the fluctuations of the inflaton, 
$N_{*} \equiv \partial N / \partial \phi_*$, 
$N_{**} \equiv \partial^2 N / \partial \phi_*^2$, 
$N_{\rm e} \equiv \partial N / \partial \phi_{\rm e}$, 
$N_{\rm ee} \equiv \partial^2 N / \partial \phi_{\rm e}^2
$, and 
$~\star~$ denotes the convolution.
Here, we assume that $\delta \phi_*$ is a Gaussian random field
and $\Braket{ \delta \phi A^\mu } = 0$.

For simplicity, we estimate the fluctuation of the vector fields in the Coulomb gauge: $\delta A_0 = 0$ and $k_i A^i = 0$. Then, the evolution equation of the fluctuations of the vector field is given by
\begin{eqnarray}
\ddot{\cal A}_i - \frac{\ddot{f}}{f} {\cal A}_i 
- a^2 \partial_j \partial^j {\cal A}_i = 0~,
\end{eqnarray}
where ${\cal A}_i \equiv f \delta A_i$, $~\dot{}~$
denotes the derivative with respect to the conformal time, and we neglect the contribution from the potential term. 
When $f \propto a, a^{-2}$ with appropriate
quantization of the fluctuations of the vector field, 
we have the scale-invariant power spectrum of $\delta A^i$ on superhorizon
scale as \cite{Yokoyama:2008xw, Dimopoulos:2009am, Martin:2007ue} 
\begin{eqnarray}
\Braket{\delta A^i_{\rm e}({\bf k_1}) \delta A^j_{\rm e}({\bf k_2})} 
= (2\pi)^3 P_\phi(k) f_{\rm e}^{-2} P^{ij}(\hat{\bf k_1})
\delta\left(\sum_{n=1}^2 {\bf k_n}\right)~, \label{eq:ini_power_A}
\end{eqnarray}
where $a$ is the scale factor, $P^{ij}(\hat{\bf k}) = \delta^{ij} -
\hat{k}^i \hat{k}^j$, $~\hat{}~$ denotes the unit vector, and $f_{\rm e}
\equiv f(t_{\rm e})$. 
Therefore, substituting this expression into Eq.~(\ref{eq:ini_power_deltaN}),
we can rewrite the power spectrum of the primordial curvature perturbations, $\zeta$, as
\begin{eqnarray}
\begin{split}
\Braket{\prod_{n=1}^2 \zeta({\bf k_n})} &\equiv (2 \pi)^3 P_{\zeta}({\bf k_1}) 
\delta\left(\sum_{n=1}^2 {\bf k_n} \right)~, \\
P_{\zeta}({\bf k}) 
&= P_\phi(k) \left[ N_*^2 + \left(\frac{N_{\rm e}}{f_{\rm e}}\right)^2 
q^i q^j P_{ij}(\hat{\bf k}) \right]~,
\end{split}
\end{eqnarray}
where $q_i \equiv d \phi_{\rm e} / d A^i, q_{ij} \equiv d^2 \phi_{\rm e} / (d A^i d A^j)$.
We can divide this expression into the isotropic part and
the anisotropic part as~\cite{Ackerman:2007nb} 
\begin{eqnarray}
P_\zeta({\bf k}) \equiv P_\zeta^{\rm iso}(k)
\left[ 1 + g_\beta \left( \hat{\bf q} \cdot \hat{\bf k} \right)^2
\right]~, \label{eq:P} 
\end{eqnarray}
with
\begin{eqnarray}
P^{\rm iso}_\zeta(k) = N_*^2 P_\phi(k) (1 + \beta)~, \ \ 
g_\beta = - \frac{\beta}{1 + \beta}~,
\end{eqnarray}
where $\beta = \left(N_{\rm e} / N_* / f_{\rm e}\right)^2 |{\bf q}|^2$.
The
bispectrum of the primordial curvature perturbation given by Eq.~(\ref{eq:ini_bis_deltaN}) 
can be written as
\begin{eqnarray}
\Braket{\prod_{n=1}^3 \zeta({\bf k_n})} &\equiv& (2 \pi)^3 F_{\zeta}({\bf k_1}, {\bf k_2}, {\bf k_3}) 
\delta\left(\sum_{n=1}^3 {\bf k_n} \right)~, \label{eq:ini_bis_general} \\
F_\zeta({\bf k_1}, {\bf k_2}, {\bf k_3}) 
&=& \left( \frac{g_\beta}{\beta} \right)^2 
P^{\rm iso}_\zeta(k_1) P^{\rm iso}_\zeta(k_2) \nonumber \\
&&\times \left[\frac{N_{**}}{N_*^2} + \beta^2 \hat{q}^a \hat{q}^b
\left(\frac{1}{N_{\rm e}} \hat{q}^{cd}  + \frac{N_{\rm ee}}{N_{\rm e}^2}\hat{q}^c \hat{q}^d \right)
P_{ac}(\hat{\bf k_1}) P_{bd}(\hat{\bf k_2}) \right] 
\nonumber \\
&&
+ 2 \ {\rm perms.}~. \label{eq:F_general}
\end{eqnarray}
Here $\hat{q}^{cd} \equiv q^{cd} / |{\bf q}|^2$ and we have assumed that
the fluctuation of the vector field $\delta A^i$ almost obeys Gaussian statistics; hence $\Braket{ \delta A_{\rm e}^\mu({\bf k_1}) \delta A_{\rm e}^\nu({\bf k_2}) \delta A_{\rm e}^\rho({\bf k_3}) } = 0$. 

Hereinafter, for calculating the CMB bispectrum explicitly, we adopt
a simple model whose potential looks like an Abelian Higgs model in the unitary gauge as~\cite{Yokoyama:2008xw}
\begin{eqnarray}
V(\phi, \chi, A^i) = \frac{\lambda}{4}(\chi^2 - v^2)^2 + \frac{1}{2} g^2
 \phi^2 \chi^2 + \frac{1}{2}m^2 \phi^2 + \frac{1}{2}h^2 A^\mu A_\mu \chi^2~,
\end{eqnarray}
where $\lambda, g$, and $h$ are the coupling constants, $m$
is the inflaton mass, and $v$ is the vacuum expectation value of $\chi$. 
Since the effective mass squared of the waterfall field is given by 
\begin{eqnarray} 
m_\chi^2 
\equiv \frac{\partial^2 V}{\partial \chi^2} 
= - \lambda v^2 + g^2 \phi_{\rm e}^2 + h^2 A^i A_i = 0~,
\end{eqnarray}
and the critical value of the inflaton $\phi_{\rm e}$ can be obtained as
\begin{eqnarray}
g^2 \phi_{\rm e}^2 = \lambda v^2 - h^2 A^iA_i~,
\end{eqnarray}
we can express $\beta$, $q^i$ and $q^{ij}$ in Eq.~(\ref{eq:F_general})
in terms of the model parameters as 
\begin{eqnarray}
\hat{q}^i = - \hat{A}^i~, \ \ \ 
\hat{q}^{ij} = - \frac{1}{\phi_{\rm e}}
\left[\left( \frac{g \phi_{\rm e}}{h A} \right)^2 \delta^{ij} + \hat{A}^i
 \hat{A}^j \right]~, \ \ \ 
\beta \simeq \frac{1}{f_{\rm e}^2} \left( \frac{h^2 A}{g^2 \phi_{\rm e}} \right)^2~,  
\end{eqnarray}
where we have used $N_* \simeq - N_e \simeq 1/ \sqrt{2\epsilon}$
with 
$\epsilon \equiv (\partial V / \partial \phi / V)^2 / 2$ being a slow-roll parameter and $|{\bf A}| \equiv A$.
Substituting these quantities into Eq.~(\ref{eq:F_general}), the bispectrum of primordial curvature
perturbations is obtained as
\begin{eqnarray}
\begin{split}
F_\zeta ({\bf k_1}, {\bf k_2}, {\bf k_3}) 
&= C P^{\rm iso}_\zeta(k_1) P^{\rm iso}_\zeta(k_2) 
\hat{A}^a \hat{A}^b
\delta^{cd}
P_{ac}(\hat{\bf k_1}) P_{bd}(\hat{\bf k_2})
+ 2 \ {\rm perms.}~, \\
C &\equiv - g^2_\beta \frac{\phi_{\rm e}}{N_{\rm e}} \left( \frac{g}{h A} \right)^2~. \label{eq:F}
\end{split}
\end{eqnarray}
Note that in the above expression, we have neglected the effect of the longitudinal polarization in the vector field for simplicity
\footnote{Owing to this treatment, we can use the quantities estimated in the Coulomb gauge as Eq.~(\ref{eq:F_general}). In a more precise discussion, we should take into account the contribution of the longitudinal mode in the unitary gauge.} and the terms that are
suppressed by a slow-roll parameter $\eta \equiv \partial^2 V /
\partial \phi ^2 / V $ because $- N_{**} / N_*^2 \simeq N_{\rm ee} /
N_{\rm e}^2 \simeq - (N_{\rm e} \phi_{\rm e})^{-1} \simeq \eta$.
Since the current CMB observations suggest $g_\beta < {\cal O}(0.1)$
(e.g., Refs.~\cite{Groeneboom:2008fz, Groeneboom:2009cb}) and $N_{\rm
e}^{-1} \simeq - \sqrt{2 \epsilon}$, the overall amplitude of the bispectrum in this model, $C$, does not seem to be sufficiently large to
be detected.    
However, even if $g_\beta \ll 1$ and $\epsilon \ll 1$, $C$ can become greater than unity for $ (g / h / A)^2 \phi_{\rm e} \gg 1$. Thus, we expect meaningful signals also in the CMB bispectrum. 
Then, in the next subsection,
we closely investigate the CMB bispectrum generated from the primordial bispectrum given by Eq.~(\ref{eq:F})
and discuss a new characteristic feature of the CMB bispectrum induced by the statistical anisotropy of the primordial bispectrum. 

\subsection{CMB statistically-anisotropic bispectrum} \label{subsec:cmb_bis_RV}

In this subsection, we give a formula of the CMB bispectrum generated from the
primordial bispectrum, which has statistical anisotropy owing to the fluctuations of the vector field,
given by Eq.~(\ref{eq:F}).
We also discuss the special signals of this CMB bispectrum, which vanish in the statistically isotropic bispectrum. 

\subsubsection{Formulation}

The CMB fluctuation can be expanded in terms of the spherical harmonic
function as
\begin{eqnarray}
\frac{\Delta X(\hat{\bf n})}{X} = \sum_{\ell m} a_{X, \ell m} Y_{\ell m}(\hat{\bf n})~,
\end{eqnarray} 
where $\hat{\bf n}$ is a unit vector pointing toward a line-of-sight direction, and $X$ denotes the intensity ($\equiv I$) and polarizations ($\equiv E, B$). According to Eq.~(\ref{eq:alm}), the coefficient, $a_{\ell m}$, generated from primordial curvature
perturbations, $\zeta$, is expressed as  
\begin{eqnarray}
a_{X, \ell m} &=& 4\pi (-i)^\ell \int_0^\infty \frac{k^2
 dk}{(2\pi)^3}
\zeta_{\ell m}(k) {\cal T}_{X, \ell}(k) \ \ \ ({\rm for} \  X = I,E) ~, \\ 
\zeta_{\ell m}(k)  &\equiv& \int d^2 \hat{\bf k} \zeta({\bf k})
Y^*_{\ell m}(\hat{\bf k})~, \label{eq:zeta_lm}
\end{eqnarray}
where ${\cal T}_{X, \ell}$ is the time-integrated transfer function of scalar
modes as described in Eq.~(\ref{eq:transfer_explicit}). Using
these equations, the CMB bispectrum generated from the bispectrum of the primordial
curvature perturbations is given by
\begin{eqnarray}
\Braket{\prod_{n=1}^3 a_{X_n, \ell_n m_n}}
= \left[\prod_{n=1}^3 4 \pi (-i)^{\ell_n} \int_0^\infty \frac{k_n^2
 dk_n}{(2\pi)^3} {\cal T}_{X_n, \ell_n}(k_n) \right]
\Braket{\prod_{n=1}^3 \zeta_{\ell_n m_n}(k_n)} , \label{eq:form_CMB_bis}
\end{eqnarray}
with
\begin{eqnarray}
\Braket{\prod_{n=1}^3 \zeta_{\ell_n m_n}(k_n)} 
= \left[ \prod_{n=1}^3 \int d^2 \hat{\bf k_n} Y^*_{\ell_n
     m_n}(\hat{\bf k_n}) \right] 
(2\pi)^3 \delta\left(\sum_{n=1}^3 {\bf k_n} \right) 
F_\zeta({\bf k_1}, {\bf k_2}, {\bf k_3})~. 
\end{eqnarray}
We expand the angular dependencies which appear in the Dirac delta function, $\delta ({\bf k_1}+{\bf k_2}+{\bf k_3})$, and
the function, $F_\zeta({\bf k_1},{\bf k_2},{\bf k_3})$, given by Eq.~(\ref{eq:F}) with respect to the spin spherical harmonics as 
\begin{eqnarray}
\delta\left( \sum_{n=1}^3 {\bf k_n} \right) 
&=& 8 \int_0^\infty y^2 dy 
\left[ \prod_{n=1}^3 \sum_{L_n M_n} 
 (-1)^{L_n/2} j_{L_n}(k_n y) 
Y_{L_n M_n}^*(\hat{{\bf k_n}}) \right] 
\nonumber \\
&&\times 
I_{L_1 L_2 L_3}^{0 \ 0 \ 0}
 \left(
  \begin{array}{ccc}
  L_1 & L_2 & L_3 \\
  M_1 & M_2 & M_3 
  \end{array}
 \right)~, \\
\hat{A}^a \hat{A}^b \delta^{cd} 
P_{ac}(\hat{\bf k_1}) P_{bd}(\hat{\bf k_2}) 
&=& - 4\left(\frac{4\pi}{3}\right)^3 
\sum_{L, L',L_A = 0,2} I_{L 1 1}^{0 1 -1} I_{L' 1 1}^{0 1
-1} 
I_{1 1 L_A}^{0 0 0} 
\left\{
  \begin{array}{ccc}
  L & L' & L_A \\
  1 & 1 & 1
  \end{array}
 \right\} \nonumber \\
&&\times \sum_{M M' M_A} Y_{L M}^*(\hat{\bf k_1}) Y_{L' M'}^*(\hat{\bf k_2}) 
Y_{L_A M_A}^*(\hat{\bf A})  
\left(
  \begin{array}{ccc}
  L & L' & L_A \\
  M & M' & M_A 
  \end{array}
 \right) ~, \label{eq:product}
\end{eqnarray}
where the $2 \times 3$ matrices of a bracket and a curly bracket denote the Wigner-$3j$ and $6j$
symbols, respectively, and 
\begin{eqnarray}
I^{s_1 s_2 s_3}_{l_1 l_2 l_3} 
\equiv \sqrt{\frac{(2 l_1 + 1)(2 l_2 + 1)(2 l_3 + 1)}{4 \pi}}
\left(
  \begin{array}{ccc}
  l_1 & l_2 & l_3 \\
   s_1 & s_2 & s_3 
  \end{array}
 \right)~.
\end{eqnarray}
Here, we have used the expressions of an arbitrary unit vector and a projection tensor as Appendices \ref{appen:wigner} and \ref{appen:polarization}. 
Note that for $Y_{00}^*(\hat{\bf A}) = 1/\sqrt{4\pi}$, the contribution
of $L_A = 0$ in Eq.~(\ref{eq:product}) is independent of the direction
of the vector field. Therefore, the statistical
anisotropy is generated from the signals of $L_A = 2$. 
By integrating these spherical harmonics over each unit vector, the angular
dependences on ${\bf k_1}, {\bf k_2}, {\bf k_3}$ can be reduced to the Wigner-$3j$ symbols as
\begin{eqnarray}
\begin{split}
\int d^2 \hat{\bf k_1} Y^*_{\ell_1 m_1} Y^*_{L_1 M_1} Y^*_{L M}
&= I_{\ell_1 L_1 L}^{0 \ 0 \ 0 }
 \left(
  \begin{array}{ccc}
  \ell_1 & L_1 & L \\
  m_1 & M_1 & M 
  \end{array}
 \right)~, \\
\int d^2 \hat{\bf k_2} Y^*_{\ell_2 m_2} Y^*_{L_2 M_2} Y^*_{L' M'}
&= I_{\ell_2 L_2 L'}^{0 \ 0 \ 0 }
 \left(
  \begin{array}{ccc}
  \ell_2 & L_2 & L' \\
  m_2 & M_2 & M' 
  \end{array}
 \right)~,  \\
\int d^2 \hat{\bf k_3} Y^*_{\ell_3 m_3} Y^*_{L_3 M_3}
&= (-1)^{m_3} \delta_{L_3, \ell_3} \delta_{M_3, -m_3} ~.
\end{split}
\end{eqnarray}
From these equations, we obtain an alternative explicit form of the
bispectrum of $\zeta_{\ell m}$ as 
\begin{eqnarray}
\Braket{\prod_{n=1}^3 \zeta_{\ell_n m_n}(k_n)}
&=& - (2\pi)^3 8 \int_0^\infty y^2 dy \sum_{L_1 L_2}
(-1)^{\frac{L_1+L_2+\ell_3}{2}} I_{L_1 L_2 \ell_3}^{0 \ 0 \ 0} \nonumber \\
&&\times P^{\rm iso}_\zeta(k_1) j_{L_1}(k_1 y) P^{\rm iso}_\zeta(k_2) j_{L_2}(k_2 y) C j_{\ell_3}(k_3
y) \nonumber \\
&&\times 
 4\left(\frac{4\pi}{3}\right)^3 (-1)^{m_3}
\sum_{L, L',L_A = 0,2} 
I_{L 1 1}^{0 1 -1} I_{L' 1 1}^{0 1 -1} \nonumber \\
&&\times I_{\ell_1 L_1 L}^{0 \ 0 \ 0 } I_{\ell_2 L_2 L'}^{0 \ 0 \ 0 }
I_{1 1 L_A}^{0 0 0} 
\left\{
  \begin{array}{ccc}
  L & L' & L_A \\
  1 & 1 & 1
  \end{array}
 \right\}
\nonumber \\
&&\times
 \sum_{M_1 M_2 M M' M_A} Y_{L_A M_A}^*(\hat{\bf A}) 
\left(
  \begin{array}{ccc}
  L_1 & L_2 & \ell_3 \\
  M_1 & M_2 & -m_3 
  \end{array}
 \right) \nonumber \\
&&\times
 \left(
  \begin{array}{ccc}
  \ell_1 & L_1 & L \\
  m_1 & M_1 & M 
  \end{array}
 \right)
 \left(
  \begin{array}{ccc}
  \ell_2 & L_2 & L' \\
  m_2 & M_2 & M' 
  \end{array}
 \right)
\left(
  \begin{array}{ccc}
  L & L' & L_A \\
  M & M' & M_A 
  \end{array}
 \right) \nonumber \\
&&+ {\rm 2 \ perms.} ~. \label{eq:zeta_lm_bis}
\end{eqnarray}
This equation implies that, owing to the vector field ${\bf A}$, 
the CMB bispectrum has a direction dependence, and hence, the dependence on $m_1, m_2, m_3$ cannot be confined 
only to a Wigner-$3j$ symbol, namely, 
\begin{eqnarray}
\Braket{\prod_{n=1}^3 \zeta_{\ell_n m_n}(k_n)} 
\neq (2\pi)^3 {\cal F}_{\ell_1 \ell_2 \ell_3}(k_1, k_2, k_3)
\left(
  \begin{array}{ccc}
  \ell_1 & \ell_2 & \ell_3 \\
  m_1 & m_2 & m_3 
  \end{array}
 \right)~. 
\end{eqnarray}
This fact truly indicates the violation of the rotational invariance in the 
bispectrum of the primordial curvature perturbations and leads to the
statistical anisotropy on the CMB bispectrum.

Let us consider the explicit form of the CMB bispectrum.
Here, we set the coordinate as $\hat{{\bf A}} = \hat{{\bf z}}$. Then, by substituting Eq.~(\ref{eq:zeta_lm_bis}) into Eq.~(\ref{eq:form_CMB_bis}) and using the
relation $Y_{L_A M_A}^*(\hat{{\bf z}}) = \sqrt{(2L_A + 1 )/ (4\pi)}
\delta_{M_A,0}$, the CMB bispectrum is expressed as 
\begin{eqnarray}
\Braket{\prod_{n=1}^3 a_{X_n, \ell_n m_n}} 
&=& - \int_0^\infty y^2 dy 
\left[ \prod_{n=1}^3 \frac{2}{\pi} \int_0^\infty k_n^2 dk_n 
{\cal T}_{X_n,\ell_n}(k_n) \right] \nonumber \\
&&\times \sum_{L_1 L_2}
(-1)^{\frac{\ell_1+\ell_2+L_1+L_2}{2} + \ell_3}
I_{L_1 L_2 \ell_3}^{0 \ 0 \ 0} 
\nonumber \\
&&\times 
P^{\rm iso}_\zeta(k_1) j_{L_1}(k_1 y) P^{\rm iso}_\zeta(k_2) j_{L_2}(k_2 y) C j_{\ell_3}(k_3
y)  \nonumber \\
&&\times 
4\left(\frac{4\pi}{3}\right)^3 (-1)^{m_3}
\sum_{L, L',L_A = 0,2}  
I_{L 1 1}^{0 1 -1} I_{L' 1 1}^{0 1 -1} \nonumber \\
&&\times I_{\ell_1 L_1 L}^{0 \ 0 \ 0 } I_{\ell_2 L_2 L'}^{0 \ 0 \ 0 }
I_{1 1 L_A}^{0 0 0} 
\left\{
  \begin{array}{ccc}
  L & L' & L_A \\
  1 & 1 & 1
  \end{array}
 \right\}
\nonumber \\
&&\times
\sqrt{\frac{2L_A +1}{4\pi}}
\sum_{M = -2}^2   
\left(
  \begin{array}{ccc}
  L_1 & L_2 & \ell_3 \\
  -m_1-M & -m_2+M & -m_3 
  \end{array}
 \right) \nonumber \\
&&\times 
 \left(
  \begin{array}{ccc}
  \ell_1 & L_1 & L \\
  m_1 & -m_1-M & M 
  \end{array}
 \right) 
 \left(
  \begin{array}{ccc}
  \ell_2 & L_2 & L' \\
  m_2 & -m_2+M & -M 
  \end{array}
 \right) \nonumber \\
&&\times 
\left(
  \begin{array}{ccc}
  L & L' & L_A \\
  M & -M & 0
  \end{array}
 \right) + {\rm 2 \ perms.}~. \label{eq:cmb_bis_RV}
\end{eqnarray}
By the selection rules of the Wigner symbols described in Appendix \ref{appen:wigner}, the ranges of $\ell_1, \ell_2, \ell_3, m_1, m_2$ and $m_3$, and the summation ranges in terms of $L_1$ and $L_2$ are limited as 
\begin{eqnarray}
\begin{split}
& \sum_{n=1}^3 \ell_n = {\rm even} ~, \ \
\sum_{n=1}^3 m_n = 0 ~, \\
& L_1 = |\ell_1 -2|, \ell_1, \ell_1 + 2~, \ \ 
L_2 = |\ell_2 -2|, \ell_2, \ell_2 + 2~,  \\
& |L_2 - \ell_3| \leq L_1 \leq L_2 + \ell_3~. \label{eq:lm_range}
\end{split}
\end{eqnarray}

\subsubsection{Behavior of the CMB statistically-anisotropic bispectrum}

On the basis of Eq.~(\ref{eq:cmb_bis_RV}), we compute the CMB bispectra for the several $\ell$'s and $m$'s. 
Then, we modify the Boltzmann Code for Anisotropies in the Microwave Background (CAMB)
\cite{Lewis:1999bs,Lewis:2004ef} and use the Common Mathematical Library SLATEC \cite{slatec}. 

\begin{figure}[t]
  \begin{tabular}{cc}
    \begin{minipage}{0.5\hsize}
  \begin{center}
    \includegraphics[width=8cm,clip]{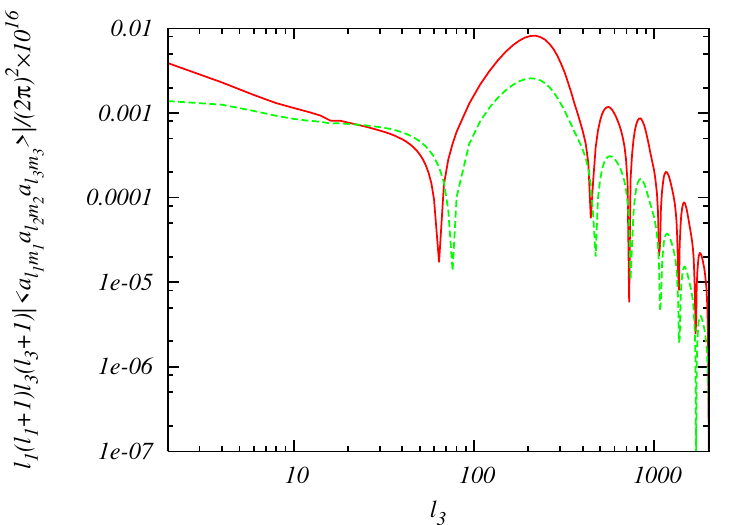}
  \end{center}
\end{minipage}
\begin{minipage}{0.5\hsize}
  \begin{center}
    \includegraphics[width=8cm,clip]{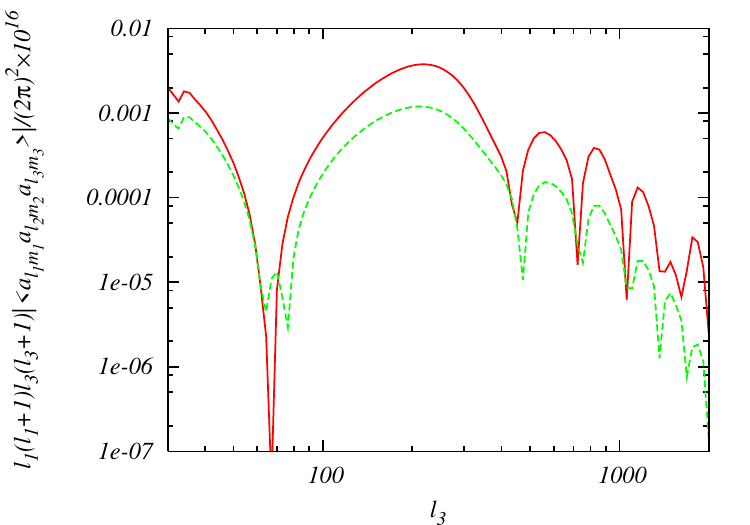}
  \end{center}
\end{minipage}
\end{tabular}
  \caption{Absolute values of the CMB statistically anisotropic
 bispectrum of the intensity mode given by Eq.~(\ref{eq:cmb_bis_RV}) with $C = 1$ (red solid line) and the statistically isotropic one given by Eq.~(\ref{eq:cmb_bis_iso}) with $f_{\rm NL} = 5$ (green dashed line) for $\ell_1 = \ell_2 = \ell_3$. The left and right figures are plotted in the configurations $(m_1, m_2, m_3) = (0,0,0), (10,20,-30)$, respectively. 
The parameters are fixed to the mean values limited from the WMAP-7yr data as reported in Ref.~\cite{Komatsu:2010fb}.} \label{fig:SSS_RV_samel.pdf}
\end{figure}

In Fig.~\ref{fig:SSS_RV_samel.pdf}, 
the red solid lines are the CMB statistically anisotropic bispectra of the intensity mode given by
Eq.~(\ref{eq:cmb_bis_RV}) with $C = 1$, and the green dashed lines are 
the statistically isotropic one sourced from the local-type non-Gaussianity of curvature perturbations given by Eq.~(\ref{eq:S_loc})
\begin{eqnarray}
\Braket{\prod_{n=1}^3 a_{X_n, \ell_n m_n}} 
&=& I_{\ell_1 \ell_2 \ell_3}^{0~0~0}
\left(
  \begin{array}{ccc}
  \ell_1 & \ell_2 & \ell_3 \\
  m_1 & m_2 & m_3
  \end{array}
 \right) 
\int_0^\infty y^2 dy 
\left[ \prod_{n=1}^3 \frac{2}{\pi} \int_0^\infty k_n^2 dk_n 
{\cal T}_{X_n,\ell_n}(k_n) j_{\ell_n}(k_n y) \right] 
\nonumber \\
&&\times 
\left( P^{\rm iso}_\zeta(k_1) P^{\rm iso}_\zeta(k_2) \frac{6}{5} f_{\rm NL}
 + {\rm 2 \ perms.} \right)~, \label{eq:cmb_bis_iso}
\end{eqnarray} 
with $f_{\rm NL} = 5$ for $\ell_1 = \ell_2 = \ell_3$ and two sets of $m_1, m_2, m_3$. 
From this figure, we can see that the red solid lines are in good agreement with the green dashed
line in the dependence on $\ell$ for both configurations of $m_1, m_2, m_3$.
This seems to be because the bispectrum of primordial curvature perturbations affected by the fluctuations of vector field given by Eq.~(\ref{eq:F})
has not only the anisotropic part but also the isotropic part and both parts have the same amplitude. 
In this sense, it is expected that the angular dependence on the vector field $\hat{{\bf A}}$ does not contribute much to a change in the shape of the CMB bispectrum. 
We also find that the anisotropic bispectrum for $C \sim 0.3$ 
is comparable in magnitude to the case with $f_{\rm NL} = 5$ for the standard local type,
which corresponds to the upper bound on the local-type non-Gaussianity expected from the PLANCK experiment \cite{:2006uk}.

In the discussion of the CMB power spectrum,
if the rotational invariance is violated in the primordial power spectrum given by Eq.~(\ref{eq:P}),
the signals in the off-diagonal configurations of $\ell$ also have
nonzero values \cite{Ackerman:2007nb, Boehmer:2007ut, Watanabe:2010bu}.
Likewise,
there are special configurations in the CMB bispectrum induced from the
statistical anisotropy on the primordial bispectrum as Eq.~(\ref{eq:F}). 
The selection rule (\ref{eq:lm_range}) suggests
that the statistically anisotropic bispectrum (\ref{eq:cmb_bis_RV}) could be nonzero in the multipole configurations given by 
\begin{eqnarray}
\ell_1 = |\ell_2 - \ell_3| - 4, |\ell_2 - \ell_3| - 2, \ell_2 + \ell_3 +
 2, \ell_2 + \ell_3 + 4~, \label{eq:special_l}
\end{eqnarray}
and two permutations of $\ell_1, \ell_2, \ell_3$. 
In contrast, in these configurations, 
the isotropic bispectrum (e.g., Eq.~(\ref{eq:cmb_bis_iso})) vanishes owing to the triangle condition of the
Wigner-$3j$ symbol 
$\left(
  \begin{array}{ccc}
  \ell_1 & \ell_2 & \ell_3 \\
  m_1 & m_2 & m_3
  \end{array}
 \right) $ and the nonzero components arise only from 
\begin{eqnarray}
|\ell_2 - \ell_3| \leq \ell_1 \leq \ell_2 + \ell_3~. \label{eq:default_l}
\end{eqnarray}
Therefore, the signals of the configurations (\ref{eq:special_l}) have the pure information of the statistical anisotropy on the CMB bispectrum. 

\begin{figure}[t]
  \begin{tabular}{cc}
    \begin{minipage}{0.5\hsize}
  \begin{center}
    \includegraphics[width=8cm,clip]{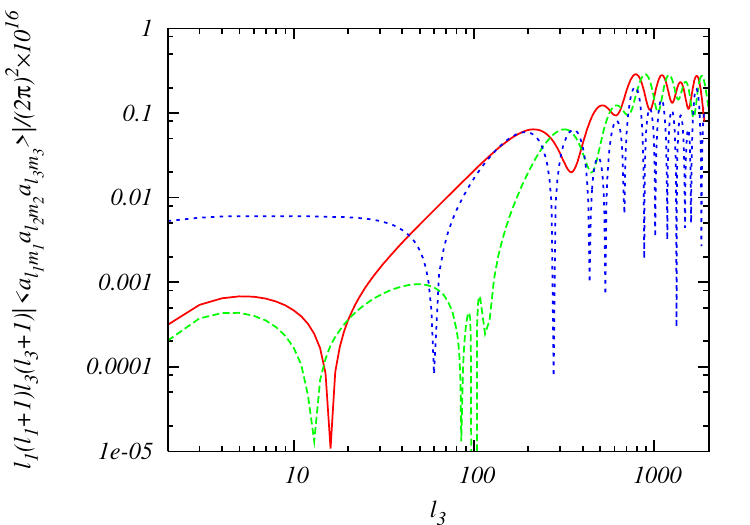}
  \end{center}
\end{minipage}
\begin{minipage}{0.5\hsize}
  \begin{center}
    \includegraphics[width=8cm,clip]{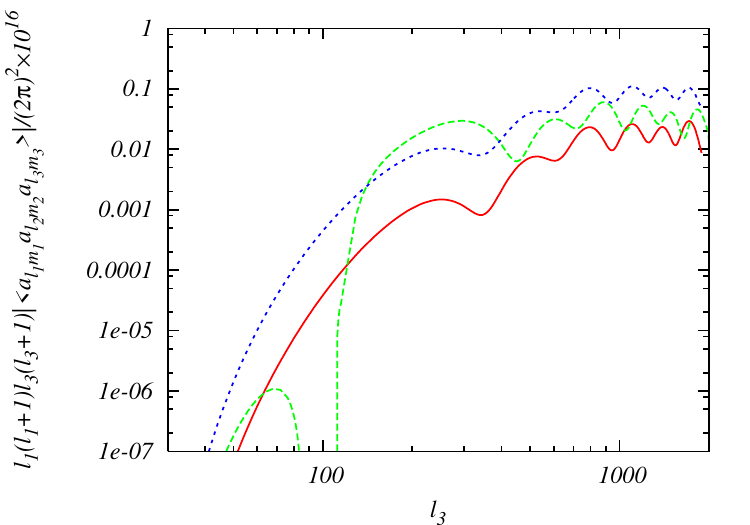}
  \end{center}
\end{minipage}
\end{tabular}
\caption{Absolute values of the CMB statistically anisotropic bispectra of the intensity mode given by Eq.~(\ref{eq:cmb_bis_RV}) for $(m_1, m_2, m_3) = (0,0,0)$ (left panel) and $(10,20,-30)$ (right one) as the function with respect to $\ell_3$. The lines correspond to
 the spectra for $(\ell_1, \ell_2) = (102 + \ell_3, 100)$ (red solid line),
 $(|100 - \ell_3| - 2, 100)$ (green dashed line) and $(100 + \ell_3, 100)$ (blue dotted line). The parameters are identical to the values defined in
 Fig.~\ref{fig:SSS_RV_samel.pdf}.}
  \label{fig:SSS_RV_C1_difl.pdf}
\end{figure}

Figure \ref{fig:SSS_RV_C1_difl.pdf} shows the CMB anisotropic bispectra of the intensity mode given by Eq.~(\ref{eq:cmb_bis_RV}) with $C = 1$ for the several configurations of $\ell$'s and $m$'s as a function of $\ell_3$. The red solid line and green dashed line satisfy the special relation (\ref{eq:special_l}), namely, $\ell_1 = \ell_2 + \ell_3 + 2, |\ell_2 - \ell_3| - 2$, and the blue dotted line obeys a configuration of Eq.~(\ref{eq:default_l}), namely, $\ell_1 = \ell_2 + \ell_3$. From this figure, we confirm that the signals in the special configuration (\ref{eq:special_l}) are comparable in magnitude to those for $\ell_1 = \ell_2 + \ell_3$. Therefore, if the rotational invariance is violated on the primordial bispectrum of curvature perturbations, the signals for $\ell_1 = \ell_2 + \ell_3 + 2, |\ell_2 - \ell_3| - 2$ can also become beneficial observables. 
Here, note that the anisotropic bispectra in the other special
configurations: $\ell_1 = \ell_2 + \ell_3 + 4, |\ell_2 - \ell_3| - 4$
are zero. It is because 
these signals arise from only the contribution of $L = L' = L_A = 2, L_1 = \ell_1 \pm 2, L_2 = \ell_2 \pm 2$ in Eq.~(\ref{eq:cmb_bis_RV}) owing to the selection rules of the Wigner symbols, and the summation of the four Wigner-$3j$ symbols over $M$ vanishes for all $\ell$'s and $m$'s. Hence, in this anisotropic bispectrum, the additional signals arise from only two configurations $\ell_1 = \ell_2 + \ell_3 + 2, |\ell_2 - \ell_3| - 2$ and these two permutations.

\subsection{Summary and discussion}

In this section, we investigated the statistical anisotropy in the CMB bispectrum by considering the modified hybrid inflation model where the waterfall field also couples with the vector field \cite{Yokoyama:2008xw}.
We calculated the CMB bispectrum sourced from the non-Gaussianity of curvature perturbations affected by the vector field. 
In this inflation model, owing to the dependence on the direction of the vector field, the correlations of the curvature perturbations violate the rotational invariance. Then, interestingly, even if the magnitude of the parameter $g_\beta$ characterizing the statistical anisotropy of
the CMB power spectrum is too small, the amplitude of the non-Gaussianity can become large depending on several coupling constants of the fields. 

Following the procedure of Sec.~\ref{sec:formula} \cite{Shiraishi:2010kd}, we formulated the statistically anisotropic CMB bispectrum and confirm that three azimuthal quantum numbers $m_1, m_2$ and $m_3$ are not confined only to the Wigner symbol 
$\left(
  \begin{array}{ccc}
  \ell_1 & \ell_2 & \ell_3 \\
  m_1 & m_2 & m_3
  \end{array}
 \right)$. 
This is evidence that the rotational invariance is violated in the CMB bispectrum and implies the existence of the signals not obeying the triangle condition of the above Wigner symbol as $|\ell_2 - \ell_3| \leq \ell_1 \leq \ell_2 + \ell_3$. 
We demonstrated that the signals of the CMB bispectrum for $\ell_1 = \ell_2 + \ell_3 + 2, |\ell_2 - \ell_3| -2$ and these two permutations do not vanish. In fact, the statistically isotropic bispectra are exactly zero for these configurations; hence, these signals have the pure information of the statistical anisotropy. 
Because the amplitudes of these intensity bispectra are comparable to those for $\ell_1 = \ell_2 + \ell_3$, it might be possible to detect these contributions of the statistical anisotropy in future experiments, which would give us novel information about the physics of the early Universe.
Of course, also for the $E$-mode polarization, we can give the same discussions and results. 

Although we assume a specific potential of inflation to show the statistical anisotropy on the CMB bispectrum explicitly, the above calculation and discussion will be applicable to other inflation models where the rotational invariance violates.

\bibliographystyle{JHEP}
\bibliography{paper}

%% file: ch8_pal.tex
\section{Parity violation of gravitons in the CMB bispectrum}\label{sec:parity_violating}


Non-Gaussian features in the cosmological perturbations include detailed 
information on the nature of the early Universe, and 
there have
been many works that attempt to extract them from the bispectrum (three-point function)
of the cosmic microwave background (CMB) anisotropies
(e.g., Refs.~\cite{Komatsu:2001rj, Bartolo:2004if, Babich:2004gb, Komatsu:2010fb}). 
However, most of these discussions are limited in the cases that the
scalar-mode contribution dominates in the non-Gaussianity and also
are based on the assumption of rotational invariance and parity conservation. 

In contrast, there are several studies on the non-Gaussianities of not
only the scalar-mode perturbations but also the vector- and
tensor-mode perturbations \cite{Maldacena:2002vr, Brown:2005kr,
Adshead:2009bz}. 
These sources produce the additional signals on the CMB bispectrum
\cite{Shiraishi:2010kd} and can give a dominant contribution by 
considering such highly non-Gaussian sources as the stochastic magnetic fields \cite{Shiraishi:2011dh}.
Furthermore, even in the CMB bispectrum induced from the scalar-mode non-Gaussianity, 
if the rotational invariance is violated in the non-Gaussianity, 
the characteristic signals appear \cite{Shiraishi:2011ph}. 
Thus, it is very important to clarify these less-noted signals to understand the precise picture of the early Universe. 

Recently, the parity violation in the graviton non-Gaussianities has 
been discussed in Refs.~\cite{Maldacena:2011nz, Soda:2011am}. 
Maldacena and
Pimentel first calculated the primordial bispectrum of the gravitons
sourced from parity-even (parity-conserving) and parity-odd (parity-violating) Weyl
cubic terms, namely, $W^3$ and $\widetilde{W}W^2$, respectively, by making use of the spinor
helicity formalism.\cite{Maldacena:2011nz} Soda $\it et \ al.$ proved that the parity-violating
non-Gaussianity of the primordial gravitational waves induced from $\widetilde{W}W^2$ emerges not in the
exact de-Sitter space-time but in the quasi de-Sitter space-time, and hence, its
amplitude is proportional to a slow-roll parameter.\cite{Soda:2011am} In these studies,
the authors assume that the coupling constant of the Weyl cubic terms is independent of time.

In this section, we estimate the primordial
non-Gaussianities of gravitons generated from $W^3$ and
$\widetilde{W}W^2$ with the time-dependent coupling parameter~\cite{Weinberg:2008hq}. We consider the case where the coupling is given by a power of the
conformal time. We show that in such a model, the parity violation in the non-Gaussianity of the primordial gravitational waves would not vanish even in the exact de-Sitter space-time. The effects of the parity violation on the CMB power spectrum have been well-studied, where an attractive result is that the cross-correlation between 
the intensity and $B$-mode polarization is generated \cite{Alexander:2004wk, Saito:2007kt, Gluscevic:2010vv, Sorbo:2011rz}. 
On the other hand, in the CMB bispectrum, owing to the mathematical property of the spherical harmonic function, 
the parity-even and parity-odd signals should arise from just the
opposite configurations of multipoles \cite{Okamoto:2002ik, Kamionkowski:2010rb}. Then, we 
formulate and numerically calculate the CMB bispectra induced by these
non-Gaussianities that contain all the correlations between the intensity ($I$)
and polarizations ($E,B$) and
show that 
the signals from $W^3$ (parity-conserving)
appear in the configuration of the multipoles
where those from $\widetilde{W}W^2$ (parity-violating) vanish and vice versa. These discussions are based on Ref.~\cite{Shiraishi:2011st}. 

This section is organized as follows. In the next subsection, we derive the
primordial bispectrum of gravitons induced by $W^3$ 
and $\widetilde{W}W^2$ with the coupling constant proportional to
the power of the conformal time. In Sec.~\ref{subsec:CMB_bis_PV}, we calculate
the CMB bispectra sourced from these non-Gaussianities, analyze their
behavior and find some peculiar signatures of the parity violation. 
The final subsection is devoted to summary and discussion. In Appendices
\ref{appen:pol_tens} and \ref{appen:weyl_in-in}, we describe the detailed calculations of the contractions of the polarization tensors and unit vectors, and of the intial bispectra by the in-in formalism. 

Throughout this section, we use $M_{\rm pl} \equiv 1/\sqrt{8 \pi G}$,
where $G$ is the Newton constant and the rule that all the Greek characters and alphabets run from $0$ to $3$ and from $1$ to $3$, respectively. 

\subsection{Parity-even and -odd non-Gaussianity of gravitons}

In this subsection, we formulate the primordial non-Gaussianity of gravitons
generated from the Weyl cubic terms with the running coupling constant as a
function of a conformal time, $f(\tau)$, whose action is given by
\begin{eqnarray}
S = \int d\tau d^3x \frac{f(\tau)}{\Lambda^2}
\left( \sqrt{-g} W^3  +  \widetilde{W}W^2 \right)~, \label{eq:action_PV}
\end{eqnarray}
with
\begin{eqnarray}
\begin{split}
W^3 &\equiv W^{\alpha \beta}{}_{\gamma \delta} W^{\gamma
 \delta}{}_{\sigma \rho} W^{\sigma \rho}{}_{\alpha \beta}~, \\
\widetilde{W}W^2 &\equiv \epsilon^{\alpha \beta \mu \nu} 
W_{\mu \nu \gamma \delta} W^{\gamma \delta}{}_{\sigma \rho} 
W^{\sigma \rho}{}_{\alpha \beta}~,
\end{split}
\end{eqnarray}
where $W^{\alpha \beta}{}_{\gamma \delta}$ denotes the Weyl tensor,
$\epsilon^{\alpha \beta \mu \nu}$ is a 4D Levi-Civita tensor
normalized as $\epsilon^{0123} = 1$, and $\Lambda$ is a scale that sets the
value of the higher derivative corrections \cite{Maldacena:2011nz}. 
Note that $W^3$ and $\widetilde{W}W^2$ have the even and odd parities,
respectively. 
In the following discussion, we assume that the coupling constant is given by
\begin{eqnarray}
f(\tau) = \left( \frac{\tau}{\tau_*} \right)^A ~,
\end{eqnarray} 
where $\tau$ is a conformal time.
Here, we have set $f(\tau_*) = 1$.
Such a coupling can be readily realized by considering a dilaton-like coupling in the slow-roll inflation as discussed in Sec.~\ref{subsubsec:coupling}. 

\subsubsection{Calculation of the primordial bispectrum}

Here, let us focus on the calculation of the primordial
bispectrum induced by $W^3$ and $\widetilde{W}W^2$ of
Eq.~(\ref{eq:action_PV}) on the exact de-Sitter space-time in a more straightforward manner than those of
Refs.~\cite{Maldacena:2011nz, Soda:2011am}. 

At first, we consider the tensor perturbations on the Friedmann-Lemaitre-Robertson-Walker metric as
\begin{eqnarray}
ds^2 = a^2 ( -d\tau^2 + e^{\gamma_{ij}} dx^i dx^j)~,
\end{eqnarray}
where $a$ denotes the scale factor and $\gamma_{ij}$ obeys the transverse traceless conditions;
$\gamma_{ii} = \partial \gamma_{ij}/ \partial x^j =0$
\footnote{$\gamma_{ij}$ is identical to $h_{ij}$ in Sec.~\ref{sec:inflation}.}. Up to the second order, even if the action includes the Weyl cubic terms
given by Eq.~(\ref{eq:action_PV}),
the gravitational wave obeys the action as
\cite{Maldacena:2011nz, Soda:2011am}
\begin{eqnarray}
S = \frac{M_{\rm pl}^2}{8} \int d\tau dx^3 a^2(\dot{\gamma}_{ij}
 \dot{\gamma}_{ij} - \gamma_{ij,k} \gamma_{ij,k})~,
\end{eqnarray}
where $~\dot~ \equiv \partial/\partial \tau$ and $_{,i} \equiv \partial/\partial x^i$. 
We expand the gravitational wave with a transverse and traceless
polarization tensor $e^{(\lambda)}_{ij}$ and
the creation and annihilation operators $a^{(\lambda) \dagger},
a^{(\lambda)}$ as 
\begin{eqnarray}
\gamma_{ij}({\bf x}, \tau) &=& \int \frac{d^3 {\bf k}}{(2\pi)^3}
 \sum_{\lambda = \pm 2} \gamma_{dS}(k,\tau) a^{(\lambda)}_{{\bf k}} e^{(\lambda)}_{ij}(\hat{{\bf k}}) e^{i {\bf k} \cdot {\bf x}} + h.c. \nonumber \\
&=& \int \frac{d^3 {\bf k}}{(2\pi)^3} \sum_{\lambda = \pm 2} 
\gamma^{(\lambda)}({\bf k}, \tau)
e^{(\lambda)}_{ij}(\hat{{\bf k}}) e^{i {\bf k} \cdot {\bf x}}~, 
\end{eqnarray}
with 
\begin{eqnarray}
\gamma^{(\lambda)}({\bf k}, \tau) 
&\equiv& \gamma_{dS}(k,\tau) a^{(\lambda)}_{{\bf k}} 
+  \gamma^*_{dS}(k,\tau) a^{(\lambda) \dagger}_{-{\bf k}}~.
\end{eqnarray}
Here, $\lambda \equiv \pm 2$ denotes the helicity of the gravitational wave and we use the polarization tensor satisfying the relations as Eq.~(\ref{eq:pol_tens_relation}). 
The creation and annihilation operators
$a^{(\lambda) \dagger}, a^{(\lambda)}$ obey the relations as
\begin{eqnarray}
\begin{split}
a^{(\lambda)}_{\bf k} \Ket{0} &= 0~, \\
\left[ a^{(\lambda)}_{\bf k}, a^{(\lambda') \dagger}_{\bf k'}
\right] &= (2\pi)^3 \delta({\bf k} - {\bf k'}) \delta_{\lambda,
\lambda'}~, \label{eq:a}
\end{split}
\end{eqnarray}
where $\ket{0}$ denotes a vacuum eigenstate. 
Then, the mode function of gravitons on the de Sitter space-time
$\gamma_{dS}$ satisfies the field equation as
\begin{eqnarray}
  \ddot{\gamma}_{dS} - \frac{2}{\tau} \dot{\gamma}_{dS} + k^2 \gamma_{dS}  = 0~,
\end{eqnarray}
and a solution is given by
\begin{eqnarray}
\gamma_{dS} = i \frac{H}{M_{\rm pl} } 
\frac{e^{- i k \tau}}{k^{3/2}} (1 + ik\tau)~,
\label{eq:sol}
\end{eqnarray}
where $H = - (a\tau)^{-1}$ is the Hubble parameter and has a constant value
 in the exact de Sitter space-time.

On the basis of the in-in formalism (see, e.g.,
Refs.~\cite{Maldacena:2002vr, Weinberg:2005vy}) and the above results, we calculate the tree-level
bispectrum of gravitons on the late-time limit. According to this
formalism, the expectation value of an operator depending on time in the interaction picture,
$O(t)$, is written as
\begin{eqnarray}
\Braket{O(t)} = \Braket{ 0 |\bar{T} e^{i \int H_{int}(t') dt'} O(t) 
T e^{-i \int H_{int}(t') dt'} | 0 } ~,
\end{eqnarray} 
where $T$ and $\bar{T}$ are respectively time-ordering and anti-time-ordering operators
and  $H_{int}(t)$ is the interaction Hamiltonian.
Applying this
equation, the primordial bispectrum of gravitons at the tree level can be
expressed as 
\begin{eqnarray}
\Braket{\prod_{n=1}^3 \gamma^{(\lambda_n)}({\bf k_n},\tau)} 
= i \int_{- \infty}^\tau d\tau' \Braket{ 0 | 
\left[: H_{int}(\tau'):, \prod_{n=1}^3
 \gamma^{(\lambda_n)}({\bf k_n},\tau) \right] | 0 }, \label{eq:in-in_formalism} 
\end{eqnarray} 
where $:~:$ denotes normal product. 

Up to the first order with respect to $\gamma_{ij}$, the nonzero components of the Weyl tensor are written as
\begin{eqnarray}
\begin{split}
W^{0i}{}_{0j} &= \frac{1}{4}(H\tau)^2 \gamma_{ij, \alpha \alpha}~, \\
W^{ij}{}_{0k} &= \frac{1}{2}(H\tau)^2(\dot{\gamma}_{ki,j} -
 \dot{\gamma}_{kj,i})~, \\
W^{0i}{}_{jk} &= \frac{1}{2}(H\tau)^2(\dot{\gamma}_{ik,j} -
 \dot{\gamma}_{ij,k})~, \\
W^{ij}{}_{kl} &= \frac{1}{4}(H\tau)^2
(-\delta_{ik} \gamma_{jl,\alpha \alpha} + \delta_{il}
\gamma_{jk,\alpha \alpha} + \delta_{jk} \gamma_{il,\alpha \alpha} -
\delta_{jl} \gamma_{ik,\alpha \alpha} )~, \label{eq:W}
\end{split}
\end{eqnarray}
where $\gamma_{ij, \alpha \alpha} \equiv \ddot{\gamma}_{ij} + \nabla^2
\gamma_{ij}$. Then $W^3$ and $\widetilde{W}W^2$ respectively reduce to 
\begin{eqnarray}
\begin{split}
W^3 &= W^{ij}{}_{kl} W^{kl}{}_{mn} W^{mn}{}_{ij} + 6 W^{0i}{}_{jk} W^{jk}{}_{lm} W^{lm}{}_{0i}  \\
&\quad + 12 W^{0i}{}_{0j} W^{0j}{}_{kl}  W^{kl}{}_{0i} + 8 W^{0i}{}_{0j}
 W^{0j}{}_{0k}  W^{0k}{}_{0i} ~, \\
\widetilde{W}W^2 
&= 4 \eta^{ijk} 
\left[ W_{jkpq} 
\left( W^{pq}{}_{lm} W^{lm}{}_{0i} + 2 W^{pq}{}_{0 m}
 W^{0m}{}_{0i} \right) \right.  \\
&\quad\qquad \left. + 2 W_{jk0p} \left( W^{0p}{}_{lm} W^{lm}{}_{0i} + 2 W^{0p}{}_{0 m} W^{0m}{}_{0i} \right) \right]~,
\end{split}
\end{eqnarray}
where $\eta^{ijk} \equiv \epsilon^{0ijk}$. 
Using the above expressions and $\int d\tau H_{int} = -S_{int}$,
up to the third order,
the interaction Hamiltonians of $W^3$ and $\widetilde{W}W^2$ are respectively given by
\begin{eqnarray}
\begin{split}
H_{W^3} &= - \int d^3x \Lambda^{-2} 
 (H \tau)^2 \left(\frac{\tau}{\tau_*}\right)^A  \\
&\quad\times \frac{1}{4} \gamma_{ij,\alpha \alpha} 
\left[ 
 \gamma_{jk,\beta \beta} \gamma_{ki,\sigma \sigma}
 + 6 \dot{\gamma}_{kl,i}\dot{\gamma}_{kl,j} 
 + 6 \dot{\gamma}_{ik,l}\dot{\gamma}_{jl,k}
 - 12 \dot{\gamma}_{ik,l}\dot{\gamma}_{kl,j}
\right]~,  \\
H_{\widetilde{W}W^2} 
&= - \int d^3 x 
 \Lambda^{-2} (H \tau)^2 \left(\frac{\tau}{\tau_*}\right)^A \\
&\quad\times
\eta^{ijk}
\left[ \gamma_{kq,\alpha \alpha} 
( - 3 
{\gamma}_{jm,\beta \beta}  \dot{\gamma}_{iq, m}
+ {\gamma}_{mi,\beta \beta } \dot{\gamma}_{mq,j} ) + 4 \dot{\gamma}_{pj,k}
\dot{\gamma}_{pm,l} ( \dot{\gamma}_{il,m} - \dot{\gamma}_{im,l} )
\right]~. \label{eq:H_int}
\end{split}
\end{eqnarray}
Substituting the above expressions into Eq.~(\ref{eq:in-in_formalism}), using the solution given by Eq.~(\ref{eq:sol}), and considering the
late-time limit as $\tau \rightarrow 0$, 
we can obtain an explicit form of the primordial bispectra: 
\begin{eqnarray}
\Braket{\prod_{n=1}^3 \gamma^{(\lambda_n)}({\bf k_n})}_{int}
=  (2\pi)^3 \delta \left(\sum_{n=1}^3 {\bf k_n}\right)
f^{(r)}_{int}(k_1, k_2, k_3) 
f^{(a)}_{int}(\hat{\bf k_1}, \hat{\bf k_2}, \hat{\bf k_3})~, \label{eq:ggg}
\end{eqnarray}
with\footnote{Here, we set that $\tau_* < 0$.}
\begin{eqnarray}
\begin{split}
f^{(r)}_{W^3} 
&= 8 \left(\frac{H}{M_{\rm pl}}\right)^6 \left( \frac{H}{\Lambda} \right)^2 
{\rm Re} \left[\tau_*^{-A} \int^0_{-\infty}d\tau' {\tau'}^{5+A} e^{-ik_t \tau'} \right] ~,  \\
f^{(a)}_{W^3}
&= e_{ij}^{(-\lambda_1)} 
\left[ \frac{1}{2} e_{jk}^{(-\lambda_2)} e_{ki}^{(-\lambda_3)} 
+ \frac{3}{4} e_{kl}^{(-\lambda_2)} e_{kl}^{(-\lambda_3)} \hat{k_2}_i \hat{k_3}_j 
\right.  \\
&\qquad\qquad \left. 
+ \frac{3}{4}  e_{ki}^{(-\lambda_2)} e_{jl}^{(-\lambda_3)} \hat{k_2}_l \hat{k_3}_k
- \frac{3}{2}  e_{ik}^{(-\lambda_2)} e_{kl}^{(-\lambda_3)} \hat{k_2}_l \hat{k_3}_j
\right] + 5 \ {\rm perms} , \\
f_{\widetilde{W}W^2}^{(r)} 
&=  
8 \left(\frac{H}{M_{\rm pl}}\right)^6 \left( \frac{H}{\Lambda} \right)^2 
{\rm Im} 
\left[\tau_*^{-A} \int^0_{-\infty}d\tau' {\tau'}^{5+A} e^{-ik_t \tau'} \right] ~, \\
f_{\widetilde{W}W^2}^{(a)}
&= i \eta^{ijk} 
\left[ e_{kq}^{(-\lambda_1)} 
\left\{- 3 e_{jm}^{(-\lambda_2)} e_{iq}^{(-\lambda_3)}
\hat{k_3}_m + e_{mi}^{(-\lambda_2)} e_{mq}^{(-\lambda_3)} \hat{k_3}_j \right\} \right. \\
&\qquad\quad \left. 
+ e_{pj}^{(-\lambda_1)} e_{pm}^{(-\lambda_2)}  
\hat{k_1}_k \hat{k_2}_l
\left\{ e_{il}^{(-\lambda_3)} \hat{k_3}_m 
- e_{im}^{(-\lambda_3)}  \hat{k_3}_l \right\} \right] + 5 \ {\rm perms}~. \label{eq:ggg_f}
\end{split}
\end{eqnarray}
Here, 
$k_t \equiv \sum_{n=1}^3 k_n$,
$int = W^3 $ and $\widetilde{W}W^2$, 
``5 perms'' denotes the five symmetric
terms under the permutations of $(\hat{{\bf k_1}}, \lambda_1),
(\hat{{\bf k_2}}, \lambda_2)$, and $(\hat{{\bf k_3}}, \lambda_3)$. 
From the above expressions,
we find that the bispectra of the primordial gravitational wave induced from $W^3$
and $\widetilde{W}W^2$ are proportional to the real and imaginary parts of 
$\tau_*^{-A} \int_{-\infty}^0 d\tau' \tau'^{5+A} e^{- i k_t \tau'}$,
respectively. 
This difference comes from the number of $\gamma_{ij ,\alpha
\alpha}$ and $\dot{\gamma}_{ij, k}$. 
$H_{W^3}$ consists of the products of an odd number of the former terms 
and an even number of the latter terms. 
On the other hand, in
$H_{\widetilde{W}W^2}$, the situation is the opposite. 
Since the former and latter terms contain 
$\ddot{\gamma}_{dS} - k^2 \gamma_{dS} 
= (2 H \tau' / M_{\rm pl}) k^{3/2} e^{-ik\tau'}$ and $\dot{\gamma}_{dS} 
= i ( H \tau' / M_{\rm pl}) k^{1/2} e^{-ik\tau'}$,
respectively, the total numbers of $i$ are different in each time
integral. Hence, the contributions of the real and imaginary parts roll upside
down in $f_{W^3}^{(r)}$ and $f_{\widetilde{W}W^2}^{(r)}$. 
Since the time integral in the bispectra can be analytically evaluated as
\begin{eqnarray}
\tau_*^{-A} \int_{-\infty}^0 d\tau' 
\tau'^{5+A} e^{- i k_t \tau'} 
= \left[ \cos\left( \frac{\pi}{2}A \right) 
+ i \sin \left(\frac{\pi}{2}A\right) \right]\Gamma(6+A) 
k_t^{-6}(-k_t \tau_*)^{-A}~,
\end{eqnarray}
$f_{W^3}^{(r)}$ and $f_{\widetilde{W}W^2}^{(r)}$ reduce to
\begin{eqnarray}
f^{(r)}_{W^3} 
&=& 8 \left(\frac{H}{M_{\rm pl}}\right)^6 
\left( \frac{H}{\Lambda} \right)^2 
\cos\left(\frac{\pi}{2}A\right) \Gamma(6+A) k_t^{-6}(-k_t \tau_*)^{-A}
~,  \label{eq:radial_w3} \\
f_{\widetilde{W}W^2}^{(r)} 
&=&  
8 \left(\frac{H}{M_{\rm pl}}\right)^6 
\left( \frac{H}{\Lambda} \right)^2 
\sin\left(\frac{\pi}{2}A\right) \Gamma(6+A) k_t^{-6}(-k_t \tau_*)^{-A}
~, \label{eq:radial_ww2}
\end{eqnarray}
where $\Gamma(x)$ is the Gamma function. For more detailed derivation of the graviton bispectrum, see Appendix \ref{appen:weyl_in-in}. 

From this equation, we can see that
in the case of the time-independent coupling, which corresponds to the $A = 0$ case,
the bispectrum from $\widetilde{W}W^2$ vanishes.
This is consistent with a claim in
Ref.~\cite{Soda:2011am}
\footnote{
In Ref.~\cite{Soda:2011am}, the authors have shown that
for $A=0$, the bispectrum from $\widetilde{W}W^2$ has a nonzero value
upward in the first order of the slow-roll parameter.}. 
On the other hand, interestingly, if $A$ deviates
 from $0$, it is possible to realize the nonzero bispectrum
 induced from $\widetilde{W}W^2$ even in the exact
 de Sitter limit. Thus, we expect the signals from $\widetilde{W}W^2$
 without the slow-roll suppression, which can be comparable to those from $W^3$
 and become sufficiently large to observe in the CMB.

\subsubsection{Running  coupling constant} \label{subsubsec:coupling}

Here, we discuss how to realize $f \propto \tau^A$ within the framework of the standard slow-roll inflation. 
During the standard slow-roll inflation,
the equation of motion of the scalar field $\phi$, which has a potential $V$,
 is expressed as 
\begin{eqnarray}
\dot{\phi} \simeq \pm \sqrt{2 \epsilon_\phi} M_{\rm pl} \tau^{-1} ~,
\end{eqnarray}
where $\epsilon_\phi \equiv [\partial V / \partial \phi /
(3 M_{\rm pl} H^2)]^2 / 2 $ is a slow-roll parameter
for $\phi$, 
$+$ and $-$ signs are taken to be for $\partial V/ \partial \phi > 0$ and $\partial V/ \partial \phi < 0$, respectively, and we have assumed that $ aH = -1/\tau$. 
The solution of the above equation is given by 
\begin{eqnarray}
\phi = \phi_* \pm \sqrt{2 \epsilon_\phi} M_{\rm pl} \ln \left( \frac{\tau}{\tau_*} \right)~.
\end{eqnarray}
Hence, if we assume a dilaton-like coupling as 
$f \equiv e^{(\phi - \phi_*) / M}$, we have 
\begin{eqnarray}
f(\tau) = \left( \frac{\tau}{\tau_*} \right)^A ~, \ \ 
A = \pm \sqrt{2 \epsilon_\phi} \frac{M_{\rm pl}}{M}~, \label{eq:coupling_moduli}
\end{eqnarray} 
where $M$ is an arbitrary energy scale.
Let us take $\tau_*$ to be a time when
the scale of the present horizon of the Universe exits
the horizon during inflation, namely,
$ \left| \tau_* \right| = k_*^{-1} \sim 14 {\rm Gpc}$.
Then, the coupling $f$, which determines the amplitude of
the bispectrum of the primordial gravitational wave induced
from the Weyl cubic terms, is on the order of unity for the
current cosmological scales.
From Eq.~(\ref{eq:coupling_moduli}),
we have $A = \pm 1/2$ with $M = \sqrt{8\epsilon_\phi} M_{\rm pl}$. 
As seen in Eqs.~(\ref{eq:radial_w3}) and (\ref{eq:radial_ww2}), this leads
 to an interesting situation that the bispectra from $W^3$ and
 $\widetilde{W}W^2$ have a comparable magnitude as $f^{(r)}_{W^3} = \pm f^{(r)}_{\widetilde{W}W^2}$. 
 Hence, we can expect that in the CMB bispectrum, the signals from these
 terms are almost the same. 

In the next subsection, we demonstrate these through the explicit calculation of the CMB bispectra. 

\subsection{CMB parity-even and -odd bispectrum} \label{subsec:CMB_bis_PV}

In this subsection, following the calculation approach discussed in
Sec.~\ref{sec:maldacena}, we formulate the CMB bispectrum induced
from the non-Gaussianities of gravitons sourced by $W^3$ and $\widetilde{W}W^2$ terms discussed in the
previous subsection.

\subsubsection{Formulation}\label{subsubsec:formulation} 

Conventionally, the CMB fluctuation is expanded with the spherical harmonics as 
\begin{eqnarray}
\frac{\Delta X (\hat{{\bf n}})}{X} = \sum_{\ell m} a_{X, \ell m} Y_{\ell
 m}(\hat{{\bf n}})~, \label{eq:cmb_anisotropy}
\end{eqnarray}
where $\hat{{\bf n}}$ is a unit vector pointing toward a line-of-sight direction, and $X$ means the intensity ($\equiv I$) and
the electric and magnetic polarization modes ($\equiv E, B$). 
By performing the line-of-sight integration, the coefficient, $a_{\ell m}$,
generated from the primordial fluctuation of gravitons, $\gamma^{(\pm 2)}$, is
given by [corresponding to Eq.~(\ref{eq:alm})]
\begin{eqnarray}
a_{X, \ell m} &=& 4\pi (-i)^\ell \int_0^\infty \frac{k^2 dk}{(2\pi)^3}
 {\cal T}_{X,\ell}(k) \sum_{\lambda = {\pm 2}} 
\left(\frac{\lambda}{2}\right)^x \gamma_{\ell m}^{(\lambda)}(k)~, \\
\gamma_{\ell m}^{(\lambda)}(k)  &\equiv& \int d^2 \hat{{\bf k}} 
\gamma^{(\lambda)}({\bf k}) 
{}_{-\lambda}Y^*_{\ell m}(\hat{{\bf k}})~, \label{eq:gamma_lm}
\end{eqnarray}
where $x$ discriminates the parity of three modes: $x = 0$ for $X = I,E$
and $x=1$ for $X = B$, and ${\cal T}_{X, \ell}$ is the time-integrated
transfer function of tensor modes (\ref{eq:transfer_explicit}). Like Eq.~(\ref{eq:cmb_bis_general}), we can obtain the CMB bispectrum generated from
the primordial bispectrum of gravitons as 
\begin{eqnarray}
\Braket{\prod_{n=1}^3 a_{X_n, \ell_n m_n}} 
&=&  \left[\prod_{n=1}^3 4\pi (-i)^{\ell_n} \int \frac{k_n^2 dk_n}{(2\pi)^3}
 {\cal T}_{X_n,\ell_n}(k_n) \sum_{\lambda_n = \pm 2} 
\left( \frac{\lambda_n}{2} \right)^{x_n} \right] \nonumber \\
&&\times
\Braket{\prod_{n=1}^3 \gamma_{\ell_n m_n}^{(\lambda_n)}(k_n) }. 
\label{eq:cmb_bis_form}
\end{eqnarray}

In order to derive an explicit form of this CMB bispectrum, at first,
 we need to express all the functions containing the angular dependence
on the wave number vectors with the spin spherical harmonics. 
Using the results of Appendix \ref{appen:pol_tens},  
$f^{(a)}_{W^3}$ and $f^{(a)}_{\widetilde{W}W^2}$ can be calculated as 
\begin{eqnarray}
f_{W^3}^{(a)}
&=&  
\left( 8\pi \right)^{3/2} 
\sum_{L', L'' = 2, 3} \sum_{M, M', M''} 
\left(
  \begin{array}{ccc}
   2 & L' & L'' \\
  M & M' & M''
  \end{array}
 \right) 
\nonumber \\
&&\times 
{}_{\lambda_1}Y_{2 M}^*(\hat{{\bf k_1}}) 
{}_{\lambda_2}Y_{L' M'}^*(\hat{{\bf k_2}}) {}_{\lambda_3}Y_{L''
M''}^*(\hat{{\bf k_3}}) 
\nonumber \\
&&\times 
\left[- \frac{1}{20}\sqrt{\frac{7}{3}} \delta_{L', 2} \delta_{L'', 2} 
+ (-1)^{L'} I_{L' 1 2}^{\lambda_2 0 -\lambda_2} I_{L'' 1 2}^{\lambda_3 0
-\lambda_3}
 \right. \nonumber \\
&&\qquad \left. 
 \times 
\left(  
- \frac{\pi}{5}
\left\{
  \begin{array}{ccc}
   2 & L' & L'' \\
   2 & 1 & 1 
  \end{array}
 \right\} 
- \pi 
\left\{
  \begin{array}{ccc}
   2 & L' & L'' \\
   1 & 1 & 2 \\
   1 & 2 & 1
  \end{array}
 \right\} \right. \right. \nonumber \\
&&\qquad\quad \left. \left. 
+ 2\pi 
\left\{
  \begin{array}{ccc}
   2 & 1 & L' \\
   2 & 1 & 1 
  \end{array}
 \right\}
\left\{
  \begin{array}{ccc}
   2 & L' & L'' \\
   2 & 1 & 1 
  \end{array}
 \right\}
\right)
\right] + 5 \ {\rm perms}
~, \label{eq:fa_w3} \\
f_{\widetilde{W}W^2}^{(a)}
&=& \left( 8\pi \right)^{3/2}  
\sum_{L', L'' = 2, 3} \sum_{M, M', M''} 
\left(
  \begin{array}{ccc}
   2 & L' & L'' \\
  M & M' & M''
  \end{array}
 \right) \nonumber \\
&&\times
{}_{\lambda_1}Y_{2 M}^*(\hat{{\bf k_1}}) 
{}_{\lambda_2}Y_{L' M'}^*(\hat{{\bf k_2}}) {}_{\lambda_3}Y_{L''
M''}^*(\hat{{\bf k_3}}) (-1)^{L''} I_{L'' 1 2}^{\lambda_3 0 -\lambda_3} 
\nonumber \\
&& \times 
\left[
\delta_{L',2} 
\left( 3 \sqrt{\frac{2\pi}{5}} 
\left\{
  \begin{array}{ccc}
   2 & 2 & L'' \\
   1 & 2 & 1 
  \end{array}
 \right\}
- 2 \sqrt{2\pi} 
\left\{
  \begin{array}{ccc}
   2 & 2 & L'' \\
   1 & 1 & 1 \\
   1 & 1 & 2
  \end{array}
 \right\} \right)
\right. \nonumber \\
&&\quad \left.  
+ \frac{\lambda_1}{2} 
I_{L' 1 2}^{\lambda_2 0 -\lambda_2} 
\left( - \frac{4 \pi}{3} 
\left\{
  \begin{array}{ccc}
   2 & L' & L'' \\
   1 & 2 & 1 \\
   1 & 1 & 2
  \end{array}
\right\}
+ \frac{2\pi}{15} \sqrt{\frac{7}{3}} 
\left\{
  \begin{array}{ccc}
   2 & L' & L'' \\
   1 & 2 & 2    
  \end{array}
\right\}
\right)
\right] \nonumber \\
&&+ 5 \ {\rm perms} ~, \label{eq:fa_ww2}
\end{eqnarray}
where the $2 \times 3$ matrix of a bracket, and the $2 \times 3$ and $3
\times 3$ matrices of a curly bracket denote the Wigner-$3j, 6j$ and $9j$
symbols, respectively, and 
\begin{eqnarray}
I^{s_1 s_2 s_3}_{l_1 l_2 l_3} 
\equiv \sqrt{\frac{(2 l_1 + 1)(2 l_2 + 1)(2 l_3 + 1)}{4 \pi}}
\left(
  \begin{array}{ccc}
  l_1 & l_2 & l_3 \\
   s_1 & s_2 & s_3 
  \end{array}
 \right)~.
\end{eqnarray} 
The delta function is also expanded as 
\begin{eqnarray}
\delta\left( \sum_{n=1}^3 {{\bf k_n}} \right) 
&=& 8 \int_0^\infty y^2 dy 
\left[ \prod_{n=1}^3 \sum_{L_n M_n} 
 (-1)^{L_n/2} j_{L_n}(k_n y) 
Y_{L_n M_n}^*(\hat{{\bf k_n}}) \right] 
\nonumber \\
&&\times 
I_{L_1 L_2 L_3}^{0 \ 0 \ 0}
 \left(
  \begin{array}{ccc}
  L_1 & L_2 & L_3 \\
  M_1 & M_2 & M_3 
  \end{array}
 \right)~.
\end{eqnarray}
Next, we integrate all the spin spherical harmonics over
$\hat{\bf k_1}, \hat{\bf k_2}, \hat{\bf k_3}$ as
\begin{eqnarray}
\begin{split}
\int d^2 \hat{\bf k_1} {}_{- \lambda_1}Y_{\ell_1 m_1}^* Y_{L_1 M_1}^* {}_{\lambda_1} Y_{2M}^* &= I_{\ell_1 L_1 2}^{\lambda_1 0 -\lambda_1}
\left(
  \begin{array}{ccc}
  \ell_1 & L_1 & 2 \\
  m_1 & M_1 & M 
  \end{array}
 \right) ~, \\
\int d^2 \hat{\bf k_2} {}_{- \lambda_2}Y_{\ell_2 m_2}^* Y_{L_2 M_2}^* {}_{\lambda_2} Y_{L' M'}^* &= I_{\ell_2 L_2 L'}^{\lambda_2 0 -\lambda_2}
\left(
  \begin{array}{ccc}
  \ell_2 & L_2 & L' \\
  m_2 & M_2 & M' 
  \end{array}
 \right) ~, \\
\int d^2 \hat{\bf k_3} {}_{- \lambda_3}Y_{\ell_3 m_3}^* Y_{L_3 M_3}^* {}_{\lambda_3} Y_{L'' M''}^* &= I_{\ell_3 L_3 L''}^{\lambda_3 0 -\lambda_3}
\left(
  \begin{array}{ccc}
  \ell_3 & L_3 & L'' \\
  m_3 & M_3 & M'' 
  \end{array}
 \right) ~.
\end{split}
\end{eqnarray}
Through the summation over the azimuthal
quantum numbers, the product of the above five Wigner-$3j$ symbols is expressed with
the Wigner-$9j$ symbols as
\begin{eqnarray}
&& \sum_{\substack{M_1 M_2 M_3 \\ M M' M''}}
\left(
  \begin{array}{ccc}
  L_1 &  L_2 & L_3 \\
   M_1 & M_2 & M_3
  \end{array}
 \right)
\left(
  \begin{array}{ccc}
  2 &  L' & L'' \\
   M & M' & M''
  \end{array}
 \right) 
\nonumber \\
&&\qquad \times
\left(
  \begin{array}{ccc}
  \ell_1 &  L_1 & 2 \\
   m_1 & M_1 & M
  \end{array}
 \right)
\left(
  \begin{array}{ccc}
  \ell_2 &  L_2 & L' \\
   m_2 & M_2 & M'
  \end{array}
 \right)
\left(
  \begin{array}{ccc}
  \ell_3 &  L_3 & L'' \\
   m_3 & M_3 & M''
  \end{array}
 \right) \nonumber \\
&& \qquad\qquad\qquad = 
\left(
  \begin{array}{ccc}
  \ell_1 & \ell_2 & \ell_3 \\
   m_1 & m_2 & m_3
  \end{array}
 \right)
\left\{
  \begin{array}{ccc}
  \ell_1 & \ell_2 & \ell_3 \\
   L_1 & L_2 & L_3 \\
   2 & L' & L'' \\
  \end{array}
 \right\}~.
\end{eqnarray} 
Finally, performing the summation over the helicities, namely $\lambda_1, \lambda_2$ and $\lambda_3$, as 
\begin{eqnarray}
\begin{split}
\sum_{\lambda = \pm 2} 
\left(\frac{\lambda}{2}\right)^{x}
I_{\ell L 2}^{\lambda 0 -\lambda} 
&=
\begin{cases}
2 I_{\ell L 2}^{2 0 -2}  & 
( \ell + L + x = {\rm even} ) \\
0 & ( \ell + L + x = {\rm odd} )
\end{cases} ~, \\
\sum_{\lambda = \pm 2} 
\left(\frac{\lambda}{2}\right)^{x}
I_{\ell L L'}^{\lambda 0 -\lambda}
I_{L' 1 2}^{\lambda 0 -\lambda}
&=
\begin{cases}
2 I_{\ell L L'}^{2 0 -2} I_{L' 1 2}^{2 0 -2} & 
( \ell + L + x = {\rm odd} ) \\
0 & ( \ell + L + x = {\rm even} ) 
\end{cases} ~, \\
\sum_{\lambda = \pm 2} 
\left(\frac{\lambda}{2}\right)^{x+1}
I_{\ell L 2}^{\lambda 0 -\lambda} 
&=
\begin{cases}
2 I_{\ell L 2}^{2 0 -2}  & 
( \ell + L + x = {\rm odd} ) \\
0 & ( \ell + L + x = {\rm even} )
\end{cases} ~,
\end{split}
\end{eqnarray}
and considering the selection rules of the Wigner symbols as described in Appendix \ref{appen:wigner}, we derive the CMB bispectrum generated from the non-Gaussianity of gravitons induced by $W^3$ as
\begin{eqnarray}
&& \Braket{\prod_{n=1}^3 a_{X_n, \ell_n m_n}}_{W^3}
= \left(
  \begin{array}{ccc}
  \ell_1 & \ell_2 & \ell_3 \\
   m_1 & m_2 & m_3
  \end{array}
 \right)
\int_0^\infty y^2 dy
\sum_{L_1 L_2 L_3} (-1)^{\frac{L_1 + L_2 + L_3}{2}} I_{L_1 L_2 L_3}^{0~0~0}
\nonumber \\
&&\qquad \times 
\left[\prod_{n=1}^3 \frac{2}{\pi} (-i)^{\ell_n} \int k_n^2 dk_n
 {\cal T}_{X_n,\ell_n}(k_n) j_{L_n}(k_n y)\right] f_{W^3}^{(r)}(k_1,
k_2, k_3) 
\nonumber \\
&&\qquad \times 
\left(8 \pi \right)^{3/2} 
\sum_{L', L'' = 2, 3} 8 I_{\ell_1 L_1 2}^{2 0 -2} I_{\ell_2 L_2 L'}^{2 0 -2} 
I_{\ell_3 L_3 L''}^{2 0 -2} 
\left\{
  \begin{array}{ccc}
  \ell_1 & \ell_2 & \ell_3 \\
   L_1 & L_2 & L_3 \\
   2 & L' & L'' \\
  \end{array}
 \right\}
\nonumber \\
&&\qquad \times  
\left[- \frac{1}{20}\sqrt{\frac{7}{3}} \delta_{L', 2} \delta_{L'', 2}
\left( \prod_{n=1}^3 {\cal D}_{L_n, \ell_n, x_n}^{(e)} \right) 
\right. \nonumber \\ 
&&\qquad\quad \left.  + (-1)^{L'} I_{L' 1 2}^{2 0 -2} I_{L'' 1 2}^{2 0 -2} 
{\cal D}_{L_1, \ell_1, x_1}^{(e)} {\cal D}_{L_2, \ell_2, x_2}^{(o)} 
{\cal D}_{L_3, \ell_3, x_3}^{(o)} \right. \nonumber \\
&&\qquad\qquad \left. \times
\left( - \frac{\pi}{5} 
\left\{
  \begin{array}{ccc}
   2 & L' & L'' \\
   2 & 1 & 1 
  \end{array}
 \right\} 
- \pi 
\left\{
  \begin{array}{ccc}
   2 & L' & L'' \\
   1 & 1 & 2 \\
   1 & 2 & 1
  \end{array}
 \right\} \right. \right. \nonumber \\
&&\qquad\qquad\quad \left. \left.
+ 2\pi 
\left\{
  \begin{array}{ccc}
   2 & 1 & L' \\
   2 & 1 & 1 
  \end{array}
 \right\}
\left\{
  \begin{array}{ccc}
   2 & L' & L'' \\
   2 & 1 & 1 
  \end{array}
 \right\}
\right)
\right]
+ 5 \ {\rm perms} ~, \label{eq:cmb_bis_w3}
\end{eqnarray}
and $\widetilde{W}W^2$ as 
\begin{eqnarray}
&&\Braket{\prod_{n=1}^3 a_{X_n, \ell_n m_n}}_{\widetilde{W}W^2}
= \left(
  \begin{array}{ccc}
  \ell_1 & \ell_2 & \ell_3 \\
   m_1 & m_2 & m_3
  \end{array}
 \right)
\int_0^\infty y^2 dy
\sum_{L_1 L_2 L_3} (-1)^{\frac{L_1 + L_2 + L_3}{2}} I_{L_1 L_2 L_3}^{0~0~0}
\nonumber \\
&&\qquad \times 
\left[\prod_{n=1}^3 \frac{2}{\pi} (-i)^{\ell_n} \int k_n^2 dk_n
 {\cal T}_{X_n,\ell_n}(k_n) j_{L_n}(k_n y)\right] f_{\widetilde{W}W^2}^{(r)}(k_1,
k_2, k_3) 
\nonumber \\
&&\qquad \times 
\left( 8\pi \right)^{3/2} 
\sum_{L', L'' = 2, 3} 8 I_{\ell_1 L_1 2}^{2 0 -2} I_{\ell_2 L_2 L'}^{2 0 -2}
I_{\ell_3 L_3 L''}^{2 0 -2} 
\left\{
  \begin{array}{ccc}
  \ell_1 & \ell_2 & \ell_3 \\
   L_1 & L_2 & L_3 \\
   2 & L' & L'' \\
  \end{array}
 \right\}
(-1)^{L''} I_{L'' 1 2}^{2 0 -2} 
\nonumber \\
&&\qquad \times 
\left[
\delta_{L',2}
{\cal D}_{L_1, \ell_1, x_1}^{(e)}
{\cal D}_{L_2, \ell_2, x_2}^{(e)} {\cal D}_{L_3, \ell_3, x_3}^{(o)}
\right.  \nonumber \\
&&\qquad\qquad 
\left. \times \left( 
3 \sqrt{\frac{2\pi}{5}}
\left\{
  \begin{array}{ccc}
   2 & 2 & L'' \\
   1 & 2 & 1 
  \end{array}
 \right\}
- 2 \sqrt{2\pi}  
\left\{
  \begin{array}{ccc}
   2 & 2 & L'' \\
   1 & 1 & 1 \\
   1 & 1 & 2
  \end{array}
 \right\} \right) \right. \nonumber \\
&&\quad \left.  
\qquad + I_{L' 1 2}^{2 0 -2} 
\left(\prod_{n=1}^3 {\cal D}_{L_n, \ell_n, x_n}^{(o)} \right) 
\right. \nonumber \\
&&\qquad\qquad \left. \times
\left( - \frac{4 \pi}{3} 
\left\{
  \begin{array}{ccc}
   2 & L' & L'' \\
   1 & 2 & 1 \\
   1 & 1 & 2
  \end{array}
\right\}
+ \frac{2\pi}{15} \sqrt{\frac{7}{3}} 
\left\{
  \begin{array}{ccc}
   2 & L' & L'' \\
   1 & 2 & 2    
  \end{array}
\right\} 
\right) \right] 
+ 5 \ {\rm perms}~. 
\label{eq:cmb_bis_ww2}
\end{eqnarray}
Here, ``5 perms'' denotes the five symmetric
terms under the permutations of $(\ell_1, m_1, x_1)$, $(\ell_2, m_2,
x_2)$, and $(\ell_3, m_3, x_3)$, and 
we introduce the filter functions as 
\begin{eqnarray}
\begin{split}
{\cal D}^{(e)}_{L, \ell, x} 
&\equiv (\delta_{L, \ell - 2} + \delta_{L, \ell} + \delta_{L, \ell + 2}) 
\delta_{x, 0}
+ (\delta_{L, \ell - 3} + \delta_{L, \ell - 1} + \delta_{L, \ell + 1} + \delta_{L, \ell + 3}) \delta_{x,1} ~, \\
{\cal D}^{(o)}_{L, \ell, x} 
&\equiv (\delta_{L, \ell - 2} + \delta_{L, \ell} + \delta_{L, \ell + 2}) 
\delta_{x, 1} 
+ ( \delta_{L, \ell - 3} + \delta_{L, \ell - 1} +
\delta_{L, \ell + 1} + \delta_{L, \ell + 3} ) \delta_{x,
0} ~,
\end{split}
\end{eqnarray}
where the superscripts $(e)$ and $(o)$ denote $L + \ell + x = {\rm even}$
and $= {\rm odd}$, respectively. 
From Eqs.~(\ref{eq:cmb_bis_w3}) and (\ref{eq:cmb_bis_ww2}), we can see
that the azimuthal quantum numbers $m_1, m_2$, and $m_3$ are confined only in a Wigner-$3j$ symbol as 
$\left(
  \begin{array}{ccc}
  \ell_1 & \ell_2 & \ell_3 \\
   m_1 & m_2 & m_3
  \end{array}
 \right)$.
This guarantees the rotational invariance of the CMB
bispectrum. Therefore, this bispectrum survives if the triangle
inequality is satisfied as $|\ell_1 - \ell_2| \leq \ell_3 \leq \ell_1 + \ell_2 $. 

Considering the products between the ${\cal D}$ functions in Eq.~(\ref{eq:cmb_bis_w3}) and the selection rules as $\sum_{n=1}^3 L_n = {\rm even}$, we can notice that the CMB bispectrum from $W^3$ does not vanish only for 
\begin{eqnarray}
\sum_{n=1}^3 (\ell_n + x_n) = {\rm even} ~.
\end{eqnarray} 
Therefore, $W^3$ contributes the $III, IIE, IEE, IBB, EEE$, and $EBB$
spectra for $\sum_{n=1}^3 \ell_n = {\rm even}$ and the $IIB, IEB, EEB$,
and $BBB$ spectra for $\sum_{n=1}^3
\ell_n = {\rm odd}$. This property can arise from any sources keeping the
parity invariance such as $W^3$. 
On the other hand, in the same manner, we understand that the CMB bispectrum from $\widetilde{W}W^2$ survives only for 
\begin{eqnarray}
\sum_{n=1}^3 (\ell_n + x_n) = {\rm odd} ~.
\end{eqnarray} 
By these constraints, we find that in reverse, $\widetilde{W}W^2$
generates the $IIB, IEB, EEB$, and $BBB$ spectra for $\sum_{n=1}^3 \ell_n =
{\rm even}$ and the $III, IIE, IEE, IBB, EEE$, and $EBB$ spectra for $\sum_{n=1}^3 \ell_n = {\rm odd}$. This
is a characteristic signature of the parity violation as mentioned in
Refs.~\cite{Okamoto:2002ik, Kamionkowski:2010rb}. Hence, if we analyze the 
information of the CMB bispectrum not only for $\sum_{n=1}^3 \ell_n =
{\rm even}$ but also for $\sum_{n=1}^3 \ell_n = {\rm odd}$, it may be
possible to check the parity violation at the level of the
three-point correlation. 

The above discussion about the multipole configurations of the CMB
bispectra can be easily understood only if one consider the parity
transformation of the CMB intensity and polarization fields in the real space (\ref{eq:cmb_anisotropy}). 
The $III$, $IIE$, $IEE$, $IBB$, $EEE$ and
 $EBB$ spectra from $W^3$, and the $IIB, IEB, EEB$, and $BBB$ spectra
 from $\widetilde{W}W^2$ have even parity, namely, 
\begin{eqnarray} 
\Braket{\prod_{i=1}^3 \frac{\Delta X_i (\hat{{\bf n_i}})}{X_i}} 
= \Braket{\prod_{i=1}^3 \frac{\Delta X_i (-\hat{{\bf n_i}})}{X_i}}~. 
\end{eqnarray}
Then, from the multipole expansion
(\ref{eq:cmb_anisotropy}) and its parity flip version as 
\begin{eqnarray}
\frac{\Delta X (- \hat{{\bf n}})}{X} = \sum_{\ell m} a_{X, \ell m} Y_{\ell
 m}(- \hat{{\bf n}}) = \sum_{\ell m} (-1)^\ell a_{X, \ell m} Y_{\ell
 m}(\hat{{\bf n}})~,
\end{eqnarray}
one can
notice that $\sum_{n=1}^3 \ell_n = {\rm even}$ must be satisfied. 
On the other hand, since the $IIB, IEB, EEB$, and $BBB$ spectra
 from $W^3$, and the $III$, $IIE$, $IEE$, $IBB$, $EEE$, and
 $EBB$ spectra from $\widetilde{W}W^2$ have odd parity, namely, 
\begin{eqnarray} 
\Braket{\prod_{i=1}^3 \frac{\Delta X_i (\hat{{\bf n_i}})}{X_i}} 
= - \Braket{\prod_{i=1}^3 \frac{\Delta X_i (-\hat{{\bf n_i}})}{X_i}}~, 
\end{eqnarray}
one can obtain $\sum_{n=1}^3 \ell_n = {\rm odd}$.

In Sec.~\ref{subsubsec:results}, we compute the CMB
bispectra (\ref{eq:cmb_bis_w3}) and (\ref{eq:cmb_bis_ww2}) 
when $A = \pm 1/2, 0, 1$, that is, the signals
from $W^3$ become as large as those from $\widetilde{W}W^2$ and either signals vanish.
  
\subsubsection{Evaluation of $f_{W^3}^{(r)}$ and $f_{\widetilde{W}W^2}^{(r)}$}

\begin{figure}[t]
  \begin{tabular}{cc}
    \begin{minipage}{0.5\hsize}
  \begin{center}
    \includegraphics[width=8.0cm,clip]{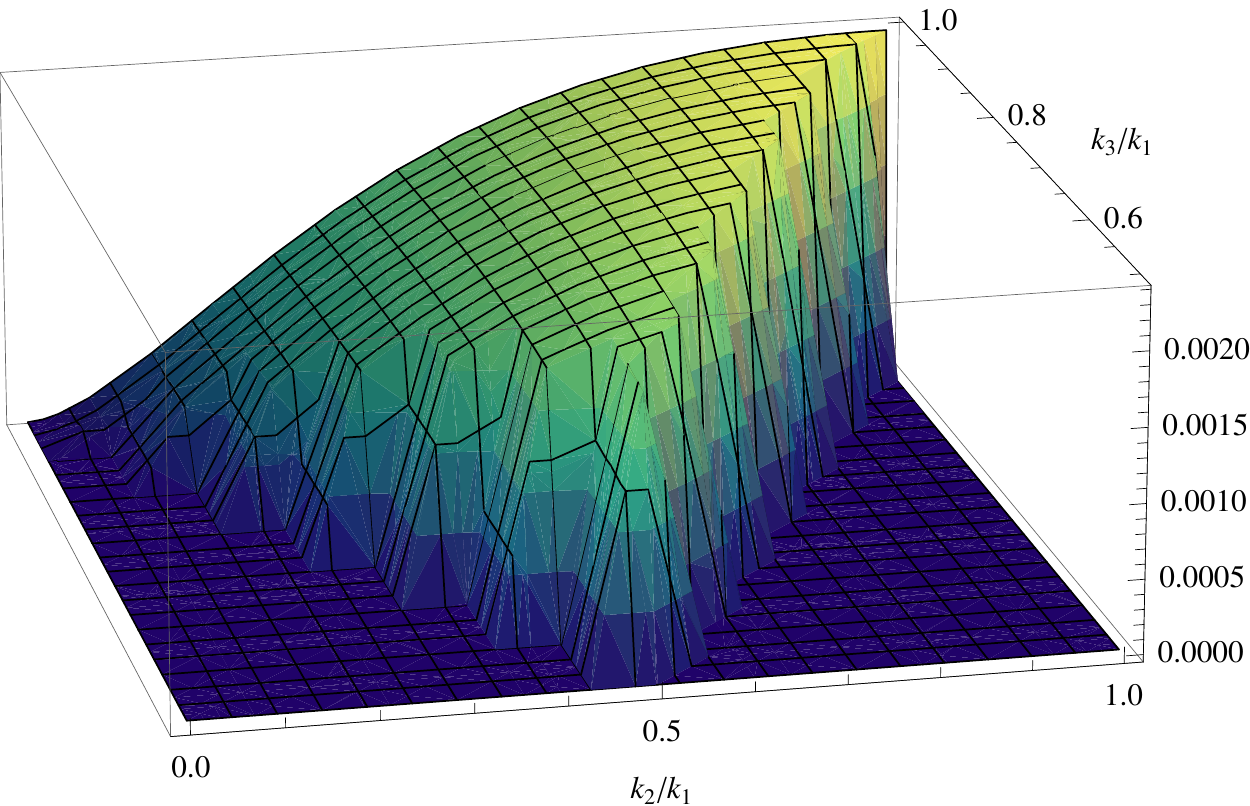}
  \end{center}
\end{minipage}
\begin{minipage}{0.5\hsize}
  \begin{center}
    \includegraphics[width=8.0cm,clip]{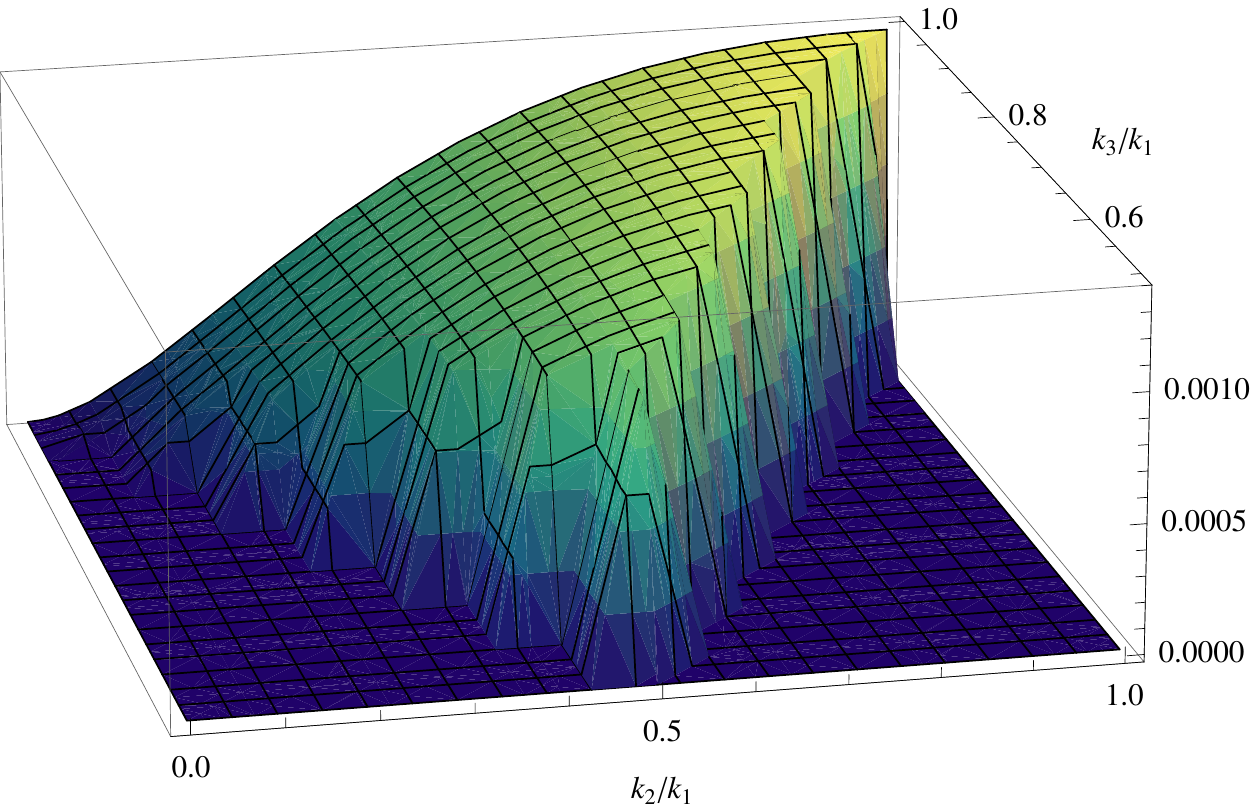}
  \end{center}
\end{minipage}
\end{tabular}
\\
  \begin{tabular}{cc}
    \begin{minipage}{0.5\hsize}
  \begin{center}
    \includegraphics[width=8.0cm,clip]{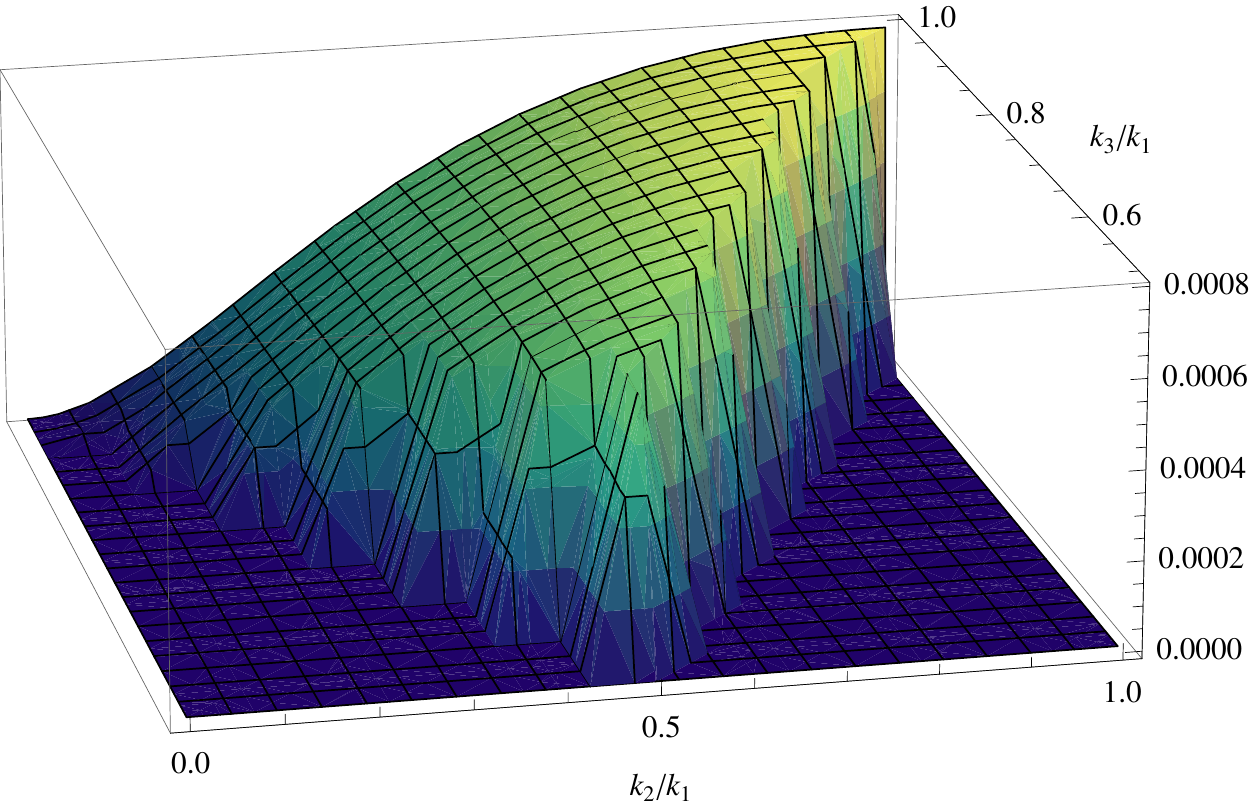}
  \end{center}
\end{minipage}
\begin{minipage}{0.5\hsize}
  \begin{center}
    \includegraphics[width=8.0cm,clip]{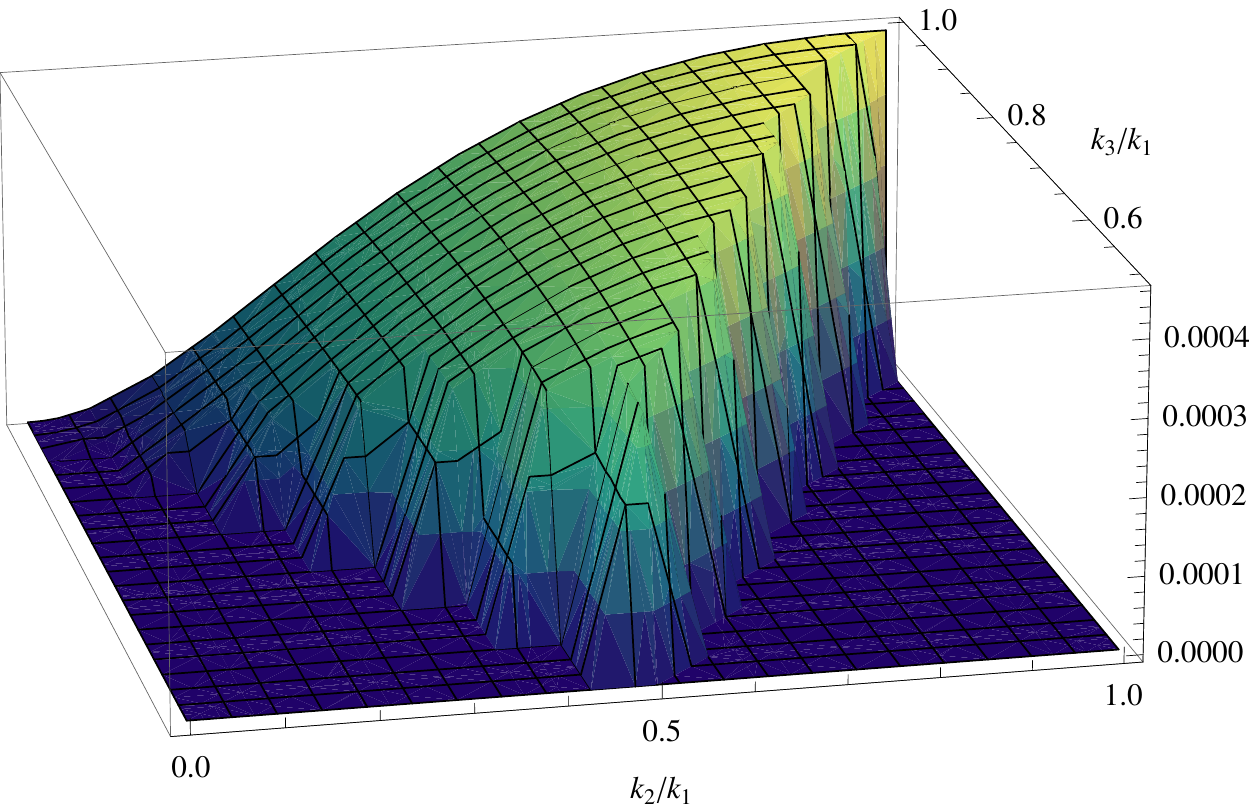}
  \end{center}
\end{minipage}
\end{tabular}
  \caption{Shape of $k_1^2 k_2^2 k_3^2 S_{A}$ for $A = -1/2$ (top left panel), $0$ (top right one), $1/2$ (bottom left one), and $1$ (bottom right one) as the function of $k_2 / k_1$ and $k_3 / k_1$.} \label{fig:S}
\end{figure}

Here, to compute the CMB bispectra (\ref{eq:cmb_bis_w3}) and
(\ref{eq:cmb_bis_ww2}) in finite time, we express the radial functions,
 $f^{(r)}_{W^3}$ and $f^{(r)}_{\widetilde{W}W^2}$, with some terms of the
 power of $k_1, k_2$, and $k_3$. 
Let us focus on the dependence on $k_1, k_2$, and $k_3$ in
Eqs.~(\ref{eq:radial_w3}) and (\ref{eq:radial_ww2}) as
\begin{eqnarray}
f_{W^3}^{(r)} \propto f_{\widetilde{W}W^2}^{(r)} &\propto& k_t^{-6}(-k_t \tau_*)^{-A} 
=  \frac{S_A(k_1, k_2, k_3)}{(k_1 k_2 k_3)^{A/3} (-\tau_*)^{A}}~, 
\end{eqnarray}
where we define $S_A$ to satisfy $S_A \propto k^{-6}$ as
\begin{eqnarray}
S_A(k_1, k_2, k_3) &\equiv&
\frac{(k_1 k_2 k_3)^{A/3}}{k_t^{6+A}}~.
\end{eqnarray}
In Fig.~\ref{fig:S}, we plot $S_{A}$ for $A = -1/2, 0, 1/2$, and $1$. From this, we
notice that the shapes of $S_A$ are similar to the equilateral-type 
configuration as Eq.~(\ref{eq:S_eq})
\begin{eqnarray}
S_{\rm equil}(k_1, k_2, k_3) 
&=& 6 \left(- \frac{1}{k_1^3 k_2^3} - \frac{1}{k_2^3 k_3^3} - \frac{1}{k_3^3 k_1^3} - \frac{2}{k_1^2 k_2^2 k_3^2} \right. \nonumber \\
&&\quad \left. 
+ \frac{1}{k_1 k_2^2 k_3^3} + \frac{1}{k_1 k_3^2 k_2^3}
+ \frac{1}{k_2 k_3^2 k_1^3} + \frac{1}{k_2 k_1^2 k_3^3}
+ \frac{1}{k_3 k_1^2 k_2^3} + \frac{1}{k_3 k_2^2 k_1^3}
 \right)~. 
\end{eqnarray}
To evaluate how a function $S$ is similar in shape to a function $S'$, we introduce a correlation function as \cite{Babich:2004gb, Senatore:2009gt}
\begin{eqnarray}
\cos(S \cdot S') \equiv \frac{S \cdot S'}
{(S \cdot S)^{1/2} (S' \cdot S')^{1/2} }~,
\end{eqnarray}
with
\begin{eqnarray}
S \cdot S' 
&\equiv& \sum_{{\bf k_i}} \frac{S(k_1, k_2, k_3) S'(k_1, k_2, k_3)}
{P(k_1) P(k_2) P(k_3)} \nonumber \\
&\propto& \int_0^1 d x_2 
\int_{1-x_2}^1 d x_3 x_2^4 x_3^4 S(1, x_2, x_3) S'(1, x_2, x_3)~,
\end{eqnarray}
where the summation is performed over all ${\bf k_i}$, which form a triangle and $P(k) \propto k^{-3}$ denotes the power spectrum. 
This correlation function gets to 1 when $S = S'$. 
In our case, this is calculated as  
\begin{eqnarray}
\cos (S_{A} \cdot S_{\rm equil}) 
&\simeq& 
\begin{cases}
0.968 ~, & (A = -1/2) \\
0.970 ~, & (A = 0) \\
0.971 ~, & (A = 1/2) \\
0.972 ~, & (A = 1) \\
\end{cases}
\end{eqnarray}
that is, an approximation that $S_A$ is proportional to $S_{\rm equil}$
seems to be valid. Here, we also calculate the correlation functions with the local- and orthogonal-type non-Gaussianities \cite{Komatsu:2010fb} and conclude that these contributions are negligible. Thus, we determine the proportionality coefficient as 
\begin{eqnarray}
S_A &\simeq& 
\frac{S_A \cdot S_{\rm equil}}{S_{\rm equil} \cdot
S_{\rm equil}}  S_{\rm equil} 
= 
\begin{cases}
4.40 \times 10^{-4}S_{\rm equil} ~, & (A = -1 / 2) \\
2.50 \times 10^{-4}S_{\rm equil} ~, & (A = 0) \\
1.42 \times 10^{-4}S_{\rm equil} ~, & (A = 1 / 2) \\
8.09 \times 10^{-5}S_{\rm equil} ~. & (A = 1)
\end{cases}
\end{eqnarray}
Substituting this into Eqs.~(\ref{eq:radial_w3}) and (\ref{eq:radial_ww2}), we obtain reasonable formulae of the radial functions
for $A = 1/2$ as
\begin{eqnarray}
f_{W^3}^{(r)} = f_{\widetilde{W}W^2}^{(r)}
\simeq  \left( \frac{\pi^2}{2} r A_S \right)^4 \left( \frac{M_{\rm pl}}{\Lambda} \right)^2 
\frac{10395}{8} \sqrt{\frac{\pi}{2}} 
\frac{1.42 \times 10^{-4} S_{\rm equil}}{ (-\tau_*)^{1/2} (k_1 k_2 k_3)^{1/6} }
~, \label{eq:fr_+0.5} 
\end{eqnarray}
and for $A = - 1/2$ as
\begin{eqnarray}
f_{W^3}^{(r)} = - f_{\widetilde{W}W^2}^{(r)}
\simeq  \left( \frac{\pi^2}{2} r A_S \right)^4 \left( \frac{M_{\rm pl}}{\Lambda} \right)^2 
\frac{945}{4} \sqrt{\frac{\pi}{2}} 
4.40 \times 10^{-4} (-\tau_*)^{1/2} (k_1 k_2 k_3)^{1/6} S_{\rm equil}~. 
\label{eq:fr_-0.5}
\end{eqnarray}
Here, we also use 
\begin{eqnarray}
\left( \frac{H}{M_{\rm pl}} \right)^2 = \frac{\pi^2}{2} r A_S~,
\end{eqnarray}
where $A_S$ is the amplitude of primordial curvature perturbations and $r$ is the tensor-to-scalar ratio \cite{Komatsu:2010fb, Shiraishi:2010kd}. 
For $A = 0$, the signals from $\widetilde{W}W^2$ disappear as
$f^{(r)}_{\widetilde{W}W^2} = 0$ and the finite radial function of $W^3$
is given by 
\begin{eqnarray}
f^{(r)}_{W^3} \simeq  
\left( \frac{\pi^2}{2} r A_S \right)^4 \left( \frac{M_{\rm pl}}{\Lambda} \right)^2 960 \times 2.50 \times 10^{-4}S_{\rm equil}~. \label{eq:fr_0}
\end{eqnarray}
In contrast, for $A = 1$, since $f^{(r)}_{W^3} = 0$, we have only the parity-violating contribution from $\widetilde{W}W^2$ as 
\begin{eqnarray}
f^{(r)}_{\widetilde{W}W^2} \simeq 
\left( \frac{\pi^2}{2} r A_S \right)^4 
\left( \frac{M_{\rm pl}}{\Lambda} \right)^2 5760 \times
\frac{8.09 \times 10^{-5} S_{\rm equil}}{(- \tau_*) (k_1 k_2 k_3)^{1/3}}~. \label{eq:fr_1}
\end{eqnarray} 

\subsubsection{Results} \label{subsubsec:results}

On the basis of the analytical formulae (\ref{eq:cmb_bis_w3}), (\ref{eq:cmb_bis_ww2}), (\ref{eq:fr_+0.5}), (\ref{eq:fr_-0.5}), (\ref{eq:fr_0}) and (\ref{eq:fr_1}), we compute the CMB bispectra 
from $W^3$ and $\widetilde{W}W^2$ for $A = - 1/2, 0 , 1/2$, and $1$.
 Then, we modify the Boltzmann Code for Anisotropies in the
Microwave Background (CAMB) \cite{Lewis:1999bs,Lewis:2004ef}. In
calculating the Wigner symbols, we use the Common
Mathematical Library SLATEC \cite{slatec} and some analytic formulae described in Appendices \ref{appen:wigner} and \ref{appen:polarization}. 
From the dependence of the radial functions $f_{W^3}^{(r)}$ and $f_{\widetilde{W}W^2}^{(r)}$ on the wave numbers, we can see that 
the shapes of the CMB
bispectra from $W^3$ and $\widetilde{W}W^2$ are similar to the
equilateral-type configuration. Then, the significant signals arise from
multipoles satisfying $\ell_1 \simeq \ell_2 \simeq \ell_3$. We confirm this by calculating the CMB bispectrum for several $\ell$'s. Hence, in the
following discussion, we give the discussion with the spectra for $\ell_1 \simeq \ell_2 \simeq \ell_3$. However, we do not focus on the spectra from $\sum_{n=1}^3 \ell_n = {\rm odd}$ for $\ell_1 = \ell_2 = \ell_3$ because these vanish due to the asymmetric nature.

\begin{figure}[!t]
  \begin{center}
    \includegraphics[width=13cm,clip]
{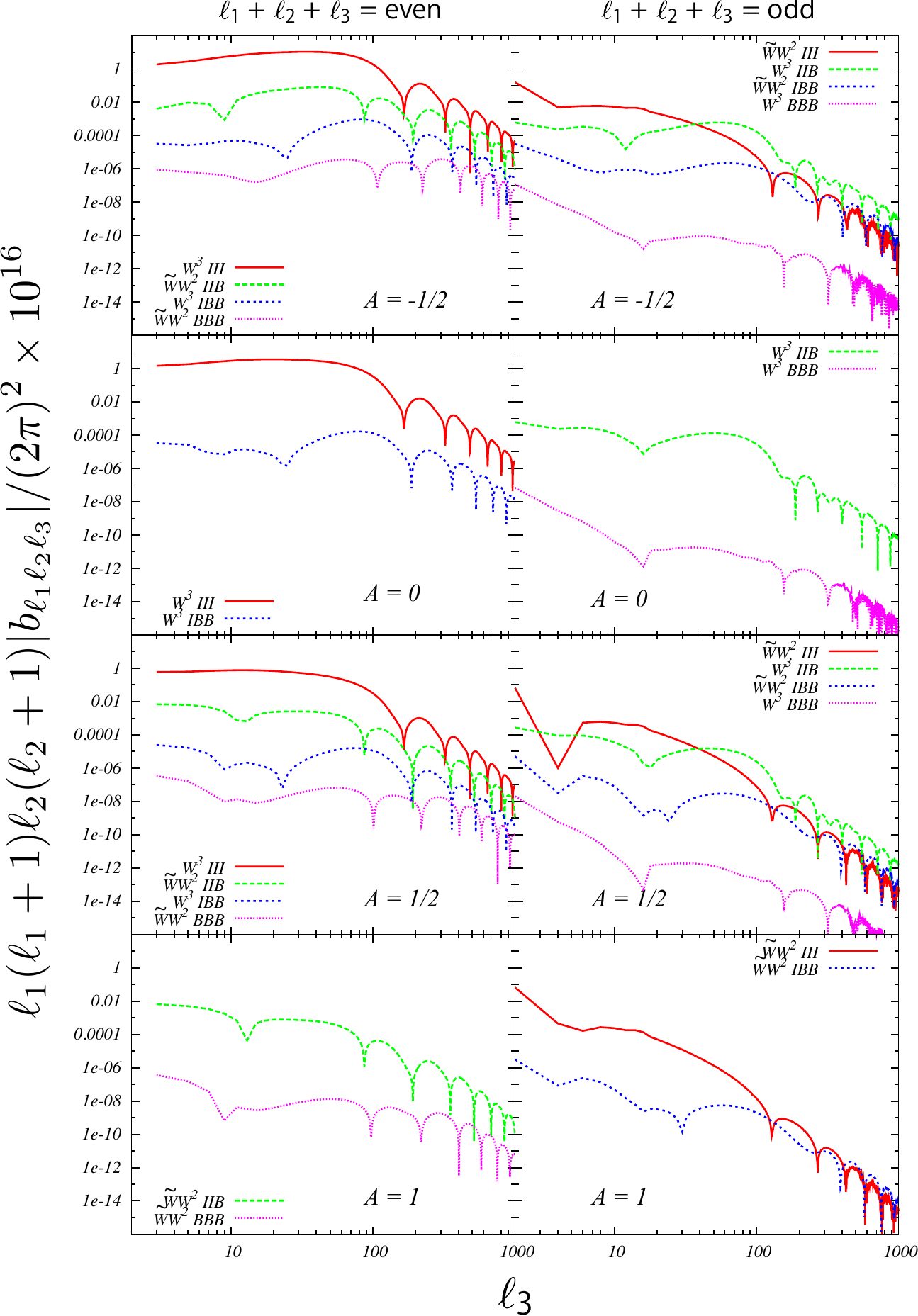}
  \end{center}
  \caption{Absolute values of the CMB $III, IIB, IBB$, and $BBB$ spectra induced by $W^3$ and $\widetilde{W}W^2$ for $A
 = -1/2, 0, 1/2$, and $1$. We set that three multipoles have identical values as $\ell_1 - 2 = \ell_2 - 1 = \ell_3$. 
The left figures show the spectra not
 vanishing for $\sum_{n=1}^3 \ell_n = {\rm even}$ (parity-even mode)
 and the right ones present the spectra for $\sum_{n=1}^3 \ell_n = {\rm
 odd}$ (parity-odd mode). Here, we fix the parameters as $ \Lambda =
 3 \times 10^6 {\rm GeV}, r = 0.1$, and $\tau_* = -k_*^{-1} = - 14 {\rm Gpc}$, and other
 cosmological parameters are fixed as the mean values limited from the
 WMAP $7$-yr data \cite{Komatsu:2010fb}.} \label{fig:TTT_A_difl}
\end{figure}

In Fig.~\ref{fig:TTT_A_difl}, we present the reduced CMB $III, IIB,
IBB$, and $BBB$ spectra given by
\begin{eqnarray}
b_{X_1 X_2 X_3, \ell_1 \ell_2 \ell_3}
 &=&  ( G_{\ell_1 \ell_2 \ell_3} )^{-1} \sum_{m_1 m_2 m_3} 
\left(
  \begin{array}{ccc}
  \ell_1 & \ell_2 & \ell_3 \\
   m_1 & m_2 & m_3
  \end{array}
 \right)
\Braket{\prod_{n=1}^3 a_{X_n, \ell_n m_n}} ~,  
\end{eqnarray}
for $\ell_1 - 2  = \ell_2 - 1 = \ell_3$. 
Here, the $G$ symbol is defined by 
\cite{Kamionkowski:2010rb}
\footnote{The conventional expression of the CMB-reduced bispectrum as 
\begin{eqnarray}
b_{X_1 X_2 X_3, \ell_1 \ell_2 \ell_3} \equiv (I_{\ell_1 \ell_2 \ell_3}^{0~0~0})^{-1} 
\sum_{m_1 m_2 m_3} 
\left(
  \begin{array}{ccc}
  \ell_1 & \ell_2 & \ell_3 \\
   m_1 & m_2 & m_3
  \end{array}
 \right)
\Braket{\prod_{n=1}^3 a_{X_n, \ell_n m_n}} 
\end{eqnarray}
breaks down for $\sum_{n=1}^{3} \ell_n = {\rm odd}$ due to the
divergence behavior of $(I_{\ell_1 \ell_2 \ell_3}^{0~0~0})^{-1}$. Here,
replacing the $I$ symbol with the $G$ symbol, this problem is avoided. Of course, for $\sum_{n=1}^{3}
\ell_n = {\rm even}$,  $G_{\ell_1 \ell_2 \ell_3}$ is identical to $I_{\ell_1 \ell_2
\ell_3}^{0~0~0}$.
}
\begin{eqnarray}
G_{\ell_1 \ell_2 \ell_3} 
&\equiv& \frac{2 \sqrt{\ell_3 (\ell_3 + 1) \ell_2 (\ell_2 +
1)}}{\ell_1(\ell_1 + 1) - \ell_2 (\ell_2 + 1) - \ell_3 (\ell_3 + 1)}
\sqrt{\frac{\prod_{n=1}^3 (2 \ell_n + 1)}{4 \pi}}
\left(
  \begin{array}{ccc}
  \ell_1 & \ell_2 & \ell_3 \\
   0 & -1 & 1
  \end{array}
 \right)~.
\end{eqnarray}
At first, from this figure, we can confirm that there are similar
features of the CMB power spectrum of tensor modes
\cite{Pritchard:2004qp, Baskaran:2006qs}. In the $III$
spectra, the dominant signals are located in $\ell_3 < 100$ due to the
enhancement of the integrated Sachs-Wolfe effect. On the other
hand, since the fluctuation of polarizations is mainly produced through Thomson scattering at around the recombination and reionization epoch,
the $BBB$ spectra have two peaks for $\ell_3 < 10$ and $\ell_3 \sim 100$,
respectively. The cross-correlated bispectra between $I$ and $B$ modes
seem to contain both these effects. These features back up the consistency of our calculation.
 
The curves in Fig.~\ref{fig:TTT_A_difl} denote the spectra for $A =
-1/2, 0, 1/2$, and $1$, respectively. 
We notice that the spectra for large $A$
become red compared with those for small $A$. 
The difference in tilt of $\ell$ between these spectra is just one corresponding to the difference in $A$.  
The curves of the left and right figures obey $\sum_{n=1}^3 \ell_n
= {\rm even}$ and $= {\rm odd}$, respectively. 
As mentioned in Sec.~\ref{subsubsec:formulation}, we stress again that in the
$\ell$ configuration where the bispectrum from $W^3$ vanishes,
the bispectrum from $\widetilde{W}W^2$ survives, and vice versa
for each correlation. This is because the parities of these terms
are opposite each other. For example, this predicts a nonzero $III$ spectrum not only for $\sum_{n=1}^3 \ell_n = {\rm even}$
due to $W^3$ but also for $\sum_{n=1}^3 \ell_n = {\rm odd}$ due to $\widetilde{W}W^2$. 

We can also see that each bispectrum induced by $W^3$ has a different shape from that induced by $\widetilde{W}W^2$ corresponding to the difference in the
primordial bispectra. 
Regardless of this, the overall amplitudes of the spectra for $A = \pm 1/2$ are almost identical. 
However, if we consider $A$ deviating from these values, the
balance between the contributions of $W^3$ and $\widetilde{W}W^2$ breaks. 
For example, if $-1/2 < A < 1/2$, the contribution of $W^3$
dominates. Assuming the time-independent coupling, namely, $A = 0$, since
$f^{(r)}_{\widetilde{W}W^2} = 0$, the CMB bispectra are generated only
from $W^3$. Thus, we will never observe the parity violation of gravitons
in the CMB bispectrum. On the other hand, when $-3/2 < A < -1/2$ or $1/2
< A < 3/2$, the contribution of $\widetilde{W}W^2$ dominates. In an
extreme case, if $A = {\rm odd}$, since $f^{(r)}_{W^3} = 0$, the CMB
bispectra arise only from $\widetilde{W}W^2$ and violate the parity invariance. Then, the information of the signals under $\sum_{n=1}^3 \ell_n = {\rm odd}$ will become more important in the analysis of the $III$ spectrum.

\begin{figure}[t]
  \begin{center}
    \includegraphics[height=8cm,clip]{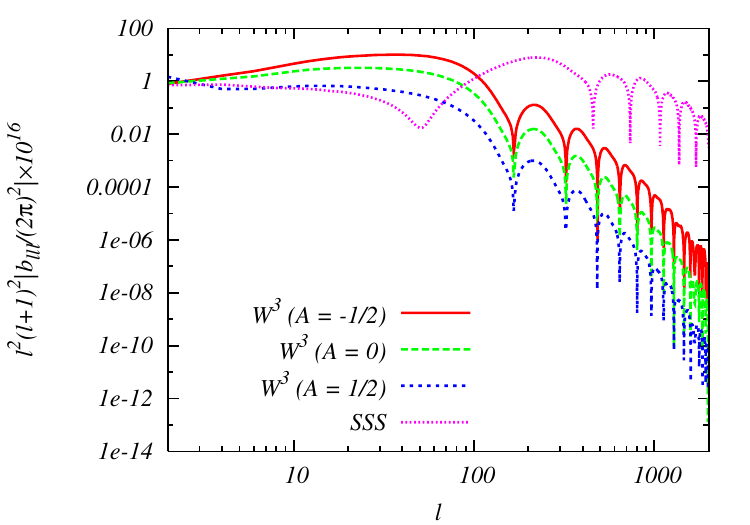}
  \end{center}
  \caption{Absolute value of the CMB $III$ spectra generated from $W^3$
 for $A = -1/2$ (red solid line), $0$ (green dashed one) and $1/2$ (blue dotted one), and generated from the equilateral-type
 non-Gaussianity given by Eq.~(\ref{eq:cmb_bis_scal}) with $f_{\rm
 NL}^{\rm equil} = 300$ (magenta dot-dashed one). 
 We set that three multipoles have identical values as $\ell_1 = \ell_2
 = \ell_3 \equiv \ell$. Here, we fix the parameters as the same values
 mentioned in Fig.~\ref{fig:TTT_A_difl}.} \label{fig:TTT_A_III_even_samel}
\end{figure}

In Fig.~\ref{fig:TTT_A_III_even_samel}, we focus on the $III$ 
spectra from $W^3$ for $\ell_1 = \ell_2 = \ell_3 \equiv \ell$ to compare these with the $III$ spectrum generated from the equilateral-type non-Gaussianity of curvature perturbations given by 
\begin{eqnarray}
b_{III, \ell_1 \ell_2 \ell_3}^{(SSS)}  
&=& \int_0^\infty y^2 dy 
\left[ \prod_{n=1}^3 \frac{2}{\pi} \int_0^\infty k_n^2 dk_n 
{\cal T}_{I,\ell_n}^{(S)}(k_n) j_{\ell_n}(k_n y) \right] 
\nonumber \\
&&\times \frac{3}{5} f_{\rm NL}^{\rm equil} (2\pi^2 A_S)^2 
S_{\rm equil}(k_1, k_2, k_3) ~, \label{eq:cmb_bis_scal}
\end{eqnarray}
where $f_{\rm NL}^{\rm equil}$ is the nonlinearity
parameter of the equilateral non-Gaussianity and ${\cal T}_{I,\ell}^{(S)}$ is the transfer function of scalar mode described in Eq.~(\ref{eq:transfer_explicit}). Note that these
three spectra vanish for $\sum_{n=1}^3 \ell_n = {\rm odd}$. 
From this figure,
we can estimate the typical amplitude of the $III$ spectra from $W^3$ at large scale as 
\begin{eqnarray}
|b_{\ell \ell \ell}| \sim \ell^{-4} \times 3.2 \times 10^{-2}
\left(\frac{{\rm GeV}}{\Lambda}\right)^2 \left(\frac{r}{0.1}\right)^4 ~.
\end{eqnarray}
This equation also seems to be applicable to the $III$ spectra from
$\widetilde{W}W^2$. 
On the other hand, the CMB bispectrum generated from the
equilateral-type non-Gaussianity on a large scale is evaluated with $f_{\rm NL}^{\rm equil}$ as
\begin{eqnarray}
|b_{\ell \ell \ell}| 
\sim \ell^{-4} \times 4 \times 10^{-15}  
\left|\frac {f_{\rm NL}^{\rm equil}}{300}\right| ~.
\end{eqnarray} 
From these estimations and ideal upper bounds on $f_{\rm NL}^{\rm equil}$
estimated only from the cosmic variance for $\ell < 100$
\cite{Creminelli:2005hu, Creminelli:2006rz, Smith:2006ud}, namely
$f_{\rm NL}^{\rm equil} \lesssim 300$ and $r \sim 0.1$, we find a rough
limit: $\Lambda \gtrsim 3 \times 10^6 {\rm GeV}$. Here, we use only the signals for $\sum_{n=1}^3 \ell_n = {\rm even}$ due to the comparison with the parity-conserving bispectrum from scalar-mode non-Gaussianity. Of course, to estimate more precisely, we will have to calculate the signal-to-noise ratio with the information of $\sum_{n=1}^3 \ell_n = {\rm odd}$ \cite{Kamionkowski:2010rb}. 

\subsection{Summary and discussion}

In this section, we have studied the CMB bispectrum generated from the graviton
non-Gaussianity induced by the parity-even and parity-odd Weyl cubic terms,
namely, $W^3$ and $\widetilde{W}W^2$,  which have a dilaton-like coupling
depending on the conformal time as $f \propto \tau^A$. Through the
calculation based on the in-in formalism, we have found that the primordial
non-Gaussianities from $\widetilde{W}W^2$ can have a magnitude comparable to that from $W^3$ even in the exact de Sitter
space-time. 

Using the explicit formulae of the primordial bispectrum, we
have derived the CMB bispectra of the intensity ($I$) and polarization ($E,B$)
modes. Then, we have confirmed that, owing to the difference in the 
transformation under parity,
the spectra from $W^3$ vanish in the $\ell$ space where those
from $\widetilde{W}W^2$ survive and vice versa. For example, 
owing to the parity-violating $\widetilde{W}W^2$ term,
the $III$ spectrum can be produced not only for $\sum_{n=1}^3 \ell_n = {\rm even}$ but
also for $\sum_{n=1}^3 \ell_n = {\rm odd}$, and the $IIB$ spectrum can
also be produced
for $\sum_{n=1}^3 \ell_n = {\rm even}$.
These signals are powerful lines of evidence the parity violation in the non-Gaussian level; hence, to
reanalyze the observational data for $\sum_{n=1}^3 \ell_n = {\rm odd}$
is meaningful work. 

When $A = -1/2, 0, 1/2$, and $1$, we have obtained reasonable numerical
results of the CMB bispectra from the parity-conserving $W^3$ and 
the parity-violating $\widetilde{W}W^2$. For $A = \pm 1/2$, we have
found that the spectra from $W^3$ and $\widetilde{W}W^2$ have almost the
same magnitudes 
even though these have a small difference in the shapes. 
In contrast, if $A = 0$ and $1$, we have confirmed that the signals from
$\widetilde{W}W^2$ and $W^3$ vanish, respectively. In the latter case,
we will observe only the parity-violating signals in the CMB bispectra
generated from the Weyl cubic terms. 
We have also found that
the shape of the non-Gaussianity from such Weyl cubic terms
is quite similar to the equilateral-type
non-Gaussianity of curvature perturbations. 
In comparison with the $III$ spectrum 
generated from the equilateral-type non-Gaussianity,
we have found that if $r = 0.1$, $\Lambda \gtrsim 3 \times 10^6 {\rm GeV}$ 
corresponds approximately to $f_{\rm NL}^{\rm equil} \lesssim 300$. 

Strictly speaking, to obtain the bound on the scale $\Lambda$, we need to
calculate the signal-to-noise ratio with the information of not only
$\sum_{n=1}^3 \ell_n = {\rm even}$ but also $\sum_{n=1}^3 \ell_n = {\rm
odd}$ for each $A$ by the application of Ref.~\cite{Kamionkowski:2010rb}. 
This will be discussed in the future. 

\bibliographystyle{JHEP}
\bibliography{paper}

%% file: ch9_pmf.tex
\section{CMB bispectrum generated from primordial magnetic fields} \label{sec:PMF}

Recent observational consequences have shown the existence of ${\cal
O}(10^{-6}) {\rm G}$ magnetic fields in galaxies and clusters of
galaxies at redshift $z \sim 0.7 - 2.0$ \cite{Bernet:2008qp,
Wolfe:2008nk, Kronberg:2007dy}. One of the scenarios to realize this is
an amplification of the magnetic fields by the galactic dynamo mechanism
(e.g. \cite{Widrow:2002ud}), which requires ${\cal O}(10^{-20}) {\rm G}$
seed fields prior to the galaxy formation.  A variety of studies have
suggested the possibility of generating the seed fields at the
inflationary epoch \cite{Martin:2007ue, Bamba:2006ga}, the cosmic phase
transitions \cite{Stevens:2010ym, Kahniashvili:2009qi}, and cosmological
recombination \cite{Ichiki:2006cd, Maeda:2008dv, Fenu:2010kh} 
and also there have been many studies about constraints on
the strength of primordial magnetic fields (PMFs) through the impact
on the cosmic microwave background (CMB) anisotropies, in particular,
the CMB power spectrum sourced from the PMFs~\cite{Subramanian:1998fn, Durrer:1998ya, Mack:2001gc, Lewis:2004ef, Paoletti:2008ck, Shaw:2009nf}. 
The PMFs excite not only the scalar fluctuation but also the vector and tensor fluctuations in the CMB fields. For example, the gravitational waves and curvature perturbations, which come from the tensor and scalar components of the PMF anisotropic stresses, produce additional CMB fluctuations at large and intermediate scales \cite{Lewis:2004ef, Shaw:2009nf}. In addition, it has been known that the magnetic vector mode may dominate the CMB small-scale fluctuations by the Doppler effect (e.g. \cite{Mack:2001gc, Lewis:2004ef}). 

The PMF anisotropic stresses depend quadratically on the magnetic seed fields. Thus, assuming the gaussianity of the PMF, the anisotropic stress and CMB fluctuation obey the highly non-Gaussian statistics \cite{Brown:2005kr, Brown:2006wv}. 
Owing to the Wick's theorem, the CMB bispectrum contains the pure non-Gaussian information. Hence, to extract the information of the PMF from the CMB fields, the analysis of the CMB bispectrum is of great utility. 
Recently, in Refs.~\cite{Seshadri:2009sy,Caprini:2009vk,Cai:2010uw,
Trivedi:2010gi}, the authors investigated the contribution of the scalar-mode anisotropic stresses of PMFs to the bispectrum of the CMB temperature fluctuations. From current CMB experimental data, some authors obtained rough limits on the PMF strength smoothed on $1 {\rm Mpc}$ scale as $B_{\rm 1 Mpc} < {\cal O}(1) {\rm nG}$. However, since in all these studies, the complicated angular dependence on the wave number vectors are neglected, there may exist any uncertainties. In addition, the authors have never considered the dependence on the vector- and tensor-mode contributions and hence more precise discussion including these concerns should be realized. 

With these motivations, we have studied the CMB scalar, vector and tensor bispectra induced from PMFs and firstly succeeded in the exact computation of them with the full-angular dependence \cite{Shiraishi:2010yk, Shiraishi:2011fi, Shiraishi:2011dh} by applying the all-sky formulae for the CMB bispectrum \cite{Shiraishi:2010kd}
\footnote{
In Ref.~\cite{Kahniashvili:2010us}, after us, the authors presented an analytic formula for the CMB temperature bispectrum generated from vector anisotropic stresses of the PMF.}. In our studies, we also updated constraints on the PMF strength. 

In this section, after reviewing the impact of PMFs on the CMB anisotropies, we present the derivation of the CMB bispectra induced from PMFs and discuss their behaviors. In addition, we put limits on the PMFs by considering the WMAP data and the expected PLANCK data \cite{Komatsu:2010fb, :2006uk}. Finally, we mention our future works. These discussions are based on our studies \cite{Shiraishi:2010yk, Shiraishi:2011fi, Shiraishi:2011dh},


\subsection{CMB fluctuation induced from PMFs} \label{subsec:alm_PMF}

\begin{figure}[t]
  \centering \includegraphics[height=12cm,clip]{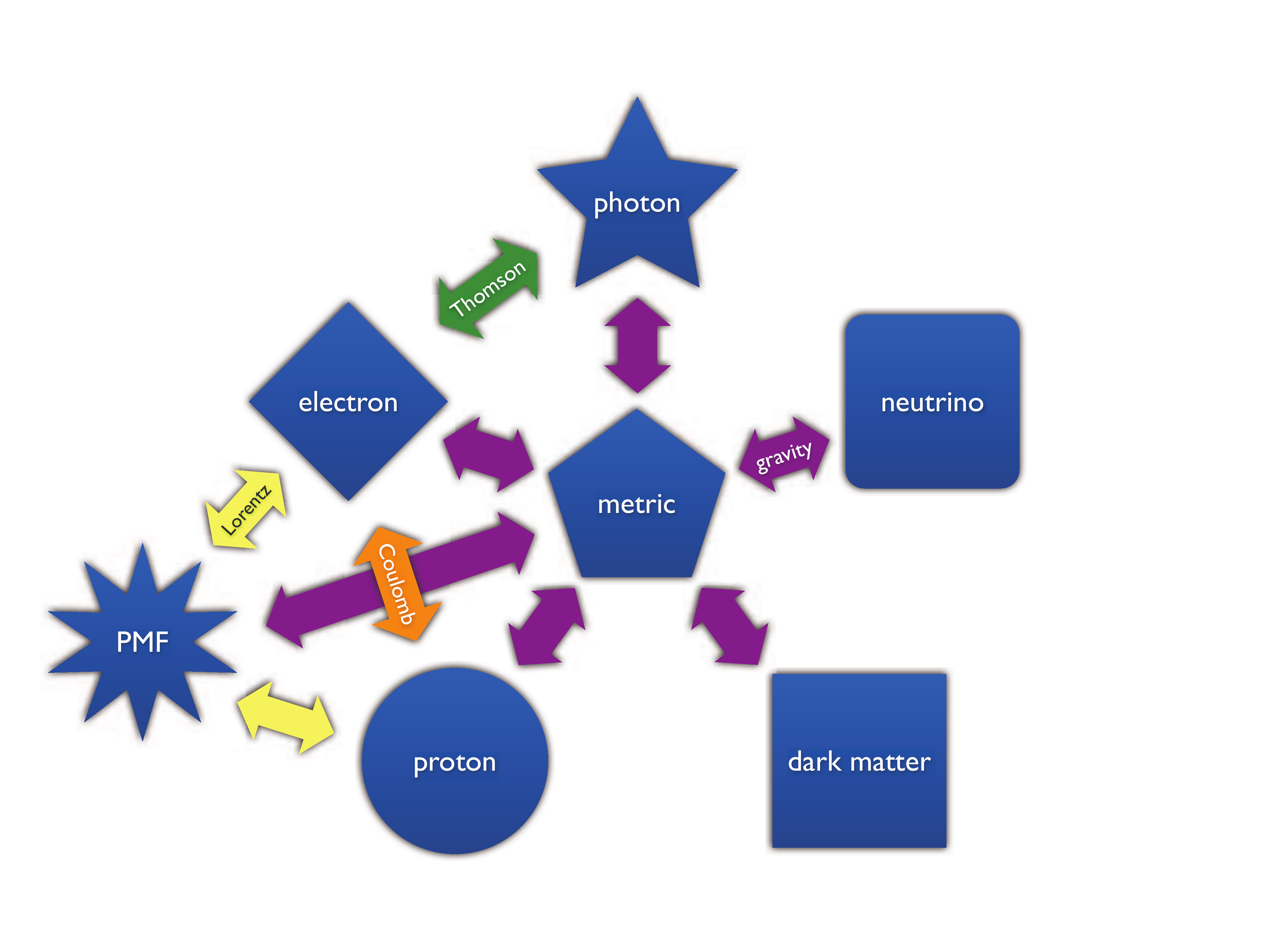}
  \caption{Interaction between several components in the Universe if the PMF exists.}
  \label{fig:CMB_PMF_w.pdf}
\end{figure}

The PMFs drive the Lorentz force and the anisotropic stress, and change the motion of baryons (protons and electrons) and the growth of the gravitational potential via the Euler and Einstein equations. Consequently, the photon's anisotropy is also affected. We illustrate this in Fig.~\ref{fig:CMB_PMF_w.pdf}. 
In the following discussion, we summarize the impacts of PMFs on the CMB fluctuations in detail and current constraints on the PMFs obtained from the CMB power spectrum. 

\subsubsection{Setting for the PMFs}

Let us consider the stochastic PMFs $B^b({\bf x}, \tau)$ on the homogeneous
background Universe which is characterized by the Friedmann-Robertson-Walker metric,
\begin{eqnarray}
ds^2 = a(\tau)^2 \left[ - d\tau^2 + \delta_{bc} dx^b dx^c \right]~,
\end{eqnarray}
where $\tau$ is a conformal time and $a (\tau)$ is a scale factor. The expansion of the Universe makes the amplitude of the magnetic fields decay as $1 / a^2$ and hence we can draw off the time dependence as
$B^b({\bf x}, \tau) = B^b({\bf x}) / a^2$. 
Each component of the energy momentum tensor (EMT) of PMFs is given by
\begin{eqnarray}
\begin{split}
T^0_{~0}(x^\mu) &= - \rho_B = - \frac{1}{8 \pi a^4} B^2({\bf x}) \equiv - \rho_\gamma(\tau) \Delta_B(x^\mu)~, \\
T^0_{~c}(x^\mu) &= T^b_{~0}(x^\mu) = 0~, \\
T^b_{~c}(x^\mu) &= \frac{1}{4\pi a^4} \left[\frac{B^2({\bf x})}{2}\delta^b_{~c}
 -  B^b({\bf x})B_c({\bf x})\right] 
\equiv \rho_\gamma(\tau) 
\left[ \Delta_B(x^\mu) \delta^b_{~c} + \Pi^b_{Bc}(x^\mu) \right]~. \label{eq:EMT_PMF} 
\end{split}
\end{eqnarray}
The Fourier components of the spatial parts are described as
\begin{eqnarray}
\begin{split}
T^b_{~c}({\bf k},\tau) &\equiv {\rho_\gamma}(\tau)
\left[ \delta^b_{~c} \Delta_B({\bf k}) + \Pi_{Bc}^b({\bf k}) \right]~, \\
\Delta_B({\bf k}) &= {1 \over 8\pi \rho_{\gamma,0}}
\int \frac{d^3 {\bf k}'}{(2\pi)^3} 
B^b({\bf k'}) B_b({\bf k} - {\bf k'}) ~, \\
\Pi^b_{Bc}({\bf k}) &=-{1 \over 4\pi \rho_{\gamma,0}} \int \frac{d^3
 {\bf k'}}{(2 \pi)^3} B^b({\bf k'}) B_c({\bf k} - {\bf k'})~, 
\label{eq:EMT_PMF_fourier} 
\end{split}
\end{eqnarray}
where we have introduced the photon energy density $\rho_{\gamma}$ in
order to include the time dependence of $a^{-4}$ and $\rho_{\gamma, 0}$
denotes the present energy density of photons.  In the following
discussion, for simplicity of calculation, we ignore the trivial
time-dependence. Hence, the index is lowered by
$\delta_{bc}$ and the summation is implied for repeated indices.

Assuming that $B^a({\bf x})$ is a Gaussian field, the
statistically isotropic power spectrum of PMFs $P_B(k)$ is defined by
\footnote{Here we neglect the helical component. This effect will be considered in Ref.~\cite{Shiraishi:2012sn}.}
\begin{eqnarray}
\langle B_a({\bf k})B_b({\bf p})\rangle
= (2\pi)^3 {P_B(k) \over 2} P_{ab}(\hat{\bf k}) \delta ({\bf k} + {\bf p})~, \label{eq:def_pmf_power}
\end{eqnarray}  
with a projection tensor
\begin{eqnarray}
P_{ab}(\hat{\bf k}) 
\equiv \sum_{\sigma = \pm 1} \epsilon^{(\sigma)}_a(\hat{\bf k}) \epsilon^{(-\sigma)}_b(\hat{\bf k}) 
 =  \delta_{ab} - \hat{k}_a \hat{k}_b~, \label{eq:projection}
\end{eqnarray}
which comes from the divergence free nature of PMFs.  Here $\hat{\bf
k}$ denotes a unit vector and $\epsilon_a^{(\pm 1)}$ is a normalized
divergenceless polarization vector which satisfies the orthogonal
condition; $\hat{k}^a \epsilon_a^{(\pm 1)} = 0$. The details of the
relations and conventions of the polarization vector are described in
Appendix \ref{appen:polarization}. Although the form of the power spectrum $P_B(k)$ is strongly dependent on the production mechanism, we assume a simple power law shape given by
\begin{eqnarray}
P_B(k) = A_B k^{n_B}~,
\end{eqnarray}
where $A_B$ and $n_B$ denote the amplitude and the spectral index of the
power spectrum of magnetic fields, respectively.  In order to
parametrize the strength of PMFs, we smooth the magnetic fields with a
conventional Gaussian filter on a comoving scale $r$:
\begin{eqnarray}
B_r^2 \equiv \int_0^\infty \frac{k^2 dk}{2 \pi^2} e^{-k^2 r^2} P_B(k) ~,
\end{eqnarray}
then, $A_B$ is calculated as
\begin{eqnarray}
{A}_B = {
\left(2 \pi \right)^{n_B + 5} B_r^2 \over \Gamma(\frac{n_B + 3}{2}) k_r^{n_B + 3} }~,
\end{eqnarray}
where $\Gamma(x)$ is the Gamma function and $k_r \equiv 2 \pi / r$.

We focus on the scalar, vector and tensor contributions induced from the PMFs, which come from the anisotropic stress of the EMT, i.e., $\Pi_{Bab}$.  Following the definition of the projection operators in Appendix \ref{appen:polarization}, the PMF anisotropic stress fluctuation is decomposed into
\begin{eqnarray}
\begin{split}
\Pi_{Bs}^{(0)}({\bf k}) 
&= \frac{3}{2} O_{ij}^{(0)}(\hat{\bf k}) \Pi_{B ij}({\bf k})  ~,\\ 
\Pi_{Bv}^{(\pm 1)}({\bf k}) 
&= \frac{1}{2} O_{ij}^{(\mp 1)}(\hat{\bf k}) 
 \Pi_{B ij}({\bf k})   ~, \\
\Pi_{Bt}^{(\pm 2)}({\bf k}) 
&= \frac{1}{2} O_{ij}^{(\mp 2)}(\hat{\bf k}) \Pi_{B ij}({\bf k})  ~. 
\label{eq:PMF_ani}
\end{split}
\end{eqnarray}
These act as sources of the CMB scalar-, vector- and tensor-mode fluctuations as follow. 

\subsubsection{Scalar and tensor modes}

If the seed magnetic fields exist in the early Universe, the scalar and tensor components of the PMF anisotropic stress are not compensated prior to neutrino decoupling \cite{Lewis:2004ef, Shaw:2009nf}, and the scalar and tensor metric perturbations generated from them survive passively. These residual metric perturbations generate the CMB anisotropies of the scalar and tensor modes. These kind of CMB anisotropies are so called ``passive mode'' and may dominate at intermediate and large scales depending on the PMF strength \cite{Shaw:2009nf}. 

To estimate curvature perturbation and gravitational wave driven by PMFs on superhorizon scales, we shall focus on the Einstein equation at the radiation dominated era. Before neutrino decoupling, the Universe is dominated by the radiative fluid. The fluid is tightly coupled to baryons and can not create any anisotropic stress. Hence, in this period, total anisotropic stress comes from only PMFs, namely, constant $\Pi_{Bs/t}^{(0 / \pm 2)}$. Until neutrino decoupling, this survives and it will be a source of metric perturbations via the Einstein equation. Then, at the superhorizon limit, the Einstein equation for scalar and tensor modes on the synchronous gauge (\ref{eq:decomposed_einstein_perturb_evolution}) reduces to same form as 
\begin{eqnarray}
\ddot{h}^{(0 / \pm 2)}({\bf k}, \tau) + \frac{2}{\tau} \dot{h}^{(0 / \pm 2)} ({\bf k}, \tau) 
= \frac{6}{\tau^2} R_{\gamma} \Pi_{Bs/t}^{(0 / \pm 2)}({\bf k})~,
\end{eqnarray}
where we have used ${\cal H} = \dot{a}/a = 1 / \tau$, ${\cal H}^2 = 8 \pi G \rho a^2 / 3$ and $R_\gamma \equiv \rho_\gamma / \rho \approx 0.6$. This is analytically solved as 
\begin{eqnarray}
h^{(0 / \pm 2)}({\bf k}, \tau) = C_1 + \frac{C_2}{\tau} 
+ 6 R_\gamma \Pi_{Bs/t}^{(0 / \pm 2)}({\bf k}) \ln \left( \frac{\tau}{\tau_B} \right) ~,
\end{eqnarray}
where $\tau_B$ is the conformal time at the generation of the PMF. On the other hand, after neutrino decoupling ($\tau > \tau_\nu = 1 {\rm MeV}^{-1}$), resultant neutrino anisotropic stress compensates the PMF anisotropic stress. Hence, right-hand side of the above Einstein equation becomes zero and the growth of metric perturbations ceases. Accordingly, for $\tau \gg \tau_\nu$, superhorizon-scale comoving curvature and tensor perturbations are evaluated as 
\begin{eqnarray}
{\cal R}({\bf k}) &=& {\cal R}({\bf k}, \tau_B) + R_\gamma \Pi_{Bs}^{(0)}({\bf k}) \ln \left( \frac{\tau_\nu}{\tau_B} \right) ~, \label{eq:ini_pmf_scalar} \\ 
h^{(\pm 2)}({\bf k}) &=& h^{(\pm 2)}({\bf k}, \tau_B) + 6 R_\gamma \Pi_{Bt}^{(\pm 2)}({\bf k}) \ln \left( \frac{\tau_\nu}{\tau_B} \right) ~, \label{eq:ini_pmf_tensor}
\end{eqnarray}
where we have used a relation of the scalar perturbations on superhorizon scales: $h^{(0)} = 6 {\cal R}$
\footnote{${\cal R}$ and $h^{(\pm 2)}$ are equal to $- \zeta$ and $- \sqrt{3} H_{T}$ of Refs.~\cite{Lewis:2004ef, Shaw:2009nf}, respectively.}. 
This logarithmic growth and saturation of metric perturbations even on superhorizon scales are caused by only the property of sources as $\rho \propto a^{-4}$, hence the above discussion is applicable to the case of general radiation fluid other than PMFs \cite{Kojima:2009gw}. These metric perturbations act as a source of the CMB fluctuations of scalar and tensor modes. 

Note that the effects of PMFs on the transfer functions ${\cal T}_{X, \ell}^{(S)}$ and ${\cal T}_{X, \ell}^{(T)}$ are inconsiderable at larger scales \cite{Shaw:2009nf}, hence it is safe to use the non-magnetic transfer functions (\ref{eq:transfer_explicit}) in the computation of the CMB spectra induced from scalar and tensor modes of the PMF anisotropic stress. 

\subsubsection{Vector mode}\label{subsubsec:pmf_vector}

The vector mode has no equivalent passive mode as the gravitational
potential of the vector mode decays away via the Einstein
equation posterior to neutrino decoupling. Thus, in the vector mode, we need to consider the impact of
the PMF on the transfer function. 
In Refs.~\cite{Durrer:1998ya, Mack:2001gc, Kahniashvili:2010us, Shiraishi:2011fi}, it is discussed that the temperature fluctuations are generated via Doppler and integrated Sachs-Wolfe effects on the CMB vector modes. On the basis of them, we derive the transfer function of the vector magnetic mode as follows.

When we decompose the metric perturbations into vector components as
\begin{eqnarray}
\begin{split}
\delta g_{0c} &= \delta g_{c0} = a^2 A_c ~, \\
\delta g_{cd} &= a^2 
\left(\partial_c h^{(V)}_{d} + \partial_d h^{(V)}_{c} \right)~,
\end{split}
\end{eqnarray}
we can construct two gauge-invariant variables, namely a vector perturbation of the extrinsic curvature and a vorticity, as
\begin{eqnarray}
\begin{split}
{\bf V} &\equiv {\bf A} - \dot{\bf h} ~, \\
{\bf \Omega} &\equiv {\bf v} - {\bf A} ~,
\end{split}
\end{eqnarray}
where ${\bf v}$ is the spatial part of the four-velocity perturbation of
a stationary fluid element and a dash denotes a partial derivative of the conformal time $\tau$.
Here, choosing a gauge as $\dot{\bf h} = 0$, we can express the Einstein equation
\begin{eqnarray}
\dot{\bf V} + 2 {\cal H} {\bf V} = - \frac{16 \pi G \rho_{\gamma,0} ({\bf \Pi_\gamma^{(V)}} + {\bf \Pi_\nu^{(V)}} + {\bf \Pi_B^{(V)}})}{a^2 k} ~, \label{eq:einstein_pmf_vec}
\end{eqnarray}
and the Euler equations for photons and baryons
\begin{eqnarray}
\dot{\bf \Omega_\gamma} + \dot{\kappa}({\bf v_\gamma} - {\bf v_b}) 
&=& 0 ~, \label{eq:euler_g}\\
\dot{\bf \Omega_b} + {\cal H} {\bf \Omega_b} - \frac{\dot{\kappa}}{R}({\bf v_\gamma} - {\bf v_b}) 
&=& \frac{\bf L^{(V)}}{a^4 (\rho_b + p_b)}~. \label{eq:euler_b}
\end{eqnarray}
Here ${\bf L^{(V)}} \equiv k \rho_{\gamma,0}{\bf \Pi_B^{(V)}}$ is the Lorentz force of vector mode and $\Pi^{(V)}_a = -i \hat{k}_b P_{ac} \Pi_{bc}$, $p$ is the isotropic pressure, the indices $\gamma, \nu$ and $b$ denote the photon, neutrino
and baryon, $\kappa$ is the optical depth, and $R \equiv (\rho_b +
p_b)/(\rho_\gamma + p_\gamma) \simeq 3 \rho_b / (4 \rho_\gamma)$.
In the tight-coupling limit as ${\bf v_\gamma} \simeq {\bf v_b}$, the photon vorticity is comparable to the baryon one: ${\bf \Omega_\gamma} \simeq {\bf \Omega_b} \equiv {\bf \Omega}$. Then, the Euler equations (\ref{eq:euler_g}) and (\ref{eq:euler_b}) are combined into
\begin{eqnarray}
(1+R)\dot{\bf \Omega} + R {\cal H} {\bf \Omega} 
= \frac{\bf L^{(V)}}{a^4 (\rho_\gamma + p_\gamma)}~, \label{eq:euler_PMF_vector_large} 
\end{eqnarray}
and this solution is given by 
\begin{eqnarray}
{\bf \Omega}({\bf k},\tau) 
\simeq \frac{\tau {\bf L^{(V)}(k)}}{(1+R)(\rho_{\gamma,0} + p_{\gamma,0})}  ~,
\end{eqnarray}
Note that Eq.~(\ref{eq:euler_PMF_vector_large}) and the above solution are valid for perturbation
wavelengths larger than the comoving Silk damping scale $L_S \equiv 2
\pi / k_S$, namely, $k < k_S$, where photon viscosity can be neglected compared to the Lorentz force. For $k > k_S$, due to the effect of the photon vorticity,
the Euler equation (\ref{eq:euler_PMF_vector_large}) is changed as \cite{Subramanian:1998fn}
\begin{eqnarray}
(1+R)\dot{\bf \Omega} 
+ R \left( {\cal H} + \frac{k^2 \chi}{a \rho_b} \right) {\bf \Omega} 
= \frac{{\bf L^{(V)}}}{a^4 (\rho_\gamma + p_\gamma)}~,
\end{eqnarray}
where $\chi = (4/15) \rho_\gamma L_\gamma a $ is the photon shear
viscosity coefficient and $L_\gamma = \dot{\kappa}^{-1}$ is the photon
comoving mean-free path. We can obtain this analytical solution: 
\begin{eqnarray}
{\bf \Omega}({\bf k},\tau) 
\simeq \frac{{\bf L^{(V)}(k)}}{(k^2 L_\gamma / 5)(\rho_{\gamma,0} + p_{\gamma,0})} ~.
\end{eqnarray}
Hence, we can summarize the vorticity of the baryon and photon fluids as
\begin{eqnarray}
\begin{split}
{\bf \Omega}({\bf k},\tau) &\simeq \beta(k,\tau) {\bf \Pi_B^{(V)}({\bf k})} ~, \\ 
\beta(k,\tau) &=
\frac{\rho_{\gamma,0}}{\rho_{\gamma,0} + p_{\gamma,0}} \times
\begin{cases}
k \tau / (1+R) & {\rm for} \ k<k_S \\
5 \dot{\kappa} / k & {\rm for} \ k>k_S
\end{cases}~.
\end{split} \label{eq:vec_transfer_pmf}
\end{eqnarray}

As mentioned above, the CMB temperature anisotropies of vector modes are produced through the Doppler and integrated Sachs-Wolfe effects as 
\begin{eqnarray}
\frac{\Delta I (\hat{\bf n})}{I}
&=& -{\bf v_\gamma} \cdot \hat{\bf n} |^{\tau_0}_{\tau_*} + \int_{\tau_*}^{\tau_0} d \tau \dot{\bf V} \cdot \hat{\bf n} ~, 
\end{eqnarray}
where $\tau_0$ is today and $\tau_*$ is the recombination epoch in
conformal time, $\mu_{k,n} \equiv \hat{\bf k} \cdot \hat{\bf n}$, $x
\equiv k(\tau_0 - \tau)$, and $\hat{\bf n}$ is a unit vector along the
line-of-sight direction. Because of compensation of the anisotropic
stresses, a solution of the Einstein equation (\ref{eq:einstein_pmf_vec})
expresses the decaying signature as ${\bf V} \propto a^{-2}$ after
neutrino decoupling. Therefore, in an integrated Sachs-Wolfe effect
term, the contribution around the recombination epoch is dominant. Furthermore, neglecting dipole contribution due to ${\bf v}$ today, we can form the coefficient of anisotropies as 
\begin{eqnarray}
a_{I, \ell m}^{(V)} &\equiv&  \int d^2 \hat{\bf n}  \frac{\Delta I(\hat{\bf n})}{I} Y^*_{\ell m} (\hat{\bf n}) \nonumber \\
&\simeq& \int \frac{d^3 {\bf k}}{(2 \pi)^3}
\int d^2 \hat{\bf n}
[{\bf \Pi_B^{(V)}}({\bf k}) \cdot \hat{\bf n}]  Y^*_{\ell m} (\hat{\bf
n}) 
\beta(k,\tau_*) e^{- i \mu_{k,n} x_*} ~. 
\end{eqnarray}
In the transformation $\hat{\bf n} \rightarrow (\mu_{k,n}, \phi_{k,n})$,
the functions are rewritten as
\begin{eqnarray}
\begin{split}
{\bf \Pi_B^{(V)} (\bf k)} \cdot \hat{\bf n}
&\rightarrow -i \sqrt{\frac{1- \mu_{k,n}^2}{2}} 
\sum_{\lambda = \pm 1} \Pi^{(\lambda)}_{Bv}({\bf k}) e^{i \lambda \phi_{k,n}} 
~, \\
Y^*_{\ell m}(\hat{\bf n}) &\rightarrow \sum_{m'} D^{(\ell)}_{m m'} 
\left( S(\hat{\bf k}) \right) Y^*_{\ell m'}(\Omega_{k,n}) ~, \\
d^2 \hat{\bf n} &\rightarrow d\Omega_{k,n}~,
\end{split}
\end{eqnarray}
where we use the relation: 
$ \Pi^{(V)}_a = \sum_{\lambda = \pm 1} - i
\Pi_{Bv}^{(\lambda)} \epsilon_a^{(\lambda)}$ and the Wigner $D$ matrix
under the rotational transformation of a unit vector parallel to $z$
axis into $\hat{\bf k}$ corresponding to Eq.~(\ref{eq:wig_D_cmb}). 
Therefore, performing the integration over $\Omega_{k,n}$ in the same
manner as Sec.~\ref{sec:CMB_anisotropy}, 
we can obtain the explicit form of $a_{I, \ell m}^{(V)}$ and express the radiation transfer function introduced in Eq.~(\ref{eq:all_alm_general}) as
\begin{eqnarray}
\begin{split}
a_{I, \ell m}^{(V)} &= 4\pi (-i)^\ell 
\int \frac{d^3 {\bf k}}{(2\pi)^3} 
\sum_{\lambda = \pm 1} \lambda
{}_{- \lambda}Y_{\ell m}^* (\hat{\bf k}) 
 \Pi^{(\lambda)}_{B v}({\bf k}) 
{\cal T}_{I, \ell}^{(V)}(k)  ~, \\
{\cal T}^{(V)}_{I,\ell}(k) 
&\simeq \left[ \frac{(\ell+1)!}{(\ell-1)!} \right]^{1/2} 
\frac{\beta(k, \tau_*)}{\sqrt{2}} \frac{j_\ell(x_*)}{x_*}~.
\end{split}
\end{eqnarray}
This is consistent with the results presented in Refs.~\cite{Lewis:2004ef, Lewis:2004sec}. 
In the same manner, the vector-mode transfer functions of polarizations are derived \cite{Mack:2001gc, Brown:2006wv}. Then, we can also express as 
\begin{eqnarray}
a_{X, \ell m}^{(V)} = 4\pi (-i)^\ell 
\int \frac{d^3 {\bf k}}{(2\pi)^3} 
\sum_{\lambda = \pm 1} \lambda^{x+1}
{}_{- \lambda}Y_{\ell m}^* (\hat{\bf k}) 
 \Pi^{(\lambda)}_{B v}({\bf k}) 
{\cal T}_{X, \ell}^{(V)}(k) ~.
\end{eqnarray}

\subsubsection{Expression of $a_{\ell m}$'s} 

From the above results and Eq.~(\ref{eq:alm}), the CMB intensity and polarization fluctuations induced from PMFs are summarized as 
\begin{eqnarray}
\begin{split}
a^{(Z)}_{X, \ell m} &= 4\pi (-i)^\ell 
\int \frac{k^2 dk}{(2\pi)^3} 
\sum_{\lambda} [{\rm sgn}(\lambda)]^{\lambda+x} \xi^{(\lambda)}_{\ell m}(k) 
{\cal T}_{X, \ell}^{(Z)}(k)  ~, \\
\xi^{(0)}_{\ell m}(k) &\approx \int d^2 \hat{\bf k} Y_{\ell m}^*(\hat{\bf k})  
\left[R_\gamma \ln\left(\frac{\tau_\nu}{\tau_B}\right) \right] \Pi^{(0)}_{Bs}({\bf k}) 
 ~,\\
\xi^{(\pm 1)}_{\ell m}(k) &\approx \int d^2 \hat{\bf k} {}_{\mp 1}Y_{\ell m}^*(\hat{\bf k})  \Pi^{(\pm 1)}_{Bv}({\bf k})   ~,\\
\xi^{(\pm 2)}_{\ell m}(k) &\approx \int d^2 \hat{\bf k} {}_{\mp 2}Y_{\ell m}^*(\hat{\bf k}) \left[6 R_\gamma \ln\left(\frac{\tau_\nu}{\tau_B}\right) \right] 
\Pi^{(\pm 2)}_{Bt}({\bf k}) ~, \label{eq:alm_PMF}
\end{split}
\end{eqnarray}
where we take ${\cal R}({\bf k}, \tau_B) = h^{(\pm 2)}({\bf k}, \tau_B) = 0$, and regard $\xi^{(0)}$ and $\xi^{(\pm 2)}$ as ${\cal R}$ and $h^{(\pm 2)}$, respectively
\footnote{In Refs.\cite{Shiraishi:2012sn, Shiraishi:2012rm}, we equate $\xi^{(0)}$ to $- {\cal R}$.}. 

\subsubsection{CMB power spectrum from PMFs} 

From the formulae (\ref{eq:alm_PMF}), the CMB power spectra from PMFs are written as
\begin{eqnarray}
\Braket{\prod_{n=1}^2 a_{X_n, \ell_n m_n}^{(Z_n)}} 
= \left[ \prod_{n=1}^2 4\pi (-i)^{\ell_n} 
\int \frac{k_n^2 dk_n}{(2\pi)^3} {\cal T}_{X_n, \ell_n}^{(Z_n)}(k_n) 
\sum_{\lambda_n} [{\rm sgn}(\lambda_n)]^{\lambda_n+x_n} \right]
\Braket{ \prod_{n=1}^2 \xi^{(\lambda_n)}_{\ell_n m_n}(k_n) }.  \label{eq:cmb_power_pmf}
\end{eqnarray}
To compute the initial power spectrum $\Braket{ \prod_{n=1}^2 \xi^{(\lambda_n)}_{\ell_n m_n}(k_n) }$, we need to deal with the power spectrum of the anisotropic stresses as 
\begin{eqnarray}
\Braket{\Pi_{Bab} ({\bf k_1}) \Pi_{Bcd} ({\bf k_2})} 
&=&\left( - 4\pi \rho_{\gamma,0} \right)^{-2}
\left[ \prod_{n=1}^2 \int \frac{d^3 {\bf k_n'}}{(2 \pi)^3} \right]
\nonumber\\
&& \times
\Braket{B_a({\bf k_1'}) B_b({\bf k_1} - {\bf k_1'}) B_c({\bf k_2'}) B_d({\bf k_2}
- {\bf k_2'})} \nonumber \\
&=&\delta\left( \prod_{n=1}^2 {\bf k_n} \right) 
\left( - 4\pi \rho_{\gamma,0} \right)^{-2}  
\int d^3{\bf k_1'} P_B(k_1') P_B(| {\bf k_1} - {\bf k_1'} |) 
  \nonumber \\
&& \times \frac{1}{4} [ P_{ad}(\hat{\bf k_1'}) P_{bc}(\widehat{{\bf k_1} - {\bf k_1'}}) 
+ P_{ac}(\hat{\bf k_1'}) P_{bd}(\widehat{{\bf k_1} - {\bf k_1'}}) ]~. \label{eq:pmf_ani_power_convolve}
\end{eqnarray}
Note that this equation includes the convolution integral and the complicated angular dependence. In Refs.~\cite{Yamazaki:2008gr, Paoletti:2008ck, Finelli:2008xh, Brown:2010ms}, the authors performed the numerical and analytical computation of this convolution integral over ${\bf k_1'}$ and provided the fitting formulae with respect to the magnitude of the wave numbers $k_1$ for each value of the magnetic spectral index $n_B$. 

\begin{figure}[t]
  \begin{center}
    \includegraphics[height=8cm]{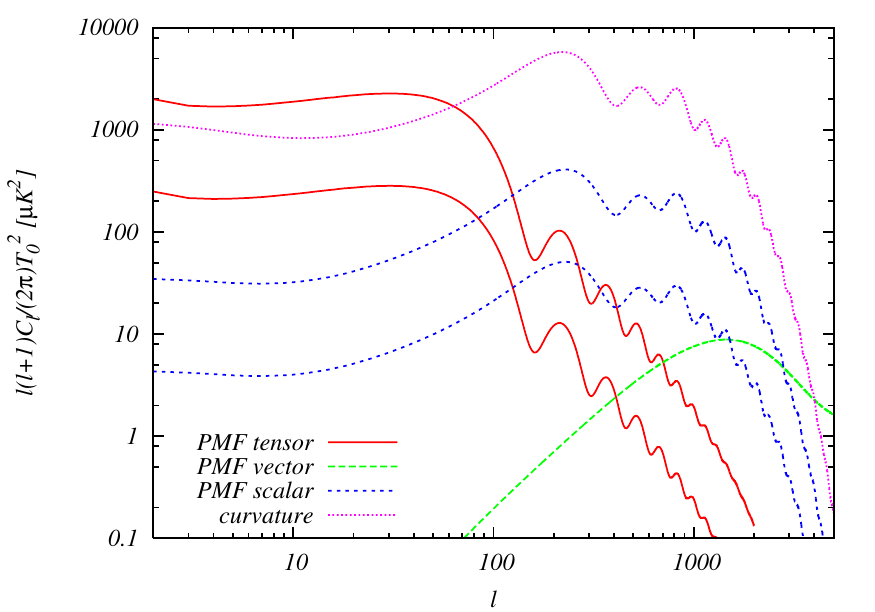}
  \end{center}
  \caption{Power spectra of the CMB intensity fluctuations. The red solid, green dashed and blue dotted lines correspond to the spectra generated from the tensor, vector and scalar components of the PMF anisotropic stress for $n_B = -2.9$, respectively. The upper (lower) line of the red solid and blue dotted ones are calculated when $\tau_{\nu}/\tau_{B} = 10^{17} (10^6)$. 
The magenta dot-dashed line expresses the spectrum sourced from the primordial curvature perturbations. The strength of PMFs is fixed to $B_{1 {\rm Mpc}} = 4.7 {\rm nG}$ and the other cosmological parameters are fixed to the mean values limited from WMAP-7yr data reported in Ref. \cite{Komatsu:2010fb}.}
  \label{fig:PMF_SVT_power.pdf}
\end{figure}

In Fig.~\ref{fig:PMF_SVT_power.pdf}, we plot the power spectra of the intensity anisotropies (\ref{eq:cmb_power_pmf}) for the scalar, vector and tensor modes when magnetic spectrum is nearly scale invariant as $n_B = -2.9$. Here, we assume that the PMFs generate from the epoch of the grand unification to that of the electroweak phase transition, i.e., $\tau_\nu / \tau_B = 10^{17} - 10^{6}$. Firstly, we will see that the shapes of the tensor and scalar power spectra are similar to those of the non-magnetic case coming from the scale-invariant primordial spectra shown in Fig.~\ref{fig:WMAP_theory_XX.pdf}. This is because PMFs impact on only the primordial gravitational waves and primordial curvature perturbations, and do not change the transfer functions of the tensor and scalar modes. For $\ell \lesssim 100$, the tensor mode dominates over the intensity signal. The scalar mode seems to dominate in the intermediate scale as $100 \lesssim \ell \lesssim 2000$. 
The vector-mode spectrum monotonically increases for larger than Silk damping scale, namely, $\ell \lesssim k_S \tau_0 \sim 2000$, and decreases for $\ell \gtrsim 2000$. These features seem to trace the scaling relation of the transfer function in terms of the wave number (\ref{eq:vec_transfer_pmf}). 
Hence, we can understand that the latter damping effect arises from the viscosity of photons. The vector mode seems to show up for very small scale, namely, $\ell \gtrsim 2000$. 

In this figure, for comparison, we also plot the CMB intensity power spectrum from the primordial curvature perturbations not depending on the PMF. In principle, comparing this spectrum with that sourced from PMFs leads to bounds on the PMF parameters. Actually, the researchers perform the parameter estimation by the Malkov Chain Monte Carlo approach \cite{Giovannini:2008yz, Yamazaki:2008gr, Finelli:2008xh, Kunze:2010ys, Paoletti:2010rx, Yamazaki:2010nf, Shaw:2010ea}. So far, the most stringent limit on the PMF strength from the CMB two-point function data of the intensity and polarizations are $B_{\rm 1Mpc} < 5 {\rm nG}$ and $n_B < -0.12$ \cite{Paoletti:2010rx}. In Refs.~\cite{Yamazaki:2008gr, Shaw:2010ea}, combining the CMB data with the information of the matter power spectrum, tighter bounds are gained.

As discussed above, conventionally, the CMB power spectra from PMFs are computed by using the fitting formulae for the power spectra of the magnetic anisotropic stresses. However, without these formulae, we can obtain the CMB power spectra by applying the mathematical tools such as the Wigner symbols \cite{Shiraishi:2011fi}. In the remaining part, focusing on the vector mode, we present this new approach and show the consistency with the conventional result. Then, we use some colors in the equations for readers to follow the complex equations more easily. 

From Eq. (\ref{eq:cmb_power_pmf}), the CMB power spectrum of the intensity
mode induced from the magnetic-vector-mode anisotropic stress is formulated as  
\begin{eqnarray}
\Braket{a^{(V)}_{I, \ell_1 m_1} a^{(V) *}_{I, \ell_2 m_2} }
&=& \Blue{ \left[ \prod_{n=1}^2 4\pi \int_0^\infty {k_n^2 dk_n \over
       (2\pi)^3} \mathcal{T}^{(V)}_{I, \ell_n}(k_n) \right] }
(-i)^{\ell_1} i^{\ell_2} \nonumber \\
&&\times
\OliveGreen{ \sum\limits_{\lambda_1, \lambda_2 = \pm 1}
\lambda_1 \lambda_2 } 
\Braket{\Pi_{Bv,\ell_1 m_1}^{(\lambda_1)}(k_1) \Pi_{Bv,\ell_2
m_2}^{(\lambda_2) *}(k_2)} \nonumber \\
&\equiv& C^{(V)}_{II, \ell_1} \delta_{\ell_1, \ell_2} \delta_{m_1, m_2} ~,
\end{eqnarray}
where the initial power spectrum, which is expanded by the spin spherical harmonics, is 
\begin{eqnarray}
\Braket{\Pi^{(\lambda_1)}_{Bv, \ell_1 m_1} (k_1) 
\Pi^{(\lambda_2) *}_{Bv, \ell_2 m_2} (k_2)}
&=& (- 4 \pi \rho_{\gamma,0})^{-2}
\Purple{\int d^2 {\hat{\bf k_1}} \int d^2 \hat{\bf k_2} {}_{- \lambda_1} 
Y^*_{\ell_1 m_1} (\hat{\bf k_1}) {}_{- \lambda_2} Y_{\ell_2 m_2}(\hat{\bf
k_2}) } \nonumber \\
&& \times 
\int_0^{k_D} k_1'^2 d k' {P}_B(k_1')  \int_0^{k_D} k_2'^2 d k_2' 
{P}_B(k_2') \nonumber \\
&&\times
\Green{ \int d^2 \hat{\bf k_1'} \int d^2 \hat{\bf k_2'} \delta({\bf k_1} -
{\bf k_1'} - {\bf k_2'}) \delta({\bf k_2} -{\bf k_2'} - {\bf k_1'}) } \nonumber \\
&& \times  
\frac{1}{4} \hat{k_1}_a \epsilon_b^{(- \lambda_1)}(\hat{\bf k_1})
\hat{k_2}_c \epsilon_d^{(\lambda_2)}(\hat{\bf k_2})
\left[ P_{ad}(\hat{\bf k_1'}) P_{bc}({\hat{\bf k_2'}}) 
+ P_{ac}({\hat{\bf k_1'}}) P_{bd}({\hat{\bf k_2'}}) \right]~, \nonumber \\
\label{eq:def_pilm_power}
\end{eqnarray}
where we use Eq.~(\ref{eq:PMF_ani}) and the definition of the vector projection operator, $O_{ab}^{(\pm 1)}(\hat{\bf k})$, in Appendix~\ref{appen:polarization}.
Note that we rewrote the power spectrum (\ref{eq:pmf_ani_power_convolve}) as more symmetric form in terms of ${\bf k_1}, {\bf k_2}$ and ${\bf k_3}$. 
Then, we should simplify this initial power spectrum. 

For the first part in two permutations, we calculate $\delta$-functions and the summations with respect to $a,b,c$ and $d$:
\begin{eqnarray}
\begin{split}
\delta({\bf k_1} - {\bf k_1'} - {\bf k_2'}) 
&= 8 \int_0^\infty A^2 dA  
\sum _{\substack{L_1 L_2 L_3 \\ \Magenta{M_1 M_2 M_3}}} 
(-1)^{\frac{L_1 + 3 L_2 + 3 L_3}{2}} 
I_{L_1 L_2 L_3}^{0~0~0} 
j_{L_1} (k_1 A) j_{L_2} (k_1' A) j_{L_3} (k_2'A) \\
& \qquad \times 
\Purple{ Y_{L_1 M_1}^*(\hat{\bf k_1}) } 
\Green{ Y_{L_2 M_2}(\hat{\bf k_1'}) Y_{L_3 -M_3}^*(\hat{\bf k_2'}) } 
\Magenta{ (-1)^{M_2} \left(
  \begin{array}{ccc}
  L_1 &  L_2 & L_3 \\
   M_1 & -M_2 & -M_3
  \end{array}
 \right) }~,  \\
\delta({\bf k_2} - {\bf k_2'} - {\bf k_1'}) 
&= 8 \int_0^\infty B^2 dB 
\sum _{\substack{L_1' L_2' L_3' \\ \Orange{M_1' M_2' M_3'} }}
(-1)^{\frac{L'_1 + 3 L'_2 + 3 L'_3}{2}}
I_{L'_1 L'_2 L'_3}^{0~0~0} 
j_{L'_1} (k_2 B) j_{L'_2} (k_2'B) j_{L'_3} (k_1'B) \\
& \qquad \times 
\Purple{ Y_{L'_1 M'_1}^*(\hat{\bf k_2}) }
\Green{ Y_{L'_2 M'_2}(\hat{\bf k_2'}) Y_{L'_3 - M'_3}^*(\hat{\bf k_1'}) }
\Orange{ (-1)^{M'_2} \left(
  \begin{array}{ccc}
  L'_1 &  L'_2 & L'_3 \\
   M'_1 & -M'_2 & -M'_3
  \end{array}
 \right) }~, \\
\hat{k_1}_a \epsilon_d^{(\lambda_2)}(\hat{\bf k_2})
P_{ad}(\hat{\bf k_1'}) 
&= \sum_{\sigma = \pm1} \sum_{\Magenta{ m_a }, \Orange{ m_d } = \pm 1, 0} 
\left(\frac{4 \pi}{3}\right)^2 \lambda_2  \\
&\quad \times
\Purple{ Y_{1 m_a}(\hat{\bf k_1})
{}_{\lambda_2} Y_{1 m_d}(\hat{\bf k_2}) }
\Green{ {}_{- \sigma} Y^*_{1 m_a}(\hat{\bf
 k_1'})
{}_{\sigma} Y^*_{1 m_d}(\hat{\bf k_1'}) } ~, \\
\hat{k_2}_c \epsilon_b^{(- \lambda_1)}(\hat{\bf k_1})
  P_{bc}({\hat{\bf k_2'}}) 
&=  \sum_{\sigma' = \pm 1} \sum_{\Orange{m_c}, \Magenta{m_b} = \pm 1, 0} 
\left(\frac{4 \pi}{3}\right)^2 (- \lambda_1) \\
&\quad \times
\Purple{ Y_{1 m_c}(\hat{\bf k_2}) {}_{-\lambda_1} Y_{1 m_b}(\hat{\bf k_1}) }
\Green{ {}_{-\sigma'} Y^*_{1 m_c}(\hat{\bf k_2'})
{}_{\sigma'} Y^*_{1 m_b}(\hat{\bf k_2'}) } ~,
\end{split}
\end{eqnarray}
perform the angular integrals of the spin spherical harmonics:
 \begin{eqnarray}
\begin{split}
&\Green{ \int d^2 \hat{\bf k_1'} {}_{- \sigma} Y^*_{1 m_a} 
Y_{L_2 M_2} {}_\sigma Y^*_{1 m_d} Y^*_{L'_3 -M'_3} } \\
&\qquad\qquad = \sum_{L \Cyan{M} S} (-1)^{\sigma + \Magenta{m_a}} I_{L_3' 1 L}^{0 -\sigma -S} I_{L_2 1
L}^{0 -\sigma -S} 
\Orange{ \left(
  \begin{array}{ccc}
  L'_3 &  1 & L \\
  - M'_3 & m_d & M 
  \end{array}
 \right) }
\Magenta{ \left(
  \begin{array}{ccc}
  L_2 &  1 & L \\
   M_2 & -m_a & M 
  \end{array}
 \right) }~, \\
&\Green{ \int d^2 \hat{\bf k_2'} 
{}_{-\sigma'} Y^*_{1 m_c} Y_{L'_2 M'_2} 
{}_{\sigma'} Y^*_{1 m_b} Y^*_{L_3 -M_3} } \\
&\qquad\qquad = \sum_{L' \Cyan{M'} S'} (-1)^{\sigma' + \Orange{m_c}} I_{L_3 1 L'}^{0 -\sigma' -S'}
I_{L'_2 1 L'}^{0 -\sigma' -S'} 
\Magenta{ \left(
  \begin{array}{ccc}
  L_3 &  1 & L' \\
   -M_3 & m_b & M' 
  \end{array}
 \right) } 
\Orange{ \left(
  \begin{array}{ccc}
  L'_2 &  1 & L' \\
   M'_2 & -m_c & M' 
  \end{array}
 \right) } ~, \\
&\Purple{ \int d^2 \hat{\bf k_1} 
{}_{-\lambda_1} Y_{1 m_b} Y_{1 m_a}
{}_{-\lambda_1} Y^*_{\ell_1 m_1} Y^*_{L_1 M_1} } \\
&\qquad\qquad = \sum_{L_k \Magenta{M_k} S_k} I_{L_1 \ell_1 L_k}^{0 \lambda_1 -S_k} I_{1 1 L_k}^{0 \lambda_1 -S_k}
\Magenta{ \left(
  \begin{array}{ccc}
  L_1 & \ell_1 & L_k \\
  M_1 & m_1 & M_k 
  \end{array}
 \right)
\left(
  \begin{array}{ccc}
  1 &  1 & L_k \\
  m_a & m_b & M_k 
  \end{array}
 \right) } ~, \\
&\Purple{ \int d^2 \hat{\bf k_2} 
{}_{\lambda_2} Y_{1 m_d} Y_{1 m_c} 
{}_{-\lambda_2} Y_{\ell_2 m_2} Y^*_{L'_1 M'_1} } \\
&\qquad\qquad
= \sum_{L_p \Orange{M_p} S_p} (-1)^{m_2 + \lambda_2 } I_{L'_1 \ell_2 L_p}^{0
\lambda_2 -S_p} I_{1 1 L_p}^{0 \lambda_2 -S_p} 
\Orange{ \left(
  \begin{array}{ccc}
  L'_1 & \ell_2 & L_p \\
  - M'_1 & m_2 & M_p 
  \end{array}
 \right)  
\left(
  \begin{array}{ccc}
  1 &  1 & L_p \\
  -m_c & -m_d & M_p 
  \end{array}
 \right) } ~, 
\end{split}
\end{eqnarray}
sum up the Wigner-$3j$ symbols over the azimuthal quantum numbers:
\begin{eqnarray}
\begin{split}
& \Magenta{ \sum_{\substack{M_1 M_2 M_3 \\ M_k m_a m_b}} 
(-1)^{M_2+m_a} 
\left(
  \begin{array}{ccc}
  1 & 1 & L_k \\
  m_a & m_b & M_k 
  \end{array}
 \right) 
\left(
  \begin{array}{ccc}
  L_1 & L_2 & L_3 \\
  M_1 & - M_2 & - M_3 
  \end{array}
 \right)  \\
&\qquad\times 
\left(
  \begin{array}{ccc}
  L_3 & 1 & L' \\
  - M_3 & m_b & M' 
  \end{array}
 \right) 
 \left(
  \begin{array}{ccc}
  L_2 &  1 & L \\
  M_2 & -m_a & M 
  \end{array}
 \right)
 \left(
  \begin{array}{ccc}
  L_1 & \ell_1 & L_k \\
  M_1 & m_1 & M_k 
  \end{array}
 \right) } \\
& \qquad\qquad = (-1)^{\Cyan{M} + \ell_1 + L_3 + L + 1} 
\Cyan{ \left(
  \begin{array}{ccc}
  L' & L & \ell_1 \\
  M' & -M & m_1 
  \end{array}
 \right) }
\left\{
  \begin{array}{ccc}
  L' & L & \ell_1 \\
  L_3 & L_2 & L_1 \\
  1 & 1 & L_k 
  \end{array}
 \right\}  ~, \\
& \Orange{ \sum_{\substack{M'_1 M'_2 M'_3 \\ M_p m_c m_d }} 
(-1)^{M'_2+m_c}  
\left(
  \begin{array}{ccc}
  1 & 1 & L_p \\
 -m_c & -m_d & M_p 
  \end{array}
 \right)
\left(
  \begin{array}{ccc}
  L'_1 & L'_2 & L'_3 \\
  M'_1 & - M'_2 & - M'_3 
  \end{array}
 \right) \\
&\qquad
\left(
  \begin{array}{ccc}
  L'_2 &  1 & L' \\
  M'_2 & -m_c & M' 
  \end{array}
 \right)
\left(
  \begin{array}{ccc}
  L'_3 & 1 & L \\
  - M'_3 & m_d & M 
  \end{array}
 \right) 
\left(
  \begin{array}{ccc}
  L'_1 & \ell_2 & L_p \\
  -M'_1 & m_2 & M_p 
  \end{array}
 \right) } \\
&\qquad\qquad = (-1)^{\Cyan{M'} + \ell_2 + L'_2 + L + 1 + L_p} 
\Cyan{ \left(
  \begin{array}{ccc}
  L' & L & \ell_2 \\
  M' & -M & m_2 
  \end{array}
 \right) } 
\left\{
  \begin{array}{ccc}
  L' & L & \ell_2 \\
  L'_2 & L'_3 & L'_1 \\
  1 & 1 & L_p 
  \end{array}
 \right\} ~,
\end{split}
\end{eqnarray}
and sum up the Wigner-$3j$ symbols over $M$ and $M'$:
\begin{eqnarray}
\Cyan{ \sum_{M M'}(-1)^{M + M'}
\left(
  \begin{array}{ccc}
  L' & L & \ell_1 \\
  M' & -M & m_1 
  \end{array}
 \right) 
\left(
  \begin{array}{ccc}
  L' & L & \ell_2 \\
  M' & -M & m_2 
  \end{array}
 \right) } 
=  \frac{(-1)^{m_2}}{2 \ell_1 + 1} \delta_{\ell_1, \ell_2} 
\delta_{m_1, m_2} ~.
\end{eqnarray}
Following the same procedures in the other permutation and calculating
the summation over $L_p$ as
\begin{eqnarray} 
\sum_{L_p} I^{0 \lambda_2 -\lambda_2}_{L'_1 \ell_2 L_p} 
I^{0 \lambda_2 -\lambda_2}_{1 1 L_p} \frac{1 + (-1)^{L_p}}{2}
\left\{
  \begin{array}{ccc}
   L' & L & \ell_2\\
  L'_2 & L'_3 & L'_1 \\
  1 & 1 & L_p
  \end{array}
 \right\}
= - \frac{3}{2 \sqrt{2 \pi}} 
I_{L'_1 \ell_2 2}^{0 \lambda_2 -\lambda_2}
 \left\{
  \begin{array}{ccc}
  L' & L & \ell_2 \\
  L'_3 & L'_2 & L'_1 \\
  1 & 1 & 2
  \end{array}
 \right\}
~, 
\end{eqnarray}
we can obtain
the exact solution of Eq.~(\ref{eq:def_pilm_power}) as 
 \begin{eqnarray}
\Braket{\Pi^{(\lambda_1)}_{Bv, \ell_1 m_1}(k_1)
 \Pi^{(\lambda_2) *}_{Bv, \ell_2 m_2}(k_2)} &=&
 - \frac{\sqrt{2\pi}}{3} 
\left(\frac{8 (2\pi)^{1/2}}{3 \rho_{\gamma,0}}\right)^2 
 / (2\ell_1 + 1) 
\delta_{\ell_1, \ell_2} \delta_{m_1,m_2}
\nonumber \\ 
&& \times \sum _{L L'}  \sum_{\substack{L_1 L_2 L_3 \\ L'_1 L'_2 L'_3}}
 (-1)^{\sum_{i=1}^3 \frac{L_i + L'_i}{2} }
I^{0~0~0}_{L_1 L_2 L_3} I^{0~0~0}_{L'_1 L'_2 L'_3}
\nonumber \\
&& \times 
\sum_{L_k} (-1)^{L'_2 + L_3} 
\left\{
  \begin{array}{ccc}
  L' & L & \ell_1 \\
  L_3 & L_2 & L_1 \\
  1 & 1 & L_k
  \end{array}
 \right\}
\left\{
  \begin{array}{ccc}
   L' & L & \ell_2\\
  L'_2 & L'_3 & L'_1 \\
  1 & 1 & 2
  \end{array}
 \right\} \nonumber \\
&& \times 
\Blue{ \int_0^\infty A^2 dA 
j_{L_1} (k_1 A)  
\int_0^\infty B^2 dB  
j_{L'_1} (k_2 B)  \nonumber \\
&& \times
\int_0^{k_D} k_1'^2 dk_1' {P}_B(k_1') j_{L_2} (k_1'A) j_{L'_3} (k_1' B) \nonumber \\
&&\times \int_0^{k_D} k_2'^2 dk_2' {P}_B(k_2') j_{L'_2} (k_2'B) j_{L_3} (k_2' A) }
 \nonumber \\ 
&& \times   
\Red{ \sum_{S,S' = \pm 1} (-1)^{L_2 + L_2' + L_3 + L'_3} I^{0 S -S}_{L'_3 1 L} I^{0 S -S}_{L_2 1 L} I^{0 S' -S'}_{L_3 1 L'} 
I^{0 S' -S'}_{L'_2 1 L'} } \nonumber \\
&&  \times 
\OliveGreen{
\lambda_1 \lambda_2 
I^{0 \lambda_1 -\lambda_1}_{L_1 \ell_1 L_k} I^{0 \lambda_1
-\lambda_1}_{1 1 L_k} I^{0 \lambda_2 -\lambda_2}_{L'_1 \ell_2 2} } ~.
\end{eqnarray}
Note that in this equation, the dependence on the azimuthal quantum
number is included only in $\delta_{m_1, m_2}$. In the similar discussion of
the CMB bispectrum, this implies that the
CMB vector-mode power spectrum generated from the magnetized anisotropic
stresses is rotationally-invariant if the PMFs satisfy the statistical
isotropy as Eq. (\ref{eq:def_pmf_power}).

Furthermore, using such evaluations as
\begin{eqnarray}
&& \Red{ \sum_{S, S' = \pm 1} (-1)^{L_2 + L'_2 + L_3 + L'_3}
I^{0 S -S}_{L'_3 1 L} I^{0 S -S}_{L_2 1 L} 
I^{0 S' -S'}_{L_3 1 L'} I^{0 S' -S'}_{L'_2 1 L'} } \nonumber \\
&&\qquad\quad =  
\begin{cases}
4 I_{L'_3 1 L}^{0 1 -1} I_{L_2 1 L}^{0 1 -1} 
I_{L_3 1 L'}^{0 1 -1} I_{L'_2 1 L'}^{0 1 -1} 
& ( L_3' + L_2, L_3 + L_2' = {\rm even} ) \\
0 & ({\rm otherwise})
\end{cases}~, \\
&& \OliveGreen{ \sum_{\lambda_1, \lambda_2 = \pm 1} 
I^{0 \lambda_1 -\lambda_1}_{L_1 \ell_1 L_k} I^{0 \lambda_1
-\lambda_1}_{1 1 L_k} I^{0 \lambda_2 -\lambda_2}_{L'_1 \ell_2 2} }
=
\begin{cases}
4 I_{L_1 \ell_1 L_k}^{0 1 -1} I_{1 1 L_k}^{0 1 -1}
I_{L_1' \ell_2 2}^{0 1 -1}  &
( L_1 + \ell_1, L_1' + \ell_2 = {\rm even} ) \\
0 & ({\rm otherwise}) 
\end{cases} ~, \\
&& 
\Blue{ \left[ \prod_{n=1}^2 4\pi \int_0^\infty {k_n^2 dk_n \over
       (2\pi)^3} \mathcal{T}^{(V)}_{I, \ell_n}(k_n) \right]
\int_0^\infty A^2 dA j_{L_1} (k_1 A)  
\int_0^\infty B^2 dB j_{L'_1} (k_2 B)  \nonumber \\
&& \qquad\qquad \times
\int_0^{k_D} k_1'^2 dk_1' {P}_B(k_1') j_{L_2} (k_1' A) j_{L'_3} (k_1' B)
\int_0^{k_D} k_2'^2 dk_2' {P}_B(k_2') j_{L'_2} (k_2' B) j_{L_3} (k_2' A) }
 \nonumber \\ 
&& \qquad \simeq  
\left[ \prod_{n=1}^2 4\pi \int_0^\infty {k_n^2 dk_n \over
       (2\pi)^3} \mathcal{T}^{(V)}_{I, \ell_n}(k_n) 
j_{\ell_n} (k_n (\tau_0-\tau_*)) \right] \nonumber \\
&& \qquad\qquad\qquad \times A_B^2 (\tau_0 - \tau_*)^4 \left( \frac{\tau_*}{5} \right)^2 
\mathcal{K}_{L_2 L'_3}^{-(n_B +1)}(\tau_0-\tau_*)
\mathcal{K}_{L'_2 L_3}^{-(n_B +1)}(\tau_0-\tau_*) ~,
\end{eqnarray}
the CMB angle-averaged power spectrum is formulated as
\begin{eqnarray}
C^{(V)}_{II,\ell}
&\simeq& - \frac{\sqrt{2\pi}}{3} \left( \frac{32 (2\pi)^{1/2}}{3\rho_{\gamma,0}} \right)^2 / (2 \ell + 1)
\left[
4\pi \int_0^\infty {k^2 dk \over (2\pi)^3}
\mathcal{T}^{(V)}_{I, \ell}(k) j_{\ell} (k (\tau_0-\tau_*))
\right]^2 \nonumber \\
&& \times
\sum_{L_1 L_1'} 
\sum_{L_k} I_{L_1 \ell L_k}^{0 1 -1} I_{1 1 L_k}^{0 1 -1}
I_{L_1' \ell 2}^{0 1 -1} 
\sum_{L L'} 
\sum_{\substack{L_2 L'_2 \\ L'_3 L_3}} 
A_B^2 (\tau_0-\tau_*)^4 \left( {\tau_* \over 5} \right)^2 \nonumber \\
&&\times \mathcal{K}_{L_2 L'_3}^{-(n_B +1)}(\tau_0-\tau_*)
\mathcal{K}_{L'_2 L_3}^{-(n_B +1)}(\tau_0-\tau_*) \nonumber \\
&& \times (-1)^{\sum_{i=1}^3 
\frac{L_i + L'_i}{2} + L'_2 + L_3} I^{0~0~0}_{L_1 L_2 L_3} 
 I^{0~0~0}_{L'_1 L'_2 L'_3} 
I_{L'_3 1 L}^{0 1 -1} I_{L_2 1 L}^{0 1 -1} 
I_{L_3 1 L'}^{0 1 -1} I_{L'_2 1 L'}^{0 1 -1} \nonumber \\
&&\times 
\left\{
  \begin{array}{ccc}
  L' & L & \ell \\
  L_3 & L_2 & L_1 \\
  1 & 1 & L_k
  \end{array}
 \right\}
\left\{
  \begin{array}{ccc}
  L' & L & \ell \\
  L'_2 & L'_3 & L'_1 \\
  1 & 1 & 2
  \end{array}
 \right\}~. \label{eq:cmb_power_pmf_vec} 
\end{eqnarray}
Here, we use the thin LSS approximation described in Sec.~\ref{subsubsec:thinLSSapp}. This has nonzero value in the configurations:
\begin{eqnarray}
\begin{split}
& (L_k, L_1) = (2, |\ell \pm 2|), (2, \ell), (1, \ell) ~, \ \ 
L'_1 = |\ell \pm 2|, \ell ~, \\
& |\ell - L| \leq L' \leq \ell + L ~, \\
& (L_2, L'_3) = (|L-1|,|L \pm 1|), (L,L), (L+1, |L \pm 1|) ~, \\
& (L'_2, L_3) = (|L'-1|,|L' \pm 1|), (L',L'), (L'+1, |L' \pm 1|)~, \\
& L_1 + L_2 + L_3 = {\rm even} ~, \ \ L_1' + L_2' + L_3' = {\rm even} ~, \\ 
& |L_1 - L_2| \leq L_3 \leq L_1 + L_2 ~, \ \ 
|L'_1 - L'_2| \leq L'_3 \leq L'_1 + L'_2 ~.
\end{split}
\end{eqnarray}

\begin{figure}[t]
  \begin{center}
    \includegraphics[height=8cm]{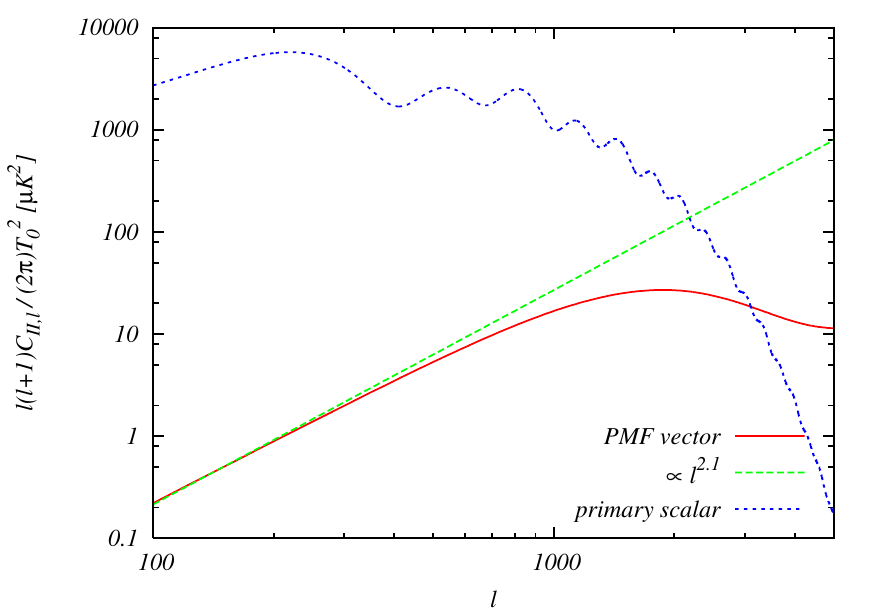}
  \end{center}
  \caption{CMB power spectra of the temperature fluctuations. The lines
 correspond to the spectra generated from vector anisotropic stress of PMFs as Eq.~(\ref{eq:cmb_power_pmf_vec}) (red solid line) and primordial curvature perturbations (blue dotted line). The green dashed line expresses the asymptotic power of the red solid one.
The PMF parameters are fixed to $n_B = -2.9$ and $B_{1 {\rm Mpc}} = 4.7 {\rm nG}$, and the other cosmological parameters are fixed to the mean values limited from WMAP-7yr data reported in Ref. \cite{Komatsu:2010fb}.
  }
  \label{fig:vec_II_check.pdf}
\end{figure}

This shape is described in Fig.~\ref{fig:vec_II_check.pdf}. 
From this figure, we confirm that the amplitude and the overall behavior of the red solid line are in broad agreement with the green dashed line of Fig.~\ref{fig:PMF_SVT_power.pdf} and the previous studies (e.g. \cite{Lewis:2004ef, Yamazaki:2006bq, Paoletti:2008ck, Shaw:2009nf}). 
For $\ell \lesssim 2000$, using the scaling relations of the Wigner
symbols at the dominant configuration $L \sim \ell, L' \sim 1$ as discussed in Sec. \ref{subsec:result_pmf}, we analytically find that $C^{(V)}_{I,\ell} \propto \ell^{n_B+3}$. This traces our numerical results as shown by the green dashed line. 

This computation approach is of great utility in the higher-order correlations. In the next subsection, in accordance with this approach, we compute the CMB bispectra sourced from PMFs. 

\subsection{Formulation for the CMB bispectrum induced from PMFs} \label{subsec:analytic_calc}

 
In this subsection, we derive the explicit form of the CMB bispectra induced from PMFs by calculating the full-angular dependence
which has never been considered in the previous studies
\cite{Seshadri:2009sy, Caprini:2009vk, Cai:2010uw, Trivedi:2010gi,
Kahniashvili:2010us}.  The following procedures are based on the
calculation rules discussed in Ref.~\cite{Shiraishi:2011fi}. 
Note that we use some colors in the following equations for readers to
follow the complex equations more easily.

\subsubsection{Bispectrum of the anisotropic stress fluctuations}

According to Eq.~(\ref{eq:EMT_PMF}), EMT of PMF at an arbitrary point,
$T^\mu{}_\nu (\bf x)$, depends quadratically on the Gaussian magnetic fields at
that point. This non-Gaussian structure is identical to the local-type
non-Gaussianity of the curvature perturbations as mentioned in Sec~\ref{subsec:NG_shape}, hence it is expected
that the statistical properties of the magnetic fields obey those of the local-type non-Gaussianity. This will be automatically shown in Sec.~\ref{subsec:optimal_pmf}. 

Using Eq. (\ref{eq:def_pmf_power}) and the Wick's theorem, the bispectrum of the anisotropic stresses is calculated as
\begin{eqnarray}
\Braket{\Pi_{Bab} ({\bf k_1}) \Pi_{Bcd} ({\bf k_2})
 \Pi_{B ef} ({\bf k_3})} 
&=&\left( - 4\pi \rho_{\gamma,0} \right)^{-3}
\left[ \prod_{n=1}^3 \int \frac{d^3 {\bf k_n'}}{(2 \pi)^3} \right]
\nonumber\\
&& \times
\Braket{B_a({\bf k_1'}) B_b({\bf k_1} - {\bf k_1'}) B_c({\bf k_2'}) B_d({\bf k_2}
- {\bf k_2'}) B_e({\bf k_3'}) B_f({\bf k_3} - {\bf k_3'})} \nonumber \\
&=&\left( - 4\pi \rho_{\gamma,0} \right)^{-3} 
\left[ \prod_{n=1}^3 
\int_0^{k_D} k_n'^2 dk_n' P_B(k_n') \int d^2 \hat{\bf k_n'} \right]
\nonumber \\
&& \times 
\delta({\bf k_1} - {\bf k_1'} + {\bf k_3'}) 
\delta({\bf k_2} - {\bf k_2'} + {\bf k_1'}) 
\delta({\bf k_3} - {\bf k_3'} + {\bf k_2'}) \nonumber \\
&& \times \frac{1}{8} [P_{ad}(\hat{\bf k_1'}) P_{be}(\hat{\bf k_3'})
P_{cf}(\hat{\bf k_2'}) + \{a \leftrightarrow b \ {\rm or} \ c \leftrightarrow d \ {\rm or} \ e \leftrightarrow f\}]~, \nonumber \\ 
\label{eq:bis_EMT}
\end{eqnarray}
where $k_D$ is the Alfv\'en-wave damping length scale
\cite{Jedamzik:1996wp, Subramanian:1997gi} as $k_D^{-1} \sim {\cal
O}(0.1)\rm Mpc$ and the curly bracket denotes the symmetric $7$ terms
under the permutations of indices: $a \leftrightarrow b$, $c
\leftrightarrow d$, or $e \leftrightarrow f$. Note that we express in a
more symmetric form than that of Ref. \cite{Brown:2005kr} to perform the
angular integrals which is described in
Sec.~\ref{subsec:analytic_calc}. To avoid the divergence of
$\Braket{\Pi_{Bab} ({\bf k_1}) \Pi_{Bcd} ({\bf k_2}) \Pi_{Bef} ({\bf
k_3})}$ in the IR limit, the value range of the spectral index is
limited as $n_B > -3$.
We note that this bispectrum depends on the Gaussian PMFs to six,
hence this is highly non-Gaussian compared with the bispectrum of
primordial curvature perturbations proportional to the Gaussian variable
to four as shown in Sec.~\ref{subsec:NG_shape}. 

\subsubsection{CMB all-mode bispectrum}

Following the general formula (\ref{eq:cmb_bis_general}) and using Eq.~(\ref{eq:alm_PMF}), the CMB bispectra induced from PMF are written as
\begin{eqnarray}
\Braket{\prod^3_{n=1} a_{X_n, \ell_n m_n}^{(Z_n)}}
= \left[ \prod^3_{n=1}
4\pi (-i)^{\ell_n}
\int_0^\infty {k_n^2 dk_n \over (2\pi)^3}
\mathcal{T}^{(Z_n)}_{X_n, \ell_n}(k_n) \sum_{\lambda_n}
 [{\rm sgn}(\lambda_n)]^{\lambda_n + x_n} \right]  
\Braket{\prod_{n=1}^3 \xi_{\ell_n m_n}^{(\lambda_n)}(k_n)}~. \nonumber \\
\label{eq:CMB_bis_PMF_general} 
\end{eqnarray}
Remember that the index $Z$ denotes the mode of perturbations: $Z = S$
(scalar), $= V$ (vector) or $=T$ (tensor) and its helicity is expressed
by $\lambda$; $\lambda = 0$ for $(Z = S)$, $= \pm 1$ for $(Z = V)$ or $=
\pm 2$ for $(Z = T)$, $X$ discriminates between intensity and two
polarization (electric and magnetic) modes, respectively, as $X = I,E,B$
and $x$ is determined by it: $x = 0$ for $X = I,E$ or $=1$ for $X = B$. 
In the following discussion, we calculate $\Braket{\prod_{n=1}^3 \xi_{\ell_n
m_n}^{(\lambda_n)}(k_n)}$ and find an explicit
formulae of the CMB bispectra corresponding to an arbitrary $Z$. 

Using Eqs. (\ref{eq:alm_PMF}) and (\ref{eq:bis_EMT}), we can write 
\begin{eqnarray}
\Braket{\prod_{n=1}^3 \xi_{\ell_n m_n}^{(\lambda_n)}(k_n)} 
&=& (-4 \pi \rho_{\gamma,0})^{-3}
\left[ \prod_{n=1}^3 \Purple{ \int d^2 \hat{\bf k_n}
	  {}_{-\lambda_n}Y_{\ell_n m_n}^* (\hat{\bf k_n}) } 
\int^{k_D}_0 k_n'^2 dk_n' P_B(k_n') \Green{\int d^2 \hat{\bf k_n'}}\right] 
 \nonumber \\
&& \times  
\delta({\bf k_1} - {\bf k_1'} + {\bf k_3'}) \delta({\bf k_2} - {\bf k_2'} + {\bf
k_1'}) \delta({\bf k_3} - {\bf k_3'} + {\bf k_2'}) \nonumber \\
&& \times  
C'_{-\lambda_1} O_{ab}^{(-\lambda_1)}(\hat{\bf k_1})
C'_{-\lambda_2} O_{cd}^{(- \lambda_2)}(\hat{\bf k_2}) 
C'_{-\lambda_3} O_{ef}^{(- \lambda_3)}(\hat{\bf k_3}) 
\nonumber \\
&& \times 
P_{ad} (\hat{\bf k_1'})  P_{be} (\hat{\bf k_3'}) P_{cf} (\hat{\bf k_2'})
~, \label{eq:xi_bis_SVT_explicit} 
\end{eqnarray}  
with
\begin{eqnarray}
C_{\lambda}' \equiv 
\begin{cases}
\frac{3}{2} R_\gamma \ln \left(\frac{\tau_\nu}{\tau_B}\right)  & (\lambda = 0)\\
\frac{1}{2} & (\lambda = \pm 1) \\
3 R_\gamma \ln \left(\frac{\tau_\nu}{\tau_B}\right) & (\lambda = \pm 2) \\
\end{cases}
~.
\end{eqnarray}
Let us consider this exact expression by expanding all the angular
dependencies with the spin-weighted spherical harmonics and rewriting the
angular integrals with the summations in terms of the multipoles and azimuthal quantum numbers. 

In the first step,
in order to perform all angular integrals, we expand each function of the wave number vectors with the spin-weighted spherical harmonics. 
By this concept, three delta functions are rewritten as
\begin{eqnarray}
\begin{split}
\delta({\bf k_1} - {\bf k_1'} + {\bf k_3'}) 
&= 8 \int_0^\infty A^2 dA
\sum _{\substack{L_1 L_2 L_3 \\ \Magenta{M_1 M_2 M_3}}} 
(-1)^{\frac{L_1 + 3 L_2 + L_3}{2}} 
I_{L_1 L_2 L_3}^{0~0~0} 
j_{L_1} (k_1 A) j_{L_2} (k_1' A) j_{L_3} (k_3' A) \\
&\quad \times \Purple{Y_{L_1 M_1}^*(\hat{\bf k_1})} 
\Green{ Y_{L_2 M_2}(\hat{\bf k_1'}) Y_{L_3 M_3}^*(\hat{\bf k_3'}) }
\Magenta{ (-1)^{M_2} \left(
  \begin{array}{ccc}
  L_1 &  L_2 & L_3 \\
   M_1 & -M_2 & M_3
  \end{array}
 \right) }~, \\
\delta({\bf k_2} - {\bf k_2'} + {\bf k_1'}) 
&= 8 \int_0^\infty B^2 dB 
\sum _{\substack{L_1' L_2' L_3' \\ \Orange{M_1' M_2' M_3'}}} 
(-1)^{\frac{L_1' + 3 L_2' + L_3'}{2}}  
I_{L'_1 L'_2 L'_3}^{0~0~0} 
j_{L'_1} (k_2 B) j_{L'_2} (k_2' B) j_{L'_3} (k_1'B) \\
&\quad \times \Purple{Y_{L'_1 M'_1}^*(\hat{\bf k_2}) }
\Green{ Y_{L'_2 M'_2}(\hat{\bf k_2'}) Y_{L'_3 M'_3}^*(\hat{\bf k_1'}) }
\Orange{ (-1)^{M'_2} \left(
  \begin{array}{ccc}
  L'_1 &  L'_2 & L'_3 \\
   M'_1 & -M'_2 & M'_3
  \end{array}
 \right) }~, \\
\delta({\bf k_3} - {\bf k_3'} + {\bf k_2'}) 
&= 8 \int_0^\infty C^2 dC 
\sum _{\substack{L_1'' L_2'' L_3'' \\ \Brown{M_1'' M_2'' M_3''}}} 
(-1)^{\frac{L''_1 + 3 L''_2 + L''_3}{2}} 
I_{L''_1 L''_2 L''_3}^{0~0~0} 
j_{L''_1} (k_3 C) j_{L''_2} (k_3' C) j_{L''_3} (k_2 'C) \\
&\quad \times \Purple{Y_{L''_1 M''_1}^*(\hat{\bf k_3}) }
\Green{ Y_{L''_2 M''_2}(\hat{\bf k_3'}) Y_{L''_3 M''_3}^*(\hat{\bf k_2'}) }
\Brown{ (-1)^{M''_2} \left(
  \begin{array}{ccc}
  L''_1 &  L''_2 & L''_3 \\
   M''_1 & -M''_2 & M''_3
  \end{array}
 \right) }~, \label{eq:delta_C}
\end{split}
\end{eqnarray}
where
\begin{eqnarray}
I^{s_1 s_2 s_3}_{\ell_1 \ell_2 \ell_3} 
\equiv \sqrt{\frac{(2 \ell_1 + 1)(2 \ell_2 + 1)(2 \ell_3 + 1)}{4 \pi}}
\left(
  \begin{array}{ccc}
  \ell_1 & \ell_2 & \ell_3 \\
   s_1 & s_2 & s_3 
  \end{array}
 \right)~.
\end{eqnarray}
The other functions in Eq.~(\ref{eq:xi_bis_SVT_explicit}), which depend on the
angles of the wave number vectors, can be also expanded in terms
of the spin-weighted spherical harmonics as 
\begin{eqnarray}
&& O_{ab}^{(-\lambda_1)}(\hat{\bf k_1}) O_{cd}^{(-\lambda_2)}(\hat{\bf k_2})
O_{ef}^{(-\lambda_3)}(\hat{\bf k_3})
P_{ad} (\hat{\bf k_1'})  P_{be} (\hat{\bf k_3'}) P_{cf} (\hat{\bf k_2'})
\nonumber \\
&&\qquad\qquad 
= \sum_{\sigma_1, \sigma_2, \sigma_3 = \pm 1} 
- C_{-\lambda_1} \sqrt{\frac{3}{8\pi}} \sum_{\Magenta{\mu_1 m_a m_b}} 
\left(\frac{4 \pi}{3}\right)^2 
\nonumber \\
&&\qquad\qquad\quad\times 
\Purple{{}_{\lambda_1} Y_{2 \mu_1}^* (\hat{\bf k_1}) }
\Magenta{
\left(
  \begin{array}{ccc}
  2 & 1 & 1 \\
  \mu_1 & m_a & m_b
  \end{array}
 \right) 
}
\Green{ {}_{- \sigma_1}Y^*_{1 m_a} (\hat{\bf k_1'}) {}_{\sigma_3}Y^*_{1 m_b} (\hat{\bf k_3'})}
\nonumber \\
&&\qquad\qquad\quad 
\times C_{-\lambda_2} \sqrt{\frac{3}{8\pi}} \sum_{\Orange{\mu_2 m_c m_d}} 
\left(\frac{4 \pi}{3}\right)^2
\Purple{ {}_{\lambda_2} Y_{2 \mu_2}^* (\hat{\bf k_2}) }
\Orange{
\left(
  \begin{array}{ccc}
  2 & 1 & 1 \\
  \mu_2 & m_c & m_d
  \end{array}
 \right)
} 
\Green{{}_{- \sigma_2}Y^*_{1 m_c} (\hat{\bf k_2'}) {}_{\sigma_1}Y^*_{1 m_d}
 (\hat{\bf k_1'}) }
\nonumber \\
&&\qquad\qquad\quad 
\times C_{-\lambda_3} \sqrt{\frac{3}{8\pi}} 
 \sum_{\Brown{ \mu_3 m_e m_f}} 
\left(\frac{4 \pi}{3}\right)^2
\Purple{ {}_{\lambda_3} Y_{2 \mu_3}^* (\hat{\bf k_3}) } 
\Brown{
\left(
  \begin{array}{ccc}
  2 & 1 & 1 \\
  \mu_3 & m_e & m_f
  \end{array}
 \right)
} 
\Green{ {}_{- \sigma_3}Y^*_{1 m_e} (\hat{\bf k_3'}) {}_{\sigma_2}Y^*_{1 m_f} (\hat{\bf k_2'}) }~, \nonumber \\
\end{eqnarray}
where we have used Eqs.~(\ref{eq:projection}), some relations in Appendices~\ref{appen:wigner} and \ref{appen:polarization}, and the definition (\ref{eq:projection_operator}) as 
\begin{eqnarray}
\begin{split}
O_{ab}^{(\lambda)}(\hat{\bf k}) 
&= C_\lambda \sqrt{\frac{3}{8\pi}}
\sum_{M m_a m_b} {}_{-\lambda}Y_{2 M}^*(\hat{\bf k})
\alpha_a^{m_a} \alpha_b^{m_b}
 \left(
  \begin{array}{ccc}
  2 & 1 & 1 \\
  M & m_a & m_b
  \end{array}
 \right) ~, \\
C_\lambda 
&= \begin{cases}
-2  & (\lambda = 0) \\
 2\sqrt{3} \lambda & (\lambda = \pm 1) \\ 
 2\sqrt{3} & (\lambda = \pm 2)  
\end{cases} ~.
\end{split}
 \end{eqnarray}

In the second step, let us consider performing all angular integrals and
replacing them with the Wigner-$3j$ symbols.  Three angular integrals with
respect to $\hat{\bf k_1'}, \hat{\bf k_2'}$ and $\hat{\bf k_3'}$ are
given as 
\begin{eqnarray}
\begin{split}
& \Green{
\int d^2 \hat{\bf k_1'} {}_{-\sigma_1} Y^*_{1 m_a} Y_{L_2 M_2} 
{}_{\sigma_1} Y^*_{1 m_d} Y^*_{L'_3 M'_3}} \\
&\qquad\qquad= - \sum_{L \Cyan{M}} \sum_{S = \pm 1} (-1)^{\Magenta{m_a}} I_{L_3' 1 L}^{0 -\sigma_1 -S} I_{L_2 1 L}^{0 -\sigma_1 -S} 
\Orange{ \left(
  \begin{array}{ccc}
  L'_3 &  1 & L \\
   M'_3 & m_d & M 
  \end{array}
 \right) } 
\Magenta{ \left(
  \begin{array}{ccc}
  L_2 &  1 & L \\
   M_2 & -m_a & M 
  \end{array}
 \right) }~, \\
&\Green{
\int d^2 \hat{\bf k_2'} {}_{- \sigma_2} Y^*_{1 m_c} Y_{L'_2 M'_2} 
{}_{\sigma_2} Y^*_{1 m_f} Y^*_{L''_3 M''_3} } \\
&\qquad\qquad= - \sum_{L' \Cyan{M'}} \sum_{S' = \pm 1} (-1)^{\Orange{m_c}} I_{L_3'' 1 L'}^{0 -\sigma_2 -S'} I_{L'_2 1 L'}^{0 -\sigma_2 -S'} 
\Brown{ \left(
  \begin{array}{ccc}
  L''_3 &  1 & L' \\
   M''_3 & m_f & M' 
  \end{array}
 \right) } 
\Orange{ \left(
  \begin{array}{ccc}
  L'_2 &  1 & L' \\
   M'_2 & -m_c & M' 
  \end{array}
 \right) }~, \\
&\Green{
\int d^2 \hat{\bf k_3'} 
{}_{- \sigma_3} Y^*_{1 m_e} Y_{L''_2 M''_2} 
{}_{\sigma_3} Y^*_{1 m_b} Y^*_{L_3 M_3}
} \\
&\qquad\qquad = - \sum_{L'' \Cyan{M''}} \sum_{S'' = \pm 1} (-1)^{\Brown{m_e}} I_{L_3 1 L''}^{0 -\sigma_3 -S''} I_{L''_2 1 L''}^{0 -\sigma_3 -S''} 
\Magenta{ \left(
  \begin{array}{ccc}
  L_3 &  1 & L'' \\
   M_3 & m_b & M'' 
  \end{array}
 \right)  }
\Brown{ \left(
  \begin{array}{ccc}
  L''_2 &  1 & L'' \\
   M''_2 & -m_e & M'' 
  \end{array}
 \right) }~, \label{eq:ang_int_kdpdqd}
\end{split}
\end{eqnarray}
where we have used a property of spin-weighted spherical harmonics given
by Eq.~(\ref{eq:product_sYlm}). 
We can also perform the angular integrals with respect to $\hat{\bf
k_1}, \hat{\bf k_2}$ and $\hat{\bf k_3}$ as
\begin{eqnarray}
\begin{split}
& \Purple{
\int d^2 \hat{\bf k_1} 
 Y^*_{L_1 M_1}
{}_{-\lambda_1}Y^*_{\ell_1 m_1} {}_{\lambda_1}Y_{2 \mu_1}^*
} = I_{L_1 \ell_1 2}^{0 \lambda_1 -\lambda_1} 
\Magenta{ \left(
  \begin{array}{ccc}
  L_1 & \ell_1 & 2 \\
  M_1 & m_1 & \mu_1
  \end{array}
 \right)  } ~,\\
\Purple{
& \int d^2 \hat{\bf k_2} 
Y^*_{L'_1 M'_1} {}_{-\lambda_2} Y^*_{\ell_2 m_2} {}_{\lambda_2} Y^*_{2 \mu_2} 
} 
=  I_{L'_1 \ell_2 2}^{0 \lambda_2 -\lambda_2} 
\Orange{ \left(
  \begin{array}{ccc}
  L'_1 & \ell_2 & 2 \\
  M'_1 & m_2 & \mu_2 
  \end{array}
 \right)   } ~,\\
& \Purple{ 
\int d^2 \hat{\bf k_3} 
Y^*_{L''_1 M''_1} {}_{- \lambda_3} Y^*_{\ell_3 m_3} {}_{\lambda_3} Y^*_{2 \mu_3} 
} 
=  I_{L''_1 \ell_3 2}^{0 \lambda_3 -\lambda_3} 
\Brown{ \left(
  \begin{array}{ccc}
  L''_1 & \ell_3 & 2 \\
  M''_1 & m_3 & \mu_3
  \end{array}
 \right) 
 }~. \label{eq:ang_int_kpq}
\end{split} 
\end{eqnarray} 
At this point, all the angular integrals in Eq.~(\ref{eq:xi_bis_SVT_explicit})
have been reduced into the Wigner-$3j$ symbols.

As the third step, we consider summing up the Wigner-$3j$ symbols
in terms of the azimuthal quantum numbers and replacing them with the
Wigner-$6j$ and $9j$ symbols, which denote Clebsch-Gordan coefficients
between two other eigenstates coupled to three and four individual
momenta \cite{Hu:2001fa, Gurau:2008yh, Jahn/Hope:1954, Shiraishi:2010kd}.
Using these properties, we can express the summation of five Wigner-$3j$
symbols with a Wigner-$9j$ symbol:
\begin{eqnarray}
\begin{split}
&\Magenta{
\sum_{\substack{M_1 M_2 M_3 \\ \mu_1 m_a m_b}} (-1)^{M_2+m_a} 
\left(
  \begin{array}{ccc}
  L_1 & L_2 & L_3 \\
  M_1 & - M_2 & M_3 
  \end{array}
 \right)
\left(
  \begin{array}{ccc}
  2 & 1 & 1 \\
  \mu_1 & m_a & m_b 
  \end{array}
 \right) \\
&\qquad\qquad\qquad \times  
\left(
  \begin{array}{ccc}
  L_3 & 1 & L'' \\
  M_3 & m_b & M'' 
  \end{array}
 \right)
 \left(
  \begin{array}{ccc}
  L_2 &  1 & L \\
  M_2 & -m_a & M 
  \end{array}
 \right)
\left(
  \begin{array}{ccc}
  L_1 & \ell_1 & 2 \\
  M_1 & m_1 & \mu_1
  \end{array}
 \right)
} \\
&\qquad\qquad = - (-1)^{\Cyan{M} + \ell_1 + L_3 + L} 
\Cyan{ \left(
  \begin{array}{ccc}
  L'' & L & \ell_1 \\
  M'' & -M & m_1 
  \end{array}
 \right) }
\left\{
  \begin{array}{ccc}
  L'' & L & \ell_1 \\
  L_3 & L_2 & L_1 \\
  1 & 1 & 2 
  \end{array}
 \right\}~, \\
&\Orange{
\sum_{\substack{M'_1 M'_2 M'_3 \\ \mu_2 m_c m_d} } (-1)^{M'_2 + m_c} 
\left(
  \begin{array}{ccc}
  L'_1 & L'_2 & L'_3 \\
  M'_1 & - M'_2 & M'_3 
  \end{array}
 \right)
\left(
  \begin{array}{ccc}
  2 & 1 & 1 \\
  \mu_2 & m_c & m_d 
  \end{array}
 \right) \\
&\qquad\qquad\qquad \times
\left(
  \begin{array}{ccc}
  L'_3 & 1 & L \\
  M'_3 & m_d & M 
  \end{array}
 \right)
 \left(
  \begin{array}{ccc}
  L'_2 &  1 & L' \\
  M'_2 & -m_c & M' 
  \end{array}
 \right)
\left(
  \begin{array}{ccc}
  L'_1 & \ell_2 & 2 \\
  M'_1 & m_2 & \mu_2
  \end{array}
 \right)
} \\
&\qquad\qquad= - (-1)^{\Cyan{M'} + \ell_2 + L'_3 + L'} 
\Cyan{ \left(
  \begin{array}{ccc}
  L & L' & \ell_2 \\
  M & -M' & m_2 
  \end{array}
 \right) }
\left\{
  \begin{array}{ccc}
  L & L' & \ell_2 \\
  L'_3 & L'_2 & L'_1 \\
  1 & 1 & 2 
  \end{array}
 \right\}~, \\
&\Brown{
\sum_{\substack{M''_1 M''_2 M''_3 \\ \mu_3 m_e m_f}} (-1)^{M''_2 + m_e} 
\left(
  \begin{array}{ccc}
  L''_1 & L''_2 & L''_3 \\
  M''_1 & - M''_2 & M''_3 
  \end{array}
 \right)
\left(
  \begin{array}{ccc}
  2 & 1 & 1 \\
  \mu_3 & m_e & m_f 
  \end{array}
 \right) \\
&\qquad\qquad\qquad \times
\left(
  \begin{array}{ccc}
  L''_3 & 1 & L' \\
  M''_3 & m_f & M' 
  \end{array}
 \right)
 \left(
  \begin{array}{ccc}
  L''_2 &  1 & L'' \\
  M''_2 & -m_e & M'' 
  \end{array}
 \right)
\left(
  \begin{array}{ccc}
  L''_1 & \ell_3 & 2 \\
  M''_1 & m_3 & \mu_3 
  \end{array}
 \right)
} \\
&\qquad\qquad= - (-1)^{\Cyan{M''} + \ell_3 + L''_3 + L''} 
\Cyan{ \left(
  \begin{array}{ccc}
  L' & L'' & \ell_3 \\
  M' & -M'' & m_3 
  \end{array}
 \right) }
\left\{
  \begin{array}{ccc}
  L' & L'' & \ell_3 \\
  L''_3 & L''_2 & L''_1 \\
  1 & 1 & 2 
  \end{array}
 \right\} ~.
\end{split} \label{eq:sum_ms}
\end{eqnarray}
Furthermore, we can also sum up the renewed Wigner-$3j$ symbols 
arising in the above equations
over $M, M'$ and $M''$ with the Wigner-$6j$ symbol as \cite{mathematica}
\begin{eqnarray}
&& \Cyan{
\sum_{M M' M''} (-1)^{M + M' + M''} 
\left(
  \begin{array}{ccc}
  L'' & L & \ell_1 \\
  M'' & -M & m_1 
  \end{array}
 \right)
\left(
  \begin{array}{ccc}
  L & L' & \ell_2 \\
  M & -M' & m_2 
  \end{array}
 \right)
\left(
  \begin{array}{ccc}
  L' & L'' & \ell_3 \\
  M' & -M'' & m_3 
  \end{array}
 \right)} \nonumber \\
&&\qquad\qquad = (-1)^{L+L'+L''}
\left(
  \begin{array}{ccc}
  \ell_1 & \ell_2 & \ell_3 \\
  m_1 & m_2 & m_3 
  \end{array}
 \right)
\left\{
  \begin{array}{ccc}
  \ell_1 & \ell_2 & \ell_3 \\
  L' & L'' & L 
  \end{array}
 \right\}~.
\end{eqnarray}   
With this prescription, one can find that the three azimuthal numbers
are confined only in the Wigner-$3j$ symbol as 
$\left(
  \begin{array}{ccc}
  \ell_1 & \ell_2 & \ell_3 \\
  m_1 & m_2 & m_3 
  \end{array}
 \right)$.
 This $3j$ symbol arises from $\Braket{\prod_{n=1}^3 \xi^{(\lambda_n)}_{\ell_n m_n}(k_n)}$
and exactly ensures the rotational invariance of the CMB
 bispectrum as pointed out above.

Consequently, we can obtain an exact form of the primordial angular bispectrum given by
\begin{eqnarray}
\Braket{\prod_{n=1}^3 \xi_{\ell_n m_n}^{(\lambda_n)}(k_n)}
&=& \left(
  \begin{array}{ccc}
  \ell_1 & \ell_2 & \ell_3 \\
  m_1 & m_2 & m_3
  \end{array}
 \right)
(-4 \pi \rho_{\gamma,0})^{-3}\left[ \prod_{n=1}^3 \int_0^{k_D} k_n'^2 dk_n' P_B(k_n') \right]
\nonumber \\ 
&& \times \sum_{L L' L''} \sum_{S, S', S'' = \pm 1} 
\left\{
  \begin{array}{ccc}
  \ell_1 & \ell_2 & \ell_3 \\
  L' & L'' & L 
  \end{array}
 \right\} \nonumber \\
&&\times 
f^{S'' S \lambda_1}_{L'' L \ell_1 }(k_3',k_1',k_1) f^{S S' \lambda_2}_{L L' \ell_2}(k_1',k_2',k_2)
f^{S' S'' \lambda_3}_{L' L'' \ell_3}(k_2',k_3',k_3),
 \label{eq:xi_bis_pmf_SVT_exact}
\end{eqnarray}
where 
\begin{eqnarray}
f^{S'' S \lambda}_{L'' L \ell}(r_3, r_2, r_1) 
&=& \sum_{L_1 L_2 L_3} \int_0^\infty y^2 dy j_{L_3}(r_3 y) j_{L_2}(r_2 y) j_{L_1}(r_1 y)  \nonumber \\
&& \times
 (-1)^{\ell + L_2 + L_3} (-1)^{\frac{L_1 + L_2 + L_3}{2}} 
I^{0~0~0}_{L_1 L_2 L_3} I^{0 S'' -S''}_{L_3 1 L''} I^{0 S -S}_{L_2 1 L} 
I_{L_1 \ell 2}^{0 \lambda -\lambda} 
 \left\{
  \begin{array}{ccc}
  L'' & L & \ell \\
  L_3 & L_2 & L_1 \\
  1 & 1 & 2
  \end{array}
 \right\} \nonumber \\
&&\times
\left\{
 \begin{array}{ll}
\frac{2}{\sqrt{3}} (8\pi)^{3/2} R_\gamma 
\ln \left(\tau_\nu / \tau_B\right)  & (\lambda = 0) \\
 \frac{2}{3} (8\pi)^{3/2} \lambda  & (\lambda = \pm 1) \\
  - 4 (8\pi)^{3/2} R_\gamma \ln \left( \tau_\nu / \tau_B \right) &
   (\lambda = \pm 2)
 \end{array}
\right. ~. \label{eq:f_exact}
\end{eqnarray}
Here, the coefficients have been calculated as
\begin{eqnarray}
C_{-\lambda} C_{-\lambda}' \sqrt{\frac{3}{8\pi}} \left(\frac{4\pi}{3}\right)^2 8
&=& \begin{cases}
- \frac{2}{\sqrt{3}} (8\pi)^{3/2} R_\gamma 
\ln \left(\tau_\nu / \tau_B\right)  & (\lambda = 0) \\
 - \frac{2}{3} (8\pi)^{3/2} \lambda & (\lambda = \pm 1) \\ 
  4 (8\pi)^{3/2} R_\gamma \ln \left( \tau_\nu / \tau_B \right) & (\lambda = \pm 2)  
\end{cases} ~. 
\end{eqnarray}

\begin{figure}[t]
\begin{tabular}{cc}
\begin{minipage}[t]{0.5\hsize}
  \begin{center}
    \includegraphics[width=8cm,clip]{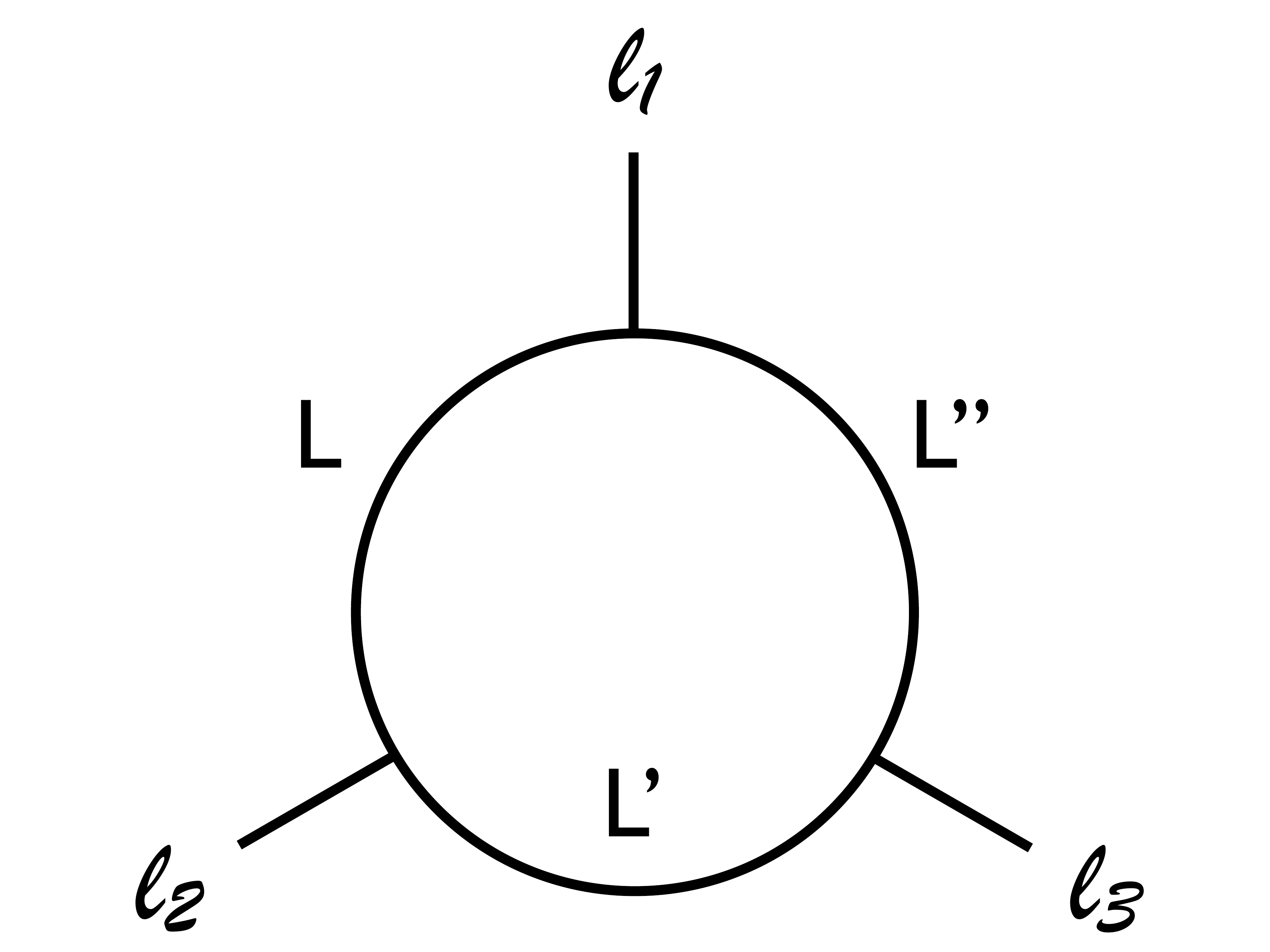}
  \end{center}
\end{minipage}
\begin{minipage}[t]{0.5\hsize}
  \begin{center}
    \includegraphics[width=8cm,clip]{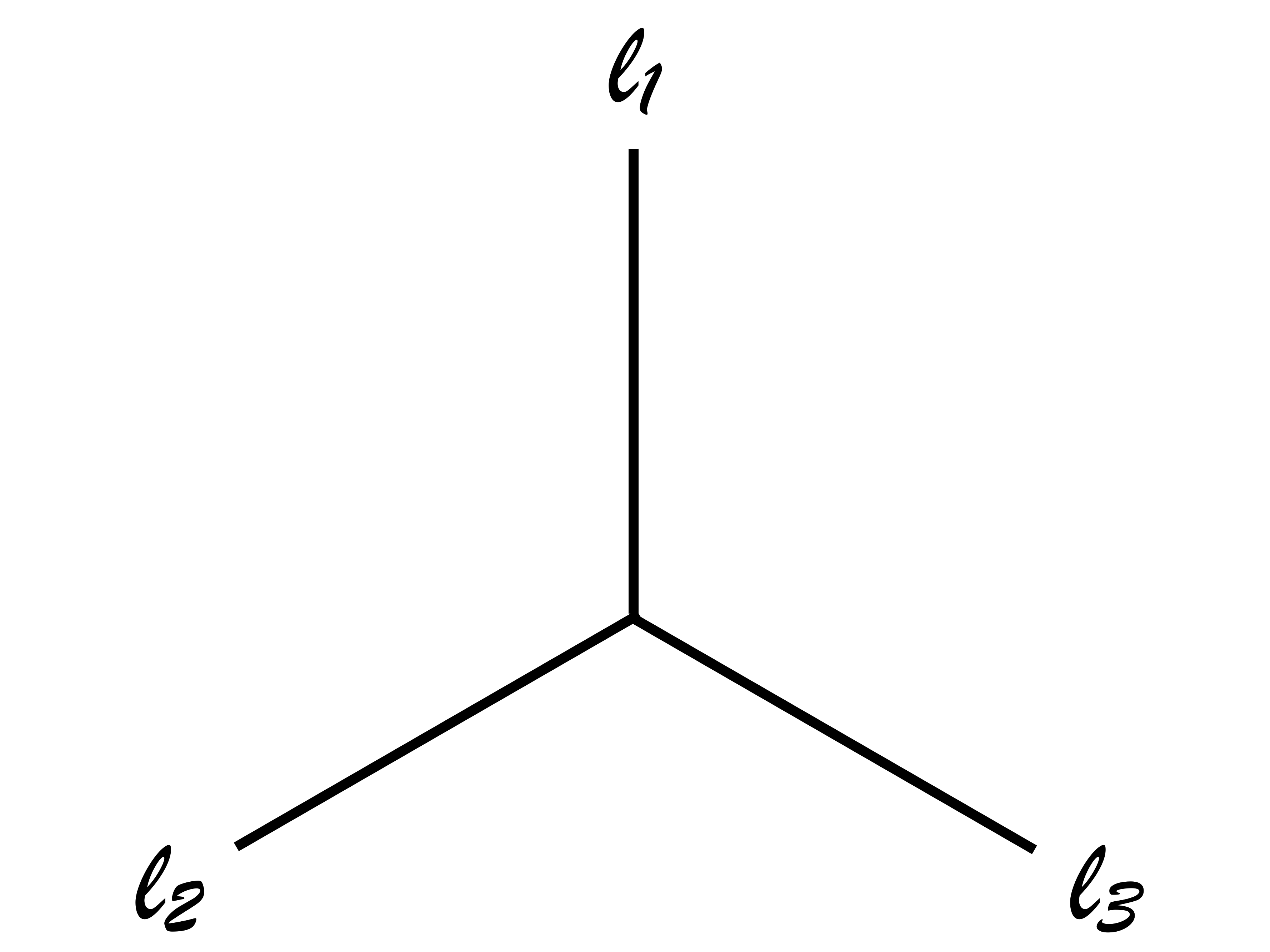}
  \end{center}
\end{minipage}
\end{tabular}
  \caption{Diagrams with respect to multipoles \cite{Shiraishi:2012rm}. The left panel corresponds to Eq.~(\ref{eq:CMB_bis_PMF_complete}). Due to the Wigner-$6j$ symbol originated with the sextuplicate dependence on the Gaussian PMFs, the true multipoles $\ell_1, \ell_2$ and $\ell_3$ are linked with the dummy ones $L, L'$, and $L''$ and the 1-loop structure is realized. The right panel represents the tree-structure diagram, which arises from the CMB bispectrum induced by the four-point function of the Gaussian fields as mentioned in the previous sections.}
\label{fig:diagram}
\end{figure}

Substituting Eq.~(\ref{eq:xi_bis_pmf_SVT_exact}) into
Eq.~(\ref{eq:CMB_bis_PMF_general}),  we can formulate the CMB bispectra
generated from arbitrary three modes such as the scalar-scalar-vector and
tensor-tensor-tensor correlations with the $f$ function as 
\begin{eqnarray}
\Braket{\prod_{n=1}^3 a_{X_n, \ell_n m_n}^{(Z_n)}}
&=& \left(
  \begin{array}{ccc}
  \ell_1 & \ell_2 & \ell_3 \\
  m_1 & m_2 & m_3
  \end{array}
 \right) 
(- 4 \pi \rho_{\gamma,0})^{-3} \nonumber \\
&&\times \left[ \prod\limits^3_{n=1}
(-i)^{\ell_n}
\int {k_n^2 dk_n \over 2\pi^2}
\mathcal{T}_{X_n, \ell_n}^{(Z_n)}(k_n)
\sum\limits_{\lambda_n} [{\rm sgn}(\lambda_n)]^{\lambda_n + x_n}
\int_0^{k_D} k_n'^2 dk_n' P_B(k_n') \right] \nonumber \\ 
&& \times 
\sum_{L L' L''} \sum_{S, S', S'' = \pm 1} 
\left\{
  \begin{array}{ccc}
  \ell_1 & \ell_2 & \ell_3 \\
  L' & L'' & L 
  \end{array}
 \right\} \nonumber \\
&& \times
f^{S'' S \lambda_1}_{L'' L \ell_1}(k_3',k_1',k_1) f^{S S' \lambda_2}_{L L' \ell_2}(k_1',k_2',k_2)
f^{S' S'' \lambda_3}_{L' L'' \ell_3}(k_2',k_3',k_3)~. \label{eq:CMB_bis_PMF_complete}
\end{eqnarray}
From this form, it can be easily seen that due to the sextuplicate dependence on the Gaussian PMFs, the Wigner-$6j$ symbol connects the true multipoles ($\ell_1, \ell_2$ and $\ell_3$) and the dummy ones ($L, L'$ and $L''$), and the 1-loop calculation with respect to these mulipoles is realized as illustrated in the left panel of Fig.~\ref{fig:diagram}. Due to the extra summations over $L, L'$ and $L''$, it takes a lot of time to compute this compared with the tree-level calculation presented in the previous sections. 

\subsection{Treatment for numerical computation}

In order to perform the numerical computation of the CMB bispectra, we
give the explicit angle-averaged form of Eq.~(\ref{eq:CMB_bis_PMF_complete}) as
\begin{eqnarray}
B^{(Z_1 Z_2 Z_3)}_{X_1 X_2 X_3, \ell_1 \ell_2 \ell_3}
&=& 
C_{Z_1} C_{Z_2} C_{Z_3} 
\left( -4 \pi \rho_{\gamma,0} \right)^{-3} 
\sum _{L L' L''} 
\left\{
  \begin{array}{ccc}
  \ell_1 & \ell_2 & \ell_3 \\
  L' & L'' & L 
  \end{array}
 \right\}
\nonumber \\
&& \times
\sum_{\substack{L_1 L_2 L_3 \\ L'_1 L'_2 L'_3 \\ L''_1 L''_2 L''_3}} 
(-1)^{\sum_{n=1}^3\frac{L_n+L'_n+L''_n+2 \ell_n}{2}} 
I^{0~0~0}_{L_1 L_2 L_3} I^{0~0~0}_{L'_1 L'_2 L'_3} I^{0~0~0}_{L''_1
L''_2 L''_3}  \nonumber \\
&& \times
\left\{
  \begin{array}{ccc}
  L'' & L & \ell_1 \\
  L_3 & L_2 & L_1 \\
  1 & 1 & 2
  \end{array}
 \right\}
\left\{
  \begin{array}{ccc}
  L & L' & \ell_2 \\
  L'_3 & L'_2 & L'_1 \\
  1 & 1 & 2
  \end{array}
 \right\}
\left\{
  \begin{array}{ccc}
  L' & L'' & \ell_3 \\
  L''_3 & L''_2 & L''_1 \\
  1 & 1 & 2
  \end{array}
 \right\} \nonumber \\
&& \times 
\Blue{
\left[ \prod\limits^3_{n=1}
(-i)^{\ell_n}
\int_0^\infty {k_n^2 dk_n \over 2 \pi^2}
\mathcal{T}^{(Z_n)}_{X_n, \ell_n}(k_n)\right] \nonumber \\
&& \times 
\int_0^\infty A^2 dA j_{L_1} (k_1 A)  
\int_0^\infty B^2 dB j_{L'_1} (k_2 B)  
\int_0^\infty C^2 dC j_{L''_1} (k_3 C) \nonumber \\ 
&& \times \int_0^{k_D} k_1'^2 dk_1' P_B(k_1') 
j_{L_2} (k_1' A) j_{L'_3} (k_1' B) 
 \int_0^{k_D} k_2'^2 dk_2' P_B(k_2')
j_{L'_2} (k_2'B) j_{L''_3} (k_2'C) \nonumber \\
&& \times 
\int_0^{k_D} k_3'^2 dk_3'  P_B(k_3') 
j_{L''_2} (k_3'C) j_{L_3} (k_3'A) }\nonumber \\
&& \times 
\Red{\sum_{S, S', S'' = \pm 1} (-1)^{L_2 + L'_2 + L''_2 + L_3 + L'_3 + L''_3}
I^{0 S -S}_{L'_3 1 L} I^{0 S -S}_{L_2 1 L} I^{0 S' -S'}_{L''_3 1 L'} 
I^{0 S' -S'}_{L'_2 1 L'} I^{0 S'' -S''}_{L_3 1 L''} I^{0 S'' -S''}_{L''_2 1 L''}} \nonumber \\
&& \times 
\OliveGreen{ \sum_{\lambda_1 \lambda_2 \lambda_3} 
[{\rm sgn(\lambda_1)}]^{x_1} I^{0 \lambda_1 -\lambda_1}_{L_1 \ell_1 2} 
[{\rm sgn(\lambda_2)}]^{x_2} I^{0 \lambda_2 -\lambda_2}_{L'_1 \ell_2 2} 
[{\rm sgn(\lambda_3)}]^{x_3} I^{0 \lambda_3 -\lambda_3}_{L''_1 \ell_3 2}
} ~,
\end{eqnarray}
with 
\begin{eqnarray}
C_Z \equiv
\left\{
 \begin{array}{ll}
\frac{2}{\sqrt{3}} (8\pi)^{3/2} R_\gamma 
\ln \left(\tau_\nu / \tau_B\right)  & (Z = S) \\
 \frac{2}{3} (8\pi)^{3/2} & (Z = V) \\
  - 4 (8\pi)^{3/2} R_\gamma \ln \left( \tau_\nu / \tau_B \right) &
   (Z = T) 
 \end{array}
\right. ~.
\end{eqnarray}
We consider performing the summations with respect to the
helicities. By considering the selection rules of the
Wigner-$3j$ symbol, the summations over $S, S'$ and $S''$ are performed as
\begin{eqnarray}
&&\Red{\sum_{S, S', S'' = \pm 1} (-1)^{L_2 + L'_2 + L''_2 + L_3 + L'_3 + L''_3}
I^{0 S -S}_{L'_3 1 L} I^{0 S -S}_{L_2 1 L} I^{0 S' -S'}_{L''_3 1 L'} 
I^{0 S' -S'}_{L'_2 1 L'} I^{0 S'' -S''}_{L_3 1 L''} I^{0 S'' -S''}_{L''_2 1 L''}} \nonumber \\
&&\qquad\qquad = I_{L'_3 1 L}^{0 1 -1} I_{L_2 1 L}^{0 1 -1} I_{L''_3 1 L'}^{0 1 -1}
I_{L'_2 1 L'}^{0 1 -1} I_{L_3 1 L''}^{0 1 -1} I_{L''_2 1 L''}^{0 1 -1}
\nonumber \\
&&\qquad\qquad\quad\times 
\left\{
 \begin{array}{ll}
 8 & (L'_3 + L_2, L''_3 + L'_2, L_3 + L''_2 = {\rm even}) \\
 0 & ({\rm otherwise}) 
 \end{array}
\right. ~.
\end{eqnarray}
By the same token, the summations over $\lambda_1, \lambda_2$ and $\lambda_3$ are given by
\begin{eqnarray}
&& \OliveGreen{ \sum_{\lambda_1 \lambda_2 \lambda_3} 
[{\rm sgn(\lambda_1)}]^{x_1} I^{0 \lambda_1 -\lambda_1}_{L_1 \ell_1 2} 
[{\rm sgn(\lambda_2)}]^{x_2} I^{0 \lambda_2 -\lambda_2}_{L'_1 \ell_2 2} 
[{\rm sgn(\lambda_3)}]^{x_3} I^{0 \lambda_3 -\lambda_3}_{L''_1 \ell_3 2}
} \nonumber \\
&&\qquad\qquad = I_{L_1~\ell_1~2}^{0 |\lambda_1| -|\lambda_1|}
 I_{L_1'~\ell_2~2}^{0 |\lambda_2| -|\lambda_2|} I_{L_1''~\ell_3~2}^{0
 |\lambda_3| -|\lambda_3|} \nonumber \\
&&\qquad\qquad\quad\times 
\left\{
 \begin{array}{ll}
 2^{3-N_S} & (L_1 + \ell_1 + x_1, L_1' + \ell_2 + x_2,  L_1'' + \ell_3 +
  x_3 = {\rm even}) \\ 
 0 & ({\rm otherwise}) 
 \end{array}
\right. ~, \label{eq:summation_helicity}
\end{eqnarray} 
where $N_S$ is the number of the scalar modes constituting the CMB
bispectrum
\footnote{Caution about a fact that $|\lambda|$ is determined by $Z$, namely, $|\lambda| = 0, 1, 2$ for $Z = S, V, T$, respectively.}. 
Thus, we rewrite the bispectrum as 
\begin{eqnarray}
B^{(Z_1 Z_2 Z_3)}_{X_1 X_2 X_3, \ell_1 \ell_2 \ell_3}
&=& 
C_{Z_1} C_{Z_2} C_{Z_3} 
\left( -4 \pi \rho_{\gamma,0} \right)^{-3} 
\sum _{L L' L''} 
\left\{
  \begin{array}{ccc}
  \ell_1 & \ell_2 & \ell_3 \\
  L' & L'' & L 
  \end{array}
 \right\}
\nonumber \\
&& \times
\sum_{\substack{L_1 L_2 L_3 \\ L'_1 L'_2 L'_3 \\ L''_1 L''_2 L''_3}} 
(-1)^{\sum_{n=1}^3\frac{L_n+L'_n+L''_n+2 \ell_n}{2}} 
I^{0~0~0}_{L_1 L_2 L_3} I^{0~0~0}_{L'_1 L'_2 L'_3} I^{0~0~0}_{L''_1
L''_2 L''_3}  \nonumber \\
&& \times
\left\{
  \begin{array}{ccc}
  L'' & L & \ell_1 \\
  L_3 & L_2 & L_1 \\
  1 & 1 & 2
  \end{array}
 \right\}
\left\{
  \begin{array}{ccc}
  L & L' & \ell_2 \\
  L'_3 & L'_2 & L'_1 \\
  1 & 1 & 2
  \end{array}
 \right\}
\left\{
  \begin{array}{ccc}
  L' & L'' & \ell_3 \\
  L''_3 & L''_2 & L''_1 \\
  1 & 1 & 2
  \end{array}
 \right\} \nonumber \\
&& \times 
\Blue{
\left[ \prod\limits^3_{n=1}
(-i)^{\ell_n}
\int_0^\infty {k_n^2 dk_n \over 2 \pi^2}
\mathcal{T}^{(Z_n)}_{X_n, \ell_n}(k_n)\right] \nonumber \\
&& \times 
\int_0^\infty A^2 dA j_{L_1} (k_1 A)  
\int_0^\infty B^2 dB j_{L'_1} (k_2 B)  
\int_0^\infty C^2 dC j_{L''_1} (k_3 C) \nonumber \\ 
&& \times \int_0^{k_D} k_1'^2 dk_1' P_B(k_1') 
j_{L_2} (k_1' A) j_{L'_3} (k_1' B) 
 \int_0^{k_D} k_2'^2 dk_2' P_B(k_2')
j_{L'_2} (k_2'B) j_{L''_3} (k_2'C) \nonumber \\
&& \times 
\int_0^{k_D} k_3'^2 dk_3'  P_B(k_3') 
j_{L''_2} (k_3'C) j_{L_3} (k_3'A) }\nonumber \\
&& \times 
8 I_{L'_3 1 L}^{0 1 -1} I_{L_2 1 L}^{0 1 -1} I_{L''_3 1 L'}^{0 1 -1}
I_{L'_2 1 L'}^{0 1 -1} I_{L_3 1 L''}^{0 1 -1} I_{L''_2 1 L''}^{0 1 -1}
{\cal Q}_{L'_3, L_2, L} {\cal Q}_{L''_3, L'_2, L'} {\cal Q}_{L_3, L''_2,
L''}
\nonumber \\
&& \times 
2^{3-N_S} I_{L_1~\ell_1~2}^{0 |\lambda_1| -|\lambda_1|}
 I_{L_1'~\ell_2~2}^{0 |\lambda_2| -|\lambda_2|} I_{L_1''~\ell_3~2}^{0
 |\lambda_3| -|\lambda_3|} 
{\cal U}_{L_1, \ell_1, x_1} {\cal U}_{L_1', \ell_2, x_2} {\cal U}_{L_1'', \ell_3, x_3} 
~, \label{eq:CMB_bis_PMF_numerical}
\end{eqnarray}
where we introduce the filter functions as 
\begin{eqnarray}
\begin{split}
{\cal Q}_{L'_3, L_2, L} &\equiv (\delta_{L'_3, L+1} + \delta_{L'_3,
 |L-1|}) (\delta_{L_2, L+1} + \delta_{L_2, |L-1|}) + \delta_{L'_3, L}
 \delta_{L_2, L} \\
{\cal U}_{L_1, \ell_1, x_1} &\equiv (\delta_{L_1, \ell_1-2} +
 \delta_{L_1, \ell_1} + \delta_{L_1, \ell_1 + 2}) \delta_{x_1,0} 
+  (\delta_{L_1, \ell_1-1} + \delta_{L_1, \ell_1 + 1}) \delta_{x_1,1} ~.
\end{split}
\end{eqnarray}
The above analytic expression seems to be quite useful to calculate the CMB
bispectrum induced from PMFs with the full-angular dependence. 
However, it is still too hard to calculate numerically, because the full
expression of the bispectrum has six integrals. In addition, when we calculate the spectra for large $\ell$'s,
this situation becomes worse since we spend a lot of time calculating the
Wigner symbols for large $\ell$'s. The CMB signals of the vector mode appear at $\ell > 2000$, hence we need the reasonable approximation in calculation of the CMB bispectra composed of the vector modes. In what follows, we introduce an approximation, the so-called thin last scattering surface (LSS) approximation to reduce the integrals. 

\subsubsection{Thin LSS approximation} \label{subsubsec:thinLSSapp}

\begin{figure}[t]
  \begin{center}
    \includegraphics[height=8cm,clip]{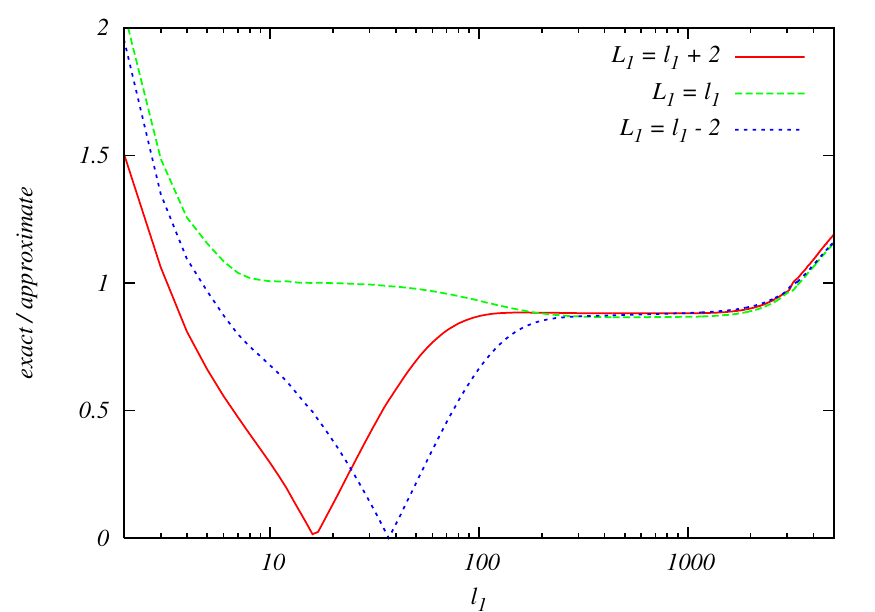}
  \end{center}
  \caption{The ratio of the left-hand side (exact solution) to the right-hand side (approximate solution) in Eq. (\ref{eq:thin_LSS_check}). The lines correspond to the case for $L_1 = \ell_1 + 2$ (red solid line), for $L_1 = \ell_1$ (green dashed one), and for $L_1 = \ell_1 - 2$ (blue dotted one).}
  \label{fig:thin_LSS_check.pdf}
\end{figure}

Let us consider the parts of the integrals with respect to $A, B, C, k',
p'$ and $q'$ in the full expression of the 
bispectrum (\ref{eq:CMB_bis_PMF_numerical}) of $B_{III, \ell_1 \ell_2 \ell_3}^{(VVV)}$. In the computation of the CMB bispectrum, the integral in terms of $k$, ($p$ and $q$) appears in the form as $\int k^2 dk {\cal
T}^{(V)}_{I,\ell_1}(k) j_{L_1}(k A)$.  We find that this integral is
sharply-peaked at $A \simeq \tau_0 -\tau_*$, where $\tau_0$ is the
present conformal time and $\tau_*$ is the conformal time of the
recombination epoch. According to Sec.~\ref{subsubsec:pmf_vector},
the vorticity of subhorizon scale sourced by magnetic fields around the
recombination epoch mostly contributes to generate the CMB vector
perturbation. On the other hand, since the vector mode in the metric
decays after neutrino decoupling, the integrated Sachs-Wolfe effect
after recombination is not observable. Such a behavior of the transfer function would be understood on the basis of the calculation in Sec.~\ref{subsec:alm_PMF} and we expect ${\cal T}^{(V)}_{I,\ell_1}(k)
\propto j_{\ell_1}(k(\tau_0 - \tau_*))$,  and the $k$-integral behaves
like $\delta(A - (\tau_0 - \tau_*))$.  By the numerical computation, we
found that
\begin{eqnarray}
\int^{\infty}_0 A^2 dA \int_0^\infty k_1^2 dk_1 {\cal T}^{(V)}_{I,\ell_1}(k_1) j_{L_1}(k_1 A)
\simeq  (\tau_0 - \tau_*)^2 \left({\tau_* \over 5}\right) \int k_1^2 dk_1 {\cal T}^{(V)}_{I,\ell_1}(k_1) j_{\ell_1}(k_1 (\tau_0 - \tau_*)), \label{eq:thin_LSS_check}
\end{eqnarray} 
is a good approximation for $L_1 = \ell_1 \pm 2, \ell_1$ as described in
Fig. \ref{fig:thin_LSS_check.pdf}. Note that only the cases $L_1 =
\ell_1 \pm 2, \ell_1$ should be considered due to the selection rules
for Wigner-3j symbols as we shall see later. From this figure, we can
find that the approximation (the right-handed term of
Eq. (\ref{eq:thin_LSS_check})) has less than $20 \%$ uncertainty for
$\ell_1 \simeq L_1 \gtrsim 100$, and therefore this approximation leads
to only less than $10 \%$ uncertainty in the bound on the strength of PMFs
if we place the constraint from the bispectrum data at $\ell_1,
\ell_2 ,\ell_3 \gtrsim 100$ 
\footnote{\Black{Of course, if we calculate the
bispectrum at smaller multipoles and the CMB bispectra are produced by other modes than the vector one, we may perform the full integration without this approximation.}}.
Using this approximation, namely $A = B = C \rightarrow \tau_0 - \tau_*$
and $\int dA = \int dB = \int dC \rightarrow \tau_*/5$, the integrals
with respect to $A,B,C,k', p'$ and $q'$ are estimated as
\begin{eqnarray}
&&\Blue{
\left[ \prod\limits^3_{n=1}
4\pi (-i)^{\ell_n}
\int_0^\infty {k_n^2 dk_n \over (2\pi)^3}
\mathcal{T}^{(V)}_{I, \ell_n}(k_n)\right] 
\int_0^\infty A^2 dA j_{L_1} (k_1 A)  
\int_0^\infty B^2 dB j_{L'_1} (k_2 B)  
\int_0^\infty C^2 dC j_{L''_1} (k_3 C) \nonumber \\ 
&& \qquad\qquad \times 
 \int_0^{k_D} k_1'^2 dk_1'  P_B(k_1') 
j_{L_2} (k_1'A) j_{L'_3} (k_1' B) 
 \int_0^{k_D} k_2'^2 dk_2' P_B(k_2')
j_{L'_2} (k_2'B) j_{L''_3} (k_2'C) \nonumber \\
&& \qquad\qquad\times 
\int_0^{k_D} k_3'^2 dk_3'  P_B(k_3') 
j_{L''_2} (k_3'C) j_{L_3} (k_3'A) } \nonumber \\
&& \qquad \simeq
\left[ \prod\limits^3_{n=1}
4\pi (-i)^{\ell_n}
\int_0^\infty {k_n^2 dk_n \over (2\pi)^3}
\mathcal{T}^{(V)}_{I, \ell_n}(k_n)
j_{\ell_n} (k_n (\tau_0-\tau_*))
\right] 
A_B^3 (\tau_0-\tau_*)^6 \left( \frac{\tau_*}{5} \right)^3
\nonumber \\
&& \qquad\qquad\qquad \times  
\mathcal{K}_{L_2 L'_3}^{-(n_B +1)}(\tau_0-\tau_*)
\mathcal{K}_{L'_2 L''_3}^{-(n_B +1)}(\tau_0-\tau_*) 
\mathcal{K}_{L''_2 L_3}^{-(n_B +1)}(\tau_0-\tau_*) ~. 
\end{eqnarray}
Here the function ${\cal K}^N_{ll'}$ is defined as 
\begin{eqnarray}
\mathcal{K}_{l l'}^{N}(y) 
&\equiv& \int_0^\infty dk k^{1 - N} j_{l} (ky) j_{l'} (ky) \nonumber \\
&=& \frac{\pi}{2y} \frac{y^{N-1}}{2^N} \frac{\Gamma(N) \Gamma(\frac{l + l' + 2 - N}{2})}{\Gamma(\frac{l - l' + 1 + N}{2}) \Gamma(\frac{-l + l' + 1 + N}{2}) 
\Gamma(\frac{l + l' + 2 + N}{2})} \ ({\rm for} \ y,N, l+l'+2-N>0)~, \nonumber \\
\end{eqnarray} 
which behaves asymptotically as ${\cal K}_{l l'}^{N}(y) \propto l^{-N}$
for $l \sim l' \gg 1$.  Here we have evaluated the $k^\prime$ integrals
by setting $k_D \rightarrow \infty$.
This is also a good approximation because the integrands are suppressed
enough for $k', p', q' < k_D \sim {\cal O}(10) \rm Mpc^{-1}$. 

\subsubsection{Selection rules of the Wigner-$3j$ symbol}

From the selection rules of the Wigner symbols as described in Appendix \ref{appen:wigner}, we can further limit
the summation range of the multipoles as   
\begin{eqnarray}
\begin{split}
& |L - \ell_2| \leq L^\prime \leq L + \ell_2 ~ , \ \ 
{\rm Max}[|L - \ell_1|, |L^\prime - \ell_3|] \leq L^{\prime\prime} \leq {\rm Min}[ L +
 \ell_1, L' + \ell_3] ~ , \\
& L_1 + L_2 + L_3 = {\rm even}~, \ \ 
L'_1 + L'_2 + L'_3 = {\rm even}~, \ \
L''_1 + L''_2 + L''_3 = {\rm even}~, \\
& |L_1 - L_2| \leq L_3 \leq L_1 + L_2 ~ , \ \ |L'_1 - L'_2| \leq  L'_3
 \leq L'_1 + L'_2 ~ , \ \ |L''_1 - L''_2| \leq  L''_3 \leq L''_1 + L''_2
~, \label{eq:range_L1Lk} 
\end{split}  
\end{eqnarray}
and from the above restrictions the multipoles in the bispectrum,
$\ell_1, \ell_2 $ and $\ell_3$, are also limited as
\begin{eqnarray}
|\ell_1 - \ell_2| \leq \ell_3 \leq \ell_1 + \ell_2 ~ .
\end{eqnarray}
Therefore, these selection rules significantly reduce the number of
calculation. In these ranges, while $L'$ and $L''$ are limited by
$L$, only $L$ has no upper bound. However, we can show that
the summation of $L$ is suppressed at $\ell_1\sim \ell_2\sim \ell_3 \ll
L$ as follows.
When the summations
with respect to $L,L'$ and $L''$ are evaluated at large $L, L'$ and
$L''$, namely $\ell_1, \ell_2, \ell_3 \ll L \sim L' \sim L'', L_2
\sim L'_3 \sim L, L'_2 \sim L''_3 \sim L'$ and $L''_2 \sim L_3 \sim
L''$, we get
\begin{eqnarray}
&&\sum_{L L' L''}
 \left\{
  \begin{array}{ccc}
  \ell_1 & \ell_2 & \ell_3 \\
  L' & L'' & L 
  \end{array}
 \right\} 
\sum_{\substack{L_2 L'_2 L''_2 \\ L'_3 L''_3 L_3}} 
\int_0^{k_D} k_1'^2 dk_1'  P_B(k_1') 
j_{L_2} (k_1'A) j_{L'_3} (k_1' B) 
 \nonumber \\
&& \qquad\qquad\times 
\int_0^{k_D} k_2'^2 dk_2' P_B(k_2')
j_{L'_2} (k_2'B) j_{L''_3} (k_2'C) 
\int_0^{k_D} k_3'^2 dk_3'  P_B(k_3') 
j_{L''_2} (k_3'C) j_{L_3} (k_3'A)  \nonumber \\
&& \qquad\qquad 
\times (-1)^{\sum_{i=1}^3 \frac{L_i + L_i' + L''_i}{2}} 
 I^{0~0~0}_{L_1 L_2 L_3} 
 I^{0~0~0}_{L'_1 L'_2 L'_3}  I^{0~0~0}_{L''_1 L''_2 L''_3} 
I_{L'_3 1 L}^{0 1 -1} I_{L_2 1 L}^{0 1 -1} 
I_{L''_3 1 L'}^{0 1 -1} I_{L'_2 1 L'}^{0 1 -1} 
I_{L_3 1 L''}^{0 1 -1} I_{L''_2 1 L''}^{0 1 -1}
\nonumber \\
&& \qquad\qquad \times 
\left\{
  \begin{array}{ccc}
  L'' & L & \ell_1 \\
  L_3 & L_2 & L_1 \\
  1 & 1 & 2
  \end{array}
 \right\}
\left\{
  \begin{array}{ccc}
  L & L' & \ell_2 \\
  L'_3 & L'_2 & L'_1 \\
  1 & 1 & 2
  \end{array}
 \right\}
\left\{
  \begin{array}{ccc}
  L' & L'' & \ell_3 \\
  L''_3 & L''_2 & L''_1 \\
  1 & 1 & 2
  \end{array}
 \right\} \nonumber \\
&&\qquad \propto \sum_{L L' L''} (L L' L'')^{n_B + 4/3} 
 ~.
\end{eqnarray}
Therefore, we may obtain a stable result with the summations
over a limited number of $L$ when we consider the magnetic power
spectrum is as red as $n_B \sim -2.9$, because the summations of $L^\prime$ and
$L^{\prime\prime}$ are limited by $L$. Here, we use the analytic
formulas of the $I$ symbols which are given by
\begin{eqnarray} 
\left\{
  \begin{array}{ccc}
  \ell_1 & \ell_2 & \ell_3 \\
  L' & L'' & L 
  \end{array}
 \right\} \propto (L L' L'')^{-1/6}~, \ \
{\cal K}_{L_2 L'_3}^{-(n_B + 1)} \propto L^{n_B+1}~, \ \
\left\{
  \begin{array}{ccc}
  L'' & L & \ell_1 \\
  L_3 & L_2 & L_1 \\
  1 & 1 & 2
  \end{array}
 \right\} \propto (L'' L)^{-1/2}~,
\end{eqnarray}
as described in detail in Appendix \ref{appen:wigner}.

Using the thin LSS approximation and the summation rules described above, we can
perform the computation of the CMB bispectrum containing full-angular
dependence in a reasonable time.

\subsection{Shape of the non-Gaussianity}\label{subsec:optimal_pmf} 

In this subsection, in order to understand the shape of the non-Gaussianities arising from PMFs, we reduce the bispectra of the PMF anisotropic stress by the pole approximation \cite{Shiraishi:2012rm}. 

Let us focus on the structure of the bispectrum of the PMF anisotropic stresses (\ref{eq:bis_EMT}). If the magnetic spectrum is enough red as $n_B \sim -3$, the integral over the wave number vectors is almost determined by the behavior of the integrand around at three poles as ${k_1'}, {k_2'}, {k_3'} \sim 0$. Considering the effects around at these poles, we can express the bispectrum of the PMF anisotropic stresses approximately as 
\begin{eqnarray}
&&\Braket{\Pi_{B ab}({\bf k_1}) \Pi_{B cd}({\bf k_2}) \Pi_{B ef}({\bf k_3})}
\sim (- 4\pi \rho_{\gamma,0})^{-3} 
\frac{\alpha A_B}{n_B + 3} k_*^{n_B + 3} \frac{8\pi}{3} 
 \delta\left(\sum_{n=1}^3 {\bf k_n}\right) \nonumber \\
&&\qquad\qquad \times 
\frac{1}{8}
\left[ P_B(k_1) P_B(k_2) \delta_{ad} P_{be}(\hat{\bf k_1}) P_{cf}(\hat{\bf k_2}) 
+ P_B(k_2) P_B(k_3) P_{ad}(\hat{\bf k_2}) P_{be}(\hat{\bf k_3}) \delta_{cf} \right. \nonumber \\
&&\qquad\qquad\qquad \left. +  P_B(k_1) P_B(k_3) P_{ad}(\hat{\bf k_1}) \delta_{be} P_{cf}(\hat{\bf k_3}) 
+ \{a \leftrightarrow b \ {\rm or} \ c \leftrightarrow d \ {\rm or} \ e \leftrightarrow f\} \right]~, 
\end{eqnarray}
where we evaluate as 
\begin{eqnarray}
  \int d^3 {\bf k'} P_B(k') P_{ab}(\hat{\bf k'})
\sim
\alpha \int_0^{k_*} k'^2 dk' P_B(k') \int d^2 \hat{\bf k'} P_{ab}(\hat{\bf k'})   = \frac{\alpha A_B}{n_B + 3} k_*^{n_B + 3} \frac{8\pi}{3} \delta_{ab} ~.
\end{eqnarray}
Note that $\alpha$ is an unknown parameter and should be determined by the comparison with the exact bispectra [(\ref{eq:CMB_bis_PMF_complete}) or (\ref{eq:CMB_bis_PMF_numerical})], and we take $k_* = 10 {\rm Mpc}^{-1}$. 

Under this approximation, the angular bispectrum of the primordial tensor perturbations ($\lambda_1,\lambda_2, \lambda_3 = \pm 2$) is given by  
\begin{eqnarray}
\Braket{\prod_{n=1}^3 \xi_{\ell_n m_n}^{(\lambda_n)}(k_n)} 
&\sim& 
\left[\prod_{n=1}^3 \int d^2 \hat{\bf k_n} {}_{-\lambda_n}Y_{\ell_n m_n}^*(\hat{\bf k_n}) \right] 
\left[\frac{R_\gamma \ln(\tau_\nu / \tau_B )}{4\pi \rho_{\gamma,0}} \right]^3 
\frac{\alpha A_B}{n_B + 3} k_*^{n_B + 3} \frac{8\pi}{3} 
 \delta\left(\sum_{n=1}^3 {\bf k_n}\right) \nonumber \\
&&\times 
\left[ P_B(k_1) P_B(k_2) \delta_{ad} P_{be}(\hat{\bf k_1}) P_{cf}(\hat{\bf k_2}) 
+ P_B(k_2) P_B(k_3) P_{ad}(\hat{\bf k_2}) P_{be}(\hat{\bf k_3}) \delta_{cf} \right. \nonumber \\
&&\quad \left. +  P_B(k_1) P_B(k_3) P_{ad}(\hat{\bf k_1}) \delta_{be} P_{cf}(\hat{\bf k_3}) \right]
 \nonumber \\
&&\times 
 (-27) e_{ab}^{(-\lambda_1)}(\hat{\bf k_1}) e_{cd}^{(-\lambda_2)}(\hat{\bf k_2})
e_{ef}^{(-\lambda_3)}(\hat{\bf k_3})~ . 
\label{eq:ini_bis_optimal}
\end{eqnarray}
Using Eq.~(\ref{eq:projection_operator}), we reduce the contraction of the subscripts in the $TTT$ spectrum to 
\begin{eqnarray}
&&O_{ab}^{(-\lambda_1)}(\hat{\bf k_1}) O_{cd}^{(-\lambda_2)}(\hat{\bf k_2})
O_{ef}^{(-\lambda_3)}(\hat{\bf k_3})
\left[ P_B(k_1) P_B(k_2) \delta_{ad} P_{be}(\hat{\bf k_1}) P_{cf}(\hat{\bf k_2}) 
 \right. \nonumber \\
&&\qquad\qquad\qquad\qquad 
\left. + P_B(k_2) P_B(k_3) P_{ad}(\hat{\bf k_2}) P_{be}(\hat{\bf k_3}) \delta_{cf}
+  P_B(k_1) P_B(k_3) P_{ad}(\hat{\bf k_1}) \delta_{be} P_{cf}(\hat{\bf k_3}) \right] \nonumber \\
&&\qquad\qquad 
= 
e_{ae}^{(-\lambda_1)}(\hat{\bf k_1}) 
e_{ef}^{(-\lambda_3)}(\hat{\bf k_3}) e_{fa}^{(-\lambda_2)}(\hat{\bf
k_2}) 
[P_B(k_1)P_B(k_2) + 2 \ {\rm perms.}]
\nonumber \\
&&\qquad\qquad = 
- \frac{( 8\pi )^{5/2}}{3} I_{2 1 1}^{0 1 -1} I_{2 1 1}^{0 1 -1}
 \left\{
  \begin{array}{ccc}
  2 & 2 & 2 \\
  1 & 1 & 1
  \end{array}
 \right\} 
[P_B(k_1)P_B(k_2) + 2 \ {\rm perms.}]
\nonumber \\
&&\qquad\qquad\quad\times 
\sum_{M, M', M''} {}_{\lambda_1} Y_{2 M}^*(\hat{\bf k_1}) 
{}_{\lambda_2}Y_{2 M'}^*(\hat{\bf k_2}) {}_{\lambda_3}Y_{2
M''}^*(\hat{\bf k_3})
\left(
  \begin{array}{ccc}
  2 & 2 & 2 \\
  M & M' & M''
  \end{array}
 \right) ~.
\end{eqnarray}
The delta function is also expanded with the spin spherical harmonics as Eq.~(\ref{eq:delta})
\begin{eqnarray}
\delta\left( \sum_{i=1}^3 {\bf k_i} \right) 
&=& 8 \int_0^\infty y^2 dy 
\left[ \prod_{i=1}^3 \sum_{L_i M_i} 
 (-1)^{L_i/2} j_{L_i}(k_i y) 
Y_{L_i M_i}^*(\hat{\bf k_i}) \right] \nonumber \\
&&\times 
I_{L_1 L_2 L_3}^{0~0~0}
\left(
  \begin{array}{ccc}
  L_1 & L_2 & L_3 \\
  M_1 & M_2 & M_3 
  \end{array}
 \right)~.
\end{eqnarray}
Then, the angular integrals are performed as   
\begin{eqnarray}
\begin{split}
\int d^2 \hat{\bf k_1} {}_{-\lambda_1}Y_{\ell_1 m_1}^* Y_{L_1 M_1}^* {}_{\lambda_1}Y_{2 M}^* &=
 I_{\ell_1 L_1 2}^{\lambda_1 0 -\lambda_1}
\left(
  \begin{array}{ccc}
  \ell_1 & L_1 & 2 \\
  m_1 & M_1 & M 
  \end{array}
 \right) ~, \\
\int d^2 \hat{\bf k_2} {}_{-\lambda_2}Y_{\ell_2 m_2}^* Y_{L_2 M_2}^* {}_{\lambda_2}Y_{2 M'}^* &= I_{\ell_2 L_2 2}^{\lambda_2 0 -\lambda_2}
\left(
  \begin{array}{ccc}
  \ell_2 & L_2 & 2 \\
  m_2 & M_2 & M' 
  \end{array}
 \right) ~, \\
\int d^2 \hat{\bf k_3} {}_{-\lambda_3}Y_{\ell_3 m_3}^* Y_{L_3 M_3}^* {}_{\lambda_3}Y_{2 M''}^* &= I_{\ell_3 L_3 2}^{\lambda_3 0 -\lambda_3}
\left(
  \begin{array}{ccc}
  \ell_3 & L_3 & 2 \\
  m_3 & M_3 & M'' 
  \end{array}
 \right) ~,
\end{split}
\end{eqnarray}
and all the Wigner-$3j$ symbols are summed up as  
\begin{eqnarray}
&& \sum_{\substack{M_1 M_2 M_3 \\ M M' M''}}
\left(
  \begin{array}{ccc}
  L_1 &  L_2 & L_3 \\
   M_1 & M_2 & M_3
  \end{array}
 \right)
\left(
  \begin{array}{ccc}
  2 &  2 & 2 \\
  M & M' & M''
  \end{array}
 \right) 
\nonumber \\
&&\qquad \times
\left(
  \begin{array}{ccc}
  \ell_1 &  L_1 & 2 \\
   m_1 & M_1 & M
  \end{array}
 \right)
\left(
  \begin{array}{ccc}
  \ell_2 &  L_2 & 2 \\
   m_2 & M_2 & M'
  \end{array}
 \right)
\left(
  \begin{array}{ccc}
  \ell_3 &  L_3 & 2 \\
   m_3 & M_3 & M''
  \end{array}
 \right) \nonumber \\
&& \qquad\qquad\qquad = 
\left(
  \begin{array}{ccc}
  \ell_1 & \ell_2 & \ell_3 \\
   m_1 & m_2 & m_3
  \end{array}
 \right)
\left\{
  \begin{array}{ccc}
  \ell_1 & \ell_2 & \ell_3 \\
   L_1 & L_2 & L_3 \\
   2 & 2 & 2 \\
  \end{array}
 \right\}~. 
\end{eqnarray}
Thus the initial bispectrum (\ref{eq:ini_bis_optimal}) is rewritten as 
\begin{eqnarray}
\Braket{\prod_{n=1}^3 \xi_{\ell_n m_n}^{(\lambda_n)}(k_n)} 
&\sim& 
\left(
  \begin{array}{ccc}
  \ell_1 & \ell_2 & \ell_3 \\
   m_1 & m_2 & m_3
  \end{array}
 \right) 
\left[\frac{R_\gamma \ln(\tau_\nu / \tau_B )}{4\pi \rho_{\gamma,0}} \right]^3 
 \frac{\alpha A_B}{n_B + 3} k_*^{n_B + 3} \frac{8\pi}{3} \nonumber \\
&&\times
8 \int_0^\infty y^2 dy 
\left[ \prod_{n=1}^3 \sum_{L_n} 
 (-1)^{L_n/2} j_{L_n}(k_n y) \right]  
I_{L_1 L_2 L_3}^{0 \ 0 \ 0} \nonumber \\
&&\times   
 \left\{
  \begin{array}{ccc}
  2 & 2 & 2 \\
  1 & 1 & 1
  \end{array}
 \right\} 
I_{2 1 1}^{0 1 -1} I_{2 1 1}^{0 1 -1} 
I_{\ell_1 L_1 2}^{\lambda_1 0 -\lambda_1} I_{\ell_2 L_2 2}^{\lambda_2 0 -\lambda_2} I_{\ell_3 L_3 2}^{\lambda_3 0 -\lambda_3}
\left\{
  \begin{array}{ccc}
  \ell_1 & \ell_2 & \ell_3 \\
   L_1 & L_2 & L_3 \\
   2 & 2 & 2 \\
  \end{array}
 \right\}
 \nonumber \\
&&\times 
9 (8\pi)^{5/2} [ P_B(k_1)P_B(k_2) + 2 {\rm perms.} ]  ~. 
\end{eqnarray}
Comparing the exact initial bispectrum of the tensor modes (\ref{eq:xi_bis_pmf_SVT_exact}) with this equation, we can see that the number of the time-integrals and summations in terms of the multipoles decreases. This means that corresponding to the pole approximation, the 1-loop calculation (the left panel of Fig.~\ref{fig:diagram}) reaches the tree-level one (the right one of that figure). This approximation seems to be applicable to the non-Gaussianity generated from the chi-squared fields without the complicated angular dependence \cite{Lyth:2006gd}. 
Note that the scaling behaviors of these initial bispectra with respect to $k_1, k_2$ and $k_3$ are in agreement with that of the local-type non-Gaussianity (\ref{eq:S_loc}). Thus, if the pole approximation is valid, we can expect that the PMFs generate the CMB bispectra coming from the local-type non-Gaussianity. 
Via the summation over $\lambda_1, \lambda_2$ and $\lambda_3$ as Eq.~(\ref{eq:summation_helicity}), 
the approximate CMB bispectra of the tensor modes are quickly formulated: 
\begin{eqnarray}
B_{X_1 X_2 X_3, \ell_1 \ell_2 \ell_3}^{{\rm app} (T T T )}(\alpha)
&=& 
\left[\frac{R_\gamma \ln(\tau_\nu / \tau_B )}{4\pi \rho_{\gamma,0}} \right]^3 
 \frac{\alpha A_B}{n_B + 3} k_*^{n_B + 3} \frac{8\pi}{3} 
\sum_{L_1 L_2 L_3} (-1)^{\frac{L_1 + L_2 + L_3}{2}} I_{L_1 L_2 L_3}^{0 \ 0 \ 0}
\nonumber \\
&&\times 
 \left\{
  \begin{array}{ccc}
  2 & 2 & 2 \\
  1 & 1 & 1
  \end{array}
 \right\} 
I_{2 1 1}^{0 1 -1} I_{2 1 1}^{0 1 -1} 
\left\{
  \begin{array}{ccc}
  \ell_1 & \ell_2 & \ell_3 \\
   L_1 & L_2 & L_3 \\
   2 & 2 & 2 \\
  \end{array}
 \right\}
\nonumber \\
&&\times 
8 I_{\ell_1 L_1 2}^{2 0 -2} I_{\ell_2 L_2 2}^{2 0 -2} I_{\ell_3 L_3 2}^{2 0 -2} 
{\cal U}_{L_1, \ell_1, x_1} {\cal U}_{L_2, \ell_2, x_2} {\cal U}_{L_3, \ell_3, x_3} 
\nonumber \\
&&\times 
8 \int_0^\infty y^2 dy 
\left[ \prod_{n=1}^3 
 (-i)^{\ell_n} \int_0^\infty \frac{k_n^2 dk_n}{2 \pi^2} 
{\cal T}_{X_n, \ell_n}^{(T)}(k_n) j_{L_n}(k_n y) \right]  \nonumber \\
&&\times 
9 (8\pi)^{5/2} [ P_B(k_1)P_B(k_2) + 2 {\rm perms.} ] \label{eq:cmb_bis_pmf_pole_app} ~, 
\end{eqnarray}
where the multipoles are limited as
\begin{eqnarray}
\sum_{n = 1}^3 L_n = {\rm even}~, \ \ |L_1 - L_2| \leq L_3 \leq L_1 + L_2 ~,
\end{eqnarray}
and the triangle inequality imposes
\begin{eqnarray}
|\ell_1 - \ell_2| \leq \ell_3 \leq \ell_1 + \ell_2 ~.
\end{eqnarray} 
 In the next subsection, we compare these approximate spectra with the exact spectra given by Eq.~(\ref{eq:CMB_bis_PMF_numerical}) and evaluate the validity of the pole approximation. 

\subsection{Analysis}\label{subsec:result_pmf}

In this subsection, we show the result of the CMB intensity-intensity-intensity spectra induced from the auto-correlations of the each-mode anisotropic stress. In order to compute numerically, we insert Eq. (\ref{eq:CMB_bis_PMF_numerical}) into the Boltzmann code for anisotropies in the microwave background (CAMB)
\cite{Lewis:2004ef, Lewis:1999bs}.  We use the transfer functions shown in Sec.~\ref{subsec:alm_PMF}. 
In the calculation of the Wigner-$3j, 6j$ and $9j$ symbols, we use a
common mathematical library called SLATEC \cite{slatec} and analytical
expressions in Appendix \ref{appen:wigner}.

In Fig. \ref{fig:PMF_SVT_III_samel}, we plot the CMB reduced bispectra of these modes defined as~\cite{Komatsu:2001rj}
\begin{eqnarray}
b^{(Z_1 Z_2 Z_3)}_{III,\ell_1 \ell_2 \ell_3}
\equiv \left( I^{0~0~0}_{\ell_1 \ell_2 \ell_3} \right)^{-1} 
B^{(Z_1 Z_2 Z_3)}_{III,\ell_1 \ell_2 \ell_3}~,
\end{eqnarray} 
for $\ell_1 = \ell_2 = \ell_3$. 
Here, for comparison, we also write the bispectrum generated from the local-type primordial non-Gaussianities of curvature perturbations given by Eq.~(\ref{eq:S_loc}). 

From the red solid lines, we can find the enhancement at $\ell \lesssim 100$ in tensor-tensor-tensor bispectra. It is because the ISW effect gives the dominant signal like in the CMB anisotropies of tensor modes \cite{Shaw:2009nf, Pritchard:2004qp}. 
From the green dashed line, one can see that the peak of the vector-vector-vector bispectrum is located at $\ell \sim 2000$ and the position is similar to
that of the angular power spectrum $C^{(V)}_{I, \ell}$ induced from the
vector mode as calculated in Sec.~\ref{subsec:alm_PMF}. At small scales, the vector mode contributes to the CMB power spectrum through
the Doppler effect. Thus, we can easily find that the Doppler
effect can also enhance the CMB bispectrum on small scale. 
From the blue dotted lines, we can see that the scalar-scalar-scalar bispectra are boosted around at $\ell \sim 200$ due to the acoustic oscillation of the fluid of photons and baryons. On the other hand, as $\ell$ enlarges, the spectra are suppressed by the Silk damping effect. These features are also observed in the non-magnetic case (the magenta dot-dashed line), however, owing to the difference of the angular dependence on the wave number vectors in the source bispectra, the location of the nodes slightly differs. 
Comparing the behaviors between the three spectra arising from PMFs, we confirm that the tensor, scalar and vector modes become effective for $\ell \lesssim 100, 100 \lesssim \ell \lesssim 2000$ and $\ell \gtrsim 2000$, respectively, like the behaviors seen in the power spectra. Thus, for $\ell < 1000$, namely the current instrumental limit of the angular resolution such as the PLANK experiment \cite{:2006uk}, we expect that the auto- and cross-correlations between the scalar and tensor modes will be primary signals of PMFs in the CMB bispectrum. 
  
The overall amplitudes of $b_{III, \ell \ell \ell}^{(SSS)}$ and $b_{III, \ell \ell \ell}^{(TTT)}$ seem to be comparable to $\left[C_{II, \ell}^{(S)}\right]^{3/2}$ and $\left[C_{II, \ell}^{(T)}\right]^{3/2}$. However, we find that the amplitude of
$b_{III, \ell \ell \ell}^{(VVV)}$ is smaller than the above expectation. 
This is because the configuration of multipoles, corresponding to the angles of
wave number vectors, is limited to the conditions placed by the Wigner
symbols. We can understand this by considering the scaling relation
with respect to $\ell$ at high $\ell$.  If the magnetic power spectrum given by
Eq. (\ref{eq:def_pmf_power}) is close to the scale-invariant shape, the
configuration that satisfies $L \sim L'' \sim \ell$ and $L' \sim 1$
contributes dominantly in the summations. Furthermore, the other
multipoles are evaluated as 
\begin{eqnarray}
L_1 \sim L'_1 \sim L''_1 \sim \ell~, \ \
L_2 \sim L''_2 \sim L_3 \sim L'_3  \sim \ell~, \ \ L'_2 \sim L''_3 \sim 1~,
\end{eqnarray}
from the triangle conditions described in Appendix
\ref{appen:wigner}. Then we can find $b^{(VVV)}_{III, \ell \ell \ell}
\propto \ell^{2 n_B + 4}$ for $\ell \lesssim 1000$, where we have also
used the following relations
\begin{eqnarray}
\begin{split}
& \int k^2 dk \mathcal{T}^{(V)}_{I, \ell_i}(k) j_{\ell_i} (k(\tau_0 - \tau_*))
 \propto \ell~, \ \ 
\left\{
  \begin{array}{ccc}
  \ell_1 & \ell_2 & \ell_3 \\
  L' & L'' & L 
  \end{array}
 \right\} 
\propto \ell^{-1}~, \ \
{\cal K}_{L_2 L'_3}^{-(n_B + 1)} \sim {\cal K}_{L''_2 L_3}^{-(n_B + 1)} 
 \propto \ell^{n_B+1}~, \\
& \left\{
  \begin{array}{ccc}
  L'' & L & \ell_1 \\
  L_3 & L_2 & L_1 \\
  1 & 1 & 2
  \end{array}
 \right\} 
\propto \ell^{-3/2}~, \ \ 
\left\{
\begin{array}{ccc}
  L & L' & \ell_1 \\
  L'_3 & L'_2 & L_1' \\
  1 & 1 & 2
  \end{array}
 \right\} \sim 
\left\{
\begin{array}{ccc}
  L' & L'' & \ell_1 \\
  L''_3 & L''_2 & L''_1 \\
  1 & 1 & 2
  \end{array}
 \right\} 
\propto \ell^{-1}~,
\end{split}
\end{eqnarray}
which, except for the first relation, are also coming from the
triangle conditions of the Wigner $3$-j symbols.  Therefore, combining
with the scaling relation of the CMB power spectrum mentioned in Sec.~\ref{subsec:alm_PMF}, we find that $b^{(VVV)}_{III, \ell \ell \ell}$ is
suppressed by a factor $\ell^{(n_B - 1)/2}$ from $C^{(V) 3/2}_{II,\ell}$. 

\begin{figure}[t]
  \begin{center}
    \includegraphics[height=8cm]{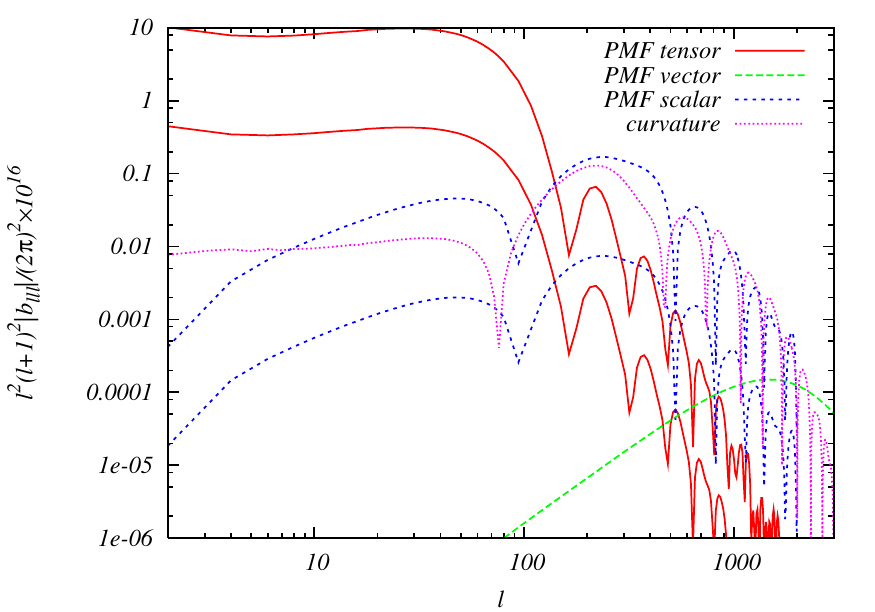}
  \end{center}
  \caption{Absolute values of the normalized reduced bispectra of temperature fluctuations for a configuration $\ell_1 = \ell_2 = \ell_3 \equiv \ell$. The red solid, green dashed, and blue dotted lines correspond to the spectra generated from the auto-correlations of the PMF tensor, vector, and scalar anisotropic stresses for $n_B = -2.9$, respectively. The upper (lower) lines of the red solid and blue dotted lines are calculated when $\tau_{\nu}/\tau_{B} = 10^{17} (10^6)$.
The magenta dot-dashed line expresses the spectrum sourced from the primordial non-Gaussianity with $f^{\rm local}_{\rm NL} = 5$. The strength of PMFs is fixed to $B_{1 {\rm Mpc}} = 4.7 {\rm nG}$ and the other cosmological parameters are fixed to the mean values limited from WMAP-7yr data reported in Ref. \cite{Komatsu:2010fb}.}
  \label{fig:PMF_SVT_III_samel}
\end{figure}

\begin{figure}[t]
  \begin{center}
    \includegraphics[height=8cm]{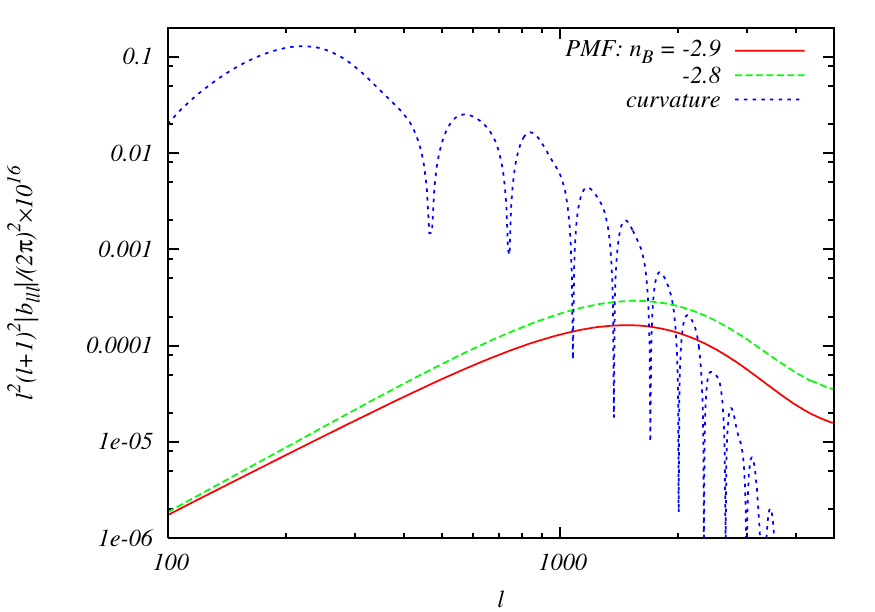}
  \end{center}
  \caption{Absolute values of the normalized reduced temperature-temperature-temperature spectra arising from the auto-correlation between the PMF vector anisotropic stresses for a configuration $\ell_1 = \ell_2 = \ell_3 \equiv \ell$. The lines correspond to the spectra generated from vector anisotropic stress for $n_B = -2.9$ (red solid line) and $-2.8$ (green dashed line), and primordial non-Gaussianity with $f^{\rm local}_{\rm NL} = 5$ (blue dotted line). The strength of PMFs is fixed to $B_{1 {\rm Mpc}} = 4.7 {\rm nG}$ and the other cosmological parameters are identical to the values used in Fig.~\ref{fig:PMF_SVT_III_samel}.}
  \label{fig:vec_III_samel_difnB.pdf}
\end{figure}

In Fig.~\ref{fig:vec_III_samel_difnB.pdf}, we show $b_{III, \ell \ell \ell}^{(VVV)}$ for $\ell_1 = \ell_2 = \ell_3$ for the different spectral index $n_B$. Red solid and green dashed lines correspond to the bispectrum with the spectral index of the power spectrum of PMFs fixed as $n_B = -
2.9$ and $- 2.8$, respectively.  
From this figure, we find that the CMB bispectrum becomes steeper
if $n_B$ becomes larger, which is similar to the case of the power
spectrum. These spectra trace the scaling relation in the above
discussion. These will lead to another constraint on the strength of PMFs.  

\begin{figure}[t]
  \begin{center}
    \includegraphics[height=8cm,clip]{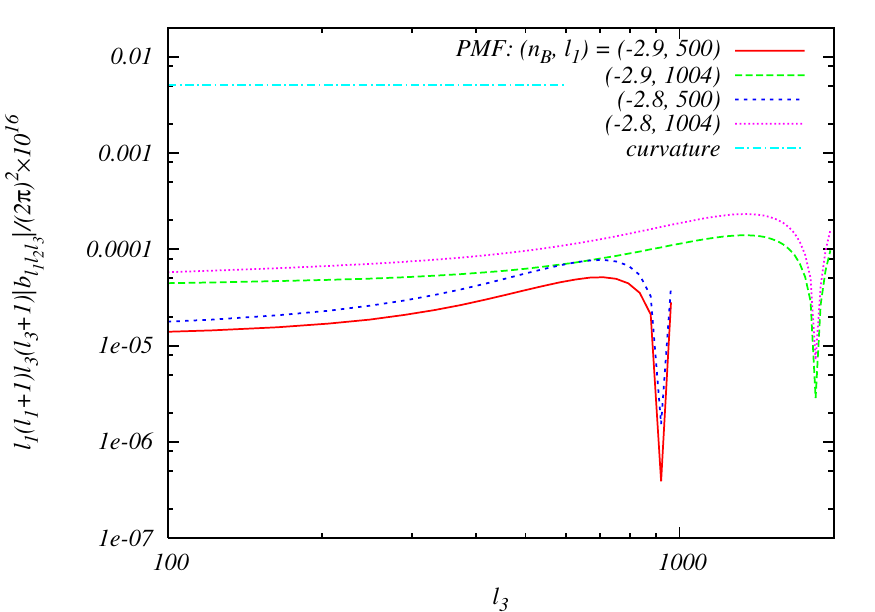}
  \end{center}
  \caption{Absolute values of the normalized reduced
 temperature-temperature-temperature bispectra induced by the auto-correlation between the PMF vector anisotropic stresses and generated by primordial
  non-Gaussianity given by Eq.~(\ref{eq:cmb_bis_local}) as a function of $\ell_3$ with $\ell_1$ and $\ell_2$ fixed to some value as indicated.
Each parameter is fixed to the same value defined in Fig.~\ref{fig:PMF_SVT_III_samel}.}
  \label{fig:vec_III_difl_difnB.pdf}
\end{figure}

\begin{figure}[t]
  \begin{center}
    \includegraphics[height=8cm,clip]{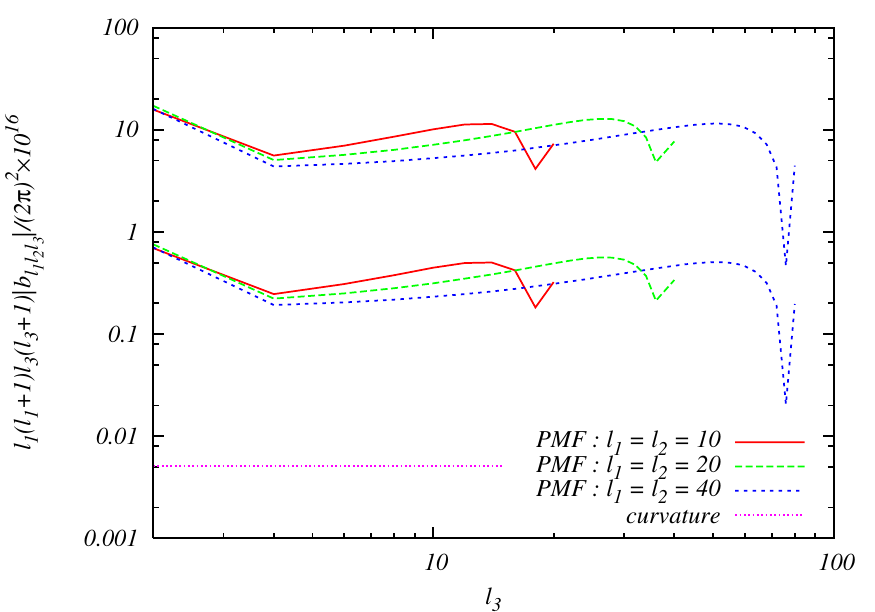}
  \end{center}
  \caption{Absolute values of the normalized reduced temperature-temperature-temperature bispectra induced by the auto-correlation between the PMF tensor anisotropic stresses and generated from primordial non-Gaussianity in curvature perturbations given by Eq.~(\ref{eq:cmb_bis_local}) as a function of $\ell_3$ with $\ell_1 = \ell_2$.
Each parameter is identical to the values defined in Fig.~\ref{fig:PMF_SVT_III_samel}.
}
  \label{fig:tens_pas_III_difl.pdf}
\end{figure}

In Figs.~\ref{fig:vec_III_difl_difnB.pdf} and \ref{fig:tens_pas_III_difl.pdf}, we show the reduced bispectrum
$b_{III, \ell_1 \ell_2 \ell_3}^{(VVV)}$ and $b_{III, \ell_1 \ell_2 \ell_3}^{(TTT)}$ with respect to $\ell_3$ with setting
$\ell_1 = \ell_2$, respectively. 
From Fig.~\ref{fig:vec_III_difl_difnB.pdf}, we can see that $b_{III, \ell_1 \ell_2 \ell_3}^{(VVV)}$ for $\ell_1,
\ell_2, \ell_3 \gtrsim 100$ is nearly flat and given as
\begin{eqnarray}
\ell_1 (\ell_1+1) \ell_3 (\ell_3+1) |b^{(VVV)}_{III, \ell_1 \ell_2 \ell_3}| 
&\sim& 2 \times 10^{-19} 
\left(\frac{B_{1 \rm Mpc}}{4.7 \rm nG} \right)^6 \label{eq:CMB_vec_bis_rough}~.
\end{eqnarray}
We can understand this by
the analytical evaluation as follows. As mentioned above, in the
summations of Eq. (\ref{eq:CMB_bis_PMF_numerical}), the configuration that
$L \sim \ell_1, L' \sim 1$ and $L'' \sim \ell_3$ contributes dominantly.
By using this and the approximations that 
\begin{eqnarray}
L_1 \sim \ell_1~, \ \ L'_1 \sim \ell_2~, \ \ L''_1 \sim \ell_3~, \ \
L_2 \sim L'_3 \sim L~, \ \ L'_2 \sim L''_3 \sim L'~, \ \ L''_2 \sim L_3 \sim L''~,
\end{eqnarray}
which again come from the triangle conditions from the Wigner symbols,
the scaling relation of $\ell_3$ at large scale is evaluated as
$b^{(VVV)}_{III,\ell_1 \ell_2 \ell_3} \propto \ell_3^{n_B + 1}$.  From
this estimation we can find that $\ell_1(\ell_1+1)\ell_3(\ell_3 +
1)b^{(VVV)}_{III, \ell_1 \ell_2 \ell_3} \propto \ell_3^{0.1}$, for $n_B
= -2.9$, and $\ell_3^{0.2}$ for $n_B = -2.8$, respectively, which match
the behaviors of the bispectra in Fig.~\ref{fig:vec_III_difl_difnB.pdf}.

From Fig.~\ref{fig:tens_pas_III_difl.pdf}, we can also see that if the
PMF spectrum obeys the nearly scale-invariant shape, $b_{III, \ell_1 \ell_2
\ell_3}^{(TTT)}$ for $\ell_1, \ell_2, \ell_3 \lesssim 100$ is given by 
\begin{eqnarray}
\ell_1 (\ell_1+1) \ell_3 (\ell_3+1) |b^{(TTT)}_{III, \ell_1 \ell_2 \ell_3}| 
&\sim& (130 - 6) \times 10^{-16} 
\left(\frac{B_{1 \rm Mpc}}{4.7 \rm nG} \right)^6 ~, \label{eq:CMB_tens_bis_rough}
\end{eqnarray}
where the factor $130$ corresponds to the $\tau_\nu / \tau_B = 10^{17}$
case and $6$ corresponds to $10^6$. 
In order to obtain a rough constraint on the magnitude of the PMF, we
compare the bispectrum induced from the PMF with that from the local-type primordial non-Gaussianity in the curvature perturbations,
which is typically estimated as \cite{Riot:2008ng}
\begin{eqnarray}
\ell_1 (\ell_1+1) \ell_3
(\ell_3+1)b_{\ell_1 \ell_2 \ell_3} \sim 4 \times 10^{-18} f^{\rm
local}_{\rm NL}~. \label{eq:cmb_bis_local}
\end{eqnarray}
By comparing this with
Eq. (\ref{eq:CMB_tens_bis_rough}), the relation between the magnitudes of
the PMF with the nearly scale-invariant power spectrum and $f^{\rm
local}_{\rm NL}$ is derived as
\begin{eqnarray}
\left( \frac{B_{1\rm Mpc}}{1\rm nG} \right) \sim (1.22 - 2.04)
|f^{{\rm local}}_{\rm NL}|^{1/6}~.
\end{eqnarray} 
Using the above equation, we can obtain the rough bound on the PMF strength.
As shown in Fig.~\ref{fig:PMF_SVT_III_samel}, because the tensor bispectrum is highly damped for $\ell \gtrsim 100$, 
we should use an upper bound on $f^{\rm local}_{\rm NL}$ obtained by the
current observational data for $\ell < 100$, namely $f^{\rm local}_{\rm
NL} < 100$ \cite{Smith:2009jr}. This value is consistent with a simple
prediction from the cosmic variance \cite{Komatsu:2001rj}. From this
value, we derive $B_{1 \rm Mpc} < 2.6 - 4.4 {\rm nG}$. 

\begin{figure}[t]
  \begin{center}
    \includegraphics[height=8cm]{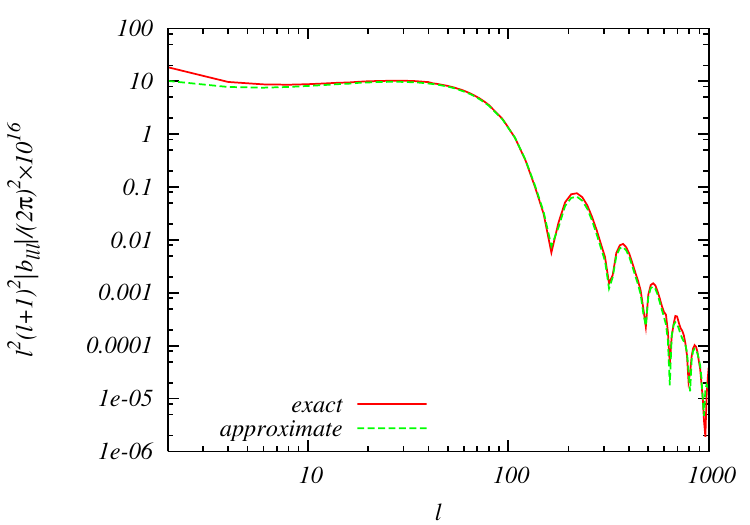}
  \end{center}
  \caption{Absolute values of the normalized reduced bispectra of temperature fluctuation for a configuration $\ell_1 = \ell_2 = \ell_3 \equiv \ell$. The red solid and green dashed lines represent the exact and approximate spectra arising from the tensor-tensor-tensor correlation of the PMF anisotropic stresses for $B_{1 {\rm Mpc}} = 4.7 {\rm nG}, n_B = -2.9$ and $\tau_{\nu}/\tau_{B} = 10^{17}$, respectively. The cosmological parameters are identical to the values defined in Fig.~\ref{fig:PMF_SVT_III_samel}.}
  \label{fig:PMF_TTT_III_pole_samel.pdf}
\end{figure}

From here, let us discuss the validity and possibility of the CMB bispectra under the pole approximation (\ref{eq:cmb_bis_pmf_pole_app}). Figure \ref{fig:PMF_TTT_III_pole_samel.pdf} shows the shapes of the CMB tensor-tensor-tensor spectra based on the exact form (\ref{eq:CMB_bis_PMF_numerical}) and approximate one (\ref{eq:cmb_bis_pmf_pole_app}). Both spectra seem to have a good agreement in the shape of the $\ell$ space. To discuss more precisely, using the correlation 
\begin{eqnarray}
b \cdot b'
\propto \sum_{\ell} b_{\ell \ell \ell} b'_{\ell \ell \ell} ~,  
\end{eqnarray}
we calculate a correlation coefficient as
\begin{eqnarray} 
\frac{b^{\rm ex} \cdot b^{\rm app}}{\sqrt{ (b^{\rm ex} \cdot b^{\rm ex})(b^{\rm app} \cdot b^{\rm app}) }} = 0.99373 ~,
\end{eqnarray} 
where $b^{\rm ex}$ and $b^{\rm app}$ are the exact and approximate reduced bispectra, respectively. This fact, which this quantity approaches unity, implies that the pole approximation can produce an almost exact copy. An unknown parameter, $\alpha$, is derived from the relation as
\begin{eqnarray}
\alpha = 
\frac{b^{\rm ex}_{\ell \ell \ell}}{b^{\rm app}_{\ell \ell \ell} (\alpha=1)}
 \approx \frac{b^{\rm ex} \cdot b^{\rm app}(\alpha = 1)}{b^{\rm app}(\alpha = 1) \cdot b^{\rm app}(\alpha = 1)} = 0.2991 ~.
\end{eqnarray}
The cases other than the tensor-tensor-tensor spectrum will be presented in Ref.~\cite{Shiraishi:2012rm}. 

As shown in the previous subsections, the CMB bispectra from PMFs arise
from the six-point correlation of the Gaussian magnetic fields and have
one-loop structure due to the summation over the additional
multipoles. Hence, it takes so long hours to estimate all $\ell$'s
contribution and it is actually impossible to compute the
signal-to-noise ratio. However, using the pole approximation, since the
summation reaches the tree-level calculation, we will obtain more
precise bound through the estimation of the signal-to-noise ratio
including the contribution of the cross-correlations between scalar and
tensor modes \cite{Shiraishi:2012rm}. 

\subsection{Summary and discussion}

In this section, on the basis of our recent works \cite{Shiraishi:2010yk,
Shiraishi:2011dh, Shiraishi:2011fi}, 
we presented the all-sky formulae for the CMB bispectra
induced by the scalar, vector, and tensor non-Gaussianities coming from
the PMFs by dealing with the full-angular dependence of the bispectrum of the PMF anisotropic
stresses. Then, expressing the angular dependence with the spin-weighted spherical harmonics and converting the angular integrals into the Wigner symbols were key points of the formulation. 
From the practical calculation, 
it is found that the CMB bispectra from the magnetic 
tensor, scalar, and vector modes dominate at large ($\ell \lesssim
100$), intermediate ($100 \lesssim \ell \lesssim 2000$), and small
($\ell \gtrsim 2000$) scales. For the discussion about the shape of the
non-Gaussianity in the PMF anisotropic stresses, we performed 
the pole approximation, which is the evaluation of the convolutions at around the divergence points of the integrands, and found that the bispectra of the PMF anisotropic stresses are classified as the local-type configuration. Owing to this, we had some significant signals of the CMB bispectra on the squeezed limit also in the multipole space.
Compared with the exact formula, the approximate one reduces the
computing time, hence we expect the use for the calculation of the
signal-to-noise ratio \cite{Shiraishi:2012rm}. 
We also investigated the dependence of the CMB bispectrum on the spectral index of the PMF power spectrum and confirmed that the CMB bispectrum induced from the PMFs is sensitive to it. Since the characteristic scale varies with the value of the spectral index, it is important to consider not only the contribution from the scalar mode, but also those from the vector and tensor modes. 

By translating the current bound on the local-type non-Gaussianity from
the CMB bispectrum into the bound on the amplitude of the magnetic
fields, we obtain a new limit: $B_{1\rm Mpc} < 2.6 - 4.4 {\rm nG}$. This is a
rough estimate coming from the large scale information of the tensor mode and a
precise constraint is expected if one considers the 
full $\ell$ contribution by using an appropriate estimator of the CMB
bispectrum induced from the primordial magnetic fields. 

Because of the complicated discussions and mathematical manipulations, here
we restrict our numerical results to the intensity bispectra of
auto-correlations between scalar, vector and tensor modes despite
the fact that our formula for the CMB bispectra
(\ref{eq:CMB_bis_PMF_complete}) contains the polarizations and the cross-correlations between scalar, vector and tensor modes. However, like the non-magnetic case \cite{Babich:2004yc}, 
the modes other than our numerical results, such as $B^{(SVT)}_{IEB,
\ell_1 \ell_2 \ell_3}$, will bring in more reasonable bounds on the
PMFs \cite{Shiraishi:2012rm}. Furthermore, the effect on the CMB four-point correlation
(trispectrum) is just beginning to be roughly discussed
\cite{Trivedi:2011vt}. Applying our studies, this should be taken into account more precisely. 

\bibliographystyle{JHEP}
\bibliography{paper}

%% file: ch10_sum.tex
\section{Conclusion}\label{sec:sum}

The main purpose of this thesis was to present the formalism for the CMB bispectrum induced by the non-Gaussianities not only in the standard scalar-mode perturbations but also in the vector- and tensor-mode ones where the violation of the rotational or parity invariance is also involved, and some attempts to prove the nature of the early Universe by applying our formalism.
To do this, we have discussed the following things. 

In Sec.~\ref{sec:intro}, we gave the introduction of this thesis. Then, we quickly summarized the history of the Universe, the paradigm in the early Universe, and the concept of this thesis. 
In Sec.~\ref{sec:inflation}, we summarized how to generate the curvature perturbations and gravitational waves and the consistency relations in the slow-roll inflation. In Sec.~\ref{sec:CMB_anisotropy}, we showed how to construct the $a_{\ell m}$'s generated from the primordial scalar, vector and tensor sources in order to formulate the CMB bispectrum easily. We also summarized the constraints on several key parameters, which characterize the nature of inflation and the dynamics of the Universe, obtained from the current CMB data. 
In Sec.~\ref{sec:PNG}, we focused on the topic of the primordial non-Gaussianities. In Sec.~\ref{sec:formula}, we gave the general formulae for the CMB bispectrum coming from not only scalar-mode but also vector- and tensor-mode perturbations, which includes both the auto- and cross-correlations between the intensity and polarizations. Next, applying this formalism, we computed the CMB bispectra from several kinds of the non-Gaussianities. 
In Sec.~\ref{sec:maldacena}, we treated the two scalars and a graviton correlator and obtained the CMB bispectrum including the tensor-mode perturbation. Here, we had a bound on the nonlinear scalar-scala-tensor coupling by the computation of the signal-to-noise ratio. 
In Sec.~\ref{sec:statistically_anisotropic}, we considered the non-Gaussianity which has the preferred direction. Through the analysis, we found that the finite signals arise from the multipoles except for the triangle inequality. We furthermore confirmed that these special signals are comparable in magnitude with the signals keeping the triangle inequality. In Sec.~\ref{sec:parity_violating}, we dealt with the graviton non-Gaussianity arising from the parity-conserving and parity-violating Weyl cubic terms. Calculating the CMB intensity and polarization bispectra, we clarified that the intensity-intensity-intensity spectrum from the parity-violating non-Gaussianity obeys the condition as $\sum_{n=1}^3 \ell_n = {\rm odd}$. These configurations will be very beneficial to check the parity violation of the Universe in the non-Gaussian level observationally. 
In Sec.~\ref{sec:PMF}, we took into account the effect of the non-Gaussianities due to the primordial magnetic fields. Depending quadratically on the magnetic fields, the magnetic anisotropic stresses obey the chi-square distributions. Since these non-Gaussian anisotropic stresses become the sources of the CMB fluctuations, their bispectra have the finite values. Computing the CMB intensity-intensity-intensity spectra, we clarified that the tensor (vector) mode dominates at large (small) scales and the scalar mode shows up at intermediate scales. 
By the pole approximation, we also found that the bispectrum of the magnetic anisotropic stresses is similar to the local-type bispectrum. 
Comparing the theoretical results with the observational limit on the local-type non-Gaussianity, we obtained a bound on the strength of the magnetic fields, $B_{\rm 1 Mpc} < 2.6 - 4.4 {\rm nG}$. We expect that this bound will be updated by considering the impacts of the cross-correlations between scalar, vector and tensor modes, and the additional information from polarizations. 

Our formalism for the CMB bispectrum is general enough to be applicable to the non-Gaussian sources other than the above ones. Moreover, this will be easily extended to the higher-order correlations. Therefore, the studies in this thesis will be very beneficial to quest for the true picture of the origin of the Universe.  


%% file: appendix.tex
\section{Spin-weighted spherical harmonic function} \label{appen:spherical_harmonics}

Here, we review the properties of the spin-weighted spherical harmonic function. In the past, this was mainly applied to the analysis of the gravitational wave (see e.g. Ref.~\cite{Thorne:1980ru}). This discussion is based on Refs.~\cite{Newman:1966ub, Goldberg:1966uu, Zaldarriaga:1996xe}. 

The spin-weighted spherical harmonic function on 2D sphere, ${}_sY_{lm}(\theta,\phi)$,  is more general expression than the ordinaly spherical harmonic function, $Y_{lm}(\theta,\phi)$, and has additional $U(1)$ symmetry characterized by a spin weight $s$. The spin-$s$ function such as ${}_sY_{lm}(\theta,\phi)$ obeys the spin raising and lowering rule as 
$(\up {}_sf)'=e^{-i(s+1)\psi} \up {}_sf$ 
and $(\down {}_sf)' = e^{-i(s-1)\psi} \down {}_sf$. Here, the spin raising and lowering operators are given by
\begin{eqnarray}
\begin{split}
\up {}_sf(\theta,\phi) &= -\sin^{s} \theta
\left[\partial_\theta + i\csc \theta
\partial_\phi \right]\sin^{-s} \theta
{}_sf(\theta,\phi)~, \\
\down {}_sf(\theta,\phi) &= -\sin^{-s} \theta 
\left[ \partial_\theta - i\csc \theta
\partial_\phi \right]\sin^{s} \theta {}_sf(\theta,\phi)~, 
\label{eq:edth}
\end{split}
\end{eqnarray}
Specifically, the spin raising and lowering operators acting twice on the
spin-$\pm 2$ function ${}_{\pm 2}f(\mu,\phi)$ such as the CMB polarization
fields can be expressed as
\begin{eqnarray} 
\begin{split}
\down^2 {}_2f(\theta,\phi)&=
\left(-\partial_\mu + {m \over 1-\mu^2}\right)^2 \left[
(1-\mu^2) {}_2f(\mu,\phi)\right] ~, \\ 
\up^2 {}_{-2}f(\theta,\phi) &=
\left(-\partial_\mu - {m \over 1-\mu^2}\right)^2 \left[(1-\mu^2) {}_{-2}
f(\mu,\phi)\right]~, 
\label{eq:operators1}
\end{split}
\end{eqnarray}
where $\mu \equiv \cos \theta$ and ${}_{\pm 2}f(\theta,\phi) = {}_{\pm 2}\tilde{f}(\mu) e^{i m \phi}$. Utilizeing these properties, we can express ${}_sY_{lm}(\theta, \phi)$ in terms of ${}_{0}Y_{l m}(\theta, \phi) = Y_{lm}(\theta, \phi)$ as 
\begin{eqnarray}
\begin{split}
{}_sY_{lm}(\theta,\phi) &=
 \left[{(l-s)!\over (l+s)!}\right]^{1\over 2}\up^s Y_{lm}(\theta,\phi)
\ \  (0 \leq s \leq l) ~,  \\ 
{}_sY_{lm}(\theta,\phi) &=
 \left[{(l+s)!\over (l-s)!}\right]^{1\over 2}(-1)^s 
\down^{-s} Y_{lm}(\theta,\phi)
\ \  (-l \leq s \leq 0) ~,
\end{split} 
\end{eqnarray}
where these equations contain 
\begin{eqnarray}
\begin{split}
\up {}_sY_{lm}(\theta,\phi) &=\left[(l-s)(l+s+1)\right]^{1\over 2}
{}_{s+1}Y_{lm}(\theta,\phi) ~, \\
\down {}_sY_{lm}(\theta,\phi) &=-\left[(l+s)(l-s+1)\right]^{1\over 2}
{}_{s-1}Y_{lm}(\theta,\phi) ~, \\
\down\up {}_sY_{lm}(\theta,\phi) &=-(l-s)(l+s+1)
{}_{s}Y_{lm}(\theta,\phi) ~.
\label{eq:propYs}
\end{split}
\end{eqnarray}
These properties reduce to an explicit expression: 
\begin{eqnarray}
{}_sY_{lm}(\theta, \phi) &=& e^{im\phi}
\left[{(l+m)!(l-m)!\over (l+s)!(l-s)!}
{2l+1\over 4\pi}\right]^{1/2}
\sin^{2l}(\theta/2) \nonumber \\
&&\times \sum_r {l-s \choose r}{l+s \choose r+s-m}
(-1)^{l-r-s+m}{\rm cot}^{2r+s-m}(\theta/2) ~.
\label{eq:expl}
\end{eqnarray} 

This holds the orthogonality and completeness conditions as 
\begin{eqnarray}
\begin{split}
\int_0^{2\pi} d\phi \int_{-1}^1 d\cos \theta {}_sY_{l' m'}^*(\theta,\phi)
{}_sY_{lm}(\theta,\phi) &= \delta_{l', l}\delta_{m', m} ~, \\
\sum_{lm} {}_sY_{lm}^*(\theta,\phi)
{}_sY_{lm}(\theta',\phi')
&= \delta(\phi-\phi')\delta(\cos\theta-\cos\theta'). 
\end{split}
\end{eqnarray}   
The reactions to complex condugate and parity transformation are given by
\begin{eqnarray}
\begin{split}
{}_sY^*_{lm}(\theta, \phi) &= (-1)^{s + m}{}_{-s}Y_{l-m}(\theta, \phi)~, \\
{}_s Y_{lm}(\pi - \theta, \phi + \pi) &= (-1)^l _{-s} Y_{l m}(\theta, \phi)~. 
\end{split}
\end{eqnarray}
Finally, we give the specific expressions for some simple cases in Tables~\ref{tab:spin1} and \ref{tab:spin2}.

\begin{table}[t]
\begin{center}
\begin{tabular}{| c | c  | c | }
\hline
$m$ & $Y_{1 m}$ & ${}_{1}Y_{1 m}$  \\ \hline
$\pm 1$ & $ -m \sqrt{ \frac{3}{8 \pi} } \sin\theta e^{m i\phi}$ 
  &
	$- \frac{1}{2}\sqrt{ \frac{3}{4 \pi} } (1 - m \cos\theta) e^{m i\phi}$ \\ \hline
0 &     $\frac{1}{2} \sqrt{\frac{3}{\pi}} \cos \theta $
  &
	$\sqrt{ \frac{3}{8\pi} } \sin \theta$
	 \\ \hline
\end{tabular}
\caption{Dipole ($l=1$) harmonics for spin-$0$ and $1$.}
\label{tab:spin1}
\end{center}
\end{table}

\begin{table}[t]
\begin{center}
\begin{tabular}{| c | c  | c | }
\hline
$m$ & $Y_{2m}$ & ${}_{2}Y_{2 m}$  \\ \hline
$\pm 2$ & ${1 \over 4} \sqrt{ 15 \over 2\pi}\, \sin^2\theta\, e^{m i\phi}$ 
  &
	${1 \over 8}\sqrt{ 5 \over \pi}\, 
\left(1 -\frac{m}{2} \cos\theta\right)^2 \,
	 e^{m i\phi}$ \\ \hline
$\pm 1$ & $-m \sqrt{15 \over 8\pi} \sin\theta\, \cos\theta \, e^{m i\phi}$ 
  &
	$- {1 \over 4}\sqrt{ 5 \over \pi} \sin\theta\, (1 - m \cos\theta)\, 
	e^{m i\phi}$ \\ \hline
0 &     ${1 \over 2}\sqrt{5 \over 4\pi}\, (3 \cos^2\theta - 1)$
  &
	${3 \over 4}\sqrt{ 5 \over 6\pi} \, \sin^2\theta$
	 \\ \hline
\end{tabular}
\caption{Quadrupole ($l=2$) harmonics for spin-$0$ and $2$.}
\label{tab:spin2}
\end{center}
\end{table}

\newpage

\section{Wigner $D$-matrix}\label{appen:wigner_d_matrix}

Here, on the basis of Refs.~\cite{Okamoto:2002ik, Weinberg:2008zzc, Goldberg:1966uu}, we introduce the properties of the Wigner $D$-matrix $D_{m m'}^{(l)}$, which is the unitary irreducible matrix of rank
$2 l + 1$ that forms a representation of the rotational group as $SU(2)$
and $SO(3)$. 
With this matrix, the change of the spin wighted spherical harmonic
function under the rotational transformation as $\hat{\bf n} \rightarrow
R \hat{\bf n}$ is expressed as
\begin{eqnarray}
_s Y_{lm}^* (R \hat{\bf n}) = \sum_{m'} D_{m m'}^{(l)} (R) 
_s Y_{lm'}^* (\hat{\bf n}) ~. \label{eq:D_transform}
\end{eqnarray} 
This satisfies the relation as 
\begin{eqnarray}
D_{m m'}^{(l) *} (R) = (-1)^{m-m'} D_{-m, m'}^{(l)} (R) = D_{m' m}^{(l)} (R^{-1}) ~.
\end{eqnarray}
When we express the rotational matrix with three Euler angles $(\alpha,
\beta, \gamma)$ under the $z-y-z$ convention as 
\begin{eqnarray}
R = \left( 
 \begin{array}{ccc}
  \cos \alpha \cos \beta \cos \gamma - \sin \alpha \sin \gamma & 
   - \cos\beta \sin\gamma \cos\alpha - \cos\gamma \sin\alpha & 
   \cos\alpha \sin\beta \\
  \cos\alpha \sin\gamma + \cos\gamma \cos\beta \sin\alpha & 
  \cos\alpha \cos\gamma - \cos\beta \sin\alpha \sin\gamma & 
  \sin\beta \sin\alpha \\
  - \cos\gamma \sin\beta & \sin\gamma \sin\beta & \cos\beta
     \end{array}
 \right)~, \label{eq:rotational_matrix_zyz}
\end{eqnarray}
we can write a general relationship between the Wigner $D$-matrix and the spin weighted spherical harmonics as 
\begin{eqnarray}
D_{m s}^{(l)}(\alpha, \beta, \gamma) = (-1)^s \sqrt{\frac{4\pi}{2 l +
 1}} {}_{-s} Y_{lm}^* (\beta, \alpha) e^{- i s \gamma} ~. \label{eq:D_Y}
\end{eqnarray} 
Like the spin weighted spherical harmonics, there also exists the
orthogonality of the Wigner $D$-matrix as
\begin{eqnarray}
\int_0^{2\pi} d \alpha \int_{-1}^1 d \cos\beta \int_0^{2\pi} d\gamma
D_{m' s'}^{(l') *} (\alpha, \beta, \gamma) D_{m s}^{(l)} (\alpha, \beta,
\gamma) = \frac{8\pi^2}{2l+1} \delta_{l',l} \delta_{m', m} \delta_{s', s}~.
\end{eqnarray}

\newpage

\section{Wigner symbols}\label{appen:wigner}

Here, we briefly review the useful properties of the Wigner-$3j, 6j$ and $9j$
symbols. The following discussions are based on Refs.~\cite{Okamoto:2002ik, Gurau:2008yh, Jahn/Hope:1954, mathematica, Hu:2001fa, Shiraishi:2010kd, Shiraishi:2011fi}.

\subsection{Wigner-$3j$ symbol}

In quantum mechanics, considering the coupling of two angular momenta as
\begin{eqnarray}
{\bf l_3} = {\bf l_1} + {\bf l_2}~,
\end{eqnarray}
the scalar product of eigenstates between the right-handed term and the left-handed one, namely, a Clebsch-Gordan coefficient, is related to the 
Wigner-$3j$ symbol:
\begin{eqnarray}
\left(
  \begin{array}{ccc}
  l_1 & l_2 & l_3 \\
   m_1 & m_2 & - m_3
  \end{array}
 \right) 
\equiv \frac{ (-1)^{l_1 - l_2 + m_3} \Braket{l_1 m_1 l_2 m_2 |
(l_1 l_2) l_3 m_3} }{\sqrt{2 l_3 + 1}}~.
\end{eqnarray}
This symbol vanishes unless the selection rules are satisfied as follows:
\begin{eqnarray}
\begin{split}
& |m_1| \leq l_1~, \ \ |m_2| \leq l_2~, \ \ |m_3| \leq l_3 ~, \ \ m_1 + m_2 = m_3 ~, \\
& |l_1 - l_2| \leq l_3 \leq l_1 + l_2 \ {\rm (the \ triangle \ condition)}~, \ \ l_1 + l_2 + l_3 \in \mathbb{Z} ~.
\end{split}
\end{eqnarray}
Symmetries of the Wigner-$3j$ symbol are given by 
\begin{eqnarray}
\left(
  \begin{array}{ccc}
  l_1 & l_2 & l_3 \\
   m_1 & m_2 & m_3
  \end{array}
 \right) 
&=& (-1)^{\sum_{i=1}^3 l_i} \left(
  \begin{array}{ccc}
  l_2 & l_1 & l_3 \\
  m_2 & m_1 & m_3
  \end{array}
 \right) 
= (-1)^{\sum_{i=1}^3 l_i} \left(
  \begin{array}{ccc}
  l_1 & l_3 & l_2 \\
  m_1 & m_3 & m_2
  \end{array}
 \right) \nonumber \\ 
&&\qquad\qquad ({\rm odd \ permutation \ of \ columns}) \nonumber \\
&=& \left(
  \begin{array}{ccc}
  l_2 & l_3 & l_1 \\
  m_2 & m_3 & m_1
  \end{array}
 \right) 
= \left(
  \begin{array}{ccc}
  l_3 & l_1 & l_2 \\
  m_3 & m_1 & m_2
  \end{array}
 \right) \nonumber \\ 
&&\qquad\qquad ({\rm even \ permutation \ of \ columns}) \nonumber \\
&=& (-1)^{\sum_{i=1}^3 l_i} 
\left(
  \begin{array}{ccc}
  l_1 & l_2 & l_3 \\
  - m_1 & - m_2 & - m_3
  \end{array}
 \right) \nonumber \\
&&\qquad\qquad ({\rm sign \ inversion \ of} \ m_1, m_2, m_3) ~ .
\end{eqnarray}
The Wigner-$3j$ symbols satisfy the orthogonality as 
\begin{eqnarray}
\begin{split}
  \sum_{l_3 m_3} (2 l_3 + 1)
\left(
  \begin{array}{ccc}
  l_1 & l_2 & l_3 \\
  m_1 & m_2 & m_3
  \end{array}
 \right)
\left(
  \begin{array}{ccc}
  l_1 & l_2 & l_3 \\
  m_1' & m_2' & m_3
  \end{array}
 \right) 
&= \delta_{m_1, m'_1} \delta_{m_2, m'_2}~, \\
 (2 l_3 + 1) \sum_{m_1 m_2} 
\left(
  \begin{array}{ccc}
  l_1 & l_2 & l_3 \\
  m_1 & m_2 & m_3
  \end{array}
 \right)
\left(
  \begin{array}{ccc}
  l_1 & l_2 & l'_3 \\
  m_1 & m_2 & m'_3
  \end{array}
 \right) 
&= \delta_{l_3, l_3'} \delta_{m_3, m'_3}~. \label{eq:Wig_3j_lllmmm}
\end{split}
\end{eqnarray}
For a special case that $\sum_{i=1}^3 l_i = {\rm even}$ and $m_1 = m_2 = m_3 = 0$, there is an analytical expression as 
\begin{eqnarray}
 \left(
  \begin{array}{ccc}
  l_1 & l_2 & l_3 \\
  0 & 0 & 0
  \end{array}
 \right) 
 = (-1)^{\sum_{i=1}^3 {- l_i \over 2}} 
\frac{ \left(\sum_{i=1}^3 \frac{l_i}{2} \right)! \sqrt{(-l_1 + l_2 + l_3)!}
\sqrt{(l_1 - l_2 + l_3)!} \sqrt{(l_1 + l_2 - l_3)!} }
{ \left(\frac{- l_1 + l_2 + l_3}{2}\right)! 
\left(\frac{l_1 - l_2 + l_3}{2}\right)! 
\left(\frac{l_1 + l_2 - l_3}{2}\right)! 
\sqrt{ \left( \sum_{i=1}^3 l_i + 1 \right)!} }.
\end{eqnarray}
This vanishes for $\sum_{i=1}^3 l_i = {\rm odd}$.
The Wigner-$3j$ symbol is related to the spin-weighted spherical harmonics as
\begin{eqnarray}
\prod_{i=1}^2 {}_{s_i} Y_{l_i m_i}(\hat{\bf n})
= \sum_{l_3 m_3 s_3} {}_{s_3} Y^*_{l_3 m_3}(\hat{\bf n}) 
I^{-s_1 -s_2
-s_3}_{l_1~l_2~l_3} 
\left(
  \begin{array}{ccc}
  l_1 & l_2 & l_3 \\
  m_1 & m_2 & m_3
  \end{array}
 \right)~, \label{eq:product_sYlm} 
\end{eqnarray}
which leads to the ``extended'' Gaunt integral including spin dependence:
\begin{eqnarray}
\int d^2 \hat{\bf n} {}_{s_1} Y_{l_1 m_1}(\hat{\bf n}) {}_{s_2} Y_{l_2 m_2}
(\hat{\bf n}){}_{s_3} Y_{l_3 m_3}(\hat{\bf n}) 
= I^{-s_1 -s_2 -s_3}_{l_1~l_2~l_3}
\left(
  \begin{array}{ccc}
  l_1 & l_2 & l_3 \\
  m_1 & m_2 & m_3
  \end{array}
 \right)~. \label{eq:gaunt}
\end{eqnarray}
Here $I^{s_1 s_2 s_3}_{l_1 l_2 l_3}
\equiv \sqrt{\frac{(2 l_1 + 1)(2 l_2 + 1)(2 l_3 + 1)}{4 \pi}}
\left(
  \begin{array}{ccc}
  l_1 & l_2 & l_3 \\
  s_1 & s_2 & s_3
  \end{array}
 \right)$.


\subsection{Wigner-$6j$ symbol}

Considering two other ways in the coupling of three angular momenta as
\begin{eqnarray}
{\bf l_5} &=& {\bf l_1} + {\bf l_2} + {\bf l_4} \\
&=& {\bf l_3} + {\bf l_4} \label{eq:l3l4} \\
&=& {\bf l_1} + {\bf l_6}~, \label{eq:l1l6}
\end{eqnarray}
the Wigner-$6j$ symbol is defined using a Clebsch-Gordan coefficient
between each eigenstate of ${\bf l_5}$ corresponding to Eqs.~(\ref{eq:l3l4}) and (\ref{eq:l1l6}) as 
\begin{eqnarray}
\left\{
  \begin{array}{ccc}
  l_1 & l_2 & l_3 \\
  l_4 & l_5 & l_6
  \end{array}
 \right\}
 \equiv \frac{ (-1)^{l_1 + l_2 + l_4 + l_5} \Braket{(l_1 l_2) l_3 ; l_4 ;
 l_5 m_5 | l_1 ; (l_2 l_4) l_6 ; l_5 m_5} }{\sqrt{(2 l_3 + 1)(2 l_6 + 1)}}~.
\end{eqnarray}
This is expressed with the summation of three Wigner-$3j$ symbols:
\begin{eqnarray}
&& \sum_{m_4 m_5 m_6} (-1)^{\sum_{i=4}^6 l_i - m_i}
\left(
  \begin{array}{ccc}
  l_5 & l_1 & l_6 \\
  m_5 & -m_1 & -m_6 
  \end{array}
 \right) 
\left(
  \begin{array}{ccc}
  l_6 & l_2 & l_4 \\
  m_6 & -m_2 & -m_4 
  \end{array}
 \right)
\left(
  \begin{array}{ccc}
  l_4 & l_3 & l_5 \\
  m_4 & -m_3 & -m_5 
  \end{array}
 \right) \nonumber \\
 &&\qquad\qquad\qquad = \left(
  \begin{array}{ccc}
  l_1 & l_2 & l_3 \\
  m_1 & m_2 & m_3 
  \end{array}
 \right) 
\left\{
  \begin{array}{ccc}
  l_1 & l_2 & l_3 \\
  l_4 & l_5 & l_6 
  \end{array}
 \right\}~; 
 \end{eqnarray}
hence, the triangle conditions are given by 
\begin{eqnarray}
\begin{split}
& |l_1 - l_2| \leq l_3 \leq l_1 + l_2, \ |l_4 - l_5| \leq l_3 \leq l_4 + l_5 ~,   \\
& |l_1 - l_5| \leq l_6 \leq l_1 + l_5, \ |l_4 - l_2| \leq l_6 \leq l_4 + l_2~. \label{eq:wig_6j_triangle}
\end{split}
\end{eqnarray}
The Wigner-$6j$ symbol obeys 24 symmetries such as
\begin{eqnarray}
\left\{
  \begin{array}{ccc}
  l_1 & l_2 & l_3 \\
  l_4 & l_5 & l_6 
  \end{array}
 \right\} 
&=& \left\{
  \begin{array}{ccc}
  l_2 & l_1 & l_3 \\
  l_5 & l_4 & l_6 
  \end{array}
 \right\} 
= \left\{
  \begin{array}{ccc}
  l_2 & l_3 & l_1 \\
  l_5 & l_6 & l_4 
  \end{array}
 \right\} \ ({\rm permutation \ of \ columns}) \nonumber \\
&=& \left\{
  \begin{array}{ccc}
  l_4 & l_5 & l_3 \\
  l_1 & l_2 & l_6 
  \end{array}
 \right\} 
= \left\{
  \begin{array}{ccc}
  l_1 & l_5 & l_6 \\
  l_4 & l_2 & l_3 
  \end{array}
 \right\} \nonumber \\ 
&& ({\rm exchange \ of \ two \ corresponding \ elements \ between \ rows}).
\end{eqnarray}
Geometrically, the Wigner-$6j$ symbol is expressed using the tetrahedron composed of four triangles obeying Eq.~(\ref{eq:wig_6j_triangle}). It is known that the Wigner-$6j$ symbol is suppressed by the square root of the volume of the tetrahedron at high multipoles.   

\subsection{Wigner-$9j$ symbol}

Considering two other ways in the coupling of four angular momenta as
\begin{eqnarray}
{\bf l_9} &=& {\bf l_1} + {\bf l_2} + {\bf l_4} + {\bf l_5} \\
&=& {\bf l_3} + {\bf l_6} \label{eq:l3l6} \\
&=& {\bf l_7} + {\bf l_8}~, \label{eq:l7l8}
\end{eqnarray}
where ${\bf l_3} \equiv {\bf l_1} + {\bf l_2}, {\bf l_6} \equiv {\bf l_4} + {\bf l_5}, {\bf l_7} \equiv {\bf l_1} + {\bf l_4}, {\bf l_8} \equiv {\bf l_2} + {\bf l_5}$,
the Wigner $9j$ symbol expresses a Clebsch-Gordan coefficient between
 each eigenstate of ${\bf l_9}$ corresponding to Eqs.~(\ref{eq:l3l6}) and
(\ref{eq:l7l8}) as 
\begin{eqnarray}
&& \left\{
 \begin{array}{ccc}
  l_1 & l_2 & l_3 \\
  l_4 & l_5 & l_6 \\
  l_7 & l_8 & l_9
 \end{array}
 \right\}  
\equiv \frac{\Braket{(l_1 l_2) l_3 ; (l_4 l_5) l_6 ; l_9 m_9 | (l_1
 l_4) l_7 ; (l_2 l_5) l_8 ; l_9 m_9}}{\sqrt{(2 l_3 + 1)(2 l_6 + 1)(2 l_7 + 1)(2 l_8
+ 1)}}~. 
\end{eqnarray}
This is expressed with the summation of five Wigner-$3j$ symbols:
\begin{eqnarray}
&& \sum_{\substack{m_4 m_5 m_6 \\ m_7 m_8 m_9}} 
\left(
  \begin{array}{ccc}
  l_4 & l_5 & l_6 \\
  m_4 & m_5 & m_6 
  \end{array}
 \right)
\left(
  \begin{array}{ccc}
  l_7 & l_8 & l_9 \\
  m_7 & m_8 & m_9 
  \end{array}
 \right) \nonumber \\
&&\qquad \times 
\left(
  \begin{array}{ccc}
  l_4 & l_7 & l_1 \\
  m_4 & m_7 & m_1 
  \end{array}
 \right)
\left(
  \begin{array}{ccc}
  l_5 & l_8 & l_2 \\
  m_5 & m_8 & m_2
  \end{array}
 \right)
\left(
  \begin{array}{ccc}
  l_6 & l_9 & l_3 \\
  m_6 & m_9 & m_3 
  \end{array}
 \right) \nonumber \\ 
&& \qquad\qquad\qquad = \left(
  \begin{array}{ccc}
  l_1 & l_2 & l_3 \\
  m_1 & m_2 & m_3
  \end{array}
 \right)
\left\{
  \begin{array}{ccc}
  l_1 & l_2 & l_3 \\
  l_4 & l_5 & l_6 \\
  l_7 & l_8 & l_9 
  \end{array}
 \right\}~, \label{eq:sum_9j_3j}
\end{eqnarray}
and that of three Wigner-$6j$ symbols:
\begin{eqnarray}
\left\{
 \begin{array}{ccc}
  l_1 & l_2 & l_3 \\
  l_4 & l_5 & l_6 \\
  l_7 & l_8 & l_9
 \end{array}
 \right\}
&=& \sum_x (-1)^{2x}(2 x + 1) 
\left\{
 \begin{array}{ccc}
  l_1 & l_4 & l_7 \\
  l_8 & l_9 & x 
 \end{array}
 \right\}
\left\{
 \begin{array}{ccc}
  l_2 & l_5 & l_8 \\
  l_4 & x & l_6 
 \end{array}
 \right\}
\left\{
 \begin{array}{ccc}
  l_3 & l_6 & l_9 \\
  x & l_1 & l_2 
 \end{array}
 \right\} ; \label{eq:sum_9j_6j}
\end{eqnarray}
hence, the triangle conditions are given by
\begin{eqnarray}
\begin{split}
& |l_1 - l_2| \leq l_3 \leq l_1 + l_2~, \ |l_4 - l_5| \leq l_6 \leq l_4 +
 l_5~, \ |l_7 - l_8| \leq l_9 \leq l_7 + l_8~,  \\
& |l_1 - l_4| \leq l_7 \leq l_1
 + l_4~, \ |l_2 - l_5| \leq l_8 \leq l_2 +
 l_5~, \ |l_3 - l_6| \leq l_9 \leq l_3 + l_6 ~.
\end{split}
\end{eqnarray}
The Wigner-$9j$ symbol obeys $72$ symmetries: 
\begin{eqnarray}
\left\{
  \begin{array}{ccc}
  l_1 & l_2 & l_3 \\
  l_4 & l_5 & l_6 \\
  l_7 & l_8 & l_9 
  \end{array}
 \right\} 
&=& (-1)^{\sum_{i = 1}^9 l_i}
\left\{
  \begin{array}{ccc}
  l_2 & l_1 & l_3 \\
  l_5 & l_4 & l_6 \\
  l_8 & l_7 & l_9 
  \end{array}
 \right\} 
= (-1)^{\sum_{i = 1}^9 l_i}
\left\{
  \begin{array}{ccc}
  l_1 & l_2 & l_3 \\
  l_7 & l_8 & l_9 \\
  l_4 & l_5 & l_6 
  \end{array}
 \right\} \nonumber \\
&&\qquad\qquad (\rm{odd \ permutation \ of \ rows \ or \ columns}) \nonumber \\
&=& \left\{
  \begin{array}{ccc}
  l_2 & l_3 & l_1 \\
  l_5 & l_6 & l_4 \\
  l_8 & l_9 & l_7 
  \end{array}
 \right\} 
= \left\{
  \begin{array}{ccc}
  l_4 & l_5 & l_6 \\
  l_7 & l_8 & l_9 \\
  l_1 & l_2 & l_3 
  \end{array}
 \right\} \nonumber \\
&&\qquad\qquad (\rm{even \ permutation \ of \ rows \ or \
 columns}) \nonumber \\
&=& \left\{
  \begin{array}{ccc}
  l_1 & l_4 & l_7 \\
  l_2 & l_5 & l_8 \\
  l_3 & l_6 & l_9 
  \end{array}
 \right\}
= \left\{
  \begin{array}{ccc}
  l_9 & l_6 & l_3 \\
  l_8 & l_5 & l_2 \\
  l_7 & l_4 & l_1 
  \end{array}
 \right\} \nonumber \\
 &&\qquad\qquad (\rm{reflection \ of \ the \ symbols})~.
\end{eqnarray}  



\subsection{Analytic expressions of the Wigner symbols}

Here, we show some analytical formulas of the Wigner symbols.

The $I$ symbols, which are defined as 
\begin{eqnarray}
I^{s_1 s_2 s_3}_{l_1 l_2 l_3}
\equiv \sqrt{\frac{(2 l_1 + 1)(2 l_2 + 1)(2 l_3 + 1)}{4 \pi}}
\left(
  \begin{array}{ccc}
  l_1 & l_2 & l_3 \\
  s_1 & s_2 & s_3
  \end{array}
 \right)~,
\end{eqnarray}
 are expressed as 
\begin{eqnarray}
I_{l_1 l_2 l_3}^{0~0~0} 
&=& 
\sqrt{\frac{\prod_{i=1}^3 (2 l_i+1)}{4 \pi}} 
(-1)^{\sum_{i=1}^3 \frac{- l_i}{2}} \nonumber \\
&&\times 
\frac{ \left(\sum_{i=1}^3 \frac{l_i}{2} \right)! \sqrt{(-l_1 + l_2 + l_3)!}
\sqrt{(l_1 - l_2 + l_3)!} \sqrt{(l_1 + l_2 - l_3)!} }
{ \left(\frac{- l_1 + l_2 + l_3}{2}\right)! 
\left(\frac{l_1 - l_2 + l_3}{2}\right)! 
\left(\frac{l_1 + l_2 - l_3}{2}\right)! 
\sqrt{ (\sum_{i=1}^3 l_i + 1)!} } 
\nonumber \\
&&\qquad\qquad\qquad\qquad\qquad\qquad\qquad\qquad\qquad\qquad\quad
   ({\rm for \ } l_1 + l_2 + l_3 = {\rm even} ) \nonumber \\
&=& 0 \ \ ({\rm for \ } l_1 + l_2 + l_3 = {\rm odd} )
~, \\
I_{l_1 ~ l_2 ~ l_3}^{0~ 1~ -1} 
&=& \sqrt{\frac{5}{8\pi}} (-1)^{l_2+1}
 \sqrt{\frac{(l_2-1)(l_2+1)}{l_2 - 1/2}} \ \
({\rm for \ } l_1 = l_2 - 2, l_3 = 2) \nonumber \\
&=& \sqrt{\frac{15}{16\pi}} (-1)^{l_2}
\sqrt{\frac{l_2 + 1/2}{(l_2 - 1/2)(l_2 + 3/2)}} \ \ 
({\rm for \ } l_1 = l_2, l_3 = 2) \nonumber \\
&=& \sqrt{\frac{5}{8\pi}} (-1)^{l_2} \sqrt{\frac{l_2(l_2+2)}{l_2 + 3/2}}
 \ \
({\rm for \ } l_1 = l_2 + 2, l_3 = 2) \nonumber \\
&=& \sqrt{\frac{3}{8\pi}} (-1)^{l_3+1}
\sqrt{l_3 + 1} \ \
({\rm for \ } l_1 = l_3 - 1, l_2 = 1) \nonumber \\
&=& \sqrt{\frac{3}{4\pi}} (-1)^{l_3+1} \sqrt{l_3 + 1/2} \ \
({\rm for \ } l_1 = l_3, l_2 = 1) \nonumber \\
&=& \sqrt{\frac{3}{8\pi}} (-1)^{l_3+1}
\sqrt{l_3} \ \
({\rm for \ } l_1 = l_3 + 1, l_2 = 1)~.
\end{eqnarray}
The Wigner-$9j$ symbols are    
calculated as 
\begin{eqnarray}
\left\{
  \begin{array}{ccc}
  l_1 & l_2 & l_3 \\
  l_4 & l_5 & l_6 \\
  1 & 1 & 2 
  \end{array}
 \right\}
&=& 
\sqrt{\frac{2(l_3 \pm 1) + 1}{5}}
\left\{
  \begin{array}{ccc}
  l_1 & l_4 & 1 \\
  l_3 \pm 2 & l_3 \pm 1 & l_5 
  \end{array}
 \right\}
\left\{
  \begin{array}{ccc}
  l_2 & l_5 & 1 \\
  l_3 \pm 1 & l_3 & l_1 
  \end{array}
 \right\} 
\ \ ({\rm for \ } l_6 = l_3 \pm 2 ) \nonumber \\
&=&
\sqrt{\frac{(2 l_3 - 1)(2 l_3 + 2)(2 l_3 + 3)}{30 (2 l_3)(2 l_3 + 1)}}
\left\{
  \begin{array}{ccc}
  l_1 & l_4 & 1 \\
  l_3 & l_3 - 1 & l_5 
  \end{array}
 \right\}
\left\{
  \begin{array}{ccc}
  l_2 & l_5 & 1 \\
  l_3-1 & l_3 & l_1 
  \end{array}
 \right\} \nonumber \\
&& + 
\sqrt{\frac{2 (2 l_3 - 1)(2 l_3 + 1)(2 l_3 + 3)}{15 (2 l_3)(2 l_3 + 2)}}
\left\{
  \begin{array}{ccc}
  l_1 & l_4 & 1 \\
  l_3 & l_3 & l_5 
  \end{array}
 \right\}
\left\{
  \begin{array}{ccc}
  l_2 & l_5 & 1 \\
  l_3 & l_3 & l_1 
  \end{array}
 \right\} \nonumber \\
&& + 
\sqrt{\frac{(2 l_3 - 1)(2 l_3)(2 l_3 + 3)}{30 (2 l_3 + 1)(2 l_3 + 2)}}
\left\{
  \begin{array}{ccc}
  l_1 & l_4 & 1 \\
  l_3 & l_3 + 1 & l_5 
  \end{array}
 \right\}
\left\{
  \begin{array}{ccc}
  l_2 & l_5 & 1 \\
  l_3 + 1 & l_3 & l_1 
  \end{array}
 \right\} \nonumber \\
&&\qquad\qquad\qquad\qquad\qquad\qquad\qquad\qquad\qquad\qquad\qquad\quad
\ \ ({\rm for \ } l_6 = l_3 ) ~, 
\end{eqnarray} 
where these Wigner-$6j$ symbols are analytically given by
\begin{eqnarray}
\left\{
  \begin{array}{ccc}
  l_1 & l_2 & 1 \\
  l_4 & l_5 & l_6 
  \end{array}
 \right\}
 &=& (-1)^{l_1+l_4+l_6+1} 
\sqrt{ \frac{ {}_{l_1+l_4+l_6+2}P_{2} ~ {}_{l_1+l_4-l_6+1}P_2 }
{ {}_{2 l_4 +3}P_3 ~ {}_{2 l_1 + 1}P_3 } } \ \ ({\rm for \ } l_2 = l_1 - 1, l_5 = l_4 + 1 ) \nonumber \\
 &=& (-1)^{l_1+l_4+l_6+1} 
\sqrt{ \frac{2 (l_1 + l_4 + l_6 +2)(l_1 + l_4 - l_6 + 1) }
{ {}_{2 l_4 +3}P_3 } }\nonumber \\
&&\times \sqrt{ \frac{(-l_1 + l_4 + l_6 + 1) (l_1 - l_4 + l_6)}
{ {}_{2 l_1 + 2}P_3 } } 
 \ \ ({\rm for \ } l_2 = l_1, l_5 = l_4 + 1 ) \nonumber \\
&=& (-1)^{l_1+l_4+l_6+1} 
\sqrt{ \frac{ {}_{-l_1+l_4+l_6+1}P_{2} ~ {}_{l_1-l_4+l_6+1}P_2 }
{ {}_{2 l_4 +3}P_3 ~ {}_{2 l_1 + 3}P_3 } } \ \ ({\rm for \ } l_2 = l_1 +
1, l_5 = l_4 + 1 ) \nonumber \\
&=& (-1)^{l_1+l_4+l_6+1} 
\left[ l_4(l_4+1) + l_1(l_1-1)(l_4+1) - l_6(l_6+1) - l_1(l_1+1)l_4
\right] \nonumber \\
&&\times \sqrt{\frac{2 (l_1 + l_4 + l_6 + 1)(l_1 + l_4 - l_6)}
{ (-l_1+l_4+l_6+1) (l_1 - l_4 +
l_6) {}_{2 l_4 + 2}P_3 ~ {}_{2 l_1 + 1}P_3 } } \ \ ({\rm for \ } l_2 = l_1 -
1, l_5 = l_4 ) \nonumber \\
&=& 2 (-1)^{l_1+l_4+l_6+1} 
\frac{ l_4(l_4+1) + l_1(l_1+1)(l_4+1) - l_6(l_6+1) - l_1(l_1+1)l_4 }
{ \sqrt{{}_{2 l_4 + 2}P_3~{}_{2 l_1 + 2}P_3 } } \nonumber \\
&&\qquad\qquad\qquad\qquad\qquad\qquad\qquad\qquad\qquad\qquad\qquad
 ({\rm for \ } l_2 = l_1, l_5 = l_4 ) \nonumber \\
&=& (-1)^{l_1+l_4+l_6+1} 
\left[ l_4(l_4+1) + (l_1+1)(l_1+2)(l_4+1) - l_6(l_6+1) - l_1(l_1+1)l_4
\right] \nonumber \\
&&\times \sqrt{\frac{2 (-l_1 + l_4 + l_6)(l_1 - l_4 + l_6 + 1)}
{(l_1+l_4+l_6+2) (l_1 + l_4 - l_6 +1) {}_{2 l_4 + 2}P_3 ~ {}_{2 l_1 +
3}P_3 } } \ \ ({\rm for \ } l_2 = l_1 + 1, l_5 = l_4 )~. \nonumber \\
\end{eqnarray}
Using these analytical formulas, one can reduce the time
cost involved with calculating the CMB bispectrum from PMFs.

\newpage

\section{Polarization vector and tensor}\label{appen:polarization}

We summarize the relations and properties of a divergenceless
polarization vector $\epsilon_a^{(\pm 1)}$ and a transverse and
traceless polarization tensor $e_{ab}^{(\pm 2)}$ \cite{Weinberg:2008zzc, Shiraishi:2010kd}~.

The polarization vector with respect to a unit
vector $\hat{\bf n}$ is expressed using two unit vectors
$\hat{\bf \theta}$ and $\hat{\bf \phi}$ perpendicular to $\hat{\bf n}$ as 
\begin{eqnarray}
\epsilon_a^{(\pm 1)}(\hat{\bf n}) = \frac{1}{\sqrt{2}}[\hat{\theta}_a(\hat{\bf n})
 \pm i~\hat{\phi}_a (\hat{\bf n}) ]~. \label{eq:pol_vec_def}
\end{eqnarray}
This satisfies the relations: 
\begin{eqnarray}
\begin{split}
\hat{n}^a \epsilon_a^{(\pm 1)}(\hat{\bf n}) &= 0~, \\
\epsilon^{(\pm 1) *}_a (\hat{\bf n}) &= \epsilon^{(\mp 1)}_a (\hat{\bf n})
 = \epsilon^{(\pm 1)}_a (-\hat{\bf n})~, \\
\epsilon^{(\lambda)}_a (\hat{\bf n}) \epsilon^{(\lambda')}_a (\hat{\bf n}) 
&= \delta_{\lambda, -\lambda'} \ \ \ ({\rm for} \ \lambda, \lambda' = \pm 1)~. 
\label{eq:pol_vec_relation}
\end{split}
\end{eqnarray} 
By defining a rotational matrix, which transforms a unit vector parallel to the $z$ axis, namely $\hat{\bf z}$, to $\hat{\bf n}$, as
\begin{eqnarray}
S(\hat{\bf n}) 
\equiv \left( 
  \begin{array}{ccc}
  \cos\theta_n \cos\phi_n & -\sin\phi_n  & \sin\theta_n \cos\phi_n \\
 \cos\theta_{n} \sin\phi_{n}  &  \cos\phi_{n} & \sin\theta_{n} \sin\phi_{n} \\
 -\sin\theta_n & 0 & \cos\theta_{n}
  \end{array}
 \right)~,
\end{eqnarray}  
we specify $\hat{\bf \theta}$ and $\hat{\bf \phi}$ as
\begin{eqnarray}
\hat{\bf \theta}(\hat{\bf n}) = S(\hat{\bf n}) \hat{\bf x}~, \ \ 
 \hat{\bf \phi}(\hat{\bf n}) = S(\hat{\bf n}) \hat{\bf y}~, \label{eq:theta_phi_def}
\end{eqnarray}
where $\hat{\bf x}$ and $\hat{\bf y}$ are unit vectors parallel
to $x$- and $y$-axes.
By using Eq.~(\ref{eq:pol_vec_def}), the polarization tensor is constructed as
\begin{eqnarray}
e^{(\pm 2)}_{ab} (\hat{\bf n}) = \sqrt{2} \epsilon^{(\pm 1)}_a(\hat{\bf n})
 \epsilon^{(\pm 1)}_b(\hat{\bf n})~. \label{eq:pol_tens_def}
\end{eqnarray}

To utilize the polarization vector and tensor in the calculation of this thesis, we need to expand Eqs.~(\ref{eq:pol_vec_def}) and (\ref{eq:pol_tens_def}) with spin spherical harmonics. 
An arbitrary unit vector is expanded with the spin-$0$ spherical harmonics as 
\begin{eqnarray}
\begin{split}
\hat{r}_a &= \sum_m \alpha_a^{m} Y_{1 m}(\hat{\bf r})~, \\
\alpha^m_a &\equiv \sqrt{\frac{2 \pi}{3}}
 \left(
  \begin{array}{ccc}
   -m (\delta_{m,1} + \delta_{m,-1}) \\
   i~ (\delta_{m,1} + \delta_{m,-1}) \\
   \sqrt{2} \delta_{m,0}
  \end{array}
\right)~. \label{eq:arbitrary_vec}
\end{split}
\end{eqnarray}
Here, note that the repeat of the index implies the summation.
The scalar product of $\alpha_a^m$ is calculated as
\begin{eqnarray}
\alpha_a^m \alpha_a^{m'} = \frac{4 \pi}{3} (-1)^m \delta_{m,-m'}~, \ \
\alpha_a^m \alpha_a^{m' *} = \frac{4 \pi}{3} \delta_{m,m'}~.
\end{eqnarray}
Through the substitution of Eq.~(\ref{eq:theta_phi_def}) into Eq.~(\ref{eq:arbitrary_vec}),
$\hat{\bf \theta}$ is expanded as 
\begin{eqnarray}
\hat{\theta}_a (\hat{\bf n}) &=& \sum_m  \alpha_a^m Y_{1m} (\hat{\bf \theta} (\hat{\bf n}))
= \sum_m  \alpha_a^m \sum_{m'} D_{m m'}^{(1) *} (S(\hat{\bf n})) Y_{1 m'}
 (\hat{\bf x}) \nonumber \\
&=& - \frac{s}{\sqrt{2}}(\delta_{s,1} + \delta_{s,-1}) \sum_m  \alpha_a^m {}_s Y_{1 m}(\hat{\bf n})~.
\end{eqnarray}
Here, we use the properties of the Wigner $D$-matrix as described in Appendix~\ref{appen:wigner_d_matrix} \cite{Shiraishi:2010sm, Goldberg:1966uu, Okamoto:2002ik, Weinberg:2008zzc}
\begin{eqnarray}
\begin{split}
Y_{\ell m}(S(\hat{\bf n})\hat{\bf x}) &= \sum_{m'} 
D^{(\ell) *}_{m m'} (S(\hat{\bf n})) Y_{\ell m'} (\hat{\bf x}) ~, \\
D_{ms}^{(\ell)} ( S(\hat{\bf n}) ) &=
\left[ \frac{4 \pi}{2\ell + 1} \right]^{1/2} (-1)^s
{}_{-s}Y_{\ell m}^*(\hat{\bf n})~.
\end{split}
\end{eqnarray}
In the same manner, $\hat{\bf \phi}$ is also calculated as
\begin{eqnarray}
\hat{\phi}_a(\hat{\bf n}) = \frac{i}{\sqrt{2}}(\delta_{s,1} + \delta_{s,-1})
\sum_m  \alpha_a^m {}_s Y_{1 m}(\hat{\bf n})~;
\end{eqnarray}
hence, the explicit form of Eq.~(\ref{eq:pol_vec_def}) is calculated as  
\begin{eqnarray}
\epsilon_a^{(\pm 1)} (\hat{\bf n}) 
= \mp \sum_m \alpha_a^m {}_{\pm 1} Y_{1 m} (\hat{\bf n})~. \label{eq:pol_vec}
\end{eqnarray}

Substituting this into Eq.~(\ref{eq:pol_tens_def}) and using the
relations of Appendix \ref{appen:wigner} and $I_{2~1~1}^{\mp 2 \pm 1 \pm
1} = \frac{3}{2 \sqrt{\pi}}$, 
the polarization tensor can also be expressed as 
\begin{eqnarray}
e^{(\pm 2)}_{ab} (\hat{\bf n})
&=& \frac{3}{\sqrt{2 \pi}}  
\sum_{M m_a m_b} {}_{\mp 2}Y_{2 M}^*(\hat{\bf n}) 
\alpha^{m_a}_{a} \alpha^{m_b}_b 
\left(
  \begin{array}{ccc}
  2 & 1 &  1\\
  M & m_a & m_b 
  \end{array}
\right)~. \label{eq:pol_tens}
\end{eqnarray}
This obeys the relations:
\begin{eqnarray}
\begin{split}
e_{aa}^{(\pm 2)}(\hat{\bf n}) &= \hat{n}_a e_{ab}^{(\pm 2)}(\hat{\bf n}) = 0~, \\
e_{ab}^{(\pm 2) *}(\hat{\bf n}) &= e_{ab}^{(\mp 2)}(\hat{\bf n}) = e_{ab}^{(\pm
2)}(- \hat{\bf n})~, \\
e_{ab}^{(\lambda)}(\hat{\bf n}) e_{ab}^{(\lambda')}(\hat{\bf n}) &= 2
\delta_{\lambda, -\lambda'} \ \ \ ({\rm for} \ \lambda, \lambda' = \pm 2)~. \label{eq:pol_tens_relation}
\end{split}
\end{eqnarray}

Using the projection operators as 
\begin{eqnarray}
\begin{split}
O_a^{(0)} e^{i {\bf k}\cdot {\bf x}} &\equiv k^{-1} \nabla_a e^{i {\bf k}\cdot {\bf x}} = i \hat{k}_a e^{i {\bf k}\cdot {\bf x}} ~, \\
O^{(0)}_{ab} e^{i {\bf k}\cdot {\bf x}} &\equiv \left( k^{-2} \nabla_a \nabla_b + \frac{\delta_{a,b}}{3} \right) e^{i {\bf k}\cdot {\bf x}}
= \left(- \hat{k}_a \hat{k}_b + \frac{\delta_{a,b}}{3} \right) e^{i {\bf k}\cdot {\bf x}} ~, \\
O_a^{(\pm 1)} e^{i {\bf k}\cdot {\bf x}} &\equiv - i \epsilon^{(\pm 1)}_a(\hat{\bf k}) e^{i {\bf k} \cdot {\bf x}} ~, \\
O^{(\pm 1)}_{ab} e^{i {\bf k}\cdot {\bf x}} 
&\equiv k^{-1} 
\left( \nabla_a O^{(\pm 1)}_b + \nabla_b O^{(\pm 1)}_a \right) e^{i {\bf k}\cdot {\bf x}}
= 
\left( \hat{k}_a \epsilon^{(\pm 1)}_b(\hat{\bf k}) + \hat{k}_b \epsilon^{(\pm 1)}_a(\hat{\bf k}) \right) e^{i {\bf k}\cdot {\bf x}} ~, \\
O_{ab}^{(\pm 2)} e^{i {\bf k}\cdot {\bf x}} &\equiv e^{(\pm 2)}_{ab}(\hat{\bf k}) e^{i {\bf k}\cdot {\bf x}}~,
\end{split}
\end{eqnarray}
the arbitrary scalar, vector and tensor are decomposed into the helicity states as 
\begin{eqnarray}
\eta({\bf k}) &=& \eta^{(0)}({\bf k}), \label{eq:scal_decompose} \\
\omega_{a}({\bf k}) &=& \omega^{(0)}({\bf k}) O^{(0)}_a + \sum_{\lambda = \pm 1} \omega^{(\lambda)}({\bf k}) O^{(\lambda)}_a~, \\
\chi_{ab}({\bf k}) &=& - \frac{1}{3} \chi_{\rm iso}({\bf k}) \delta_{a,b} + \chi^{(0)}({\bf k}) O^{(0)}_{ab} + \sum_{\lambda = \pm 1} \chi^{(\lambda)}({\bf k}) O^{(\lambda)}_{ab} + \sum_{\lambda = \pm 2} \chi^{(\lambda)}({\bf k}) O^{(\lambda)}_{ab}~.
\end{eqnarray}
Then, using Eq.~(\ref{eq:pol_vec_relation}) and (\ref{eq:pol_tens_relation}), we can find the inverse formulae as
\begin{eqnarray}
\omega^{(0)}({\bf k}) &=& -O_a^{(0)} \omega_a({\bf k})~, \\ 
\omega^{(\pm 1)}({\bf k}) &=& -O_a^{(\mp 1)}(\hat{\bf k}) \omega_{a}({\bf k})~, \\
\chi^{(0)}({\bf k}) &=& \frac{3}{2} O_{ab}^{(0)}(\hat{\bf k}) \chi_{ab}({\bf k})~, \\
\chi^{(\pm 1)}({\bf k}) &=& \frac{1}{2} O_{ab}^{(\mp 1)}(\hat{\bf k}) \chi_{ab}({\bf k})~, \\
\chi^{(\pm 2)}({\bf k}) &=& \frac{1}{2} O_{ab}^{(\mp 2)}(\hat{\bf k}) \chi_{ab}({\bf k})~. \label{eq:tens_decompose} 
\end{eqnarray}

From these, we can derive the relations of several projection operators as
\begin{eqnarray}
\begin{split}
O_{ab}^{(0)}(\hat{\bf k}) &= - \hat{k}_a \hat{k}_b + \frac{1}{3} \delta_{ab} \\
&= - \sqrt{\frac{3}{2\pi}} \sum_{M m_a m_b} 
Y_{2 M}^*(\hat{\bf k}) \alpha_a^{m_a} \alpha_b^{m_b} 
\left(
  \begin{array}{ccc}
  2 & 1 & 1 \\
  M & m_a & m_b 
  \end{array}
 \right) ~, \\
O_{ab}^{(\pm 1)}(\hat{\bf k}) 
& =  
\hat{k}_a {\epsilon}_b^{(\pm 1)}(\hat{\bf k}) 
+ \hat{k}_b {\epsilon}_a^{(\pm 1)}(\hat{\bf k}) \\
&= \pm \frac{3}{\sqrt{2 \pi}} \sum_{M m_a m_b} 
{}_{\mp 1} Y^*_{2 M}(\hat{\bf k}) \alpha_a^{m_a} \alpha_b^{m_b} 
\left(
  \begin{array}{ccc}
  2 & 1 & 1 \\
  M & m_a & m_b
  \end{array}
 \right)
~, \\
O_{ab}^{(\pm 2)}(\hat{\bf k}) 
&= e_{ab}^{(\pm 2)}(\hat{\bf k}) \\
&= \frac{3}{\sqrt{2 \pi}}
\sum_{M m_a m_b} {}_{\mp 2}Y_{2 M}^*(\hat{\bf k})
\alpha_a^{m_a} \alpha_b^{m_b}
 \left(
  \begin{array}{ccc}
  2 & 1 & 1 \\
  M & m_a & m_b
  \end{array}
 \right) ~, \\
P_{ab}(\hat{\bf k}) &\equiv \delta_{ab} - \hat{k}_a \hat{k}_b \\
&= -2 \sum_{L=0,2} I_{L 1 1}^{0 1 -1} \sum_{M m_a m_b} 
Y^*_{L M}(\hat{\bf k}) \alpha_a^{m_a} \alpha_b^{m_b} 
\left(
  \begin{array}{ccc}
  L & 1 & 1 \\
  M & m_a & m_b
  \end{array}
 \right) ~, \\
O_{ab}^{(0)}(\hat{\bf k}) P_{bc}(\hat{\bf k}) 
&= \frac{1}{3} P_{ac}(\hat{\bf k}) 
~, \\
O_{ab}^{(\pm 1)}(\hat{\bf k}) P_{bc}(\hat{\bf k}) 
&= \hat{k}_a \epsilon_{c}^{(\pm 1)}(\hat{\bf k}) 
~, \\
O_{ab}^{(\pm 2)}(\hat{\bf k}) P_{bc}(\hat{\bf k}) 
&= e_{ac}^{(\pm 2)}(\hat{\bf k}) ~, \\ 
\hat{k}_c &= i \eta^{abc} \epsilon_a^{(+1)}(\hat{\bf k}) 
\epsilon_b^{(-1)}(\hat{\bf k}) ~, \\ 
\eta^{abc} \hat{k}_a \epsilon_b^{(\pm 1)}(\hat{\bf k}) 
&= \mp i \epsilon_c^{(\pm 1)}(\hat{\bf k})
~. \label{eq:projection_operator}
\end{split}
\end{eqnarray}

\newpage

\section{Calculation of $f_{W^3}^{(a)}$ and $f_{\widetilde{W}W^2}^{(a)}$} \label{appen:pol_tens}

Here, we calculate each product between the wave number vectors and the
polarization tensors of $f_{W^3}^{(a)}$ and $f_{\widetilde{W}W^2}^{(a)}$
mentioned in Sec.~\ref{subsubsec:formulation} \cite{Shiraishi:2011st}. 

Using the relations discussed in Appendix \ref{appen:polarization}, the all terms of $f_{W^3}^{(a)}$ are written as
\begin{eqnarray}
\begin{split}
e_{ij}^{(-\lambda_1)} e_{jk}^{(-\lambda_2)} e_{ki}^{(-\lambda_3)} 
&= - ( 8\pi )^{3/2} 
\sum_{M, M', M''} {}_{\lambda_1}Y_{2 M}^*(\hat{{\bf k_1}}) 
{}_{\lambda_2}Y_{2 M'}^*(\hat{{\bf k_2}}) {}_{\lambda_3}Y_{2 M''}^*(\hat{{\bf k_3}}) \\
&\quad \times 
\frac{1}{10} \sqrt{\frac{7}{3}}
\left(
  \begin{array}{ccc}
   2 & 2 & 2 \\
  M & M' & M''
  \end{array}
 \right)~, \\
e_{ij}^{(-\lambda_1)} e_{kl}^{(-\lambda_2)} e_{kl}^{(-\lambda_3)} \hat{k_2}_i \hat{k_3}_j 
&= - (8\pi)^{3/2} 
\sum_{L', L'' = 2, 3} \sum_{M, M', M''} {}_{\lambda_1}Y_{2 M}^*(\hat{{\bf k_1}}) 
{}_{\lambda_2}Y_{L' M'}^*(\hat{{\bf k_2}}) {}_{\lambda_3}Y_{L'' M''}^*(\hat{{\bf k_3}})  \\
&\quad \times 
\frac{4\pi}{15}(-1)^{L'} I_{L' 1 2}^{\lambda_2 0 -\lambda_2} I_{L'' 1 2}^{\lambda_3 0 -\lambda_3}
\left(
  \begin{array}{ccc}
   2 & L' & L'' \\
  M & M' & M''
  \end{array}
 \right)
\left\{
  \begin{array}{ccc}
   2 & L' & L'' \\
  2 & 1 & 1
  \end{array}
 \right\}
~, \\
e_{ij}^{(-\lambda_1)} e_{ki}^{(-\lambda_2)} e_{jl}^{(-\lambda_3)} \hat{k_2}_l \hat{k_3}_k 
&= -( 8\pi )^{3/2} 
\sum_{L', L'' = 2, 3} \sum_{M, M', M''} 
{}_{\lambda_1}Y_{2 M}^*(\hat{{\bf k_1}}) 
{}_{\lambda_2}Y_{L' M'}^*(\hat{{\bf k_2}}) {}_{\lambda_3}Y_{L'' M''}^*(\hat{{\bf k_3}})  \\
&\quad \times 
\frac{4\pi}{3} (-1)^{L'} I_{L' 1 2}^{\lambda_2 0 -\lambda_2} I_{L'' 1 2}^{\lambda_3 0 -\lambda_3}
\left(
  \begin{array}{ccc}
   2 & L' & L'' \\
  M & M' & M''
  \end{array}
 \right)
\left\{
  \begin{array}{ccc}
   2 & L' & L'' \\
   1 & 1 & 2 \\
   1 & 2 & 1
  \end{array}
 \right\}~, \\
e_{ij}^{(-\lambda_1)} e_{ik}^{(-\lambda_2)} e_{kl}^{(-\lambda_3)} \hat{k_2}_l \hat{k_3}_j 
&= - (8\pi)^{3/2} 
\sum_{L', L'' = 2, 3} \sum_{M, M', M''} {}_{\lambda_1}Y_{2 M}^*(\hat{{\bf k_1}}) 
{}_{\lambda_2}Y_{L' M'}^*(\hat{{\bf k_2}}) {}_{\lambda_3}Y_{L'' M''}^*(\hat{{\bf k_3}})  \\
&\quad \times 
\frac{4\pi}{3} (-1)^{L'} I_{L' 1 2}^{\lambda_2 0 -\lambda_2} I_{L'' 1 2}^{\lambda_3 0 -\lambda_3}
\left(
  \begin{array}{ccc}
   2 & L' & L'' \\
  M & M' & M''
  \end{array}
 \right)  \\
&\quad \times
\left\{
  \begin{array}{ccc}
   2 & 1 & L' \\
   2 & 1 & 1 
  \end{array}
 \right\}
\left\{
  \begin{array}{ccc}
   2 & L' & L'' \\
   2 & 1 & 1 
  \end{array}
 \right\}
~.
\end{split}
\end{eqnarray}

In the calculation of $f_{\widetilde{W}W^2}^{(a)}$, we also need to
consider the dependence of the tensor contractions on $\eta^{ijk}$. Making use of the relation:
\begin{eqnarray}
\eta^{abc} \alpha^{m_a}_a \alpha^{m_b}_b \alpha^{m_c}_c 
= -i \left(\frac{4\pi}{3}\right)^{3/2} \sqrt{6}
\left(
  \begin{array}{ccc}
   1 & 1 & 1 \\
  m_a & m_b & m_c
  \end{array}
 \right) ~,
\end{eqnarray}
the first two terms of $f_{\widetilde{W}W^2}^{(a)}$ reduce to 
\begin{eqnarray}
\begin{split}
i \eta^{ijk} e_{kq}^{(-\lambda_1)} e_{jm}^{(-\lambda_2)} e_{iq}^{(-\lambda_3)} \hat{k_3}_m 
&= - ( 8\pi)^{3/2} 
\sum_{L'' = 2, 3} \sum_{M, M', M''} {}_{\lambda_1}Y_{2 M}^*(\hat{{\bf k_1}}) 
{}_{\lambda_2}Y_{2 M'}^*(\hat{{\bf k_2}}) {}_{\lambda_3}Y_{L'' M''}^*(\hat{{\bf k_3}}) \\
&\quad \times \sqrt{\frac{2\pi}{5}} (-1)^{L''}  
I_{L'' 1 2}^{\lambda_3 0 -\lambda_3} 
\left(
  \begin{array}{ccc}
   2 & 2 & L'' \\
  M & M' & M''
  \end{array}
 \right)
\left\{
  \begin{array}{ccc}
   2 & 2 & L'' \\
  1 & 2 & 1
  \end{array}
 \right\} ~, \\
 i \eta^{ijk} e_{kq}^{(-\lambda_1)} e_{mi}^{(-\lambda_2)}
e_{mq}^{(-\lambda_3)} \hat{k_3}_j 
&= -( 8\pi )^{3/2} 
\sum_{L'' = 2, 3} \sum_{M, M', M''} {}_{\lambda_1}Y_{2 M}^*(\hat{{\bf k_1}}) 
{}_{\lambda_2}Y_{2 M'}^*(\hat{{\bf k_2}}) {}_{\lambda_3}Y_{L'' M''}^*(\hat{{\bf k_3}}) \\
&\quad \times 2\sqrt{2\pi} (-1)^{L''} I_{L'' 1 2}^{\lambda_3 0 -\lambda_3} 
\left(
  \begin{array}{ccc}
   2 & 2 & L'' \\
  M & M' & M''
  \end{array}
 \right)
\left\{
  \begin{array}{ccc}
   2 & 2 & L'' \\
  1 & 1 & 1 \\
  1 & 1 & 2
  \end{array}
 \right\}~.
\end{split}
\end{eqnarray}
For the other terms, by using the relation 
\begin{eqnarray}
\eta^{abc} \hat{k}_a e_{bd}^{(\lambda)}(\hat{{\bf k}}) 
&=& -\frac{\lambda}{2} i e_{cd}^{(\lambda)}(\hat{{\bf k}}) ~, \label{eq:eigen}
\end{eqnarray}
 we have 
\begin{eqnarray}
\begin{split}
i \eta^{ijk} e_{pj}^{(-\lambda_1)} e_{pm}^{(-\lambda_2)}  
\hat{k_1}_k \hat{k_2}_l e_{il}^{(-\lambda_3)} \hat{k_3}_m  
&= - \frac{\lambda_1}{2} (8\pi)^{3/2} \sum_{L', L'' = 2, 3} \sum_{M, M', M''} \\
&\quad \times {}_{\lambda_1}Y_{2 M}^*(\hat{{\bf k_1}}) {}_{\lambda_2}Y_{L' M'}^*(\hat{{\bf k_2}}) {}_{\lambda_3}Y_{L'' M''}^*(\hat{{\bf k_3}}) \\
&\quad \times \frac{4\pi}{3}(-1)^{L''} I_{L' 1 2}^{\lambda_2 0 -\lambda_2}
I_{L'' 1 2}^{\lambda_3 0 -\lambda_3} \\
&\quad \times 
\left(
  \begin{array}{ccc}
   2 & L' & L'' \\
  M & M' & M''
  \end{array}
 \right)
\left\{
  \begin{array}{ccc}
   2 & L' & L'' \\
  1 & 2 & 1 \\
  1 & 1 & 2
  \end{array}
 \right\} ~,  \\
 i \eta^{ijk} e_{pj}^{(-\lambda_1)} e_{pm}^{(-\lambda_2)}  
\hat{k_1}_k \hat{k_2}_l e_{im}^{(-\lambda_3)}  \hat{k_3}_l 
&=  - \frac{\lambda_1}{2} (8\pi)^{3/2}
 \sum_{L', L'' = 2, 3} \sum_{M, M', M''} \\
&\quad \times {}_{\lambda_1}Y_{2 M}^*(\hat{{\bf k_1}}) 
{}_{\lambda_2}Y_{L' M'}^*(\hat{{\bf k_2}}) {}_{\lambda_3}Y_{L'' M''}^*(\hat{{\bf k_3}}) \\
&\quad \times \frac{2\pi}{15} \sqrt{\frac{7}{3}} (-1)^{L''} 
I_{L' 1 2}^{\lambda_2 0 -\lambda_2}
I_{L'' 1 2}^{\lambda_3 0 -\lambda_3} \\
&\quad \times 
\left(
  \begin{array}{ccc}
   2 & L' & L'' \\
  M & M' & M''
  \end{array}
 \right)
\left\{
  \begin{array}{ccc}
   2 & L' & L'' \\
  1 & 2 & 2 
  \end{array}
 \right\}~. 
\end{split}
\end{eqnarray}

\newpage

\section{Graviton non-Gaussianity from the Weyl cubic terms} \label{appen:weyl_in-in}

Here, let us derive the bispectra of gravitons coming from the parity-even and parity-odd Weyl cubic terms, namely, Eqs.~(\ref{eq:ggg}) and (\ref{eq:ggg_f}) \cite{Shiraishi:2011st}. 
For convenience, we decompose the interaction Hamiltonians of $W^3$ and $\widetilde{W}W^2$ (\ref{eq:H_int}) into 
\begin{eqnarray}
H_{int} = \sum_{n=1}^4 H_{int}^{(n)}.
\end{eqnarray}
Depending on this, we also split the graviton non-Gaussianity as
\begin{eqnarray}
\Braket{\prod_{n=1}^3 \gamma^{(\lambda_n)}({\bf k_n})}_{int} 
= \sum_{m=1}^4 \Braket{\prod_{n=1}^3 \gamma^{(\lambda_n)}({\bf k_n})}_{int}^{(m)} ~. 
\end{eqnarray}
In what follows, we shall show the computation of each fraction. 

\subsection{$W^3$}

The bracket part of Eq.~ (\ref{eq:in-in_formalism}) in terms of $H_{W^3}^{(1)}$
is expanded as
\begin{eqnarray}
&& \Braket{0| \left[: H^{(1)}_{W^3}(\tau'):, \prod_{n=1}^3 
\gamma^{(\lambda_n)}({\bf k_n}, \tau) \right] |0} \nonumber \\
&&\qquad= 
\Braket{0| : H^{(1)}_{W^3}(\tau'): \prod_{n=1}^3 
\gamma^{(\lambda_n)}({\bf k_n}, \tau) |0} - 
\Braket{0| 
\left[ \prod_{n=1}^3 \gamma^{(\lambda_n)}({\bf k_n}, \tau) \right] 
: H^{(1)}_{W^3}(\tau'): |0}
 \nonumber \\
&&\qquad= - \int d^3x' \Lambda^{-2} (H \tau')^2 \left(\frac{\tau'}{\tau_*}\right)^A  \frac{1}{4}
\left[ \prod_{n=1}^3  \int \frac{d^3 {\bf k_n'}}{(2 \pi)^3} 
e^{i {\bf k'_n} \cdot {\bf x'} } 
 \sum_{\lambda'_n = \pm 2} \right]
e_{ij}^{(\lambda'_1)}(\hat{\bf k'_1}) e_{jk}^{(\lambda'_2)}(\hat{\bf k'_2}) e_{ki}^{(\lambda'_3)}(\hat{\bf k'_3})
\nonumber \\
&&\qquad\quad\times 
\left[ : \Braket{0|  
\left\{ \prod_{n=1}^3 (\ddot{\gamma}_{dS} - k'^{2}_n \gamma_{dS} )
 (k_n', \tau') a^{(\lambda'_n)}_{k'_n} \right\} 
\left\{ \prod_{m=1}^3 
\gamma^*_{dS}(k_m, \tau)  
a^{(\lambda_m) \dagger}_{- k_m} \right\}  |0} : \right. \nonumber \\
&&\qquad\quad\qquad \left. - : \Braket{0| 
\left\{ \prod_{m=1}^3 \gamma_{dS} (k_m, \tau) 
a^{(\lambda_m)}_{k_m} \right\}
\left\{ \prod_{n=1}^3   
(\ddot{\gamma}^*_{dS} - k'^{2}_n \gamma^*_{dS} ) (k_n', \tau')  
 a^{(\lambda'_n) \dagger}_{-k'_n} \right\} 
|0} :  \right] \nonumber \\
&&\qquad= - \Lambda^{-2} (H \tau')^2 \left(\frac{\tau'}{\tau_*}\right)^A  \frac{1}{4}
(2\pi)^3 \delta\left(\sum_{n=1}^3 {\bf k_n} \right)
e_{ij}^{(\lambda_1)}(- \hat{\bf k_1}) e_{jk}^{(\lambda_2)}(- \hat{\bf k_2}) e_{ki}^{(\lambda_3)}(- \hat{\bf k_3})
\nonumber \\
&&\qquad\quad \times 
6  \left[  \left\{ \prod_{n=1}^3
(\ddot{\gamma}_{dS} - k^{2}_n \gamma_{dS} ) (k_n, \tau') \gamma^*_{dS}(k_n, \tau) \right\}
- \left\{ \prod_{n=1}^3 \gamma_{dS} (k_n, \tau)
(\ddot{\gamma}^*_{dS} - k^{2}_n \gamma^*_{dS} ) (k_n, \tau') \right\} \right] \nonumber \\
&&\qquad= - \frac{3}{2} \Lambda^{-2} (H \tau')^2 
\left(\frac{\tau'}{\tau_*}\right)^A
(2\pi)^3 \delta\left(\sum_{n=1}^3 {\bf k_n} \right)
e_{ij}^{(- \lambda_1)}(\hat{\bf k_1}) e_{jk}^{(- \lambda_2)}(\hat{\bf k_2}) e_{ki}^{(- \lambda_3)}(\hat{\bf k_3}) \nonumber \\
&&\qquad\quad \times  
2i {\rm Im} 
\left[ \prod_{n=1}^3  (\ddot{\gamma}_{dS} - k^{2}_n \gamma_{dS} ) (k_n, \tau') \gamma^*_{dS}(k_n, \tau) \right]~.  
\end{eqnarray}
Here, we use 
\begin{eqnarray}
\begin{split}
\Braket{0|: \prod_{n=1}^3 a^{(\lambda'_n)}_{k'_n} a^{(\lambda_n) \dagger}_{-k_n} :|0}
&= (2\pi)^9 \delta({\bf k_1} + {\bf k'_3}) \delta_{\lambda_1, \lambda_3'}
\delta({\bf k'_1} + {\bf k_3}) \delta_{\lambda'_1, \lambda_3}
\delta({\bf k_2} + {\bf k'_2}) \delta_{\lambda_2, \lambda_2'} + 5 \ {\rm perms.} \\
&= \Braket{0|: \prod_{n=1}^3 a^{(\lambda_n)}_{k_n} a^{(\lambda'_n) \dagger}_{-k'_n} :|0} \\
e_{ij}^{(-\lambda) }(\hat{\bf k}) &= e_{ij}^{(\lambda)}(- \hat{\bf k}) ~.
\end{split}
\end{eqnarray}
Furthermore, since 
\begin{eqnarray}
\begin{split}
\ddot{\gamma}_{dS} - k^{2} \gamma_{dS} &= \frac{2 H \tau'}{M_{\rm pl}} k^{3/2} e^{-ik\tau'} ~, \\ 
\prod_{n=1}^3 \gamma^*_{dS}(k_n,\tau) &\xrightarrow{\tau \rightarrow 0}
i \frac{H^3}{M_{\rm pl}^3} 
(k_1 k_2 k_3 ) ^{-3/2} ~, 
\end{split}
\end{eqnarray}
the time integral at $\tau \to 0$ is performed as 
\begin{eqnarray}
&& {\rm Im} \left[ \int_{-\infty}^\tau d\tau' (H \tau')^2 
\left(\frac{\tau'}{\tau_*}\right)^A 
\prod_{n=1}^3 (\ddot{\gamma}_{dS} - k^{2}_n \gamma_{dS} )(k_n, \tau') 
\gamma^*_{dS}(k_n, \tau) 
\right] \nonumber \\ 
&&\qquad\qquad= \frac{8 H^5}{M_{\rm pl}^3} \sqrt{k_1^3 k_2^3 k_3^3} 
{\rm Im} 
\left[ \left( \prod_{n=1}^3 \gamma_{dS}^*(k_n, \tau) \right) \tau_*^{-A}\int_{-\infty}^\tau d\tau' \tau'^{5+A}
e^{- i k_t \tau'}  \right] \nonumber \\
&&\qquad\qquad \xrightarrow{\tau \rightarrow 0}
\frac{8 H^8}{M_{\rm pl}^6}  
{\rm Re} \left[ \tau_*^{-A}\int_{-\infty}^0 d\tau' \tau'^{5+A}
e^{- i k_t \tau'} \right]~,
\end{eqnarray}
where $k_t \equiv \sum_{n=1}^3 k_n$. Thus, the graviton non-Gaussianity in the late time limit arising from $H_{W^3}^{(1)}$ is 
\begin{eqnarray}
\Braket{\prod_{n=1}^3 \gamma^{(\lambda_n)}({\bf k_n})}^{(1)}_{W^3}
&=&  (2\pi)^3 \delta\left(\sum_{n=1}^3 {\bf k_n} \right) 
8 \left(\frac{H}{M_{\rm pl}}\right)^6 \left( \frac{H}{\Lambda} \right)^2 
{\rm Re} \left[ \tau_*^{-A}\int_{-\infty}^0 d\tau' \tau'^{5+A}
e^{- i k_t \tau'} \right] \nonumber \\
&&\times 3 e_{ij}^{(-\lambda_1)}(\hat{\bf k_1}) e_{jk}^{(-\lambda_2)}(\hat{\bf k_2}) e_{ki}^{(-\lambda_3)}(\hat{\bf k_3}) ~.
\end{eqnarray}
The bracket part in terms of $H_{W^3}^{(2)}$ is given by
\begin{eqnarray}
&& \Braket{0| \left[: H^{(2)}_{W^3}(\tau'):, \prod_{n=1}^3 
\gamma^{(\lambda_n)}({\bf k_n}, \tau) \right] |0} \nonumber \\
&&\qquad= 
\Braket{0| : H^{(2)}_{W^3}(\tau'): \prod_{n=1}^3 
\gamma^{(\lambda_n)}({\bf k_n}, \tau) |0} - 
\Braket{0| 
\left[ \prod_{n=1}^3 \gamma^{(\lambda_n)}({\bf k_n}, \tau) \right] 
: H^{(2)}_{W^3}(\tau'): |0}
 \nonumber \\
&&\qquad= \frac{3}{2} \Lambda^{-2} (H \tau')^2 
\left(\frac{\tau'}{\tau_*}\right)^A k_2 k_3
(2\pi)^3 \delta\left(\sum_{n=1}^3 {\bf k_n} \right)
 \hat{k_2}_i \hat{k_3}_j 
e_{ij}^{(- \lambda_1)}(\hat{\bf k_1}) e_{kl}^{(- \lambda_2)}(\hat{\bf
k_2}) e_{kl}^{(- \lambda_3)}(\hat{\bf k_3}) \nonumber \\
&&\qquad\quad \times  
2i {\rm Im} 
\left[  (\ddot{\gamma}_{dS} - k^{2}_1 \gamma_{dS} ) (k_1, \tau')
 \dot{\gamma}_{dS} (k_2, \tau') \dot{\gamma}_{dS} (k_3, \tau') 
\prod_{n=1}^3\gamma^*_{dS}(k_n, \tau) \right]  + 5 \ {\rm perms.}~.  
\end{eqnarray}
Using 
\begin{eqnarray}
\dot{\gamma}_{dS} = i \frac{H \tau}{M_{\rm pl}} \sqrt{k} e^{-i k \tau'} ~,
\end{eqnarray}
we can reduce the time integral to  
\begin{eqnarray}
&& {\rm Im} 
\left[ \int_{-\infty}^\tau d\tau' (H \tau')^2 
\left( \frac{\tau'}{\tau_*} \right)^A k_2 k_3  
(\ddot{\gamma}_{dS} - k^{2}_1 \gamma_{dS} )(k_1, \tau')  
\dot{\gamma}_{dS}(k_2, \tau')
\dot{\gamma}_{dS}(k_3, \tau') \prod_{n=1}^3 \gamma^*_{dS}(k_n,\tau)
\right] 
\nonumber \\
&&\qquad\qquad 
\xrightarrow{\tau \rightarrow 0} 
 - \frac{2 H^8}{M_{\rm pl}^6} {\rm Re} 
\left[ \tau_*^{-A}\int_{-\infty}^0 d\tau' \tau'^{5+A}
e^{- i k_t \tau'} \right] ~, 
\end{eqnarray}
and obtain 
\begin{eqnarray}
\Braket{\prod_{n=1}^3 \gamma^{(\lambda_n)}({\bf k_n})}^{(2)}_{W^3}
&=& (2\pi)^3 \delta \left(\sum_{n=1}^3 {\bf k_n}\right) 
8\left( \frac{H}{M_{\rm pl}} \right)^6 \left( \frac{H}{\Lambda} \right)^2 
{\rm Re} \left[ \tau_*^{-A}\int_{-\infty}^0 d\tau' \tau'^{5+A}
e^{- i k_t \tau'} \right] 
\nonumber \\
&&\times \frac{3}{4} \hat{k_2}_i e_{ij}^{(-\lambda_1)}(\hat{\bf k_1}) \hat{k_3}_j
e_{kl}^{(-\lambda_2)}(\hat{\bf k_2}) e_{kl}^{(-\lambda_3)}(\hat{\bf k_3}) + 5 \ {\rm perms.} ~.
\end{eqnarray}
The graviton non-Gaussianities from $H_{W^3}^{(3)}$ and $H_{W^3}^{(4)}$ are derived in the same manner as that from $H_{W^3}^{(2)}$:
\begin{eqnarray}
\sum_{m = 3}^4 \Braket{\prod_{n=1}^3 \gamma^{(\lambda_n)}({\bf k_n})}^{(m)}_{W^3}
&=&  (2\pi)^3 \delta \left(\sum_{n=1}^3 {\bf k_n}\right) 
8\left( \frac{H}{M_{\rm pl}} \right)^6 \left( \frac{H}{\Lambda} \right)^2 
{\rm Re} \left[ \tau_*^{-A}\int_{-\infty}^0 d\tau' \tau'^{5+A}
e^{- i k_t \tau'} \right] 
\nonumber \\
&&\times \left[ \frac{3}{4}
\hat{k_3}_k e_{ki}^{(-\lambda_2)}(\hat{\bf k_2}) e_{ij}^{(-\lambda_1)}(\hat{\bf k_1})
e_{jl}^{(-\lambda_3)}(\hat{\bf k_3}) \hat{k_2}_l 
\right. \nonumber \\
&&\quad \left. 
- \frac{3}{2}
\hat{k_3}_j e_{ji}^{(-\lambda_1)}(\hat{\bf k_1}) e_{ik}^{(-\lambda_2)}(\hat{\bf k_2})
e_{kl}^{(-\lambda_3)}(\hat{\bf k_3}) \hat{k_2}_l + 5 \ {\rm perms.}\right] ~.
\end{eqnarray}

\subsection{$\widetilde{W}W^2$}

At first, we shall focus on the contribution of $H_{\widetilde{W}W^2}^{(1)}$. The bracket part is computed as 
\begin{eqnarray}
&& \Braket{0| 
\left[: H^{(1)}_{\widetilde{W}W^2}(\tau'):, \prod_{n=1}^3 \gamma^{(\lambda_n)}({\bf k_n}, \tau) \right] |0} \nonumber \\
&&\qquad= \Braket{0| : H^{(1)}_{\widetilde{W}W^2}(\tau'): \prod_{n=1}^3 
\gamma^{(\lambda_n)}({\bf k_n}, \tau) |0} 
- \Braket{0| 
\left[ \prod_{n=1}^3 \gamma^{(\lambda_n)}({\bf k_n}, \tau) \right]
: H^{(1)}_{\widetilde{W}W^2}(\tau'): |0}  \nonumber \\ 
&&\qquad= - \int d^3x' \Lambda^{-2} (H \tau')^2 
\left(\frac{\tau'}{\tau_*}\right)^A (-3)
\left[ \prod_{n=1}^3 
\int \frac{d^3 {\bf k_n'}}{(2 \pi)^3} 
e^{i {\bf k'_n} \cdot {\bf x'} } \sum_{\lambda'_n = \pm 2} \right] \nonumber \\
&&\qquad\quad\times \eta^{ijk} e_{kq}^{(\lambda'_1)}(\hat{\bf k'_1})
e_{jm}^{(\lambda'_2)}(\hat{\bf k'_2}) e_{iq}^{(\lambda'_3)}(\hat{\bf
k'_3})(i {k'_3}_m)
\nonumber \\
&&\qquad\quad\times
\left[   
\left( \ddot{\gamma}_{dS} - k'^2_1 {\gamma}_{dS}\right) (k'_1, \tau') 
\left( \ddot{\gamma}_{dS} - k'^2_2 {\gamma}_{dS}  \right)
(k'_2, \tau') 
\dot{\gamma}_{dS}(k'_3, \tau') \right. \nonumber \\
&&\qquad\quad\qquad \left. 
\times : \Braket{0| 
\left\{ \prod_{m=1}^3 a^{(\lambda'_m)}_{k'_m} \right\}
\left\{ \prod_{n=1}^3 \gamma^*_{dS}(k_n, \tau) a^{(\lambda_n) \dagger}_{- k_n} \right\}
  |0} : \right. \nonumber \\
&&\qquad\qquad \left. -  
\left( \ddot{\gamma}^*_{dS} - k'^2_1 {\gamma}^*_{dS}\right) (k'_1, \tau') 
\left( \ddot{\gamma}^*_{dS} - k'^2_2 {\gamma}^*_{dS}  \right)
(k'_2, \tau') 
\dot{\gamma}^*_{dS}(k'_3, \tau') \right. \nonumber \\
&&\qquad\qquad\qquad \left. \times
: \Braket{0| 
\left\{ \prod_{n=1}^3 \gamma_{dS}(k_n, \tau)
a^{(\lambda_n)}_{k_n} \right\} 
\left\{ \prod_{m=1}^3 a^{(\lambda'_m) \dagger}_{- k'_m} \right\} |0} :
\right] \nonumber \\
&&\qquad= \Lambda^{-2} (H \tau')^2 
\left(\frac{\tau'}{\tau_*}\right)^A
3 (2\pi)^3 \delta\left( \sum_{n=1}^3 {\bf k_n} \right)
\eta^{ijk} e_{kq}^{(\lambda_1)}(- \hat{\bf k_1})
e_{jm}^{(\lambda_2)}(- \hat{\bf k_2}) 
e_{iq}^{(\lambda_3)}(- \hat{\bf k_3})(- i {k_3}_m) 
\nonumber \\
&&\qquad\quad\times
\left[   
\left( \ddot{\gamma}_{dS} - k^2_1 {\gamma}_{dS}\right) (k_1, \tau') 
\left( \ddot{\gamma}_{dS} - k^2_2 {\gamma}_{dS}  \right) (k_2, \tau') 
\dot{\gamma}_{dS}(k_3, \tau') 
\left\{ \prod_{n=1}^3 \gamma^*_{dS}(k_n, \tau) \right\} \right. \nonumber \\
&&\qquad\quad\qquad \left. -  
\left( \ddot{\gamma}^*_{dS} - k^2_1 {\gamma}^*_{dS}\right) (k_1, \tau') 
\left( \ddot{\gamma}^*_{dS} - k^2_2 {\gamma}^*_{dS} \right) (k_2, \tau') 
\dot{\gamma}^*_{dS}(k_3, \tau') 
\left\{ \prod_{n=1}^3 \gamma_{dS}(k_n, \tau) \right\}
\right] \nonumber \\
&&\qquad\quad + 5 \ {\rm perms.} \nonumber \\
&&\qquad= \Lambda^{-2} (H \tau')^2 
\left(\frac{\tau'}{\tau_*}\right)^A (-3i) k_3 
(2\pi)^3 \delta\left( \sum_{n=1}^3 {\bf k_n} \right)
\eta^{ijk} e_{kq}^{(- \lambda_1)}(\hat{\bf k_1})
e_{jm}^{(- \lambda_2)}(\hat{\bf k_2}) 
e_{iq}^{(- \lambda_3)}(\hat{\bf k_3}) \hat{k_3}_m 
\nonumber \\
&&\qquad\quad\times 2i {\rm Im}\left[ \left( \ddot{\gamma}_{dS} - k_1^2 {\gamma}_{dS} \right)(k_1, \tau')
 \left( \ddot{\gamma}_{dS} - k_2^2 {\gamma}_{dS} \right)(k_2, \tau')  \dot{\gamma}_{dS}(k_3, \tau') 
\left\{ \prod_{n=1}^3 \gamma^*_{dS}(k_n, \tau) \right\} \right] \nonumber \\
&&\qquad\quad + 5 \ {\rm perms.}
~. 
\end{eqnarray}
Via the time integral: 
\begin{eqnarray}
&& 
{\rm Im} \left[ \int_{-\infty}^\tau d\tau' (H \tau')^2  
\left(\frac{\tau'}{\tau_*}\right)^A k_3
\left( \ddot{\gamma}_{dS} - k_1^2 {\gamma}_{dS} \right)(k_1, \tau')
 \left( \ddot{\gamma}_{dS} - k_2^2 {\gamma}_{dS} \right)(k_2, \tau') 
\right. \nonumber \\
&&\qquad \left. \times
\dot{\gamma}_{dS}(k_3, \tau')  
\left\{ \prod_{n=1}^3 \gamma^*_{dS}(k_n, \tau) \right\}
 \right] \nonumber \\
&&\qquad\qquad 
\xrightarrow{\tau \rightarrow 0} 
- \frac{4 H^8}{M_{\rm pl}^6} {\rm Im} 
\left[ \tau_*^{-A} \int_{-\infty}^0 d\tau' \tau'^{5+A}
e^{-i k_t \tau'} \right] ~,
\end{eqnarray}
we have 
\begin{eqnarray}
\Braket{\prod_{n=1}^3 \gamma^{(\lambda_n)}({\bf k_n})}^{(1)}_{\widetilde{W}W^2}
&=&  (2\pi)^3 \delta\left(\sum_{n=1}^3 {\bf k_n}\right) 
8 \left(\frac{H}{M_{\rm pl}}\right)^6 \left( \frac{H}{\Lambda} \right)^2 
{\rm Im} 
\left[ \tau_*^{-A}\int_{-\infty}^0 d\tau' \tau'^{5+A}
e^{- i k_t \tau'} \right] \nonumber \\
&& \times 
(-3i) \left[ \eta^{ijk} e_{kq}^{(-\lambda_1)}(\hat{\bf k_1}) e_{jm}^{(-\lambda_2)}(\hat{\bf k_2}) e_{iq}^{(-\lambda_3)}(\hat{\bf k_3}) \hat{k_3}_m + {\rm 5 \ perms.}\right] ~.
\end{eqnarray}
Like this, we can gain the second counterpart:
\begin{eqnarray}
\Braket{\prod_{n=1}^3 \gamma^{(\lambda_n)}({\bf k_n})}^{(2)}_{\widetilde{W}W^2}
&=&  (2\pi)^3 \delta\left(\sum_{n=1}^3 {\bf k_n}\right) 
8 \left(\frac{H}{M_{\rm pl}}\right)^6 \left( \frac{H}{\Lambda} \right)^2 
{\rm Im} 
\left[ \tau_*^{-A}\int_{-\infty}^0 d\tau' \tau'^{5+A}
e^{- i k_t \tau'} \right]  \nonumber \\
&&\times 
i \left[ \eta^{ijk} e_{kq}^{(-\lambda_1)}(\hat{\bf k_1}) e_{mi}^{(-\lambda_2)}(\hat{\bf k_2}) e_{mq}^{(-\lambda_3)}(\hat{\bf k_3}) \hat{k_3}_j + {\rm 5 \ perms.}\right] ~.
\end{eqnarray}

The bracket part with respect to $H_{\widetilde{W}W^2}^{(3)}$ is 
\begin{eqnarray}
&&\Braket{0| 
\left[: H^{(3)}_{\widetilde{W}W^2}(\tau'):, \prod_{n=1}^3 
\gamma^{(\lambda_n)}({\bf k_n}, \tau) \right] |0} \nonumber \\
&&\qquad= \Braket{0| : H^{(3)}_{\widetilde{W}W^2}(\tau'): \prod_{n=1}^3 
\gamma^{(\lambda_n)}({\bf k_n}, \tau) |0} -  
\Braket{0| 
\left[ \prod_{n=1}^3 
\gamma^{(\lambda_n)}({\bf k_n}, \tau) \right]: H^{(3)}_{\widetilde{W}W^2}(\tau'):  |0}
\nonumber \\
&&\qquad= -  \int d^3x' \Lambda^{-2} (H \tau')^2  
 \left( \frac{\tau'}{\tau_*} \right)^A
4 \left[ \prod_{n=1}^3 
\int \frac{d^3 {\bf k_n'}}{(2 \pi)^3} 
e^{i {\bf k'_n} \cdot {\bf x'} }
\sum_{\lambda'_n = \pm 2} \right] \nonumber \\
&&\qquad\quad\times  \eta^{ijk} e_{pj}^{(\lambda'_1)}(\hat{\bf k'_1})
e_{pm}^{(\lambda'_2)}(\hat{\bf k'_2}) e_{il}^{(\lambda'_3)}(\hat{\bf
k'_3})(i {k'_1}_k)(i {k'_2}_l)(i {k'_3}_m) 
\nonumber \\
&&\qquad\quad\times  
\left[  
: \Braket{0| 
\left\{ \prod_{n=1}^3 \dot{\gamma}_{dS}(k'_n, \tau')
a^{(\lambda'_n)}_{k'_n}
 \right\}
\left\{ \prod_{m=1}^3 \gamma^*_{dS}(k_m, \tau)  a^{(\lambda_m)
\dagger}_{- k_m} \right\}
 |0}: \right. \nonumber \\
&& \qquad\quad\quad \left. - : \Braket{0| 
\left\{ \prod_{m=1}^3 \gamma_{dS}(k_m, \tau) a^{(\lambda_m)}_{k_m} \right\}
\left\{ \prod_{n=1}^3 \dot{\gamma}^*_{dS}(k'_n, \tau')
  a^{(\lambda'_n) \dagger}_{- k'_n} \right\}|0} :
\right]  \nonumber \\
&&\qquad= \Lambda^{-2} (H \tau')^2 \left( \frac{\tau'}{\tau_*} \right)^A 
(-4) (2\pi)^3\delta\left(\sum_{n=1}^3 {\bf k_n}\right) \nonumber \\
&&\qquad\quad\times \eta^{ijk} e_{pj}^{(\lambda_1)}(- \hat{\bf k_1})
e_{pm}^{(\lambda_2)}(- \hat{\bf k_2}) e_{il}^{(\lambda_3)}
(- \hat{\bf k_3})(- i {k_1}_k)(- i {k_2}_l)(- i {k_3}_m) 
\nonumber \\
&&\qquad\quad\times  
\left[  
\left\{ \prod_{n=1}^3 \dot{\gamma}_{dS}(k_n, \tau')
\gamma^*_{dS}(k_n, \tau) \right\}
 - \left\{ \prod_{n=1}^3 \dot{\gamma}^*_{dS}(k_n, \tau')
\gamma_{dS}(k_n, \tau) \right\}
\right] + 5 \ {\rm perms.}
 \nonumber \\
&&\qquad= \Lambda^{-2} (H \tau')^2 
\left( \frac{\tau'}{\tau_*} \right)^A (-4) (-i)^3 k_1 k_2 k_3  
(2\pi)^3\delta\left(\sum_{n=1}^3 {\bf k_n}\right) \nonumber \\
&&\qquad\quad\times \eta^{ijk} e_{pj}^{(- \lambda_1)}(\hat{\bf k_1})
e_{pm}^{(- \lambda_2)}(\hat{\bf k_2}) e_{il}^{(- \lambda_3)}
(\hat{\bf k_3}) \hat{k_1}_k \hat{k_2}_l \hat{k_3}_m  
\nonumber \\
&&\qquad\quad\times  
2i {\rm Im} 
\left[ \prod_{n=1}^3 \dot{\gamma}_{dS}(k_n, \tau')
\gamma^*_{dS}(k_n, \tau)
\right] + 5 \ {\rm perms.}
~. 
\end{eqnarray}
The time integral is 
\begin{eqnarray}
&&{\rm Im} \left[ \int_{-\infty}^\tau d\tau' (H \tau')^2 
\left( \frac{\tau'}{\tau_*} \right)^A \prod_{n=1}^3 k_n
 \dot{\gamma}_{dS}(k_n, \tau') \gamma^*_{dS}(k_n, \tau) \right]
\nonumber \\
&&\qquad \xrightarrow{\tau \rightarrow 0} 
\frac{H^8}{M_{\rm pl}^6} 
{\rm Im} \left[ \tau_*^{-A}\int_{-\infty}^0 d\tau' \tau'^{5+A}
e^{- i k_t \tau'} \right] ~, 
\end{eqnarray}
so that the bispectrum of gravitons becomes
\begin{eqnarray}
\Braket{\prod_{n=1}^3 \gamma^{(\lambda_n)}({\bf k_n})}^{(3)}_{\widetilde{W}W^2} 
&=& (2\pi)^3 \delta \left(\sum_{n=1}^3 {\bf k_n}\right) 
8 \left(\frac{H}{M_{\rm pl}}\right)^6 \left( \frac{H}{\Lambda} \right)^2
{\rm Im} \left[ \tau_*^{-A}\int_{-\infty}^0 d\tau' \tau'^{5+A}
e^{- i k_t \tau'} \right]
\nonumber \\
&&\times 
i \left[ \eta^{ijk} e_{pj}^{(-\lambda_1)}(\hat{\bf k_1})
e_{pm}^{(-\lambda_2)}(\hat{\bf k_2}) e_{il}^{(-\lambda_3)}(\hat{\bf
k_3}) \hat{k_1}_k \hat{k_2}_l \hat{k_3}_m + {\rm 5 \ perms.}\right] ~.
\end{eqnarray}
Through the same procedure, the bispectrum from $H_{\widetilde{W}W^2}^{(4)}$ is estimated as
\begin{eqnarray}
\Braket{\prod_{n=1}^3 \gamma^{(\lambda_n)}({\bf k_n})}^{(4)}_{\widetilde{W}W^2} 
&=& (2\pi)^3 \delta \left(\sum_{n=1}^3 {\bf k_n}\right) 
8 \left(\frac{H}{M_{\rm pl}}\right)^6 \left( \frac{H}{\Lambda} \right)^2
{\rm Im} \left[ \tau_*^{-A}\int_{-\infty}^0 d\tau' \tau'^{5+A}
e^{- i k_t \tau'} \right]
\nonumber \\
&&\times 
(-i) \left[ \eta^{ijk} e_{pj}^{(-\lambda_1)}(\hat{\bf k_1})
e_{pm}^{(-\lambda_2)}(\hat{\bf k_2}) e_{im}^{(-\lambda_3)}(\hat{\bf
k_3}) \hat{k_1}_k \hat{k_2}_l \hat{k_3}_l + {\rm 5 \ perms.}\right]~. \nonumber \\ 
\end{eqnarray}


\bibliographystyle{JHEP}
\bibliography{paper}